\documentclass[sn-mathphys-ay,pdflatex]{sn-jnl}

\usepackage{graphicx}%
\usepackage{multirow}%
\usepackage{amsmath,amssymb,amsfonts}%
\usepackage{amsthm}%
\usepackage{mathrsfs}%
\usepackage[title]{appendix}%
\usepackage{xcolor}%
\usepackage{textcomp}%
\usepackage{manyfoot}%
\usepackage{booktabs}%
\usepackage{algorithm}%
\usepackage{algorithmicx}%
\usepackage{algpseudocode}%
\usepackage{listings}%

\usepackage{amsmath}
\usepackage{float}
\usepackage[utf8]{inputenc}
\usepackage{booktabs}
\usepackage{tabularx}
\usepackage{xcolor}

\numberwithin{equation}{section}

\usepackage[english]{babel}
\usepackage{amsthm}
\usepackage{amssymb}

\theoremstyle{plain}
\newtheorem{theorem}{Theorem}[section]
\newtheorem{proposition}{Proposition}[section]
\newtheorem{corollary}{Corollary}[proposition]
\newtheorem{lemma}[theorem]{Lemma}
\newtheorem{property}[theorem]{Property}
\newtheorem{remark}[theorem]{Remark}
\newtheorem{note}[theorem]{Note}
\newtheorem{definition}[theorem]{Definition}

\usepackage{soul}
\usepackage{mathtools}

\usepackage{tocloft}
\usepackage{setspace}
\usepackage{pbox}
\usepackage{makecell}

\usepackage{tikz}
\usetikzlibrary{positioning}

\providecommand{\tightlist}{%
  \setlength{\itemsep}{0pt}\setlength{\parskip}{0pt}}

\begin{document}

\title[]{Construction, Transformation and Structures of 2x2
Space-Filling Curves}


\author[1]{\fnm{Zuguang} \sur{Gu}}\email{guzuguang@suat-sz.edu.cn}

  \affil*[1]{\orgname{Shenzhen University of Advanced Technology, Institute of Cell and Gene Technology}, \orgaddress{\city{Shenzhen}, \country{China}}}

\abstract{The 2x2 space-filling curve is a type of generalized
space-filling curve characterized by basic units in ``U-shapes'' that
traverse 2x2 grids. One of the most well-known forms of such curves is
the Hilbert curve. In this work, we proposed a universal framework for
constructing general 2x2 curves where self-similarity is not strictly
required. The construction is based on a novel set of grammars that
define the expansion of curves from level 0 (single points) to level 1
(units in U-shapes), which ultimately determines all \(36
\times 2^k\) possible forms of curves on any level \(k\) initialized
from single points. We further developed an encoding system in which
each unique form of the curve is associated with a specific combination
of an initial seed and a sequence of code that sufficiently describes
both the global and local structures of the curve. We demonstrated that
this encoding system can be a powerful tool for studying 2x2 curves and
we established comprehensive theoretical foundations from the following
three key aspects. 1) We provided a deterministic encoding for any unit
on any level and any position on the curve, enabling the study of curve
generation across arbitrary parts on the curve and ranges of iterations;
2) We gave deterministic encodings for various curve transformations,
including rotations, reflections, reversals and reductions; 3) We
provided deterministic forms of curve families exhibiting specific
structures, including homogeneous curves, curves with identical shapes,
partially identical shapes, and completely distinct shapes. We also
explored families of recursive curves, subunit identically or
differently shaped curves, completely non-recursive curves, symmetric
curves and closed curves. Finally, we proposed a method to calculate the
location of any point on the curve arithmetically, within a time
complexity linear to the level of the curve. This framework and the
associated theories can be seamlessly applied to more general 2x2 curves
initialized from seed sequences represented as orthogonal paths,
allowing it to fill spaces with a much greater variety of shapes.}



\maketitle

\tableofcontents
\newpage

\section{Introduction}\label{introduction}

A space-filling curve is a continuous curve that traverses every point
in a space. In most cases, the curve is generated by repetitive patterns
in a recursive way. When the number of iterations of the generation
reaches infinity, the curve completely fills the space
\citep{sagen_hans}. In this work, we studied the construction process of
a type of space-filling curves, namely the 2x2 (2-by-2) curve, that
fills a two-dimensional space, where the basic repetitive unit is
represented as a list of three connected segments in a ``U-shape''
traversing a square of 2x2 points. In the curve generation, it starts
from a single point and after \(k\) iterations\footnote{Or the curve
starts from a 2x2 ``U-unit'' and after $k-1$ iterations.}, the curve is
represented as a list of \(4^k - 1\) segments connecting \(4^k\) points
located in a square region partitioned by \(2^k \times 2^k\) grids. This
curve is called on \textit{level $k$} or in \textit{order $k$} generated
from level 0. Mathematically, a space-filling curve is defined for
\(k \rightarrow \infty\). In this work, we only consider the
space-filling curve where \(k\) is a finite integer, i.e., the
``finite'' or ``pseudo'' space-filling curve.

\begin{figure}[t]
\centering{
\includegraphics[width=1\linewidth]{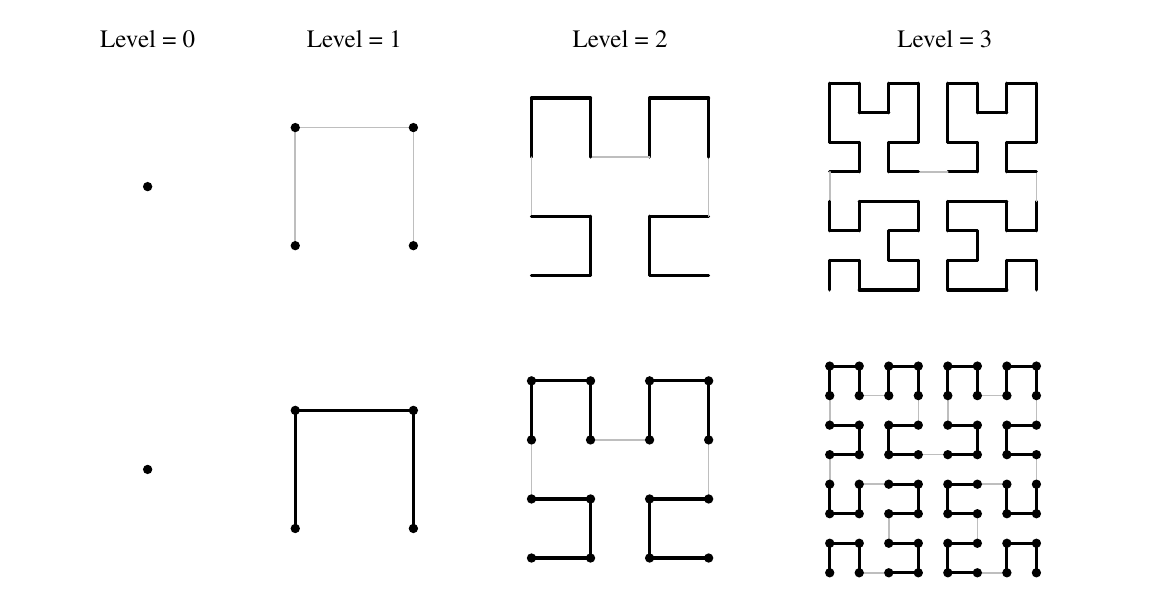}
\caption{Generation of 2x2 curves from level 0 to level 3. First row: generation under the copy-paste mode; Second row: generation under the expansion mode. The curves belong to a special form, the Hilbert curve.}\label{fig:first_four_levels}
}
\end{figure}

Current studies on the 2x2 curve mainly focus on one of its special
forms, the Hilbert curve \citep{hilbert1891}. In it, self-similarity is
required as an important attribute where a curve on a higher level is
composed of replicates of the curve from its lower levels. The
construction of the curve is normally described in a
\textit{copy-paste mode} where a curve on level \(k\) is composed of
four copies of itself on level \(k-1\), positioned in the four quadrants
of the curve region with specific orientations that are consistent in
the curve generation. As an example in the first row in Figure
\ref{fig:first_four_levels} which illustrates the generation of a
Hilbert curve from level 0 to level 3, four subcurves from the previous
level are positioned in a clockwise order of lower left, upper left,
upper right and lower right, with facing leftward, downward, downward
and rightward\footnote{If we assume the shape ``U'' is facing upward.}.
If the curve is considered as directional, traversing from its lower
left corner to its lower right corner, the first subcurve is applied
with a horizontal reflection then a 90-degree rotation clockwise, and
the fourth subcurve is applied with a horizontal reflection then a
90-degree rotation counterclockwise. There are also variants of the
Hilbert curve where self-similarity only exists in each of its four
subcurves, but not globally on the complete curve. A typical form is the
Moore curve \citep{moore1900} where a curve on level \(k\) includes four
copies of Hilbert curves on level \(k-1\) with facing right, right, left
and left. The facings of the four subcurves are defined on level 2 of
the curve. Other four variants were proposed by \citet{LIU2004741} where
the facings of the Hilbert subcurves are defined by other different
combinations of facings on level 2.

The construction under the copy-paste mode preserves the global
structure of the curve which ensures the self-similarity between levels,
but it is limited in displaying more rich types of curve structures. In
this work, from the opposite viewpoint of the copy-paste mode, we
proposed a new framework for generating general 2x2 curves where
self-similarity is not a required attribute any more. The construction
of the curve is applied from a local aspect, which we named the
\textit{expansion mode} or the \textit{division
mode}. Instead of treating the curve as four copies of its subcurves, we
treat the curve on level \(k-1\) as a list of \(4^{k-1}\) points
associated with their specific entry and exit directions. Then the
generation of the curve to level \(k\) is described as expansions of
every single point to its corresponding 2x2 unit. It is easy to see, the
curve on level \(k-1\) determines the structure of the curve on level
\(k\). To compare, on the \(k\)-th iteration of the curve generation,
the copy-paste mode only needs to adjust four identical subcurves to
connect them properly, while the expansion mode has to adjust all
\(4^{k-1}\) 2x2 units. The expansion mode, although increases the
complexity of analyzing curve structures, provides flexibility to
generate more types of curves. Similar idea of describing curve
generations as expansions is also seen in \citet{gips}.

For both copy-paste and expansion modes, the curve is generated by
applying certain rules recursively. Such rules are often called
``\textit{grammars}'' in literatures. For various grammars, the curve is
treated as directional and the generation of a curve on level \(k\) is
described as a drawing process from a start point moving to an end
point. In these grammars, a group of \textit{base patterns} on level 1,
i.e., 2x2 units, are selected and denoted as a set of symbols, then a
complete set of rules is established where curves on level 2 are purely
composed of the selected level-1 units and assigned with the same symbol
as the level-1 units if they share the same orientation or facing. In
this way, the curve can be recursively generated to any level \(k\) from
its lower levels and sufficiently expressed only by the symbols in the
set. Additionally, a set of ``\textit{commands}'' are also defined which
specify how two neighbouring units are connected, represented as a
second group of symbols. Eventually, the curve on level \(k\) is
expressed as a long sequence of symbols of base patterns and commands
that determine how the curve traverses in the space.

The most well-known grammars for constructing Hilbert curves are based
on the \textit{L}-systems \citep{prusinkiewicz_synthesis_1991}, which
include two horizontally reflected 2x2 base patterns in the U-shapes and
three commands of moving forward, turning left and turning right that
determine the location and rotation of the next unit on the curve. The
directions and rotations of units described by the \textit{L}-systems
are relative metrics because they are determined by their preceding
units. \citet{bader} described a set of static grammars which include a
set of four 2x2 base patterns in their specific orientations and four
commands for connecting neighbouring units in absolute directions, i.e.,
up, down, left and right. Also using the four absolute directional
movements, the grammars defined by \citet{SFCGen} described the 2x2 base
units by the orientations of their first two segments, which yields
eight different base patterns (corresponding to four facings of a 2x2
unit and four facings on the reflected versions). Nevertheless, these
grammars are not general and they are mainly designed for the Hilbert
curve. A new set of grammars needs to be defined for other forms of 2x2
curves. For example, the \(\beta\Omega\)-curve \citep{betaomega} is also
a type of 2x2 curves but with a very different structure from the
Hilbert curve. \citet{bader} defined a set of grammars for the
\(\beta\Omega\)-curve, but they are different and more complex than the
grammars for the Hilbert curve.

Majority of current works on 2x2 curves focus on the Hilbert curve,
which has a recursive structure on all its levels. In this work, we
extended the study to more general 2x2 curves where self-similarity is
not strictly required. We proposed a universal framework that is capable
to generate all possible forms of 2x2 curves in a unified process. There
are two major differences of our curve construction method compared to
current ones. First, instead of using level-1 units as the base
patterns, in our grammar, we use the level-0 units additionally
associated with their entry and exit directions. Such design is natural
because if we treat the curve generation as a drawing process, the pen
moves forward, rightward or leftward from the current position, thus it
implies the entry and exit directions are important attributes of every
point on the curve. Once they are determined, the final structure of the
curve is completely determined. We then defined a full set of expansion
rules from level 0 to level 1. Second, we use the expansion mode to
expand the curve to the next level. As a curve on level \(k-1\) can be
expressed as a sequence of base patterns on level 0, with the full set
of expansion rules from level 0 to level 1 defined, the form of the
curve on level \(k\) can be fully determined. On the other hand,
expanding the curve from the lowest level allows more flexibility to
tune the structures of curves. In this framework, we demonstrated, the
expansion of the complete curve is solely determined by the expansion of
its first base pattern. Additionally, integrating the entry and exit
directions into the level-0 base patterns gets rid of using a second set
of command symbols to define how units are connected, as the connections
have already been implicitly determined by the entry and exit directions
of neighbouring points on the curve. This provides compact and unified
expressions of 2x2 curves compared to other grammars.

As a companion of the construction process, we further developed an
encoding system which assigns each form of the curve a unique symbolic
expression represented as a specific combination of an initial seed and
a sequence of expansion code, where the expansion code determines how
the curve is expanded to the next levels. This provides a standardized
way to denote and distinguish all possible 2x2 curves. The construction
framework and encoding system can be seamlessly extended to more general
2x2 curves initialized by a seed sequence, not only restricted to a
single seed base, allowing to generate curves that fill arbitrary shapes
from initial orthogonal paths.

Based on the construction framework and the curve encoding system, we
established a comprehensive theoretical foundation for studying 2x2
curves. This article can be split into the following three parts from
the aspects of construction, transformation, and structures of 2x2
curves.

\begin{itemize}
\tightlist
\item
  In the first part, we introduced the framework for constructing 2x2 curves.
  We first introduced the complete expansion rules from level 0 to level 1 in
  Section \ref{expansion-rules}. Then in Section \ref{expansion-to-level-k},
  we demonstrated that the rules defined on 0-to-1 expansions are sufficient
  for generating curves to any level. We further discussed the conditions for
  properly connecting all level-1 units where we proved the curve expansion is
  solely determined by the expansion of its first base pattern. Based on this
  attribute, in Section \ref{encode-the-curve}, we developed an encoding
  system which uniquely encodes every possible form of 2x2 curves. In Section
  \ref{the-expansion-code-sequence}, We provided the forms of the expansions
  of square subcurves in any size on any locations and on any level of the
  curve. We further demonstrated the symbolic expression encodes information
  of both global and local structures of the curve.
\item
  In the second part, we studied various transformations of curves and the
  corresponding forms of their symbolic expressions. In Section
  \ref{transformation}, we discussed transformations of rotation, reflection
  and reversals. In Section \ref{reduction}, we discussed the reduction of
  curves and demonstrated how to infer the encoding of a curve by stepwise
  curve reduction.
\item
  In the third part, we explored various types of curve structures. In Section
  \ref{geometry}, we studied the geometric attributes of the entry and exit
  points on the curve. In particular, we proved a curve can be uniquely
  determined only by its entry and exit points. In Section \ref{shapes}, we
  gave the encodings for families of curves that are homogeneous, identically
  shaped, partially identically shaped, or completely distinct. In Section
  \ref{def-hc}, we provided alternative definitions for the Hilbert curve and
  the $\beta\Omega$-curve based on their structural attributes. In Section
  \ref{structures}, we studied more types of curves in their specific
  structures including recursive curves, subunit identical or different
  curves, completely non-recursive curves, symmetric curves and closed curves.
  Finally, in Section \ref{arithmetic}, we demonstrated how to arithmetically
  obtain the coordinate of any point on the curve in a linear time complexity
  to the level of the curve.
\end{itemize}

\section{Expansion rules}\label{expansion-rules}

\begin{figure}
\centering{
\includegraphics[width=1\linewidth]{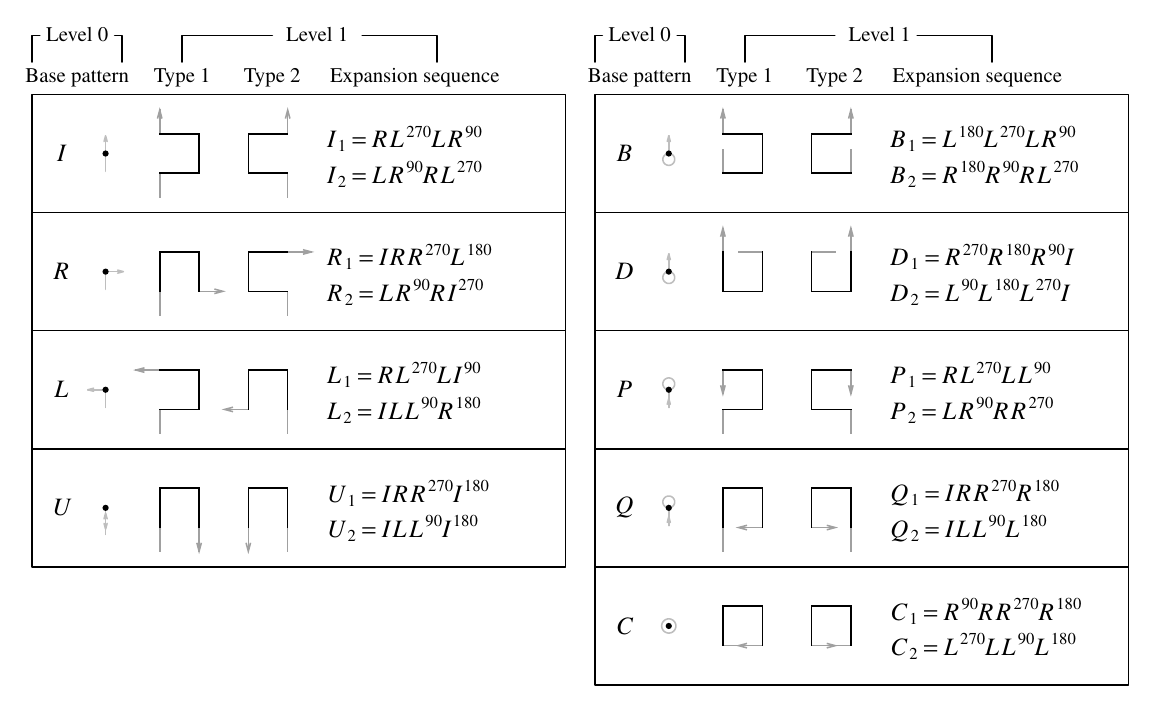}
\caption{The complete set of expansion rules from level 0 to level 1. Grey segments and arrows represent entry and exit directions of corresponding units.}\label{fig:expansion_rule}}
\end{figure}

\subsection{Level 0-to-1 expansion}\label{level-0-to-1-expansion}

A 2x2 curve is generated by recursively repeating patterns from its
sub-structures, which means, low-level structures determine high-level
structures of the curve. A 2x2 curve is normally initialized from its
lowest level represented as a single point, i.e., on level 0. In Figure
\ref{fig:expansion_rule}, we defined a complete set of nine level-0
patterns which are composed of single points associated with their
corresponding entry and exit directions. They are described as follows:

\begin{itemize}
\tightlist
\item
  $I$: bottom-in and top-out.
\item
  $R$: bottom-in and right-out.
\item
  $L$: bottom-in and left-out.
\item
  $U$: bottom-in and bottom-out.
\item
  $B$ and $D$: entry-closed and top-out. 
\item
  $P$ and $Q$: bottom-in and exit-closed.
\item
  $C$: both entry-closed and exit-closed.
\end{itemize}

\noindent We call these nine level-0 patterns \textit{base patterns}.
They serve as the basic lowest-level structures for the curve
construction.

A level-1 curve (or unit) is composed of four base patterns and is
expanded from a specific level-0 unit. The level-1 curve is also
associated with an entry direction and an exit direction where the entry
direction is the same as that of its first base pattern and the exit
direction is the same as that of its last base pattern. In the 0-to-1
expansion, each base pattern can be expanded to its corresponding
level-1 curve in two ways. Figure \ref{fig:expansion_rule} lists all
combinations of 0-to-1 expansions for the nine base patterns. All
level-1 curves have ``U-shapes'' in their specific facings.

In the diagram in Figure \ref{fig:expansion_rule}, each level-1 curve is
expressed as a sequence of four base patterns with their rotations. For
example, base pattern \(I\) is on level 0, explicitly denoted as
\(I^{(0)}\). When it is expanded to level 1 while keeping the
orientation of the unit, there are two options as listed in the diagram.
As an example here, we choose the first option (type = 1) and denote
this level-1 expansion as \(I^{(1)}_1\). Now for the following
expansion:

\begin{equation}
I^{(0)} \rightarrow I^{(1)}_1,
\notag
\end{equation}

\noindent we can describe the curve generation in four steps (Figure
\ref{fig:unit_traverse}):

\begin{itemize}
\tightlist
\item
  Step 1: bottom-in and right-out. This is the base pattern $R$ without
  rotation. We denote it as $R^{(0)}$.
\item
  Step 2: left-in and top-out. This is the base pattern $L$ with a rotation of
  $-90$ degrees \footnote{Positive values for couterclockwise rotations.}. We
  denote it as $L^{(0),-90}$ or $L^{(0),270}$.
\item
  Step 3: bottom-in and left-out. This is the base pattern $L$ without
  rotation. We denote it as $L^{(0),0}$ or simply $L^{(0)}$.
\item
  Step 4: right-in and top-out. This can be denoted as $R^{(0),90}$.
\end{itemize}

Then the expansion is written as a sequence of four base patterns with
their corresponding rotations:

\begin{figure}
\centering{
\includegraphics[width=0.2\linewidth]{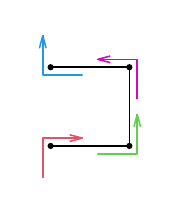}
\caption{Traversal on the level-1 unit of $I^{(1)}_1$.}\label{fig:unit_traverse}}
\end{figure}

\begin{equation}
I^{(1)}_1 = R^{(0)}L^{(0),270}L^{(0)}R^{(0),90}.
\notag
\end{equation}

In the diagram in Figure \ref{fig:expansion_rule}, notations of levels
are removed from the equations, since they can be easily inferred as the
left side of the equation always corresponds to the unit on level 1 and
the right side always corresponds to the four-base expansion from level
0. Then \(I^{(1)}_1\) can be simplified to:

\begin{equation}
I_1 = RL^{270}LR^{90}.
\notag
\end{equation}

For the nine base patterns listed in Figure \ref{fig:expansion_rule},
\(B\) and \(D\) are \textit{entry-closed} in the same structure.
However, their structures are distinguishable from level 1 of the curve.
\(P\) and \(Q\) are \textit{exit-closed} in the same structure. Their
structures are distinguishable also from level 1.

\subsection{Rotation}\label{rotation}

Each base pattern on level 0 listed in Figure \ref{fig:expansion_rule}
is associated with a rotation of zero degree. We call it in its
\textit{base
rotation state}. We can easily calculate the rotation of a base pattern
or a sequence of base patterns.

Denote \(X^\theta\) as a base pattern where
\(X \in \{I, R, L, U, B, D, P, Q,
C\}\) and \(\theta\) as a counterclockwise rotation, then we have

\begin{equation}
(X^{\theta_1})^{\theta_2} = X^{\theta_1 + \theta_2},
\notag
\end{equation}

\noindent which means rotating the base pattern twice is identical to
rotating the pattern once but by the sum of the two rotations.

Rotating a curve which is represented as a sequence of base patterns is
identical to rotating its individual base patterns separately.

\begin{equation}
\label{eq:rot_seq}
\begin{aligned}
(X_1^{\theta_1}X_2^{\theta_2}...X_n^{\theta_n})^\theta  &= (X_1^{\theta_1})^\theta(X_2^{\theta_2})^\theta ... (X_n^{\theta_n})^\theta \\
    &= X_1^{\theta_1 + \theta}X_2^{\theta_2 + \theta} ... X_n^{\theta_n + \theta} \\
\end{aligned}
\end{equation}

Above equation is obvious if we treat the curve as a rigid object where
every point on it has the same rotation as the rigid object itself.

If a sequence is composed of several subsequences (or a curve is
composed of several subcurves), rotating the sequence is identical to
rotating each subsequence separately. Denote a sequence as
\(\mathcal{S}\) composed of \(w\) subsequences, there is

\begin{equation}
\begin{aligned}
\mathcal{S}^\theta = (\mathcal{S}_1\mathcal{S}_2...\mathcal{S}_w)^\theta &= \left((X_{11}^{\theta_{11}}...)(X_{21}^{\theta_{21}}...)...(X_{w1}^{\theta_{w1}}...)\right)^\theta \\
 &= (X_{11}^{\theta_{11}+\theta}...)(X_{21}^{\theta_{21}+\theta}...)...(X_{w1}^{\theta_{w1}+\theta}...) \\
  &= (X_{11}^{\theta_{11}}...)^{\theta}(X_{21}^{\theta_{21}}...)^\theta...(X_{w1}^{\theta_{w1}}...)^\theta \\
  &= \mathcal{S}_1^\theta \mathcal{S}_2^\theta ... \mathcal{S}_w^\theta \\
\end{aligned} .
\end{equation}

When studying 2x2 curves, we only consider \(\theta\) whose modulus is
in \(\{0,
90, 180, 270\}\).

\subsection{Design of the expansion
rules}\label{design-of-the-expansion-rules}

The diagram in Figure \ref{fig:expansion_rule} lists the full set of
expansions from level 0 to level 1 for all base patterns on their base
rotation states. There are the following criterions for constructing the
expansion rules:

\begin{enumerate}
\tightlist
\item
   The entry and exit directions of the level-1 units should be the same as their
   corresponding base patterns.
\item
   For the two types of level-1 expansions of each base pattern, in the first
   type, entry point is located on the lower left corner of the square; and in
   the second type, entry point is located on the lower right corner. $D$ is
   an exception, but we require its entry point to be located on the upper
   right corner for type-1 expansion and on the upper left corner for the
   type-2 expansion.
\end{enumerate}

The third criterion is not mandatory but recommended. It simplifies the
analysis in this work.

\begin{enumerate}
\tightlist
\setcounter{enumi}{2}
\item
   All the base patterns should have the same entry direction. If some of them
   do not have entry directions, they should have the same exit direction as
   base pattern $I$. In Figure \ref{fig:expansion_rule}, we set the entry
   directions of $I$/$R$/$L$/$U$/$P$/$Q$ to vertically bottom-in (i.e., 90
   degrees), and we set the exit directions of $B$ and $D$ to vertically
   top-out (i.e., 90 degrees) since they are entry-closed.
\end{enumerate}

It is easy to see from Figure \ref{fig:expansion_rule} that a level-1
unit can be a reflected or reversed version of some other units. E.g.,
\(R_1\) is a horizontal reflection of \(L_2\), or \(U_1\) is a reversal
of \(U_2\). In our framework, we require a curve to be generated from
low-level units \emph{only by rotations} (in-plane transformation),
while we do not allow out-of-plane transformation (reflection) or
modification on the curve (reversal).

In the nine base patterns, particularly, \(I\), \(R\) and \(L\) are
called \textit{primary base patterns} because all the level-1 units are
only composed of these three ones and they represent the three basic
movements of moving forward, turning right and left. \(B\) and \(D\)
have the same structure on level 0 but they are different on level-1
where the last base patterns in \(D^{(1)}_1\) and \(D^{(1)}_2\) are
always \(I\). \(P\) and \(Q\) have the same structure on level 0 but
they are different on level-1 where the first base patterns in
\(Q^{(1)}_1\) and \(Q^{(1)}_2\) are always \(I\).

By also including all four rotations of the base patterns, the diagram
in Figure \ref{fig:expansion_rule} includes the complete set of
\(9 \times 2
\times 4 = 72\) different expansions from level 0 to level 1 for the 2x2
curves.

In the remaining part of this article, we may also refer base patterns
to \textit{bases} for simplicity. Without explicit clarification, a base
pattern \(X\) is always from the complete base set
\(\{I, R, L, U, B, D, P, Q, C\}\), and the modulus of a rotation
\(\theta\) is always from the complete rotation set
\(\{0, 90, 180, 270\}\). If there is no explicit clarification, \(X\)
always refers to \(X^\theta\) for simplicity, i.e., a base associated
with a specific rotation.

\section{\texorpdfstring{Expansion to level
\(k\)}{Expansion to level k}}\label{expansion-to-level-k}

\subsection{Recursive expansion}\label{recursive-expansion}

The diagram in Figure \ref{fig:expansion_rule} only defines the
expansion of a curve from level 0 to level 1, i.e., the 0-to-1
expansion. Nevertheless, that is sufficient for generating a curve to
any level \(k > 1\). For simplicity, we take a curve initialized from a
single base (\(\mathcal{P}_0 = X\)) as an example. Denote
\(\mathcal{P}_i\) as a curve on level \(i\) and let \((X)_n =
X_1...X_n\) be a sequence of \(n\) bases where each base \(X_i\) is
implicitly associated with its corresponding rotation. The expansion
process can be described in the following steps:

\begin{enumerate}
\tightlist
\item
   Level 0 $\rightarrow$ level 1: $\mathcal{P}_1 = (X)_4$. It generates a
   level-1 curve of 4 bases.
\item
   Level 1 $\rightarrow$ level 2: $\mathcal{P}_2 = (X)_{4^2}$. For each base
   in $\mathcal{P}_1$, we replace it with its level-1 expansion. This
   generates a curve of $4^2$ bases.
\item
   Level $k-1$ $\rightarrow$ level $k$ ($k \ge 3$): $\mathcal{P}_k =
   (X)_{4^k}$. Note the curve $\mathcal{P}_{k-1}$ on level $k-1$ is already
   represented as a sequence of $4^{k-1}$ bases. Then for each base in
   $\mathcal{P}_{k-1}$, we replace it with its level-1 expansion. This
   generates a curve of $4^k$ bases.
\end{enumerate}

It is easy to see, we only need to apply the 0-to-1 expansion repeatedly
to expand the curve to level \(k\). Let's write \(\mathcal{P}_1\) as a
sequence of four base patterns:

\begin{equation}
\mathcal{P}_1 = X_1 X_2 X_3 X_4 .
\notag
\end{equation}

\noindent When expanding \(\mathcal{P}_1\) to \(\mathcal{P}_2\) on the
next level, we replace, e.g., \(X_1\) with its level-1 unit denoted as
\(X^{(1)}_{<i>,1}\) (\(i=1\) or \(2\), i.e., the expansion type in
Figure \ref{fig:expansion_rule}), then

\begin{equation}
\mathcal{P}_2 = X^{(1)}_{<i_1>,1} X^{(1)}_{<i_2>,2} X^{(1)}_{<i_3>,3} X^{(1)}_{<i_4>,4} .
\notag
\end{equation}

One issue arises where \(i_* \in \{i_1, ..., i_4\}\) may take value of 1
or 2 for each base expansion. Then, we need a criterion here for picking
the correct expansion types for the four bases to ensure all their
level-1 units are properly connected. Let's take the first two level-1
units as an example. If \(X^{(1)}_{<i_1>,1}\) and \(X^{(1)}_{<i_2>,2}\)
are properly connected, since the last base in \(X^{(1)}_{<i_1>,1}\)
denoted as \(Z_a\) has an exit direction associated, it determines the
location of its next base which is the first base in
\(X^{(1)}_{<i_2>,2}\) denoted as \(Z_b\). On the other hand, \(Z_b\) in
\(X^{(1)}_{<i_2>,2}\) has an entry direction associated, which can also
validate the location of \(Z_a\) in \(X^{(1)}_{<i_1>,1}\). Thus \(Z_a\)
and \(Z_b\) should be compatible and there are the following statements.

\begin{note}
\label{note}
The criterion for properly connecting two level-1 units can be stated in
either of the following two ways:

1. $Z_b$ is located on one of the top, left, bottom or right of $Z_a$.

2. The segment connecting $Z_a$ and $Z_b$ is either horizontal or vertical.
\end{note}

A curve on level \(k\) can be eventually expressed as 0-to-1 expansions
from \(\mathcal{P}_{k-1}\):

\begin{equation}
\mathcal{P}_k = X^{(1)}_{<i_1>,1} X^{(1)}_{<i_2>,2} ...  X^{(1)}_{<i_{4^{k-1}}>,4^{k-1}} .
\notag
\end{equation}

\noindent Then, the criterion also ensures all the \(4^{k-1}\) 2x2 units
are properly connected and makes the final curve in the correct form. We
will discuss the solution in the next two subsections.

\subsection{Corners}\label{corners}

\begin{figure}
\centering{
\includegraphics[width=0.2\linewidth]{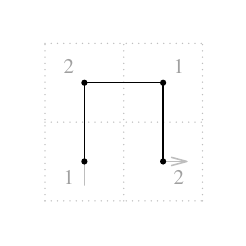}
\caption{Corners of a 2x2 unit. The lower left and upper right corners have values of 1 and the lower right and upper left corners have values of 2. The corner-tuple of the 2x2 unit is composed of the values of the entry and exit corners. The level-1 unit in this example is $R_1$.}\label{fig:corner_value}}
\end{figure}

A level-1 unit traverses through 2x2 grids starting from its entry
corner and ended at its exit corner. Let's set the lower left and upper
right corners to have values of 1 and the upper left and lower right
corners to have values of 2 (Figure \ref{fig:corner_value}). We define
the \textit{corner-tuple} denoted as \(\tau\) of the level-1 unit
\(X^{(1)}\) as a 2-tuple \((c_1, c_2)\) where \(c_1\) is the value of
the entry corner and \(c_2\) is the value of the exit corner of
\(X^{(1)}\).

\begin{equation}
\tau_{X^{(1)}} = (c_1, c_2) \quad c_1, c_2 \in \{1, 2\}
\notag
\end{equation}

All level-1 units are in the ``U-shapes'', thus they, including their
rotated versions, have either \(\tau_{X^{(1)}} = (1, 2)\) or
\(\tau_{X^{(1)}} =
(2,1)\). We define the \textit{complement} of a corner value \(c\)
denoted as \(\hat{c}\) as:

\begin{equation}
\hat{c} = \begin{cases}
2 \quad \text{if }c = 1\\ 
1 \quad \text{if }c = 2\\ 
\end{cases} .
\notag
\end{equation}

\noindent Then we rewrite the corner-tuple of \(X_{(1)}\) as:

\begin{equation}
\tau_{X^{(1)}} = (c, \hat{c}), \quad c \in \{1, 2\} .
\notag
\end{equation}

Rotating \(X^{(1)}\) by 90 degrees or its odd multiples changes the two
values in the corner-tuple, while rotating by its even multiples does
not.

\begin{equation}
\label{eq:tao}
\tau_{X^{(1), \theta}} = \begin{cases}
 (c, \hat{c}) & \quad \text{if }\theta \bmod 180 = 0 \\
 (\hat{c}, c) & \quad \text{if }\theta \bmod 180 = 90\\
\end{cases}
\end{equation}

\subsection{Connect level-1 units}\label{connect-level-1-units}

For two level-1 units \(X^{(1),\theta_1}\) and \(Y^{(1),\theta_2}\), to
properly connect them on a curve, according to the criterions in Note
\ref{note}, the two units represented as two squares can only be
connected horizontally or vertically. This results in that, if the first
unit has an exit corner value of \(\hat{c}\), the entry corner value of
the second unit must be \(c\). Then there are the following two
combinations of corner-tuples of the two units:
\(\tau_{X^{(1),\theta_1}} = (1, 2)\),
\(\tau_{Y^{(1),\theta_2}} = (1,2)\), or
\(\tau_{X^{(1),\theta_1}} = (2, 1)\),
\(\tau_{Y^{(1),\theta_2}} = (2,1)\). We write it in the general form:

\begin{equation}
\label{eq:tao2}
\tau_{X^{(1),\theta_1}} =  \tau_{Y^{(1),\theta_2}} .
\end{equation}

Notice
\((X^{(1),\theta_1}Y^{(1),\theta_2})^{-\theta_2} = X^{(1),\theta_1 -
\theta_2}Y^{(1),0}\) (Section \ref{rotation}). Then with Equation
\ref{eq:tao2},
\(\tau_{X^{(1),\theta_1 - \theta_2}} = \tau_{Y^{(1),0}}\). Assume
\(X^{(1),0}\) is associated with a corner-tuple \((c, \hat{c})\), then
with Equation \ref{eq:tao}, we obtain the solution of
\(\tau_{Y^{(1), 0}}\):

\begin{equation}
\label{eq:corner_y}
\tau_{Y^{(1), 0}} = \begin{cases}
(c, \hat{c}) &\quad \text{if } \theta_1 - \theta_2 \bmod 180 = 0 \\
(\hat{c}, c) &\quad \text{if } \theta_1 - \theta_2 \bmod 180 = 90 \\
\end{cases} .
\end{equation}

\subsection{Expansion code}\label{expansion-code}

According to the 0-to-1 expansion rules listed in Figure
\ref{fig:expansion_rule}, each base \(X\) has two types of level-1
expansions. The \textit{expansion code} of a base \(X\) encodes which
type of the level-1 unit is selected from the expansion diagram.

Corner-tuples of the two level-1 units of a base \(X\) are always
mutually complementary. In the complete set of expansion rules in Figure
\ref{fig:expansion_rule}, we require all level-1 units in type-1
expansion should have entry corner values of 1 (with the corresponding
corner-tuples \((1, 2)\)), and all level-1 units in type-2 expansion
should have entry corner values of 2 (with the corresponding
corner-tuples \((2, 1)\)). Then, denote the expansion code as \(\pi\)
(\(\pi \in \{1,2\}\)), the corner-tuple of \(X^{(1),0}\) from
type-\(\pi\) expansion is \((\pi, \hat{\pi})\).

Now we can calculate which types of level-1 expansions (i.e., the
expansion code) should be selected for bases in a sequence when expanded
to the next level. Let's still take \(X^{(1),\theta_1}Y^{(1),\theta_2}\)
as an exampple. First we should pre-select the expansion code \(\pi\)
for \(X\), then \(\tau_{X^{(1),0}_{<\pi>}} = (\pi, \hat{\pi})\). With
Equation \ref{eq:corner_y}, we can obtain the expansion code \(\pi_*\)
of \(Y\) (note the expansion code is the first value in
\(\tau_{Y^{(1),0}_{\pi_*}}\)):

\begin{equation}
\label{eq:code_xy}
\pi_* = \begin{cases}
\pi & \quad \text{if } \theta_1 - \theta_2 \bmod 180 = 0 \\
\hat{\pi} & \quad \text{if } \theta_1 - \theta_2 \bmod 180 = 90 \\
\end{cases} .
\end{equation}

Equation \ref{eq:code_xy} implies, when the expansion code of the first
base \(X\) is determined, the expansion code of the second base \(Y\) is
also determined, which determines the exact form of
\(Y^{(1),\theta_2}\).

For a sequence with more than two bases, with knowing the expansion code
of its first base, the expansion code of the remaining bases can be
calculated by repeatedly applying Equation \ref{eq:code_xy}. This yields
the followng proposition.

\begin{proposition}
\label{prop:3.2}
The form of $\mathcal{P}_k$ is only determined by the expansion of the first
base in $\mathcal{P}_{k-1}$ ($k \ge 1$).
\end{proposition}

\begin{proof}
When $k = 1$, the form of $\mathcal{P}_1$ can be uniquely selected from Figure
\ref{fig:expansion_rule} if knowing the expansion code of the
base $\mathcal{P}_0$ (assume its rotation is already included in $\mathcal{P}_0$).

When $k \ge 2$, $\mathcal{P}_{k-1}$ is expressed as a list of $4^{k-1}$
bases. We have already known that with knowing the expansion code and rotation
of the first base in $\mathcal{P}_{k-1}$, the expansion code for the remaining
bases are all determined. Then for a base $X_i$ in the sequence associated
with $\pi_i$ as its expansion code, we replace it with its type-$\pi_i$
expansion from Figure \ref{fig:expansion_rule} and apply the corresponding
rotation. We apply such process to all bases in $\mathcal{P}_{i-1}$ to
generates a deterministic sequence of $4^k$ bases. Thus the form of
$\mathcal{P}_k$ is completely determined.
\end{proof}

In the remaining part of this article, we use the form
\(X^{(1)}_{<\pi>}\) to represent a level-1 unit from type-\(\pi\)
expansion. If \(X\) is associated with a rotation \(\theta\), the
notation of level-1 unit \(X^{(1),\theta}_{<\pi>}\) should be read in a
way of \(\left(X^{(1),0}_{<\pi>}\right)^\theta\), i.e., first picking
type-\(\pi\) level-1 expansion of \(X\), expanding it, then applying a
rotation of \(\theta\). \(X^{(1)}_{<\pi>}\) is also written as
\(X_{<\pi>}\) for simplicity. If the expansion code \(\pi\) is not of
interest, \(X^{(1)}_{<\pi>}\) is written as \(X^{(1)}\).

\subsection{Example}\label{example}

We demonstrate how to expand a base \(R^{90}\) to a level-2 curve. First
let's expand it to level 1. This can be done by simply preselecting one
expansion type from Figure \ref{fig:expansion_rule}. Here we choose the
first one, i.e., taking expansion code of 1 (\(\pi_1 = 1\)). Then we
have the sequence of the level-1 curve denoted as \(\mathcal{P}_1\) as
follows.

\begin{equation}
\mathcal{P}_1 = R^{(1),90}_{<1>} = R_1^{90} = (IRR^{270}L^{180})^{90} = I^{90}R^{90}RL^{270}
\notag
\end{equation}

\begin{note}
In this article, as a convention, when we explicitly use specific base types, we
simplify the form, e.g., $R^{(1),\theta}_{<1>}$ to $R^\theta_1$ where the
integer subscript always corresponds to the expansion code. When $\theta$ is
missing, it always corresponds to the base state with zero rotation. This
convention only applies to the notations of level-1 units.
\end{note}

Next, to extend \(\mathcal{P}_1\) to level 2, we have to assign the
expansion code to each of \(IRRL\). We start from the first base \(I\)
and we preselect an expansion code for it. As there are two options, we
use \(I_1\) as an example (\(\pi_2 = 1\)). Then according to the
criterions defined in Equation \ref{eq:code_xy}, the expansion code for
the remaining bases can be calculated from their preceding bases.

\begin{equation}
\label{eq:p1}
\mathcal{P}_2 = I_1^{90}R_1^{90}R_2 L_1^{270}
\end{equation}

\noindent Then we replace each base in \(\mathcal{P}_2\) with its
corresponding level-1 expression and we obtain the final base sequence
of the level-2 curve:

\begin{equation}
\label{eq:p2}
\begin{aligned}
\mathcal{P}_2 &= (RL^{270}LR^{90})^{90} (IRR^{270}L^{180})^{90} (LR^{90}RI^{270}) (RL^{270}LI^{90})^{270} \\
    &= R^{90}LL^{90}R^{180} I^{90}R^{90}RL^{270} LR^{90}RI^{270} R^{270}L^{180}L^{270}I \\
\end{aligned} .
\end{equation}

This process can be applied repeatedly to any level \(k\), where we
always first preselect an expansion code for the first base in
\(\mathcal{P}_{k-1}\), calculate the code for the remaining bases and
expand the curve to level \(k\) by replacing each base to its
corresponding level-1 expansion.

A curve is a sequence of bases where each base is associated with a
specific rotation as well as an entry direction and an exit direction.
This means, the location and rotation of the next base are already
determined by the current base. Then, with knowing the location and
rotation of the first base, the locations and rotations of all the
remaining bases can be deterministically calculated with only knowing
the type of the bases, while their absolute rotations are not
necessarily known in advance. Then the long expression of
\(\mathcal{P}_2\) in Equation \ref{eq:p2} can be rewritten as a sequence
with only its first base associated with a rotation:

\begin{equation}
\mathcal{P}_2 = R(90)LLRIRRLLRRIRLLI .
\notag
\end{equation}

Given two connected bases \(X_1^{\theta_1}X_2^{\theta_2}\), the value of
\(\theta_2\) depends on the specific base type of its preceding base
\(X_1\) (Let's only restrict it to the three primary bases).

\begin{equation}
\label{eq:theta}
\theta_2 = \begin{cases}
\theta_1 & \quad \text{if } X_1 = I \\
\theta_1 - 90 & \quad \text{if } X_1 = R \\
\theta_1 + 90 & \quad \text{if } X_1 = L \\
\end{cases}
\end{equation}

However, when studying the expansion and transformation of a curve or
its subcurves, we still use the representation where rotations of all
bases are implicitly or explicitly added.

\subsection{Expansion path}\label{expansion-path}

When expanding a curve to the next level, each base on the curve needs
to be associated with an expansion code, which is recursively determined
by the code of its first base. Such list of expansion code along a base
sequence is called the \textit{expansion path}. In Equation \ref{eq:p1},
the expansion path of \(\mathcal{P}_1\) denoted as \(p_1\) is:

\begin{equation}
p_1 = (1, 1, 2, 1) .
\notag
\end{equation}

There exists a second expansion path denoted as \(p'_1\) if we assign
code 2 to the first base on \(\mathcal{P}_1\):

\begin{equation}
p'_1 = (2, 2, 1, 2) .
\notag
\end{equation}

Similarly, if we expand \(\mathcal{P}_2\) to the next level, there are
the following two expansion paths denoted as \(p_2\) and \(p'_2\). The
expansion path can be calculated with Equation \ref{eq:code_xy}, or even
faster with Equation \ref{eq:pipi} which we will introduce later.

\begin{equation}
\begin{array}{ll}
p_2 &= (1, 2, 1, 2, 1, 1, 2, 1, 2, 1, 2, 1, 1, 2, 1, 2) \\
p'_2 &= (2, 1, 2, 1, 2, 2, 1, 2, 1, 2, 1, 2, 2, 1, 2, 1) \\
\end{array}
\notag
\end{equation}

\begin{figure}[t]
\centering{
\includegraphics[width=1\linewidth]{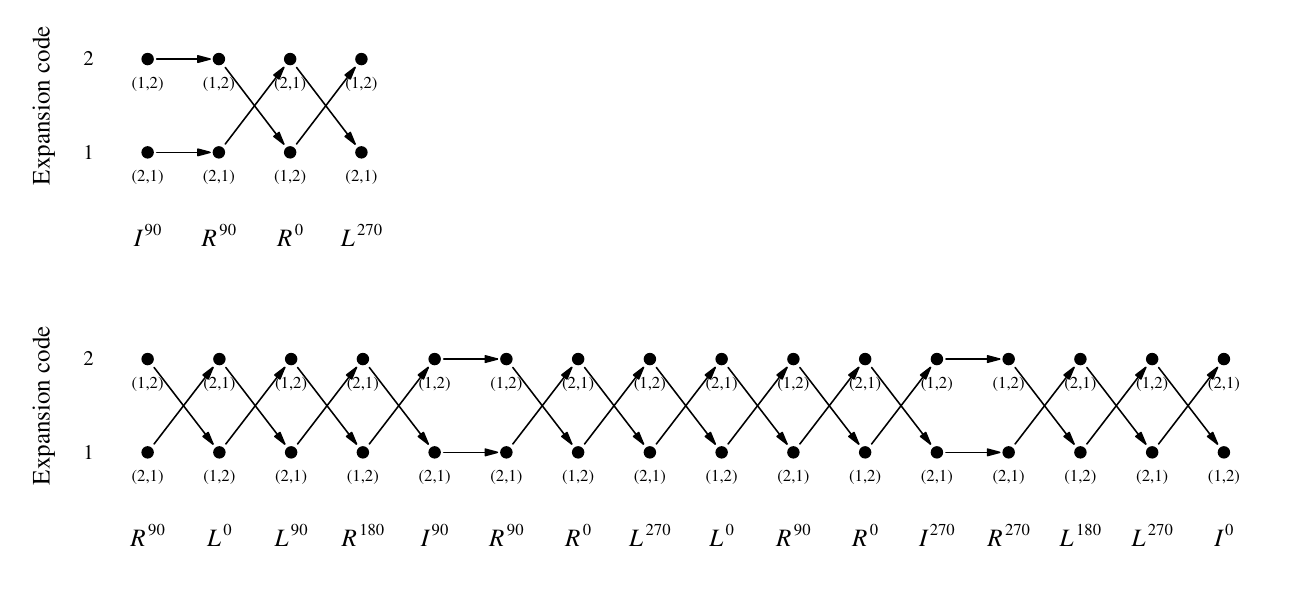}
\caption{Expansion paths of two curves. Top: the two expansion paths of $\mathcal{P}_1 = R^{90}|1$; Bottom: the two expansion paths of $\mathcal{P}_2 = R^{90}|11$. The meaning of $R^{90}|1$ and $R^{90}|11$ will be explained in later sections. The 2-tuple under each point is the corner-tuple for each base (rotation included).}\label{fig:traverse_path}}
\end{figure}

\begin{proposition}
\label{prop:expansion_path}
When a curve $\mathcal{P}_{k-1}$ is expanded to $\mathcal{P}_k$ ($k \ge 1$), there are only 
two expansion paths that are complementary and only determined by the expansion
code of the first base in $\mathcal{P}_{k-1}$.
\end{proposition}

\begin{proof}
When $k = 1$, $\mathcal{P}_0$ is a single base. The expansion path of $\mathcal{P}_0$ is just
its expansion code. Then choosing either of the two expansion code for $\mathcal{P}_0$ makes
the two expansion paths of $\mathcal{P}_0$ complementary.

When $k \ge 2$, let $\mathcal{P}_{k-1} = X_1...X_n$ where $n = 4^{k-1}$. Write
Equation \ref{eq:code_xy} as a function $\pi_{i} = f(\pi_{i-1}, \theta_{i-1} -
\theta_i)$ ($2 \le i \le n$) where $\pi_i$ is the expansion code of $X_i$ and
$\theta_i$ is the rotation associated with $X_i$. We assign $\pi_1$ to $X_1$ and
we can calculate all the remaining expansion code $\pi_i$ by $f()$, then
we have the first expansion path $(\pi_1, \pi_2, ..., \pi_n)$. Next we change $\pi_1$
to its complement $\hat{\pi}_1$. With the form of $f()$, we can easily see
$\hat{\pi}_{i} = f(\hat{\pi}_{i-1}, \theta_{i-1} - \theta_i)$ since all $\theta_i$ are not changed.
Thus we have the second expansion path $(\hat{\pi}_1, \hat{\pi}_2, ..., \hat{\pi}_n)$
which is complementary from the first expansion path. The two expansion paths 
are only determined by the code of $X_1$.
\end{proof}

The expansion paths of \(\mathcal{P}_1\) and \(\mathcal{P}_2\) are
visualized in Figure \ref{fig:traverse_path}. For a 2x2 curve, its
level-1 units with their rotations included have corner-tuples either
all \((1, 2)\) or all \((2, 1)\), which ensures the expansion path is
fully determined by the first unit (also see Proposition
\ref{prop:3.2}). However, for more complex curves such as 3x3 curves
(not included in this study), the level-1 unit can also have other
corner-tuples of \((1, 1)\) or \((2, 2)\), which makes the combinations
of different expansion paths
huge\footnote{A qiuck example of the expansion paths
in 3x3 curves can be found from
\url{https://jokergoo.github.io/sfcurve/articles/all_3x3_curve.html}.}.
The visualization of expansion paths helps to study the complexity of
the curve generation.

Here, in the expansion path, expansion code for the \(i\)-th base is
calculated from its preceding base recursively according to Equation
\ref{eq:code_xy} or \ref{eq:pipi}. In Section
\ref{expansion-code-from-the-second-base}, we will demonstrate the
expansion code can be directly calculated from the first base of the
curve sequence.

\section{Encode the curve}\label{encode-the-curve}

\begin{figure}
\centering{
\includegraphics[width=1\linewidth]{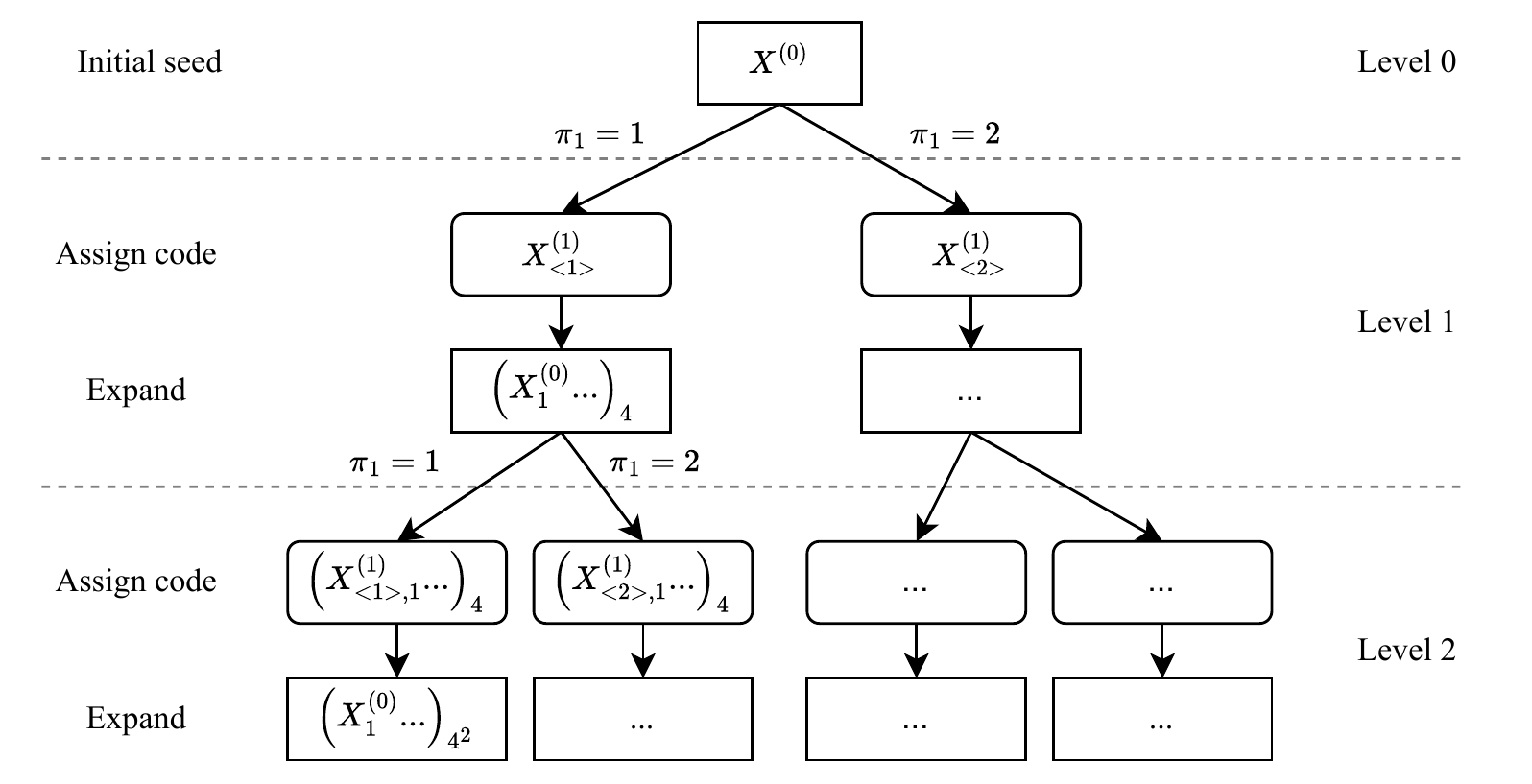}
\caption{Expansion of the curve. Expansions only till level 2 are illustrated in the figure, but it can be applied recursively to any level $k$. $X^{(0)}$: a base on level 0; $X^{(1)}_{<1>}$ or $X^{(1)}_{<2>}$: a level-1 expansion of $X$ with expansion code 1 or 2.}\label{fig:figure_expansion_chart}}
\end{figure}

\subsection{The encoding system}\label{the-encoding-system}

From level 0 which corresponds to the initial pattern of the curve, on
each level of the curve expansion, there always involve two steps: to
determine the expansion code for bases in the sequence and to replace
each base with its corresponding level-1 unit. In the previous section,
we have demonstrated that, from level \(k-1\) to level \(k\), the
expansion code of a base is determined by its preceding base, which is
eventually determined by the first base in the sequence (Proposition
\ref{prop:3.2}). Then the expansion of the curve from level 0 to level
\(k\) can be described in a binary tree schema illustrated in Figure
\ref{fig:figure_expansion_chart}. The curve expansions can be briefly
described in the following steps.

\begin{enumerate}
\tightlist
\item
   Level 0 $\rightarrow$ level 1: Pick one expansion code for the level-0 base
   and expand it into four bases.
\item
   Level 1 $\rightarrow$ level 2: Only select the expansion code for the first
   base of $\mathcal{P}_1$, and calculate the expansion code for the other
   bases, then expand the four bases into 16 bases.
\item
   Level $k-1$ $\rightarrow$ level $k$ ($k \ge 3$): Only select the expansion
   code for the first base of $\mathcal{P}_{k-1}$, and calculate the expansion
   code for the other bases, then expand the $4^{k-1}$ bases into $4^k$ bases.
\end{enumerate}

With knowing the initial base and the expansion code of the first base
in each iteration, the curve is fully determined. Then we can encode a
2x2 curve on level \(k\) denoted as \(\mathcal{C}_k\) as:

\begin{equation}
\label{eq:09}
\mathcal{C}_k = X^{(0)}|\pi_1 \pi_2 ... \pi_k \quad \pi_i \in \{1, 2\}, 1 \le i \le k ,
\end{equation}

\noindent where \(\pi_i\) is the expansion code of the first base in the
sequence when expanded from level \(i-1\) to level \(i\).

In the expansion code sequence of \(\pi_1 \pi_2 ... \pi_k\), if
\(\pi_i\) is more to the left of the sequence, it corresponds more to
the early stage of the expansion, and if the code is more to the right
side of the sequence, it corresponds more to the late stage of the
expansion.

We can remove the level-0 notation in Equation \ref{eq:09} because
apparently by definition the initial base is on level 0. We also add the
rotation to the base to have a final encoding of a 2x2 curve:

\begin{remark}
\label{remark:curve_single}
A 2x2 curve $\mathcal{C}_k$ initialized from a single base $X^{\theta}$ is encoded as: 

\begin{equation}
  \label{eq:def}
\mathcal{C}_k = X^{\theta}|\pi_1 \pi_2 ... \pi_k \quad \pi_i \in \{1, 2\} .
\end{equation}

\end{remark}

Next we prove the symbolic expression in Equation \ref{eq:def} uniquely
encodes a curve.

\begin{definition}[Identical curves]
\label{def:identical_curve}
Two curves on level $k$ denoted as $\mathcal{P}_k$ and $\mathcal{Q}_k$ are
identical when their corresponding base sequences are identical. Write
$\mathcal{P}_k = X_1^{\alpha_1} ... X_n^{\alpha_n}$ and $\mathcal{Q}_k =
Y_1^{\beta_1} ... Y_n^{\beta_n}$, then $\mathcal{P}_k = \mathcal{Q}_k$ iff
$\forall i \in \{1, ..., n\}: X_i = Y_i \text{ and } \alpha_i = \beta_i$.
\end{definition}

\begin{lemma}
\label{lemma:p_inequality}
If $\mathcal{P}_{i} \ne \mathcal{Q}_{i}$ ($0 \le i < k$), then $\mathcal{P}_{k} \ne \mathcal{Q}_{k}$.
\end{lemma}

\begin{proof}

We first prove for $k = i+1$. There are two scenarios that cause
$\mathcal{P}_{i} \ne \mathcal{Q}_{i}$.

First, there exists a base $X_j$ in the base sequence of $\mathcal{P}_{i}$
being different from the corresponding base $Y_j$ in $\mathcal{Q}_{i}$.
According to all level 0-to-1 expansions in Figure \ref{fig:expansion_rule},
we can always have $X^{(1)}_j \ne Y^{(1)}_j$ if $X_j \ne Y_j$ regardless of
which expansion code they take. Also this inequality is not affected by the
rotations associated with $X_j$ and $Y_j$. Since $X^{(1)}_j$ and $Y^{(1)}_j$
are subsequences of $\mathcal{P}_{i+1}$ and $\mathcal{Q}_{i+1}$, this results
in $\mathcal{P}_{i+1} \ne \mathcal{Q}_{i+1}$ (Definition
\ref{def:identical_curve}).

Second, for all $j \in \{1, ..., 4^i\}$, $X_j = Y_j$, but these exist a base
$X_j$ whose rotation $\alpha_j$ is different from the rotation $\beta_j$ of
its corresponding base $Y_j$ in $\mathcal{Q}_{i}$. According to Equation
\ref{eq:theta}, rotation of a base is determined by the type of its preceding
base in the sequence. Since all $X_j = Y_j$, thus $\alpha_1 \ne \beta_1$. If
$\mathcal{P}_{i}$ is expanded to the next level via code $\pi$ and
$\mathcal{Q}_{i}$ is expanded to the next level via code $\sigma$, according
to the discussion in this section, $\pi$ and $\sigma$ are also for $X_1$ and
$Y_1$ respectively. We already have $X_1 = Y_1$ in this category. If $\pi =
\sigma$, there is $X^{(1),0}_{<\pi>} = Y^{(1),0}_{<\sigma>}$, but since
$\alpha_1 \ne \beta_1$, there is $X^{(1),\alpha_1}_{<\pi>} \ne
Y^{(1),\beta_1}_{<\sigma>}$, which in turn results in $\mathcal{P}_{i+1} \ne
\mathcal{Q}_{i+1}$ (Definition \ref{def:identical_curve}). If $\pi \ne \sigma$, 
then according to Figure \ref{fig:expansion_rule}, two different expansion
code on identical bases always give two different level-1 units, thus
$X^{(1)}_{<\pi>} \ne Y^{(1)}_{<\sigma>}$ and in turn $\mathcal{P}_{i+1} \ne
\mathcal{Q}_{i+1}$ (Definition \ref{def:identical_curve}).

Now we have proven that when $\mathcal{P}_{i} \ne \mathcal{Q}_{i}$, there is 
$\mathcal{P}_{i+1} \ne \mathcal{Q}_{i+1}$. By applying it repeatedly, we can 
eventually have $\mathcal{P}_{k} \ne \mathcal{Q}_{k}$ for any $k > i$.

\end{proof}

\begin{proposition}
\label{prop:unique_curve}
For two curves on level $k$ encoded as $\mathcal{P}_k =
X^\alpha|\pi_1...\pi_k$ and $\mathcal{Q}_k = Y^\beta|\sigma_1...\sigma_k$,
$\mathcal{P}_k \ne \mathcal{Q}_k$ iff 1. $X \ne Y$, or 2. $\alpha \ne \beta$,
or 3. $\exists i \in \{1, ..., k\}: \pi_i \ne \sigma_i$. In other words, if
$\mathcal{P}_k$ and $\mathcal{Q}_k$ have the same encoding, they are
the same curve; and if they have different encodings, they are different curves.
\end{proposition}

\begin{proof}
We first discuss the case where two encodings are the same. According to Proposition
\ref{prop:expansion_path}, for $\mathcal{P}_i$ and $\mathcal{Q}_i$ expanded
from $\mathcal{P}_{i-1}$ and $\mathcal{Q}_{i-1}$ ($1 \le i \le k$), if
$\mathcal{P}_{i-1} = \mathcal{Q}_{i-1}$ and $\pi_i = \sigma_i$, the expansion
paths denoted as $p_{i-1}$ and $q_{i-1}$ of $\mathcal{P}_{i-1}$ and
$\mathcal{Q}_{i-1}$ are also the same, which makes $\mathcal{P}_{i-1}$ and
$\mathcal{Q}_{i-1}$ expanded into identical $\mathcal{P}_i$ and
$\mathcal{Q}_i$. Apparently in this category,
$\mathcal{P}_0 = \mathcal{Q}_0$. Then according to the discussion, we can
sequentially have $\mathcal{P}_1 = \mathcal{Q}_1$, ..., $\mathcal{P}_k = \mathcal{Q}_k$.
Thus two identical encodings result in two identical curves.

Next we discuss the case where two encodings are different. There are three scenarios.

1. When $X \ne Y$, there is $X^\alpha \ne Y^\beta$ for any values of $\alpha$
   and $\beta$. This means $\mathcal{P}_0 \ne \mathcal{Q}_0$. According to
   Lemma \ref{lemma:p_inequality}, we have $\mathcal{P}_k \ne \mathcal{Q}_k$.

2. When $X = Y$, $\alpha \ne \beta$ also results in $X^\alpha \ne Y^\beta$.
   We can similarly have $\mathcal{P}_k \ne \mathcal{Q}_k$. 

3. When $X = Y$ and $\alpha = \beta$, let $i$ be the first index in $\{1, ...,
   k\}$ that makes $\pi \ne \sigma$, i.e., $\pi_j = \sigma_j$ for all $1 \le j \le i-1$ and
   $\pi_i \ne \sigma_i$, then $\mathcal{P}_{i-1} =
   \mathcal{Q}_{i-1}$ because the two symbolic expressions of $\mathcal{P}_{i-1}$ and $\mathcal{Q}_{i-1}$ are identical. Let $Z$
   be the first base in the base sequence of $\mathcal{P}_{i-1}$ and $W$ be
   the first base in $\mathcal{Q}_{i-1}$, then apparently $Z = W$. When
   $\mathcal{P}_{i-1}$ is expanded to level $i$ via code $\pi_i$, $\pi_i$ is
   also the expansion code for $Z$, thus the first 2x2 unit in $\mathcal{P}_i$
   is $Z^{(1)}_{<\pi_i>}$. Similarly, the first 2x2 unit in $\mathcal{Q}_i$ is
   $W^{(1)}_{<\sigma_i>}$. With $Z = W$ and $\pi_i \ne \sigma_i$, we have
   $Z^{(1)}_{<\pi_i>} \ne W^{(1)}_{<\sigma_i>}$, and this inequality is not
   affected by the rotations associated with $Z$ and $W$. This results in
   $\mathcal{P}_{i} \ne \mathcal{Q}_{i}$ and eventually $\mathcal{P}_k \ne \mathcal{Q}_k$ (Lemma \ref{lemma:p_inequality}).

\end{proof}

Equation \ref{eq:def} and Proposition \ref{prop:unique_curve} imply
that, by fixing the base \(X\) and its rotation \(\theta\), there are
\(2^k\) different forms of curves on level \(k\), so the total number of
the forms by also considering all 9 base types and 4 rotations is

\begin{equation}
\label{eq:total_number}
4 \times 9 \times 2^k = 36 \times 2^k .
\end{equation}

\begin{figure}
\centering{
\includegraphics[width=1\linewidth]{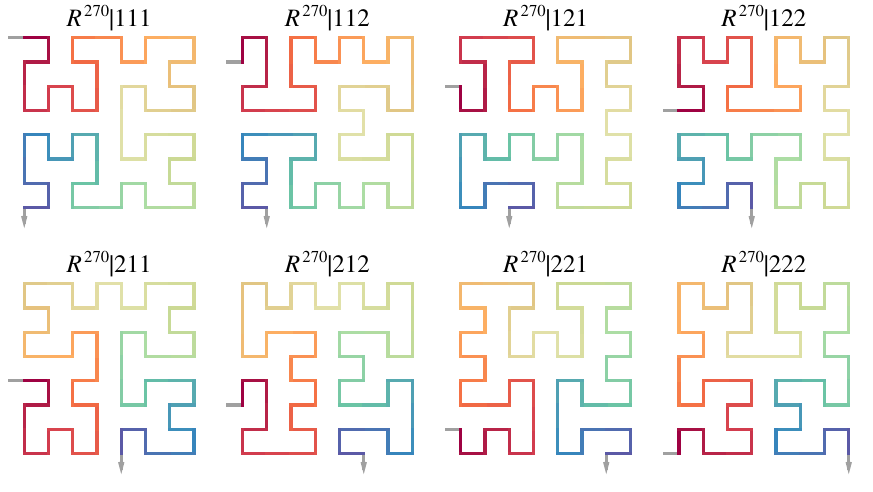}
\caption{All level-3 curves initialized by $R^{270}$. Each curve is associated with an entry direction and an exit direction.}\label{fig:R_level3}}
\end{figure}

As an example, taking \(R^{270}\) (horizontally left-in and vertically
bottom-out) as the initial base, the complete set of all 8 level-3
curves induced by \(R^{270}\) is listed in Figure \ref{fig:R_level3}.
Please note, \(\mathcal{C}_k\) is a directional curve also associated
with an entry direction and an exit direction. The number of different
forms of \(\mathcal{C}_k\) in Equation \ref{eq:total_number} also
distinguishes these factors.

The expansion code only takes value of 1 or 2, thus the code sequence
can be thought of as a sequence of binary bits, and each individual
curve can be associated with an unique integer, e.g.,

\begin{equation}
\begin{aligned}
X|111 &= X|1^{(3)} \\
X|121 &= X|3^{(3)} \\
X|222 &= X|8^{(3)}
\end{aligned}
\notag
\end{equation}

\noindent where the superscript ``$(3)$'' implies the level of the
curve. More generally, denote the integer representation of a curve on
level \(k\) as \(\delta^{(k)}\), i.e.,
\(X|\pi_1...\pi_k = X|\delta^{(k)}\), \(\delta\) can be calculated as:

\begin{equation}
\label{eq:integer}
\delta = 1 + \sum_{i=1}^k{2^{k-i}(\pi_i-1)} .
\end{equation}

The integer representation of the expansion code sequence will be used
in Section \ref{geometry}, \ref{shapes} and \ref{arithmetic} for
calculating locations of points on the curve.

\subsection{Special curves}\label{special-curves}

\begin{figure}
\centering{
\includegraphics[width=1\linewidth]{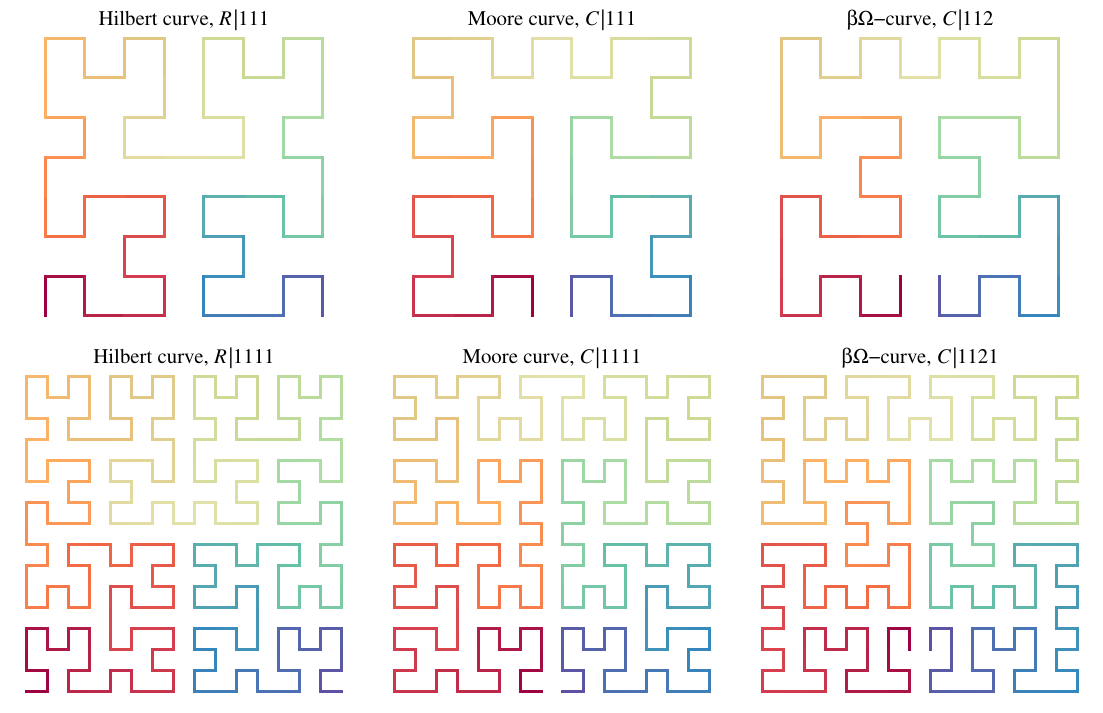}
\caption{The Hilbert curve, the Moore curve and the $\beta\Omega$-curve. The first row: three types of curves on level 3; The second row: three types of curves on level 4. Entry and exit directions are not included. Note the encoding in the title of each curve is just one from multiple possible forms.}\label{fig:standard_curve}
}
\end{figure}

Besides the Hilbert curve, there are other two types of 2x2 curves that
have been studied in literatures, the Moore curve \citep{moore1900} and
the \(\beta\Omega\)-curve \citep{betaomega}. These three types of curves
are just 2x2 curves in special encodings in our system. They can be
constructed by special initial bases and expansion code sequences. Let's
consider forms of curves starting from the lower left quadrant and
ending at the lower right quadrant (Figure \ref{fig:standard_curve}).
The Hilbert curve on level \(k\) can be encoded as:

\begin{equation}
R|(1)_k
\notag
\end{equation}

\noindent where \((1)_k\) is a sequence of \(k\) digits of 1. Since the
entry direction and exit direction of the curve are normally ignored in
current studies, there are other encodings for the Hilbert curve such as
\(R^{270}|(2)_k\), \(I^{270}|(2)_k\) or \(U|(1)_k\). These identical
curves ignoring their entry and exit directions are called
``homogeneous curves'' in this article and they will be further
discussed in Section \ref{homogeneous}.

The Moore curve is a ``closed Hilbert curve''. Its form on level \(k\)
can be encoded as:

\begin{equation}
C|(1)_k .
\notag
\end{equation}

The Moore curve is closed on the bottom-center of the curve region. It
has other homogeneous curves such as \(U|1(2)_{k-1}\) or
\(Q|1(2)_{k-1}\). Similar as the Moore curve, \citet{LIU2004741}
introduced four more variants of Hilbert curves denoted as \(L_1\) to
\(L_4\). They can be encoded by our system as \(L_1=C|1(2)_{k-1}\),
\(L_2=I^{270}|2(1)_{k-1}\), \(L_3=P^{270}|(2)_k\), and \(L_4
=B^{270}|2(1)_{k-1}\).

The Moore curve and the four Liu-variants all belong to a class of
curves, namely the order-1 Hilbert variants, which are composed of four
Hilbert curves on level \(k-1\) but in specific combinations of
orientations. Their structures will be further discussed in Section
\ref{recursive-curves}.

Last, the \(\beta\Omega\)-curve is also a closed curve. One of its
encodings on level \(k\) (\(k \ge 2\)) is:

\begin{equation}
\label{eq:betaomega}
\begin{array}{ll}
C|1\pi_2...\pi_k \quad \text{where }\pi_2 = 1, \text{ and } \pi_{i} = \hat{\pi}_{i-1} \text{ for } 3 \le i \le k
\end{array} .
\end{equation}

In Section \ref{def-hc}, we will give definitions of the Hilbert curve
and the \(\beta\Omega\)-curve as well as their variants bases on their
structural attributes. In particular, we will demonstrate the curve with
the form in Equation \ref{eq:betaomega} which are often used in
literatures is not a strict \(\beta\Omega\)-curve.

\subsection{Seed as a sequence}\label{seed-as-a-sequence}

We have demonstrated using a single base as the seed to induce the
curve. There is no restriction on the length of the seed sequence. We
can still follow the expansion steps in Section
\ref{the-encoding-system} but with small modifications. Denote the seed
sequence as \(\mathcal{S} = X_1...X_n\), the expansion steps are:

\begin{enumerate}
\tightlist
\item
   Level 0 $\rightarrow$ level 1: Pick the expansion code only for $X_1$, then
   the code for the remaining bases in $\mathcal{S}$ can be deterministically
   obtained by Equation \ref{eq:code_xy} or \ref{eq:pipi}. Replace each with its
   corresponding level-1 units. This generates a level-1 curve with $4 \times n$ bases.
\item
   Level $k-1$ $\rightarrow$ level $k$ ($k \ge 2$): Only select the expansion
   code for the first base, and calculate the expansion code for all other
   bases, then expand the $4^{k-1}\times n$ bases into $4^k \times n$ bases.
\end{enumerate}

The seed sequence represents the seed curve. The seed curve should be
continuous and have no intersection, i.e., it should be represented as
an orthogonal path. The seed sequence is normally composed of the three
primary bases of \(I\), \(R\) and \(L\). Nevertheless, to make it
general, other base types are also allowed for constructing the seed
sequence, but with the following restrictions:

\begin{enumerate}
\tightlist
\item
   $U$ can only be used as the first base or the last base in a seed sequence.
\item
   $B$ and $D$ are entry-closed, so they can only be used as the first base in
   a seed sequence.
\item
   $P$ and $Q$ are exit-closed, so they can only be used as the last base in a
   seed sequence.
\item
   $C$ is both entry-closed and exit-closed, thus it can only be used as a
   singleton while cannot be connected to other bases.
\end{enumerate}

\begin{remark}[2x2 space-filling curve]
\label{remark:curve_sequence}
A general 2x2 curve $\mathcal{C}_k$ initialized by a seed curve $\mathcal{S}=X_1 ... X_n$ ($n \ge 1$) is encoded as:

\begin{equation}
\mathcal{C}_k = \mathcal{S}|\pi_1 \pi_2 ... \pi_k ,
\notag
\end{equation}

\noindent where $\pi_1$ is the expansion code of $X_1$ from level 0 to level 1.

\end{remark}

\begin{figure}
\centering{
\includegraphics[width=1\linewidth]{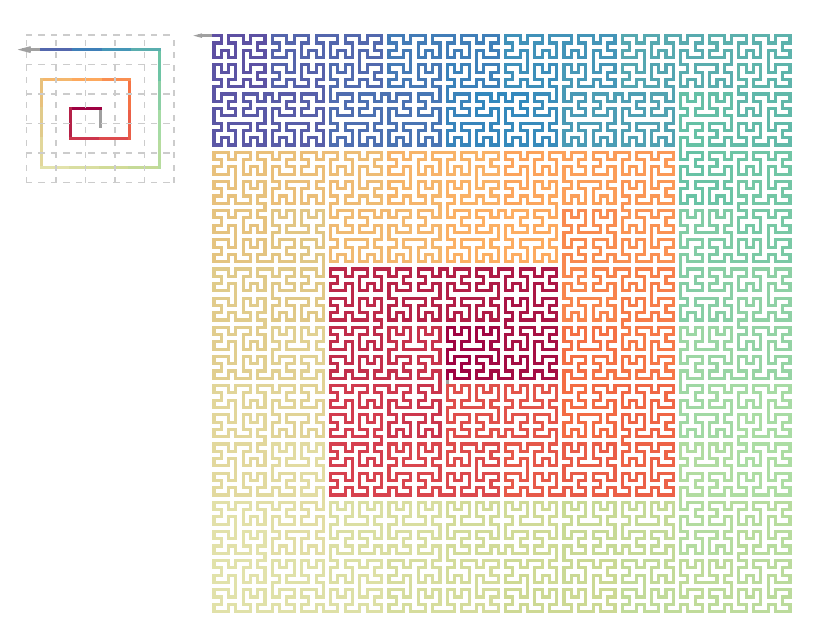}
\caption{A space-filling curve initialized from a spiral seed curve. Left: the spiral seed  curve; Right: two induced curves on level 4 with encoding $\mathcal{S}|1111$ and $\mathcal{S}|2121$.}\label{fig:sequence_seed}
}
\end{figure}

As an example, the following sequence represents a spiral seed curve
(Figure \ref{fig:sequence_seed}, left panel).

\begin{equation}
\begin{aligned}
\mathcal{S} &= L(0)LLILILIILIILIIILIIILIIII \\
  &= LL^{90}L^{180}I^{270}L^{270}ILI^{90}I^{90}L^{90}I^{180}I^{180} L^{180}I^{270}I^{270}I^{270}L^{270}IIILI^{90}I^{90}I^{90}I^{90} \\
\end{aligned}
\notag
\end{equation}

Note in the second line in the above equation, rotations from the second
base can be calculated by Equation \ref{eq:theta}. Figure
\ref{fig:sequence_seed} (right panel) illustrates the expansion of
\(\mathcal{P}_4 = \mathcal{S}|1111\) and
\(\mathcal{P}_4 = \mathcal{S}|2121\) (level-4 curves).

Fixing the seed sequence, the total number of different forms of the
induced curves on level \(k\) is \(2^k\).

\subsection{Other attributes of 2x2
curves}\label{other-attributes-of-2x2-curves}

In the remaining part of this article, if there is no explicit
clarification, we use the form \(\mathcal{P}_k\) to represent a general
level-\(k\) curve initialized from a seed sequence, i.e.,
\(\mathcal{P}_k = \mathcal{S}|\pi_1...\pi_k\). In this section, we
discuss several attributes of 2x2 curves that will be used in other
sections of this article.

\begin{remark}
$\mathcal{P}_k$ can eventually be expressed as a long sequence of bases. If the
construction of $\mathcal{P}_k$ is treated as a drawing process,
since each base has an entry direction and an exit directon associated, then
each base describes how the pen moves through the corresponding point on the
curve. Thus the sequential expression of $\mathcal{P}_k$ exactly describes
its representation as an orthogonal path.
\end{remark}

\begin{proposition}
\label{prop:IRL}
$\mathcal{P}_k$ ($k \ge 1$) only contains primary bases from $\{I, R, L\}$,
and it must contain $R$ or $L$.
\end{proposition}

\begin{proof}
According to Figure \ref{fig:expansion_rule}, all level-1 units in the
U-shapes are only composed of $I$, $R$ and $L$. Since the curve is generated
in the expansion mode, $\mathcal{P}_k$ ($k \ge 1$) can be represented as a
list of level-1 units, then $\mathcal{P}_k$ only contains based from $\{I, R,
L\}$. The second and the third bases in a level-1 units always represent both
turning right, or both turning left if the orientation of the unit is counterclockwise,
thus a level-1 unit must contain $R$ or $L$, then $\mathcal{P}_k$ ($k \ge 1$)
must contain $R$ or $L$.
\end{proof}

\begin{proposition}
\label{prop:full_IRL}
$\mathcal{P}_k$ ($k \ge 2$) contains the full set of $\{I, R, L\}$.
\end{proposition}

\begin{proof}
According to Figure \ref{fig:expansion_rule}, level-1 expansions of $R$ and
$L$ contain all the three primary bases. This results in that, expansion to
any level from $R$ or $L$ will also contain the full set of the three primary
bases. According to Proposotion \ref{prop:IRL}, $\mathcal{P}_1$ must contain
$R$ or $L$, then $\mathcal{P}_k$ ($k \ge 2$) must contain the full set of
$\{I, R, L\}$.
\end{proof}

As bases \(U\), \(B\) and \(D\) can also be the first base of a seed
sequence, we first extend Equation \ref{eq:theta} to:

\begin{equation}
\label{eq:theta2}
\theta_2 = \begin{cases}
\theta_1   & \quad \text{if } X_1 \in \{I, B, D\} \\
\theta_1 - 90   & \quad \text{if } X_1  = R  \\
\theta_1 + 90   & \quad \text{if } X_1  = L  \\
\theta_1 + 180   & \quad \text{if } X_1  = U  \\
\end{cases} .
\end{equation}

\noindent With Equation \ref{eq:code_xy} and \ref{eq:theta2}, we can
have the following proposition of assigning expansion code to all bases
in a sequence without considering their rotations.

\begin{proposition}
\label{prop:3.5}
For a curve expressed as a sequence of $X_1...X_n$ which will be expanded to
the next level, the expansion code denoted as $\pi_i$ for base $X_i$ in the
sequence is determined by the type of its preceding base $X_{i-1}$ ($2
\le i \le n$) as:

\begin{equation}
\label{eq:pipi}
\pi_i = \begin{cases}
\pi_{i-1} \quad & \textnormal{if } X_{i-1} \in \{I, U, B, D\} \\
\hat{\pi}_{i-1} \quad & \textnormal{if } X_{i-1} \in \{R, L\} \\
\end{cases} .
\end{equation}

\end{proposition}

\begin{proof}
With Equation \ref{eq:theta2}, we have $\theta_{i} - \theta_{i-1} \bmod 180 =
0$ if $X_{i-1} \in \{I, U, B, D\}$, and $\theta_{i} - \theta_{i-1} \bmod 180 =
90$ if $X_{i-1} \in \{R, L\}$. Then with Equation \ref{eq:code_xy}, we can
obtain the solution in Equation \ref{eq:pipi}.
\end{proof}

\begin{proposition}
\label{prop:entry-exit-directions}
Denote $\varphi_s()$ as a function which returns the entry direction of a
curve, and $\varphi_e()$ as a function which returns the exit direction of a
curve. The entry and exit directions keep unchanged during the curve
expansion, written as:

\begin{equation}
\begin{aligned}
\varphi_s(\mathcal{P}_i) &=  \varphi_s(\mathcal{P}_j) \\
\varphi_e(\mathcal{P}_i) &=  \varphi_e(\mathcal{P}_j) \\
\end{aligned} \quad\quad 1 \le i, j \le k ,
\notag
\end{equation}

\noindent and the equalities extend to $\mathcal{P}_0$ when the corresponding
$\varphi_s(\mathcal{P}_0)$ or $\varphi_e(\mathcal{P}_0)$ exists.

\end{proposition}

\begin{proof}

When a curve $\mathcal{P}_k$ is expanded from $\mathcal{P}_{k-1}$ ($k \ge 1$),
the first base in $\mathcal{P}_{k-1}$ denoted as $X_s^\theta$ is expanded to
$X^{(1),\theta}_s$. If $k-1 = 0$, we only consider entry-opened bases, i.e., $X_s
\in \{I, R, L, U, P, Q\}$; and if $k-1 \ge 1$, there is always $X_s \in \{I, R,
L\}$ (Proposition \ref{prop:IRL}). For both scenarios, according to the
expansion rules in Figure \ref{fig:expansion_rule}, entry direction of $X_s$
is always the same as that of its both level-1 expansions $X^{(1),0}_s$, and
in turn we can have

\begin{equation}
\varphi_s(\mathcal{P}_{k-1}) = \varphi_s(X_s^\theta) = \varphi_s(X^{(1),\theta}_s) = \varphi_s(\mathcal{P}_k) .
\notag
\end{equation}

This relation can be repeatedly applied to have:

\begin{equation}
\varphi_s(\mathcal{P}_{k}) = ... = \varphi_s(\mathcal{P}_1)
\notag
\end{equation}

\noindent and till $\varphi_s(\mathcal{P}_0)$ if it exists.

Similarly, the last base in $\mathcal{P}_{k-1}$ denoted as $X_e^\xi$ is
expanded to $X^{(1),\xi}_e$. If $k-1 = 0$, we only consider exit-opened bases,
i.e., $X_e \in \{I, R, L, U, B, D\}$; and if $k-1 \ge 1$, there is always $X_e
\in \{I, R, L\}$ (Proposition \ref{prop:IRL}). For both scenarios, according
to the expansion rules in Figure \ref{fig:expansion_rule}, exit direction of
$X_e$ is always the same as that of its both level-1 expansions $X^{(1),0}_e$,
and in turn we can have

\begin{equation}
\varphi_e(\mathcal{P}_{k-1}) = \varphi_e(X_e^\xi) = \varphi_e(X^{(1),\xi}_e) = \varphi_e(\mathcal{P}_k) .
\notag
\end{equation}

This relation can be repeatedly applied to have:

\begin{equation}
\varphi_e(\mathcal{P}_{k}) = ... = \varphi_e(\mathcal{P}_1)
\notag
\end{equation}

\noindent and till $\varphi_e(\mathcal{P}_0)$ if it exists.
\end{proof}

\begin{corollary}
\label{coro:first_rotation}
Denote $\theta_s()$ as a function which returns the rotation of the first base
in a sequence. For a seed sequence $\mathcal{S} = X_1 ... X_n$, the first base
$X_1$ is associated with a rotation $\theta$ and the rotation of the first
base of $X^{(1),0}_{<\pi_1>,1}$ is $\alpha$, then for $\mathcal{P}_k =
\mathcal{S}|\pi_1...\pi_k$, there is always

\begin{equation}
\theta_s(\mathcal{P}_{k}) = \theta_s(\mathcal{P}_1) = \theta + \alpha \quad k \ge 1 .
\notag
\end{equation}
\end{corollary}

\begin{proof}
First, the entry direction of a curve $\mathcal{P}$ is also the entry
direction of its first base $X_s$. The entry direction is a component of
$X_s$, thus it is rotated in the same amount as $X_s$ itself:

\begin{equation}
\label{eq:entry}
\varphi_s(\mathcal{P}) - \theta_s(\mathcal{P}) = \gamma_{X_s}
\end{equation}

\noindent where $\gamma_{X_s}$ is only determined by the type of $X_s$. According to
Proposition \ref{prop:IRL}, $\mathcal{P}_k$ ($k \ge 1$) only contains primary
bases. Then according to Figure \ref{fig:expansion_rule}, we have $\gamma_I =
\gamma_R = \gamma_L = 90$.

Based on Proposition \ref{prop:entry-exit-directions} and Equation \ref{eq:entry}, we have:

\begin{equation}
\begin{aligned}
\varphi_s(\mathcal{P}_k) &= \varphi_s(\mathcal{P}_1) \\
\theta_s(\mathcal{P}_k) + 90 &= \theta_s(\mathcal{P}_1) + 90 \\
\theta_s(\mathcal{P}_k)& = \theta_s(\mathcal{P}_1) \\
\end{aligned} ,
\notag
\end{equation}

\noindent then

\begin{equation}
\begin{aligned}
\theta_s(\mathcal{P}_k)& = \theta_s(\mathcal{P}_1)= \theta_s(\mathcal{S}|\pi_1) = \theta_s\left(\left(X^{(1),0}_{<\pi_1>,1}\right)^\theta ...\right) \\
    &= \theta_s \left((Z_1^\alpha ...)^\theta ...\right) \\
    &= \theta + \alpha \\
\end{aligned} 
\notag
\end{equation}

\noindent where $Z_1$ is the first base in $X^{(1),0}_{<\pi_1>,1}$.

\end{proof}

\begin{corollary}
\label{coro:first_rotation_IRL}
For a curve initialized from a single primary base, i.e., $\mathcal{P}_k = Z^\theta|\pi_1...\pi_k$ where
$Z \in \{I, R, L\}$, there is always $\theta_s(\mathcal{P}_k) = \theta$ for any $k \ge 0$.
\end{corollary}

\begin{proof}
When $k = 0$, $\theta_s(\mathcal{P}_0) = \theta_s(Z^\theta) = \theta$. When $k
\ge 1$, using Corollary \ref{coro:first_rotation}, with $Z \in \{I, R, L\}$,
there is always $\alpha = 0$, then there is $\theta_s(\mathcal{P}_k) =
\theta$.
\end{proof}

\subsection{\texorpdfstring{\(\mathcal{P}\) is mathematically a
space-filling
curve}{\textbackslash mathcal\{P\} is mathematically a space-filling curve}}\label{mathcalp-is-mathematically-a-space-filling-curve}

In this study, we focus on \(\mathcal{P}_k\) after finite iterations.
However, when \(k\) reaches infinity, the limit of \(\mathcal{P}_k\) is
a mathematically strict space-filling curve regardless how it is
constructed from the \(36 \times 2^k\) forms.

\begin{proposition}
For a general 2x2 curve $\mathcal{P}_k = \mathcal{S}|\pi_1...\pi_k$, its limit 
$\mathcal{P} = \lim_{k \to \infty } \mathcal{P}_k$ is a space-filling curve.
\end{proposition}

\begin{proof}
$\mathcal{S}$ is composed of $n$ bases. Let the base $X_i$ has a coordinate of
$(x_i, y_i)$ and it is located on the left, right, top or bottom of its
preceding $X_{i-1}$ with a distance of 1. Let $\mathcal{I} := [0, n]$ as an
one-dimensional interval and $\mathcal{Q} := \bigcup_{i=1}^n ([x_i - x_1, x_i
- x_1+1] \times [y_i - y_1, y_i - y_1+1])$ as the region determined by
$\mathcal{S}$. $\mathcal{P}$ defines a mapping $h:\mathcal{I} \rightarrow
\mathcal{Q}$.

To prove that the mapping $h$ defines a space-filling cuve, we can use the
same proof as in Theorem 2.1 in \citet{sagen_hans} or Section 2.3.3 in
\citet{bader} only with very small adjustment. The two mentioned proofs are
on the Hilbert curve which is initialized from a single base. In there, $\mathcal{I}$ is
recursively partitioned into four subintervals, and corresponding
$\mathcal{Q}$ is recursively partitioned into four subsquares. A point
$p$ on $\mathcal{I}$ can be uniquely determined by a sequence of nested
intervals and its mapping $h(p)$ on $\mathcal{Q}$ is also uniquely determined
by a sequence of nested subsquares.

There are the following two additional notes when adjusting to prove this proposition: 

\begin{enumerate}
\tightlist
\item
  The original proof is applied to a curve initialized from a single base.
  However, to extend it to the curve initialized from a seed sequence
  $\mathcal{S}$, we only need to add a pre-partitioning step where
  $\mathcal{I}$ is first partitioned into $n$ unit-intervals and $\mathcal{Q}$
  is pre-partitioned into $n$ unit-squares, where the interval and subsquare that $p$ is located on
  are inserted before the sequences of nested intervals and nested subsquares.
  This won't affect the use of the two nested sequences when the iteration reaches infinity.
\item
  Assume $p$ is located in the interval $I^{(i-1)}$ on level $i-1$ from nested partitioning on $\mathcal{I}$,
  and in square $Q^{(i-1)}$ on the corresponding level $i-1$ nested partitioning on $\mathcal{Q}$.
  Partition $I^{(i-1)}$ into four subintervals where one of them contains $p$. $Q^{(i-1)}$
  is also partitioned into four subsquares. Different selections of the expansion code $\pi_i$
  only affects how the four subsquares are arranged in $Q^{(i-1)}$ which will not affect
  that the fact that one of them contains $h(p)$.
\end{enumerate}

Now we have the same conditions as the original proofs.
Then $h:\mathcal{I} \rightarrow \mathcal{Q}$ is a surjective mapping and
$\mathcal{P}$ is a continous curve, thus $\mathcal{P}$ is a space-filling
curve.

\end{proof}

\section{The expansion code sequence}\label{the-expansion-code-sequence}

\subsection{Combinations}\label{combinations}

Let's go back to the notation of a curve on level \(k\) initialized by a
single base \(X\):

\begin{equation}
\mathcal{P}_k = X^{(0)}|\pi_1 \pi_2 ... \pi_k .
\notag
\end{equation}

The encoding represents the curve started from the initial seed
\(X^{(0)}\) (associated with a certain rotation) and expanded for \(k\)
times. We can merge \(X^{(0)}\) and the first expansion to form a new
initial seed sequence, and later expand the curve for \(k-1\) times,
written as:

\begin{equation}
\begin{aligned}
\mathcal{P}_k &= \left(X^{(0)}|\pi_1 \right) | \pi_2 ... \pi_k \\
              &=  X^{(1)}_{<\pi_1>} | \pi_2 ... \pi_k = \mathcal{P}_1| \pi_2 ... \pi_k \\
\end{aligned} .
\notag
\end{equation}

Note here \(X^{(1)}_{<\pi_1>}\) is a sequence of four bases. Similarlly,
we can move any amount of \(\pi_i\) to the left side of \(|\):

\begin{equation}
\label{eq:comb}
\begin{aligned}
\mathcal{P}_k &= \left(X^{(0)}|\pi_1 ... \pi_i \right) | \pi_{i+1} ... \pi_k \\
              &=  X^{(i)}_{<\pi_1 ... \pi_i>} | \pi_{i+1} ... \pi_k = \mathcal{P}_i | \pi_{i+1} ... \pi_k \\
\end{aligned}
\notag
\end{equation}

\noindent where \(X^{(i)}_{<\pi_1 ... \pi_i>}\) represents a level-\(i\)
curve expanded via the code sequence
\(\pi_1 ... \pi_i\)\footnote{If there is no ambiguity,
$X^{(i)}_{<\pi_1 ... \pi_i>}$ can be simply written as $X_{<\pi_1 ...\pi_i>}$.}.
The equation means a curve on level \(k\) can be generated from a
level-\(i\) curve as the seed by expanding \(k-i\) times.

We can expand the curve level-by-level where on each level, a new curve
is generated and used as the seed for the next-level expansion:

\begin{equation}
\mathcal{P}_k = \left( \left(\left(X^{(0)}|\pi_1 \right)| \pi_2 \right) | ...\right)| \pi_k ,
\notag
\end{equation}

\noindent which can be simply written as:

\begin{equation}
\mathcal{P}_k = X^{(0)}|\pi_1 | \pi_2 | ... | \pi_k .
\notag
\end{equation}

These combinations are the same if using a seed sequence
\(\mathcal{S}\).

\subsection{Expansion code from the second
base}\label{expansion-code-from-the-second-base}

In the form in Remark \ref{remark:curve_sequence}, if the seed sequence
is expanded for \(k\) iterations, i.e., to \(\mathcal{P}_k =
\mathcal{S}|\pi_1...\pi_k\), every code \(\pi_i\) in the code sequence
always corresponds to the first base of the curve on the previous level
\(\mathcal{P}_{i-1}\) which is eventually expanded from \(X_1\). In this
section, we study the form of the expansion code sequence from the
second base in \(\mathcal{S}\). Notice a curve \(\mathcal{P}_k\) can be
expressed as a curve (or a sequence) induced by \(\mathcal{P}_i\) and
expanded for \(k-i\) times (Section \ref{combinations}), thus the
analysis in this section helps to study the expansion of any base from
any level in the curve generation.

\subsubsection{One expansion}\label{one-expansion}

\paragraph{Two bases}\label{two-bases}

We first consider the following simplest form where the seed sequence
only includes two bases and expanded in one iteration:

\begin{equation}
X_1^{\theta_1} X_2^{\theta_2} | \pi = X_{<\pi>,1}^{\theta_1} X_{<\pi_*>,2}^{\theta_2}
\notag
\end{equation}

\noindent where
\(\pi_*\)\footnote{In this article, if a symbol is associated with an
asterisk, it means the symbol represents a variable whose value is going to be
solved, or just a wildcard symbol whose exact value is not of interest.}
is the expansion code for the second base \(X_2\). It can be easily
calculated based on Equation \ref{eq:code_xy}.

\begin{equation}
\pi_* = \begin{cases}
\pi \quad & \text{if }\theta_2 - \theta_1 \bmod 180 = 0\\ 
\hat{\pi} \quad & \text{if } \theta_2 - \theta_1 \bmod 180 = 90\\ 
\end{cases}
\notag
\end{equation}

The equation implies that the value of \(\pi_*\) depends on the value of
\(\theta_2 - \theta_1\). To simplify the description in this section, we
use a helper function \(s()\) to denote the solution for \(\pi_*\).
\(s()\) returns the original code sequence or its complement:

\begin{equation}
s(\pi_1...\pi_i|\theta_2 - \theta_1) = \begin{cases}
\pi_1...\pi_i \quad & \text{if }\theta_2 - \theta_1 \bmod 180 = 0\\ 
\hat{\pi}_1...\hat{\pi}_i \quad & \text{if } \theta_2 - \theta_1 \bmod 180 = 90\\ 
\end{cases} .
\notag
\end{equation}

\noindent Then we can write the solution of \(\pi_*\) as:

\begin{equation}
\label{eq:code21}
\pi_* = s(\pi|\theta_2 - \theta_1) .
\end{equation}

Straightforwardly from the definition of \(s()\), we have the following
three attributes for \(s()\).

\begin{remark}
\label{remark:s}
If two code sequences have the same condition in $s()$, i.e.,

\begin{equation}
\begin{aligned}
\pi_{*,1}...\pi_{*,i} &= s(\pi_1...\pi_i|\theta_2 - \theta_1) \\
\pi_{*,j}...\pi_{*,k} &= s(\pi_j...\pi_k|\theta_2 - \theta_1) \\
\end{aligned}
\notag
\end{equation}

\noindent where, e.g., $\pi_{*,i}$ represents a variable for the $i$-th code that is going to be
solved, then they can be concatenated to:

\begin{equation}
\begin{aligned}
\pi_{*,1}...\pi_{*,i}\pi_{*,j}...\pi_{*,k} &= s(\pi_1...\pi_i|\theta_2 - \theta_1)s(\pi_j...\pi_k|\theta_2 - \theta_1)\\
     &= s(\pi_1...\pi_i\pi_j...\pi_k|\theta_2 - \theta_1) \\
\end{aligned} .
\notag
\end{equation}

\end{remark}

\begin{remark}
\label{remark:s2}

\begin{equation}
s(\pi_1...\pi_k|\theta_2 - \theta_1 + \alpha) = \begin{cases}
s(\pi_1...\pi_k|\theta_2 - \theta_1) & \quad \textnormal{if }\alpha \bmod 180 = 0 \\
s(\hat{\pi}_1...\hat{\pi}_k|\theta_2 - \theta_1) & \quad \textnormal{if }\alpha \bmod 180 = 90 \\
\end{cases}
\notag
\end{equation}

\end{remark}

\begin{remark}
\label{remark:s3}

\begin{equation}
s(s(\pi_1...\pi_k|\theta_1)|\theta_2) = s(\pi_1...\pi_k|\theta_1 + \theta_2)
\notag
\end{equation}
\end{remark}

\paragraph{\texorpdfstring{\(n\) bases}{n bases}}\label{n-bases}

Next we extend the seed sequence to \(n\) bases and prove the following
lemma:

\begin{lemma}
\label{lemma:code1}
For a seed sequence of $n$ bases ($n \ge 2$) after one expansion with the code $\pi$,
\begin{equation}
X_1^{\theta_1} X_2^{\theta_2}... X_n^{\theta_n} | \pi = X_{<\pi>,1}^{\theta_1} X_{<\pi_{*,2}>,2}^{\theta_2} ... X_{<\pi_{*,n}>,n}^{\theta_n}
\notag
\end{equation}

\noindent where $\pi_{*,i}$ is the expansion code of the $i$-th base,
the solution is

\begin{equation}
\pi_{*,i} = s(\pi|\theta_i - \theta_1) \quad 2 \le i \le n .
\notag
\end{equation}

\end{lemma}

\begin{proof}

The scenario of $n = 2$ has already been proven in Equation \ref{eq:code21}.
For the scenario of $n \ge 3$, we first consider the first three bases. With
Equation \ref{eq:code21}, we can calculate $\pi_{*,2}$ and $\pi_{*,3}$ from
their respective preceding bases as:

\begin{equation}
\begin{aligned}
\pi_{*,2} &= s(\pi|\theta_2 - \theta_1) \\
\pi_{*,3} &= s(\pi_{*,2}|\theta_3 - \theta_2) \\
\end{aligned} .
\notag
\end{equation}

Table \ref{tab:code_expansion} enumerates all combinations of $\theta_2 -
\theta_1 \bmod 180$ and $\theta_3 - \theta_2 \bmod 180$. Values in the column
``$\pi_{*,2}$'' are directly from the definition of $s()$. Values in the
column ``$\pi_{*,3}$'' are based on the definition of $s()$ and the
values of $\pi_{*,2}$. By merging the last two columns in Table
\ref{tab:code_expansion}, we can have the solution for $\pi_{*,3}$:

\begin{equation}
\pi_{*,3} = s(\pi|\theta_3 - \theta_1) .
\notag
\end{equation}

\begin{table}
\centering
\begin{tabular}{rr|rr|r}
\toprule
 $\theta_{2}-\theta_{1} \bmod 180$ & $\pi_{*,2}$ & $\theta_{3}-\theta_{2} \bmod 180$ & $\pi_{*,3}$ & $\theta_{3}-\theta_{1} \bmod 180$  \\
\midrule
0 & $\pi$ & 0 & $\pi$ & 0 \\
90 & $\hat{\pi}$ & 0 & $\hat{\pi}$ & 90 \\
0 & $\pi$ & 90 & $\hat{\pi}$ & 90 \\
90 & $\hat{\pi}$ & 90 & $\pi$ & 0 \\
\bottomrule
\end{tabular}
\vspace*{5mm}
\caption{Calculate $\pi_{*,3}$.}
\label{tab:code_expansion}
\end{table}

By applying the same strategy repeatedly, we can extend it to any $i$ ($i \ge 3$):

\begin{equation}
\begin{aligned}
\pi_{*,i-1} &= s(\pi|\theta_{i-1} - \theta_1) \\
\pi_{*,i} &= s(\pi_{*,i-1}|\theta_i - \theta_{i-1}) \\
\end{aligned}
\notag
\end{equation}

\noindent to have the general form:

\begin{equation}
\pi_{*,i} = s(\pi|\theta_i - \theta_1) .
\notag
\end{equation}

\end{proof}

Compared to Equation \ref{eq:code_xy} where the expansion code of
\(X_i\) is calculated from \(X_{i-1}\), here the expansion code is
directly calculated from \(X_1\).

\subsubsection{\texorpdfstring{\(k\)
expansions}{k expansions}}\label{k-expansions}

Next we consider the general form. For a seed sequence of \(n\) bases
(\(n \ge
2\)) after \(k\) (\(k \ge 1\)) expansions with code sequence
\(\pi_1 ...\pi_k =
(\pi)_k\)\footnote{In this article, we always use $(\pi)_k$ to represent a
sequence of code where individual values of code are independently assigned.
This notation is only for the case when a Greek letter is used as the symbol.
If all the code in the sequence have the same value, we use the notation
$(a)_k$ or $(b)_k$.},

\begin{equation}
\label{eq:code_full}
\mathcal{P}_k = X_1^{\theta_1} X_2^{\theta_2}... X_n^{\theta_n} | (\pi)_k = X_{<(\pi)_k>,1}^{\theta_1} X_{<(\pi_{*,2})_k>,2}^{\theta_2} ... X_{<(\pi_{*,n})_k>,n}^{\theta_n} ,
\end{equation}

\noindent we want to find the solution of \((\pi_{*,i})_k =
\pi_{1*,i}...\pi_{k*,i}\) for \(2 \le i \le n\).

\paragraph{Two expansion code}\label{two-expansion-code}

We first consider the scenario of \(k = 2\). A level-2 curve can be
treated as a curve after one expansion taking the level-1 curve as the
seed sequence:

\begin{equation}
\label{eq:5.2.2-9}
\begin{aligned}
X_1^{\theta_1} ... X_i^{\theta_i}... | \pi_1 \pi_2 &= \left( X_1^{\theta_1} ... X_i^{\theta_i}... | \pi_1 \right)| \pi_2 \\
 &= X^{\theta_1}_{<\pi_1>,1}...X^{\theta_i}_{<\pi_{1*,i}>,i}...|\pi_2 \\
\end{aligned}
\end{equation}

\noindent where \(\pi_{1*,i}\) is the expansion code for \(X_i\) from
the first expansion, which can be directly calculated by Lemma
\ref{lemma:code1}:

\begin{equation}
\label{eq:5.2.2-10}
\pi_{1*,i} = s(\pi_1|\theta_i - \theta_1) .
\end{equation}

Next we continue to expand the curve to level 2. Starting from the
second line in Equation \ref{eq:5.2.2-9}, there is

\begin{equation}
\label{eq:kk2}
\begin{aligned}
\mathcal{P}_1|\pi_2 &= X^{\theta_1}_{<\pi_1>,1}...X^{\theta_i}_{<\pi_{1*,i}>,i}...|\pi_2 \\
  &=\left(X^{\theta_1}_{<\pi_1>,1}|\pi_2 \right)... \left(X^{\theta_i}_{<\pi_{1*,i}>,i}|\pi_{2*,i}\right) ... \\
  &= X^{\theta_1}_{<\pi_1 \pi_2>,1} ... X^{\theta_i}_{<\pi_{1*,i}\pi_{2*,i}>,i} ...\\
\end{aligned} .
\end{equation}

Notice \(\pi_2\) is the expansion code of the first base in
\(\mathcal{P}_1\), and \(\pi_{2*,i}\) is the expansion code of the first
base in \(X^{\theta_i}_{<\pi_{1*,i}>,i}\). To calculate \(\pi_{2*,i}\)
with Lemma \ref{lemma:code1}, we additionally need the rotation of the
first base in \(\mathcal{P}_1\) and the rotation of the first base in
\(X^{\theta_i}_{<\pi_{1*,i}>,i}\).

Denote the first base in \(\mathcal{P}_1\) as \(X_{11}^{\theta_{11}}\).
There is
\(\theta_{11} = \theta_s(\mathcal{P}_1) = \theta_s(X^{\theta_1}_{<\pi_1>,1})\).
According to Corollary \ref{coro:first_rotation},
\(\theta_s(X^{\theta_1}_{<\pi_1>,1}) = \theta_1 + \alpha_1\) where
\(\alpha_1\) is the rotation of the first base in
\(X^{(1),0}_{<\pi_1>,1}\). We have
\(\theta_{11} = \theta_1 + \alpha_1\).

Denote the first base in \(X^{\theta_i}_{<\pi_{1*,i}>,i}\) as
\(X_{i1}^{\theta_{i1}}\). There is also \(\theta_{i1} =
\theta_s(X^{\theta_i}_{<\pi_{1*,i}>,i}) = \theta_i + \alpha_i\) where
\(\alpha_i\) is the rotation of the first base in
\(X^{(1),0}_{<\pi_{1*,i}>,i}\).

Now with Lemma \ref{lemma:code1}, the expansion code of \(X_{i1}\) on
\(\mathcal{P}_1\) is:

\begin{equation}
\label{eq:pi_b}
\pi_{2*,i} = s(\pi_2|\theta_{i1} - \theta_{11}) = s(\pi_2|\theta_{i}+\alpha_i - \theta_1 - \alpha_1) .
\end{equation}

Together with Equation \ref{eq:5.2.2-10} and \ref{eq:pi_b}, we have the
solution of \(\pi_{1*,i}\pi_{2*,i}\). We can first use Equation
\ref{eq:5.2.2-10} to calculate \(\pi_{1*,i}\), then we know the form of
\(X^{(1),0}_{<\pi_{1*,i}>,i}\) and in turn we can know the value of
\(\alpha_i\). Finally, we apply Equation \ref{eq:pi_b} to obtain the
solution for \(\pi_{2*,i}\).

\paragraph{\texorpdfstring{\(k\) expansion
code}{k expansion code}}\label{k-expansion-code}

Now for the general scenario of \(k \ge 3\), we write \(\mathcal{P}_k\)
as a one-level expansion from \(\mathcal{P}_{k-1}\).

\begin{equation}
\label{eq:kcode}
\begin{aligned}
X_1^{\theta_1} ... X_i^{\theta_i}... | (\pi)_k &= \left( X_1^{\theta_1} ... X_i^{\theta_i}... | (\pi)_{k-1} \right)| \pi_k \\
 &= X^{\theta_1}_{<(\pi)_{k-1}>,1}...X^{\theta_i}_{<(\pi_{*,i})_{k-1}>,i}...|\pi_k \\
 &= \left(X^{\theta_1}_{<(\pi)_{k-1}>,1}|\pi_k\right)...\left(X^{\theta_i}_{<(\pi_{*,i})_{k-1}>,i}|\pi_{k*,i}\right)... \\
 &= X^{\theta_1}_{<(\pi)_{k}>,1}...X^{\theta_i}_{<(\pi_{*,i})_{k-1}\pi_{k*,i}>,i}... \\
\end{aligned}
\notag
\end{equation}

Similarly, to solve \(\pi_{k*,i}\), we need the rotation of the first
base in \(X^{\theta_1}_{<(\pi)_{k-1}>,1}\) and the rotation of the first
base in \(X^{\theta_i}_{<(\pi_{*,i})_{k-1}>,i}\). We still denote these
two rotations as \(\theta_{11}\) and \(\theta_{i1}\) for convenience.
Then \(\theta_{11} =
\theta_s(X^{\theta_1}_{<(\pi)_{k-1}>,1})\), and \(\theta_{i1} =
\theta_s(X^{\theta_i}_{<(\pi_{*,i})_{k-1}>,i})\). According to Corollary
\ref{coro:first_rotation}, there are:

\begin{equation}
\begin{aligned}
\theta_{11} = \theta_s(X^{\theta_1}_{<(\pi)_{k-1}>,1}) &= \theta_s(X^{\theta_1}_{<\pi_1>,1}) = \theta_1 + \alpha_1 \\
\theta_{i1} = \theta_s(X^{\theta_i}_{<(\pi_{*,i})_{k-1}>,i}) &= \theta_s(X^{\theta_i}_{<\pi_{1*,i}>,i}) = \theta_i + \alpha_i \\
\end{aligned}
\notag
\end{equation}

\noindent where \(\alpha_1\) and \(\alpha_i\) have the same meaning as
in Equation \ref{eq:pi_b}. Then we can obtain the solution of
\(\pi_{k*,i}\) as:

\begin{equation}
\label{eq:pi_all}
\pi_{k*,i} = s(\pi_k|\theta_{i1} - \theta_{11}) = s(\pi_k|\theta_{i}+\alpha_i - \theta_1 - \alpha_1) .
\end{equation}

The complete solution for \((\pi_{*,i})_k\) is in the next proposition.

\begin{proposition}
\label{prop:code_sequence}
The solution of the expansion code sequence in Equation \ref{eq:code_full} is
split into two parts:

\begin{equation}
(\pi_{*,i})_k = (\pi_{1*,i})(\pi_{2*,i}...\pi_{k*,i}) ,
\notag
\end{equation}

\noindent and the solution for each part is:

\begin{equation}
\label{eq:pi_general}
\begin{aligned}
\pi_{1*,i} &= s(\pi_1|\theta_i - \theta_1) \\
\pi_{2*,i}...\pi_{k*,i} &= s(\pi_2...\pi_k|\theta_{i}+\alpha_i - \theta_1 - \alpha_1) \\
\end{aligned}
\end{equation}

\noindent where $\alpha_1 = \theta_s(X^{(1),0}_{<\pi_1>,1})$ and $\alpha_i =
\theta_s(X^{(1),0}_{<\pi_{1*,i}>,i})$.

\end{proposition}

\begin{proof}
The solution for $\pi_{1*,i}$ is in Lemma \ref{lemma:code1}, solution for
$\pi_{2*,i}$ is in Equation \ref{eq:pi_b}, and solution for $\pi_{k*,i}$ ($k
\ge 3$) can be obtained by repeatedly applying Equation \ref{eq:pi_all}.

$\alpha_i$ only depends on $\pi_{1*,i}$ (i.e., the code for $X_i$ from the
first expansion), thus it is a constant when calculating each of $\pi_{2*,i},
..., \pi_{k*,i}$. Then in the following expansion code sequence, conditions in
all $s()$ are the same.

\begin{equation}
\pi_{2*,i}...\pi_{k*,i} = s(\pi_2|\theta_{i}+\alpha_i - \theta_1 - \alpha_1)...s(\pi_k|\theta_{i}+\alpha_i - \theta_1 - \alpha_1)
\notag
\end{equation}

According to Remark \ref{remark:s}, all $s()$ can be merged to:

\begin{equation}
\pi_{2*,i}...\pi_{k*,i} = s(\pi_2...\pi_k|\theta_{i}+\alpha_i - \theta_1 - \alpha_1)
\notag
\end{equation}

\end{proof}

\(\mathcal{S}\) is a sequence with at least two bases. \(X_1\) is the
first base in \(\mathcal{S}\), thus \(X_1 \in \{I, R, L, U, B, D\}\).
\(X_i\) (\(i \ge 2\)) is the base from the second one in
\(\mathcal{S}\), thus \(X_i \in \{I, R, L\}\) if \(X_i\) is also not the
last base. By enumerating all base types for \(X_1\) and \(X_i\), we can
simplify the solution in Proposition \ref{prop:code_sequence} to:

\begin{corollary}
\label{coro:5.6.1}
\begin{equation}
(\pi_{*,i})_k = \begin{cases}
s(\pi_1...\pi_k|\theta_i - \theta_1) & \quad \textnormal{if } X_1 \in \{I, R, L, U, B\} \\
s(\pi_1|\theta_i - \theta_1)s(\hat{\pi}_2...\hat{\pi}_k|\theta_i - \theta_1) & \quad \textnormal{if } X_1 = D \\
\end{cases}
\notag
\end{equation}

\end{corollary}

\begin{proof}
According to Figure \ref{fig:expansion_rule}, for all possible bases of $X_i
\in \{I, R, L\}$, rotation of their first bases in its level-1 expansions are
all zero. Thus, it is always $\alpha_i = 0$.

Bases of $X_1$ can be put into three groups:

\begin{enumerate}
\tightlist
\item
   When $X_1 \in \{I, R, L, U\}$, the first bases in their level-1 expansions
   all have rotations of zero, i.e., $\alpha_1 = 0$. This results in that 
   the condition in $\pi_{2*,i}...\pi_{k*,i} = s(\pi_2...\pi_k|\theta_i - \theta_1)$
   is the same as $\pi_{1*,i}$.
   According to Remark \ref{remark:s}, $\pi_{1*,i}$ and
   $\pi_{2*,i}...\pi_{k*,i}$ can be concatenated into a single $s()$.
\item
   When $X_1 = B$, the first bases in its two level-1 expansions all have
   rotations of 180 degrees, i.e., $\alpha_1 = 180$. According to Remark
   \ref{remark:s2}, $s(\pi_2...\pi_k|\theta_i - \theta_1 - 180) =
   s(\pi_2...\pi_k|\theta_i - \theta_1)$. We can also concatenate $\pi_{1*,i}$
   and $\pi_{2*,i}...\pi_{k*,i}$ into a single $s()$.
\item
   When $X_1 = D$, the first base in its two level-1 expansions are 270 or 90,
   i.e., $\alpha_1 = 90$ or $270$. According to Remark \ref{remark:s2},
   $s(\pi_2...\pi_k|\theta_i - \theta_1 - \alpha_1) =
   s(\hat{\pi}_2...\hat{\pi}_k|\theta_i - \theta_1)$.
\end{enumerate}

\end{proof}

\paragraph{Solution that does not rely on
rotations}\label{solution-that-does-not-rely-on-rotations}

Proposition \ref{prop:3.5} shows a single expansion code of a base can
be directly inferred from the type of its preceding base in the
sequence, without considering its rotation. It can be extended to the
code sequence as well.

We first consider the expansion code for the second base. According to
Corollary \ref{coro:5.6.1},

\begin{equation}
(\pi_{*,2})_k = \begin{cases}
s(\pi_1...\pi_k|\theta_2 - \theta_1) & \quad \text{if } X_1 \in \{I, R, L, U, B\} \\
s(\pi_1|\theta_2 - \theta_1)s(\hat{\pi}_2...\hat{\pi}_k|\theta_2 - \theta_1) & \quad \text{if } X_1 = D \\
\end{cases} .
\notag
\end{equation}

With Equation \ref{eq:theta2}, we can have different values of
\(\theta_2 -
\theta_1\) for different base types. Then we directly get the value from
\(s()\) and we can have a new form of solution for \((\pi_{*,2})_k\)
without rotations:

\begin{equation}
\label{eq:511}
(\pi_{*,2})_k = \begin{cases}
\pi_1\pi_2...\pi_k & \quad \text{if } X_1 \in \{I, U, B\} \\
\hat{\pi}_1\hat{\pi}_2...\hat{\pi}_k & \quad \text{if } X_1 \in \{R, L\} \\
\pi_1\hat{\pi}_2...\hat{\pi}_k & \quad \text{if } X_1 = D \\
\end{cases} .
\end{equation}

Next we consider two neighbouring bases
\(X_{i-1}^{\theta_{i-1}}X_{i}^{\theta_i}\) (\(i \ge 3\)) from the third
base in the sequence. Notice \(X_{i-1}\) is a base in the middle of a
sequence (since \(i-1
\ge 2\)), thus it can only be one of \(I\), \(R\) and \(L\). If we treat
\(X_{i-1}^{\theta_{i-1}}X_{i}^{\theta_i}\) as a sequence of two bases
and \((\pi_{*,i-1})_k\) is the code sequence of \(X_{i-1}\), then with
Corollary \ref{coro:5.6.1}, the code sequence for \(X_i\) is:

\begin{equation}
(\pi_{*,i})_k = s\left( (\pi_{*,i-1})_k | \theta_{i} - \theta_{i-1} \right) .
\notag
\end{equation}

With Equation \ref{eq:theta2}, if \(X_{i-1} = I\),
\(\theta_i - \theta_{i-1} =
0\); if \(X_{i-1} \in \{R, L\}\), \(\theta_i - \theta_{i-1} = \pm 90\),
then

\begin{equation}
\label{eq:512}
(\pi_{*,i})_k = \begin{cases}
(\pi_{*,i-1})_k & \quad \text{if } X_{i-1} = I \\
(\hat{\pi}_{*,i-1})_k & \quad \text{if } X_{i-1} \in \{R, L\} \\
\end{cases} .
\end{equation}

Let's summarize it into the following corollary:

\begin{corollary}
\label{coro:5.6.2}
We omit the rotations in the Equation \ref{eq:code_full} for simplicity. For
the following curve

\begin{equation}
\label{eq:code_full2}
\mathcal{P}_k = X_1 X_2 ... X_n| (\pi)_k = X_{<(\pi)_k>,1} X_{<(\pi_{*,2})_k>,2} ... X_{<(\pi_{*,n})_k>,n} \quad n \ge 2 ,
\notag
\end{equation}

\noindent the expansion code sequence of the $i$-th ($i \ge 2$) base is determined by its preceding
base. When $i = 2$, the solution is in Equation \ref{eq:511}, and when $i \ge
3$, the solution is in Equation \ref{eq:512}.

\end{corollary}

\subsection{Global structure and local
unit}\label{global-structure-and-local-unit}

A curve on level \(k\) can be written as:

\begin{equation}
\mathcal{P}_k = (\mathcal{S}|\pi_1...\pi_i)|\pi_{i+1}...\pi_k .
\notag
\end{equation}

\noindent This implies the curve can be treated as taking
\(\mathcal{P}_i =
\mathcal{S}|\pi_1...\pi_i\) as the seed and expanded for \(k-i\) times.
According to the process of the expansion mode of curve generation, the
seed sequence determines the global structure of the final curve. In
other words, the expansion of each base is only performed on the curve
of \(\mathcal{P}_i\). Thus, the expansion code sequence
\(\pi_1 ... \pi_i\) determines the global structure on level \(i\) of
the curve.

\begin{figure}
\centering{
\includegraphics[width=1\linewidth]{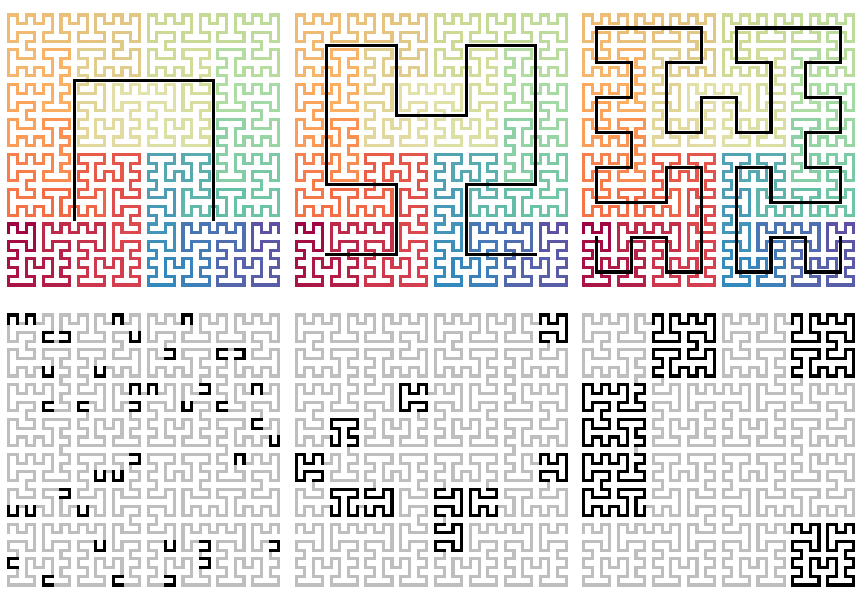}
\caption{Global structures and local units of $I^{270}|22112$. Top: Global structures on level 1 ($I^{270}|2$), level 2 ($I^{270}|22$) and level 3 ($I^{270}|221$); Bottom: Local units on the lowest 1 level (2), 2 levels (12), and 3 levels (112). 40, 10 and 5 random units are highlighted in black.}\label{fig:structure}}
\end{figure}

\begin{figure}
\centering{
\includegraphics[width=1\linewidth]{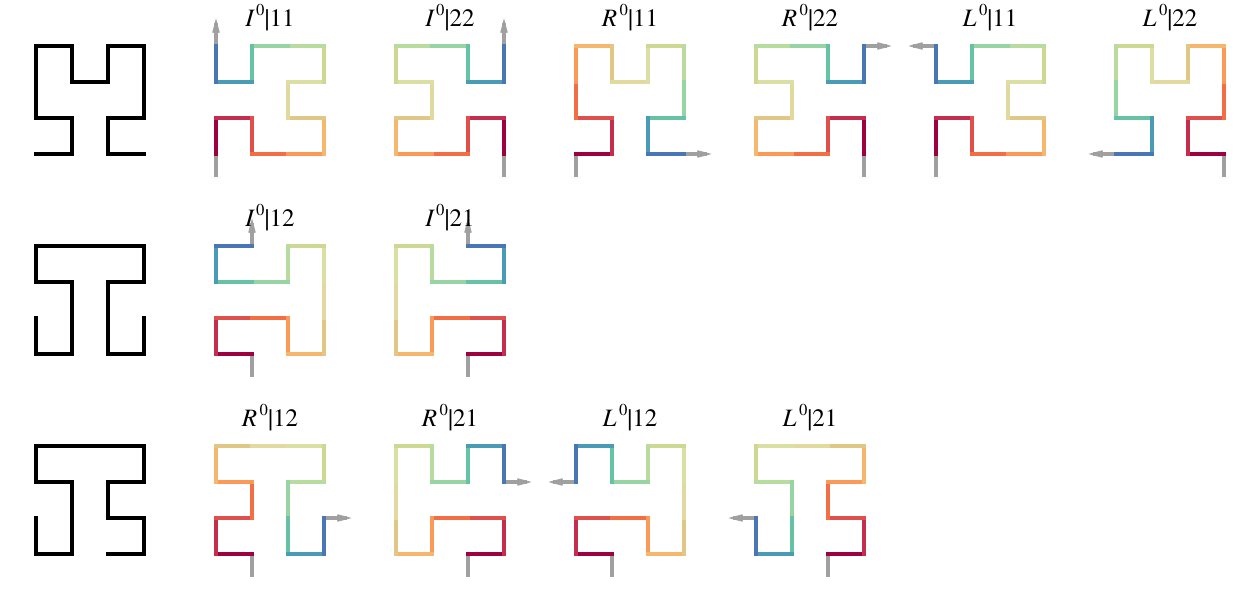}
\caption{All forms of level-2 units on their base rotations. These units are on curves with levels $\ge$ 3. The first curve on each row represents the common shape of units in the corresponding row.}\label{fig:level_2_units}}
\end{figure}

The level-\(i\) seed \(\mathcal{P}_i\) is basically a sequence of bases
denoted as \(\mathcal{P}_i = X_s ... X_e\) of \(4^i \times n\) bases
where \(n\) is the number of bases in \(\mathcal{S}\). We can express
\(\mathcal{P}_k\) as:

\begin{equation}
\begin{aligned}
\mathcal{P}_k &= \mathcal{P}_i|\pi_{i+1}...\pi_k \\
    &= X_s ... X_e|\pi_{i+1}...\pi_k \\
    &= X_{<\pi_{i+1}...\pi_{k}>,s} ... X_{<\pi_{i+1*}...\pi_{k*}>,e} \\
    &= (X_s|\pi_{i+1}...\pi_k) ... (X_e|\pi_{i+1*}...\pi_{k*}) \\
\end{aligned} .
\notag
\end{equation}

This implies the curve is composed a list of \(4^i \times n\) units on
level \(k-i\), where each unit is generated from a base on
\(\mathcal{P}_i\). The expansion code sequence from the second base in
\(\mathcal{S}\) can be calculated according to Corollary
\ref{coro:5.6.1} or \ref{coro:5.6.2} and the value depends on
\(\pi_{i+1}...\pi_k\). Thus, we could say the code sequence
\(\pi_{i+1}...\pi_k\) determines the form of local units on the curve.
An example is in Figure \ref{fig:structure}.

In particular, the lowest level-2 units are the most identifiable on the
curve. For curves with level \(k \ge 3\) written as
\(\mathcal{P}_{k-2}|\pi_{k-1}\pi_k\), according to Proposition
\ref{prop:IRL}, all bases on \(\mathcal{P}_{k-2}\) only include \(I\),
\(R\) and \(L\). Thus all level-2 units are in the form of
\(Z|\pi_{k-1}\pi_k\), where \(Z \in
\{I,R,L\}\). Figure \ref{fig:level_2_units} lists all
\(3 \times 2^2 = 12\) forms of 2x2 units in their base rotation states.
They can be classified into two groups.

\begin{enumerate}
\tightlist
\item
   $\pi_{k-1} = \pi_k$, the first row in Figure \ref{fig:level_2_units}. This
   group contains six units on their base rotations: $I|11$, $I|22|$, $L|11$,
   $L|22$, $R|11$, $R|22$. By also considering the four rotations, there are
   24 different forms.
\item
   $\pi_{k-1} \ne \pi_k$, the second and third rows in Figure
   \ref{fig:level_2_units}. They can be further split into two subgroups:
   \begin{itemize}
   \tightlist
   \item
      $X = I$. This includes $I|12$, $I|21$. There are 8 different forms
      considering the four rotations.
    \item
      $X \in \{R,L\}$. This includes $L|12$, $L|21$, $R|12$, $R|21$. There are
      16 different forms considering the four rotations.
    \end{itemize}
\end{enumerate}

Thus, these \(24 + 8 + 16 = 48\) forms construct the complete set of
level-2 units in 2x2 curves with level \(\ge\) 3.

As shown in the first curve on each row in Figure
\ref{fig:level_2_units}, all units in the same group or subgroup have
the same shape where orientations of the units are ignored. We name
level-2 units in the first group the \textit{Hilbert units}, units in
the second group, first subgroup the \textit{$\Omega$-units} and the
units in the second group, second subgroup the
\textit{$\beta$-units}\footnote{Because these two types of units have shapes
of the letters $\Omega$ and $\beta$.}.

\begin{proposition}
\label{prop:unit}
For $\mathcal{P}_k$ ($k \ge 3$), its lowest level-2 units are all
Hilbert units iff the last two code are the same in the code sequence; or a
combination of the $\beta$-units and $\Omega$-units iff the last two code are
different in the code sequence.
\end{proposition}

\begin{proof}
$\mathcal{P}_k = \mathcal{P}_{k-2}|\pi_{k-1}\pi_k = X_s...X_e|\pi_{k-1}\pi_k$
and $\mathcal{P}_{k-2}$ only includes $I$, $R$ or $L$ (Proposition
\ref{prop:IRL}). According to Corollary \ref{coro:5.6.1}, for all level-2
units of $\mathcal{P}_k$ denoted as $X_i|\pi_{k-1*}\pi_{k*}$,
where $\pi_{k-1*}\pi_{k*} = \pi_{k-1}\pi_{k}$ or $\hat{\pi}_{k-1}\hat{\pi}_{k}$.
Then iff $\pi_{k-1} = \pi_k$, all level-2 units are Hilbert units; and iff
$\pi_{k-1} \ne \pi_k$, all level-2 units are $\beta$- or $\Omega$-units.
\end{proof}

By observing the code sequence of standard curves in Section
\ref{special-curves}, for all these curves on level \(\ge\) 3, the
Hilbert curve, the Moore curve and the four other Liu-variants are only
composed of Hilbert untis. The \(\beta\Omega\)-curve is composed of
\(\beta\)-units and \(\Omega\)-units. In Section \ref{def-hc}, we will
give definitions for the Hilbert curve and the \(\beta\Omega\)-curve
based on the Hilbert unit, the \(\beta\)-unit and the \(\Omega\)-unit.

\section{Transformation}\label{transformation}

In this section, we study the forms of the symbolic expressions of
curves after various types of transformations, including rotations,
reflections, reversals and their combinations.

\subsection{Transformation on a single
base}\label{transformation-on-a-single-base}

A base is a point together with an entry direction and an exit
direction. Transformations defined in this section are applied to the
three components simultaneously. Rotation on single base has already
been discussed in Section \ref{rotation}, here we only discuss
reflection and reversal.

\subsubsection{Horizontal reflection}\label{ref_single}

Based on the expansion rules in Figure \ref{fig:expansion_rule},
horizontal reflection denoted as \(h()\) of the primary base patterns is
calculated as:

\begin{equation}
\label{eq:ref}
\begin{matrix}
\begin{aligned}
h(I^\theta) &= I^{\theta+\alpha} \\
h(R^\theta) &= L^{\theta+\alpha} \\
h(L^\theta) &= R^{\theta+\alpha} \\
\end{aligned} &
\quad
\text{where }\alpha = \begin{cases}
0  & \text{if } \theta \bmod 180 = 0\\ 
180 & \text{if } \theta \bmod 180 = 90\\ 
\end{cases} .
\end{matrix}
\end{equation}

For the non-primary base patterns \(X \in \{U, B, D, P, Q, C\}\), there
is

\begin{equation}
h(X^\theta) = X^{\theta + \alpha} .
\notag
\end{equation}

Horizontally reflecting a sequence is identical to reflecting its
individual bases.

\begin{equation}
\label{eq:reflect_seq}
h(X_1 X_2 ... X_n) = h(X_1)h(X_2)...h(X_n)
\end{equation}

We rewrite \(\alpha\) as a function \(\alpha(\theta)\) since its value
depends on \(\theta\). \(\alpha()\) returns a value of 0 or 180. There
is the following arithmetic attribute on \(\alpha()\):

\begin{remark}
\label{remark:alpha}
\begin{equation}
\alpha(\theta_1 + \theta_2) = \begin{cases}
\alpha(\theta_1) & \textnormal{if } \theta_2 \bmod 180 = 0\\
\alpha(\theta_1) + 180 & \textnormal{if } \theta_2 \bmod 180 = 90\\
\end{cases}
\notag
\end{equation}
\end{remark}

\begin{proposition}
\label{prop:6.1}
The combination of rotation and horizontal reflection on a single base has the
following relation:

\begin{equation}
\label{eq:reflection1}
\left(h(X^{\theta_1})\right)^{\theta_2} = h(X^{\theta_1 + \theta_2 + \alpha(\theta_2)}) = \left( h(X) \right)^{\theta_1 + \theta_2 + \alpha(\theta_1)} .
\end{equation}

\end{proposition}

\begin{proof}

We first write the horizontal reflection as $h(X^\theta) = \hat{X}^{\theta +
\alpha(\theta)}$ where $\hat{X}$ is the corresponding reflected base type. The
exact base type of $\hat{X}$ is not used in this proof.
We expand the first two parts of Equation \ref{eq:reflection1} separately as:

\begin{equation}
\label{eq:h01}
\begin{aligned}
\left(h(X^{\theta_1})\right)^{\theta_2} &= (\hat{X}^{\theta_1 + \alpha(\theta_1)})^{\theta_2} \\
  &= \hat{X}^{\theta_1 + \alpha(\theta_1) + \theta_2} \\
\end{aligned} ,
\notag
\end{equation}

\noindent and

\begin{equation}
h(X^{\theta_1 + \theta_2 + \alpha(\theta_2)}) = \hat{X}^{\theta_1 + \theta_2 + \alpha(\theta_2) + \alpha(\theta_1 + \theta_2 + \alpha(\theta_2))} .
\notag
\end{equation}

Note $\alpha(\theta_2)$ is always 0 or 180, then with Remark \ref{remark:alpha},

\begin{equation}
\label{eq:h02}
h(X^{\theta_1 + \theta_2 + \alpha(\theta_2)}) = \hat{X}^{\theta_1 + \theta_2 + \alpha(\theta_2) + \alpha(\theta_1 + \theta_2)}
\end{equation}

Next we expand the third part in Equation \ref{eq:reflection1}:

\begin{equation}
\label{eq:h03}
\begin{aligned}
\left( h(X) \right)^{\theta_1 + \theta_2 + \alpha(\theta_1)} &= \hat{X}^{0+\alpha(0) + \theta_1 + \theta_2 + \alpha(\theta_1)}  \\
    &= \hat{X}^{\theta_1 + \theta_2 + \alpha(\theta_1)}
\end{aligned} .
\notag
\end{equation}

Let's further simplify Equation \ref{eq:h02}. If $\theta_2 \bmod 180 = 0$,
$\alpha(\theta_1 + \theta_2) = \alpha(\theta_1)$ (Remark \ref{remark:alpha})
and $\alpha(\theta_2) = 0$, then the right side of Equation \ref{eq:h02}
becomes $\hat{X}^{\theta_1 + \theta_2 + \alpha(\theta_1)}$. If $\theta_2 \bmod
180 = 90$, $\alpha(\theta_1 + \theta_2) = \alpha(\theta_1) + 180$ (Remark
\ref{remark:alpha}), and $\alpha(\theta_2) = 180$, then the right side of
Equation \ref{eq:h02} is

\begin{equation}
\begin{aligned}
\hat{X}^{\theta_1 + \theta_2 + \alpha(\theta_2) + \alpha(\theta_1 + \theta_2)} &= \hat{X}^{\theta_1 + \theta_2 + 180 + \alpha(\theta_1) + 180} \\
  &= \hat{X}^{\theta_1 + \theta_2 + \alpha(\theta_1)} \\
\end{aligned} .
\notag
\end{equation}

So the three parts of Equation \ref{eq:reflection1} are all identical.

\end{proof}

\subsubsection{Other types of
reflections}\label{other-types-of-reflections}

Vertical reflection denoted as \(v()\) can be simply constructed by
first rotating the base by 180 degrees then by a horizontal reflection.

\begin{equation}
v(X^\theta) = h(X^{\theta+180})
\notag
\end{equation}

There are two types of diagonal reflections: the one against the
diagonal line with slop of 1 (lower left to upper right) denoted as
\(d^1()\) and the other one with slop of \(-1\) (lower right to upper
left) denoted as \(d^{-1}()\). They can be constructed by rotating the
base by 90 or \(-90\) degrees then by a horizontal reflection.

\begin{equation}
\begin{aligned}
d^1(X^\theta) &= h(X^{\theta + 90})    \\
d^{-1}(X^\theta) &= h(X^{\theta - 90}) \\
\end{aligned}
\notag
\end{equation}

\noindent Note the entry and exit directions of the curve are also
adjusted after reflections.

We will not discuss vertical and diagonal reflections in this article
because they can be simply constructed by rotations and horizontal
reflections.

\subsubsection{Reversal}\label{reversal}

According to the patterns of bases in Figure \ref{fig:expansion_rule},
the reversals of the nine bases are listed as folllows. We denote the
reversal of a base \(X\) as \(X'\), then there are:

\begin{equation}
\label{eq:rev_base}
\begin{aligned}
I' &= I^{180} & \ \ \ \ \ \ \ \  B' &= P^{180} \\
R' &= L^{90} & D' &= Q^{180} \\
L' &= R^{-90} & P' &= B^{180} \\
U' &= U  & Q' &= D^{180}\\
  &   & C' &= C  \\
\end{aligned} .
\end{equation}

\noindent The relations for reversing \(B\), \(D\), \(P\) and \(Q\) are
based on their level-1 forms.

When a sequence is reversed, the order of its individual bases are also
reversed accordingly. The reversal on a sequence is denoted as \(r()\).

\begin{equation}
\label{eq:rev_seq}
r(X_1 X_2 ... X_n) = X'_n ... X'_2 X'_1
\end{equation}

\begin{proposition}
\label{prop:6.2}
The combination of rotation and reversal on $X$ has the following relation:

\begin{equation}
\label{eq:rev_x}
(X')^\theta = (X^\theta)' .
\end{equation}

\end{proposition}

\begin{proof}
The base pattern $X$ can be written as a two-tuple $X = (\varphi_s,
\varphi_e)$ where $\varphi_s$ is its entry direction and $\varphi_e$ is its
exit direction. Reversing $X$ switches the entry and exit direction and also
reverses the orientations of the two directions.

\begin{equation}
X' = (\varphi_e + 180, \varphi_s + 180)
\notag
\end{equation}

Entry and exit directions of a base have the same amount of rotation as the
base itself.

\begin{equation}
(X')^\theta = (\varphi_e + 180 + \theta, \varphi_s + 180 + \theta)
\notag
\end{equation}

We then expand the right side of Equation \ref{eq:rev_x}:

\begin{equation}
\begin{aligned}
X^\theta &= (\varphi_s + \theta, \varphi_e + \theta) \\
(X^\theta)' &= (\varphi_e + \theta + 180, \varphi_s + \theta + 180) \\ 
\end{aligned}
\notag
\end{equation}

\noindent which results in 

\begin{equation}
(X')^\theta = (X^\theta)' .
\notag
\end{equation}

\end{proof}

\begin{lemma}
\label{lemma:reflection_entry}
Write $X = (\varphi_s, \varphi_e)$ and $h(X) = (\varphi'_s, \varphi'_e)$, then 
$\varphi'_s = \varphi_s + \alpha(\varphi_s + 90)$ and $\varphi'_e = \varphi_e + \alpha(\varphi_e + 90)$,
where $\alpha()$ is defined in Section \ref{horizontal-reflection}.
\end{lemma}

\begin{proof}
If the entry or exit direction of $X$ is vertical (with a degree of 90 or
270), it is not changed after horizontal reflection, while if it is horizontal
(with a degree of 0 or 180), the direction is reversed (by a rotation of $\pm 180$)
after horizontal reflection. We can write as followings, taking $\varphi_s$ as an example:

\begin{equation}
\varphi'_s = \begin{cases}
\varphi_s \quad & \textnormal{if } \varphi_s \bmod 180 = 90 \\
\varphi_s + 180 \quad & \textnormal{if } \varphi_s \bmod 180 = 0 \\
\end{cases} ,
\notag
\end{equation}

\noindent and it is equivalent if using $\alpha()$:

\begin{equation}
\varphi'_s = \varphi_s + \alpha(\varphi_s + 90)
\notag
\end{equation}

The calculation is the same for $\varphi'_e$ and $\varphi_e$.
\end{proof}

\begin{proposition}
\label{prop:6.2-2}
The combination of horizontal reflection and reversal on $X$ has the following
relation:

\begin{equation}
\label{eq:rev_x2}
h(X') = h(X)' .
\end{equation}

\end{proposition}

\begin{proof}
We still denote $X = (\varphi_s, \varphi_e)$ and expand $h(X)$ as:

\begin{equation}
\begin{aligned}
h(X) &= h( (\varphi_s, \varphi_e) ) \\
     &= (\varphi_s + \alpha(\varphi_s + 90), \varphi_e + \alpha(\varphi_e + 90)) \\
     &= (\varphi_s + \alpha(\varphi_s) + 180, \varphi_e + \alpha(\varphi_e) + 180) \\
\end{aligned} .
\notag
\end{equation}

\noindent where Line 2 is based on Lemma \ref{lemma:reflection_entry} and Line
3 is based on Remark \ref{remark:alpha}. We next expand the two sides in Equation
\ref{eq:rev_x2}:

\begin{equation}
\begin{aligned}
h(X') &= h( (\varphi_e + 180, \varphi_s + 180) ) \\
      &= (\varphi_e + 180 + \alpha(\varphi_e + 180) + 180, \varphi_s + 180 + \alpha(\varphi_s + 180) + 180) \\
      &= (\varphi_e + 180 + \alpha(\varphi_e) + 180, \varphi_s + 180 + \alpha(\varphi_s) + 180)  \\
      &= (\varphi_e + \alpha(\varphi_e), \varphi_s + \alpha(\varphi_s))  \\
h(X)' &= (\varphi_s + \alpha(\varphi_s) + 180, \varphi_e + \alpha(\varphi_e) + 180)' \\
      &= (\varphi_e + \alpha(\varphi_e) + 180 + 180, \varphi_s + \alpha(\varphi_s) + 180 + 180) \\
      &= (\varphi_e + \alpha(\varphi_e), \varphi_s + \alpha(\varphi_s))  \\
\end{aligned}
\notag
\end{equation}

\noindent which results in $h(X') = h(X)'$.

\end{proof}

Propositions \ref{prop:6.2} and \ref{prop:6.2-2} imply reversal is
independent to the rotation or reflection on a base.

\subsubsection{Comments}\label{comments}

As shown in this section, a base type can be generated by reflection,
reversal, or their combinations from other base types. It seems the nine
base patterns as well as their level-1 expansions listed in Figure
\ref{fig:expansion_rule} are redundant. However, allowing more
transformations while restricting the amount of base patterns makes the
forms of the curves on higher level complex, which significantly
increases the difficulty of interpretation. For example, on level-1, the
relation of \(R^{(1)}_1\) and \(R^{(1)}_2\) can be written in a complex
form of \(R^{(1)}_2 =
r(h(R^{(1),270}_1))\). Thus, we only allow rotations when building the
expansion rules, which makes the theory compact and consistent.

\subsection{Transformation on the base sequence and
subsequences}\label{transformation-on-the-base-sequence-and-subsequences}

Based on the transformation on single base, we can extend it to a base
sequence. The single transformation on the sequence has already been
introduced in Section \ref{rotation} (rotation), Section
\ref{horizontal-reflection} (horizontal reflection) and Section
\ref{reversal} (reversal). In this section, we only discuss combinations
of transformations on the base sequence.

\begin{proposition}
\label{prop:trseq_1}
The combination of rotation and horizontal reflection on a sequence
$\mathcal{S} = X_1^{\theta_1}...X_n^{\theta_n}$ has the following relation:

\begin{equation}
(h(\mathcal{S}))^{\theta} = h(\mathcal{S}^{\theta + \alpha(\theta)})
\notag
\end{equation}

\noindent where $\alpha()$ is defined in Section \ref{horizontal-reflection}.

\end{proposition}

\begin{proof}

We expand the left side of the equation

\begin{equation}
\begin{aligned}
(h(\mathcal{S}))^\theta &= \left( h(X_1^{\theta_1} ... X_n^{\theta_n}) \right)^\theta \\
  &= \left( h(X_1^{\theta_1}) ... h(X_n^{\theta_n}) \right)^\theta \\
  &= (h(X_1^{\theta_1}))^\theta ... (h(X_n^{\theta_n}))^\theta \\
  &= h(X_1^{\theta_1 + \theta + \alpha(\theta)})...h(X_n^{\theta_n + \theta+ \alpha(\theta)}) \\
  &= h(X_1^{\theta_1 + \theta+ \alpha(\theta)} ... X_n^{\theta_n + \theta+ \alpha(\theta)}) \\
  &= h((X_1^{\theta_1} ... X_n^{\theta_n})^{\theta+\alpha(\theta)}) \\
  &= h(\mathcal{S}^{\theta + \alpha(\theta)})
\end{aligned} .
\notag
\end{equation}

Explanations are:

\begin{itemize}
\tightlist
\item
 Line 2: With Equation \ref{eq:reflect_seq}, reflecting the whole sequence is
  identical to reflecting individual bases.
\item
 Line 3: With Equation \ref{eq:rot_seq}, rotating the whole sequence is
  identical to rotating individual bases.
\item
  Line 4: We apply the transformation in Proposition \ref{prop:6.1}.
\item
  Line 5: With Equation \ref{eq:reflect_seq}, reflecting individual bases is
  identical to reflecting the complete sequence.
\item
  Line 6: With Equation \ref{eq:rot_seq}, if each base is rotated by the same
  amount, the rotation can be applied to the complete sequence directly.
\end{itemize}

\end{proof}

\begin{proposition}
\label{prop:trseq_2}
The combination of rotation and reversal on a sequence $\mathcal{S} =
X_1...X_n$ has the following relation:

\begin{equation}
(r(\mathcal{S}))^\theta = r(\mathcal{S}^\theta) .
\notag
\end{equation}

\end{proposition}

\begin{proof}

We expand the left side of the equation

\begin{equation}
\begin{aligned}
(r(\mathcal{S}))^\theta &= (r(X_1...X_n))^\theta \\
  &= (X'_n ... X'_1)^\theta \\
  &= (X'_n)^\theta ... (X'_1)^\theta \\
  &= (X_n^\theta)' ... (X_1^\theta)' \\
  &= r(X_1^\theta...X_n^\theta) \\
  &= r((X_1...X_n)^\theta) \\
  &= r(\mathcal{S}^\theta) \\
\end{aligned} .
\notag
\end{equation}

Explanations of the key steps are:

\begin{itemize}
\tightlist
\item
  Line 2: With Equation \ref{eq:rev_seq}, reversing the whole sequence is
  identical to reversing individual bases in a reversed order.
\item
  Line 3: With Equation \ref{eq:rot_seq}, rotating the whole sequence is
  identical to rotating individual bases.
\item
  Line 4: Apply Proposition \ref{prop:6.2} to switch rotation and reversal
  transformations on each base.
\end{itemize}

\end{proof}

\begin{proposition}
\label{prop:trseq_3}
The combination of horizontal reflection and reversal on a sequence
$\mathcal{S} = X_1...X_n$ has the following relation:

\begin{equation}
h(r(\mathcal{S})) = r(h(\mathcal{S})) .
\notag
\end{equation}

\end{proposition}

\begin{proof}
The proof is basically the same as for Proposition \ref{prop:trseq_2} except
Proposition \ref{prop:6.2-2} is used in Lines 3 and Line 4 instead.
\end{proof}

\begin{proposition}
\label{prop:subsequence}
A sequence $\mathcal{P}$ is composed of a list of subsequences, denoted as
$\mathcal{P}=\mathcal{S}_1 ... \mathcal{S}_w$ ($w \ge 1$), there are

\begin{equation}
\begin{aligned}
h(\mathcal{P}) &= h(\mathcal{S}_1 ... \mathcal{S}_w) = h(\mathcal{S}_1)...h(\mathcal{S}_w) \\
r(\mathcal{P}) &= r(\mathcal{S}_1 ... \mathcal{S}_w) = r(\mathcal{S}_w)...r(\mathcal{S}_1) \\
\end{aligned} .
\notag
\end{equation}

\end{proposition}

\begin{proof}
Write $\mathcal{S}_i$ as $X_{i,s}...X_{i,e}$, then with
Equation \ref{eq:reflect_seq}, there are:

\begin{equation}
\begin{aligned}
h(\mathcal{S}_1 ... \mathcal{S}_w) &= h(X_{1,s}...X_{1,e}... X_{w,s}...X_{w,e}) \\
      &= h(X_{1,s})...h(X_{1,e}) ... h(X_{w,s}) ...h(X_{w,e}) \\
h(\mathcal{S}_1)...h(\mathcal{S}_w) &= h(X_{1,s}...X_{1,e}) ... h(X_{w,s}...X_{w,e}) \\
      &= h(X_{1,s})...h(X_{1,e}) ... h(X_{w,s}) ...h(X_{w,e}) \\
\end{aligned} .
\notag
\end{equation}

Thus

\begin{equation}
h(\mathcal{S}_1 ... \mathcal{S}_w) = h(\mathcal{S}_1)...h(\mathcal{S}_w)
\notag
\end{equation}

For reversal, with Equation \ref{eq:rev_seq},

\begin{equation}
\begin{aligned}
r(\mathcal{S}_1 ... \mathcal{S}_w) &= r(X_{1,s}...X_{1,e}... X_{w,s}...X_{w,e}) \\
      &= X'_{w,e}...X'_{w,s}...X'_{1,e}...X'_{1,s}\\
r(\mathcal{S}_w)...r(\mathcal{S}_1) &= r(X_{w,s}...X_{w,e})...r(X_{1,s}...X_{1,e}) \\
      &= X'_{w,e}...X'_{w,s}...X'_{1,e}...X'_{1,s}\\
\end{aligned} .
\notag
\end{equation}

Thus

\begin{equation}
r(\mathcal{S}_1 ... \mathcal{S}_w) = r(\mathcal{S}_w)...r(\mathcal{S}_1) .
\notag
\end{equation}

\end{proof}

\subsection{Transformation on the
curve}\label{transformation-on-the-curve}

After rotation, reflection or reversal on the curve, it is still a 2x2
curve, thus, there must be a symbolic expression associated with it
after the transformation. In this section, we will explore the forms of
the symbolic expressions of 2x2 curves after various transformations. We
consider a general curve \(\mathcal{P}_k = X_1...X_n|\pi_1...\pi_k\) on
level \(k\).

\subsubsection{Rotation}\label{rotation-1}

We first consider the scenario where there is only one expansion on the
curve.

\begin{lemma}
\label{lemma:rot_single}
For a sequence $\mathcal{S} = X_1^{\theta_1}...X_n^{\theta_n}$, there is
$(\mathcal{S}|\pi)^\theta = \mathcal{S}^\theta|\pi$.

\end{lemma}

\begin{proof}

We expand the two sides of the equation.

\begin{equation}
\begin{aligned}
(\mathcal{S}|\pi)^\theta &= (X_1^{\theta_1}...X_n^{\theta_n}|\pi)^\theta \\
   &=(X_{<\pi>,1}^{\theta_1}...X_{<\pi_{*,i}>,i}^{\theta_i}...X_{<\pi_{*,n}>,n}^{\theta_n})^\theta \\
   &= X_{<\pi>}^{\theta_1+\theta} ...X_{<\pi_{*,i}>}^{\theta_i+\theta} ...X_{<\pi_{*,n}>}^{\theta_n+\theta}  \\
\end{aligned}
\notag
\end{equation}

\begin{equation}
\begin{aligned}
\mathcal{S}^\theta|\pi &= (X_1^{\theta_1}...X_n^{\theta_n})^\theta|\pi \\
   &= X_1^{\theta_1 +\theta}...X_n^{\theta_n+\theta}|\pi \\
   &= X_{<\pi>}^{\theta_1 + \theta} ...X_{<\pi'_{*,i}>}^{\theta_i + \theta} ...X_{<\pi'_{*,n}>}^{\theta_n + \theta}  \\
\end{aligned}
\notag
\end{equation}

$\pi_{*,i}$ and $\pi'_{*,i}$ can be calculated based on Lemma \ref{lemma:code1}:

\begin{equation}
\begin{aligned}
\pi_{*,i} &= s(\pi|\theta_i - \theta_1) \\
\pi'_{*,i} &= s(\pi|\theta_i + \theta - \theta_1 - \theta) \\
\end{aligned} .
\notag
\end{equation}

We can see for $2 \le i \le k$, it is always $\pi_{*,i} = \pi'_{*,i}$. Then it
is easy to see $(\mathcal{S}|\pi)^\theta = \mathcal{S}^\theta|\pi$.

\end{proof}

\begin{proposition}
\label{prop:6.3}
Rotating a curve only rotates its seed sequence while the expansion code is
not changed.

\begin{equation}
\mathcal{P}_k^\theta = (X_1...X_n|\pi_1 ... \pi_k)^\theta = (X_1 ... X_n)^\theta|\pi_1 ... \pi_k
\notag
\end{equation}

\end{proposition}

\begin{proof}

With Lemma \ref{lemma:rot_single},

\begin{equation}
\mathcal{P}_k^\theta = (\mathcal{P}_{k-1}|\pi_k)^\theta = \mathcal{P}_{k-1}^\theta|\pi_k .
\notag
\end{equation}

We can apply it recursively:

\begin{equation}
\begin{aligned}
\mathcal{P}_k^\theta &= \mathcal{P}_{k-1}^\theta|\pi_k \\
             &= \mathcal{P}_{k-2}^\theta|\pi_{k-1}|\pi_k \\
             &= ... \\
             &= \mathcal{P}_0^\theta|\pi_1|...|\pi_k \\
             &= (X_1...X_n)^\theta|\pi_1...\pi_k \\
\end{aligned} .
\notag
\end{equation}

\end{proof}

\begin{remark}
Proposition \ref{prop:6.3} implies that the global rotation of the complete
curve is controlled by the rotation of its initial seed sequence. Equation
\ref{eq:theta2} implies the rotations of bases in a seed sequence are in turn
only determined by the rotation of the first base. Thus the rotation of the
complete curve is merely determined by the first base in the seed sequence.
\end{remark}

\subsubsection{Horizontal reflection}\label{horizontal-reflection}

Similar as rotations, we first consider the scenario where there is only
one expansion on the curve.

\begin{lemma}
\label{lemma:reflect_single}
For a sequence $\mathcal{S} = X_1^{\theta_1}...X_n^{\theta_n}$, there is
$h(\mathcal{S}|\pi) = h(\mathcal{S})|\hat{\pi}$.
\end{lemma}

\begin{proof}

First, by enumerating all possible forms of level-1 units in Figure
\ref{fig:expansion_rule}, we have (note $h(X)$ is also a single base):

\begin{equation}
\label{eq:hpi}
h(X_{<\pi>}) = h(X)|\hat{\pi} = h(X)_{<\hat{\pi}>} .
\notag
\end{equation}

We expand the two sides of the equation in this lemma.

\begin{equation}
\begin{aligned}
h(\mathcal{S}|\pi) &= h(X_1...X_n|\pi) \\
  &= h(X_{<\pi>,1} ... X_{<\pi_{*,i}>,i} ... X_{<\pi_{*,n}>,n}) \\
  &= h(X_{<\pi>,1}) ... h(X_{<\pi_{*,i}>,i}) ... h(X_{<\pi_{*,n}>,n}) \\
  &= h(X_1)_{<\hat{\pi}>} ... h(X_i)_{<\hat{\pi}_{*,i}>} ... h(X_n)_{<\hat{\pi}_{*,n}>}) \\
\end{aligned}
\notag
\end{equation}

\begin{equation}
\begin{aligned}
h(\mathcal{S})|\hat{\pi} &= h(X_1...X_n)|\hat{\pi} \\
  &= h(X_1)...h(X_i)...h(X_n)|\hat{\pi} \\
  &= h(X_1)_{<\hat{\pi}>} ... h(X_i)_{<\pi'_{*,i}>} ...h(X_n)_{<\pi'_{*,n}>} \\
\end{aligned}
\end{equation}

Let $\theta_i$ and $\theta_1$ be rotations associated with $X_i$ and $X_1$, there are:

\begin{equation}
\label{eq:hoz_eq1}
\begin{aligned}
\hat{\pi}_{*,i} &= s(\hat{\pi}|\theta_i - \theta_1) \\
\end{aligned}
\end{equation}

Note here $\pi'_{*,i}$ is calculated from the encoding
$h(X_1)...h(X_i)...h(X_n)|\hat{\pi}$, then

\begin{equation}
\label{eq:hoz_eq2}
\pi'_{*,i} = s(\hat{\pi}|\xi_i - \xi_1)
\end{equation}

\noindent where $\xi_i$ and $\xi_1$ are rotations of $h(X_i)$ and $h(X_1)$. According
to Lemma \ref{lemma:reflection_entry}, there is $\xi_i = \theta_i +
\alpha(\theta_i + 90)$ and $\xi_1 = \theta_1 + \alpha(\theta_1 + 90)$. Then

\begin{equation}
\xi_i - \xi_1 = \theta_i + \alpha(\theta_i + 90) - \theta_1 - \alpha(\theta_1 + 90) .
\notag
\end{equation}

Notice $\alpha(\theta_i + 90)$ and $\alpha(\theta_1 + 90)$ return 0 or 180, thus

\begin{equation}
\xi_i - \xi_1 \bmod 180 = \theta_i - \theta_1 \bmod 180 .
\notag
\end{equation}

This results in the identical conditions in Equations \ref{eq:hoz_eq1} and
\ref{eq:hoz_eq2}, thus $\hat{\pi}_{*,i} = \pi'_{*,i}$, and eventually
$h(\mathcal{S}|\pi) = h(\mathcal{S})|\hat{\pi}$.

\end{proof}

\begin{proposition}
\label{prop:6.5}
Horizontally reflecting a curve reflects the seed also change the expansion
code to the complement.

\begin{equation}
h(\mathcal{P}_k) = h(X_1...X_n|\pi_1 ... \pi_k) = h(X_1...X_n)|\hat{\pi}_1 ... \hat{\pi}_k
\notag
\end{equation}

\end{proposition}

\begin{proof}

With Lemma \ref{lemma:reflect_single}, there is:

\begin{equation}
h(\mathcal{P}_k) = h(\mathcal{P}_{k-1}|\pi_k) = h(\mathcal{P}_{k-1})|\hat{\pi}_k .
\notag
\end{equation}

The above equation can be repeatedly extended till level 1 and we can finally
have

\begin{equation}
  \label{eq:61}
h(\mathcal{P}_k) = h(\mathcal{P}_0)|\hat{\pi}_1...\hat{\pi}_k = h(X_1...X_n)|\hat{\pi}_1...\hat{\pi}_k .
\notag
\end{equation}

\end{proof}

\begin{corollary}
Combining rotation and reflection, there is

\begin{equation}
\left(h(X_1...X_n|\pi_1 ... \pi_k)\right)^\theta = \left(h(X_1...X_n)\right)^\theta|\hat{\pi}_1 ... \hat{\pi}_k .
\notag
\end{equation}

\end{corollary}

\begin{proof}
This can be proven by first applying Proposition \ref{prop:6.5} then
Proposition \ref{prop:6.3}.

\end{proof}

\subsubsection{Reversal}\label{reversal-1}

We first have the following relation by enumerating all level-1 units in
Figure \ref{fig:expansion_rule}.

\begin{equation}
\label{eq:6.3.5-1}
\begin{array}{ll}
r(X_{<\pi>}) = X'|\pi'
& \quad \text{where }
\pi' = \begin{cases}
\pi \quad & \text{if } X \in \{R,L\} \\
\hat{\pi} \quad & \text{if } X \notin \{R,L\}
\end{cases}
\end{array}
\end{equation}

In the form \(\pi'\), the superscript ``$'$'' should be better read as
an operator that is applied on an expansion code and returns the code or
its complement depending on the base type of \(X\).

\paragraph{One expansion code}\label{one-expansion-code}

The symbolic form of the reversal of the general curve
\(X_1...X_n|\pi_1...\pi_k\) is complex. We first start the analysis on a
seed curve in one expansion.

\begin{equation}
\begin{aligned}
r(X_1 ... X_n|\pi) &= r(X_{<\pi>,1} ... X_{<\pi_*>,n}) \\
  &= r(X_{<\pi>,1} ... X_{< s_n >,n}) \\
  &= r(X_{< s_n >,n}) ... r(X_{<\pi>,1}) \\
  &= (X'_n|s'_n) ... (X'_1|\pi') \\
  &= X'_{< s'_n >,n}...X'_{<\pi'>,1} \\
  &= X'_n ... X'_1|s'_n \\
  &= r(X_1...X_n)|s'_n \\
\end{aligned}
\notag
\end{equation}

Explanations for the key steps are:

\begin{itemize}
\tightlist
\item
  Line 2: $s_n$ is the expansion code for $X_n$ inferred from $X_1$ based on
  Lemma \ref{lemma:code1} (i.e., solution for $\pi_*$). The value is $s_n =
  s(\pi|\theta_n - \theta_1)$ where $\theta_1$ and $\theta_n$ are rotations
  associated with $X_1$ and $X_n$.
\item
  Line 3: According to Proposition \ref{prop:subsequence}, the reversal on the
  complete sequence is split into a list of reversed subsequences (level-1
  units).
\item
  Line 4: The reversal of the level-1 unit is applied according to Equation
  \ref{eq:6.3.5-1}.
\item
  Line 6: The expansion code $s'_n$ of the first unit is moved to the right
  side of $|$ as it controls the expansion of the whole sequence.
\end{itemize}

According to Equation \ref{eq:6.3.5-1}, the value of \(s'_n\) depends on
the base type of \(X_n\) and the value of \(s_n\). We enumerate all
combinations of \(\theta_n - \theta_1 \bmod 180\) (for calculating
\(s_n\)) and \(X_n\) to solve \(s'_n\), as listed in Table \ref{tab:sn}.

\begin{table}
\centering
\begin{tabular}{ccc|c}
\toprule
$\theta_n-\theta_1 \bmod 180$ & $s_n$ & $X_n \in \{L, R\}$ & $s'_n$ \\
\midrule
0 & $\pi$ & yes & $\pi$ \\
0 & $\pi$ & no & $\hat{\pi}$ \\
90 & $\hat{\pi}$ & yes & $\hat{\pi}$ \\
90 & $\hat{\pi}$ & no & $\pi$ \\
\bottomrule
\end{tabular}
\vspace*{5mm}
\caption{\label{tab:sn}Solve $s'_n$. Value of $s_n$ is based on the definition of $s()$ which returns $\pi$ or $\hat{\pi}$ based on $\theta_n - \theta_1 \bmod 180$. Value of $s'_n$ depends on $s_n$ and $X_n$.}
\end{table}

If we write

\begin{equation}
\label{eq:rcode1}
r(X_1...X_n|\pi) = r(X_1...X_n)|\pi^\#,
\end{equation}

\noindent then the solution of \(\pi^\#\) is exactly \(s'_n\). Then
according to Table \ref{tab:sn}, \(\pi^\#\) takes value in
\(\{\pi, \hat{\pi}\}\) depending on the value of \(\theta_n - \theta_1\)
and the base type of \(X_n\). We rewrite solutions in Table \ref{tab:sn}
to:

\begin{equation}
\label{eq:6.3.5-3}
\pi^\# = \begin{cases}
\pi \quad \text{if } \theta_n - \theta_1 \bmod 180 = 0 \text{ and } X_n \in \{L, R\} \\
\hat{\pi} \quad \text{if } \theta_n - \theta_1 \bmod 180 = 0 \text{ and } X_n \notin \{L, R\} \\
\hat{\pi} \quad \text{if } \theta_n - \theta_1 \bmod 180 = 90 \text{ and } X_n \in \{L, R\} \\
\pi \quad \text{if } \theta_n- \theta_1 \bmod 180 = 90 \text{ and } X_n \notin \{L, R\} \\
\end{cases} .
\end{equation}

We simplify the expression of Equation \ref{eq:6.3.5-3} by a helper
function \(u()\) written as:

\begin{equation}
\label{eq:6.3.5-4}
\pi^\# = u(\pi|\theta_n - \theta_1, X_n) .
\end{equation}

\paragraph{\texorpdfstring{\(k\) expansion
code}{k expansion code}}\label{k-expansion-code-1}

Next we extend to general \(k\) expansions (\(k \ge 2\)).

\begin{equation}
\begin{aligned}
r(X_1 ... X_n|\pi_1 ...\pi_k) &= r\left((X_1 ... X_n|\pi_1...\pi_{k-1})| \pi_k\right) \\
    &= r(X_1 ... X_n|\pi_1...\pi_{k-1})|\pi^\#_k \\
    &= r(X_1 ... X_n|\pi_1...\pi_{k-2})|\pi^\#_{k-1}|\pi^\#_k \\
    &= ... \\
    &= r(X_1 ... X_n)|\pi^\#_1|\pi^\#_2|...|\pi^\#_{k-1}|\pi^\#_k \\
  &= r(X_1 ... X_n)|\pi^\#_1 ...\pi^\#_k \\
\end{aligned}
\notag
\end{equation}

In the equation expansion, reversal is applied from level \(k\) to level
1 level-by-level. On each step \(i\), we treat the curve
\(\mathcal{P}_{i-1}\) as a seed sequence to be expanded to
\(\mathcal{P}_i\) (\(i \ge 2\)), then with Equation \ref{eq:rcode1}, we
always have \(r(\mathcal{P}_i) =
r(\mathcal{P}_{i-1}|\pi_i) = r(\mathcal{P}_{i-1})|\pi^\#_i\). The
sequence of \(\pi^\#_1 ...\pi^\#_k\) is going to be solved.

\(\pi^\#_1\) is already solved in Equation \ref{eq:6.3.5-4}. We look at
the reversal on \(\mathcal{P}_i\) for \(2 \le i \le k\). We expand
\(r(\mathcal{P}_i)\):

\begin{equation}
\begin{aligned}
r(\mathcal{P}_i) = r(\mathcal{P}_{i-1}|\pi_i) &= r( (X_1...X_n|\pi_1...\pi_{i-1})|\pi_i ) \\
   &=r(X_{<(\pi)_{i-1}>,1} ... X_{<(\pi_*)_{i-1}>,n} | \pi_i) \\
   &=r(X_s ... X_e|\pi_i) \\
   &= r(X_s ... X_e)|\pi^\#_i \\
\end{aligned} .
\notag
\end{equation}

In the above equation, \(\mathcal{P}_{i-1}\) is represented as list of
level \(i-1\) units and in turn as a base sequence denoted as
\(X_s...X_e\). Let \(X_s\) be associated with a rotation \(\theta_s\)
and \(X_e\) be associated with a rotation of \(\theta_e\). With Equation
\ref{eq:6.3.5-3}, \(\pi^\#_i\) can be solved as
\(\pi^\#_i = u(\pi_i|\theta_e - \theta_s, X_e)\), however, \(\theta_e -
\theta_s\) and \(X_e\) are internal variables and their values change on
each level \(k\). We want to find a deterministic solution of
\(\pi^\#_i\) which is only based on the initial seed sequence.

First notice \(\varphi_s(X) = \theta_s(X) + \gamma_X\) where
\(\varphi_s(X)\) is the entry direction of \(X\) and \(\gamma_X\) is the
difference between the entry direction and rotation of \(X\). We rewrite
\(\theta_e - \theta_s\) as
follows\footnote{We use $\theta_s(X)$ (as a function) and $\theta_s$ (as a
variable), or $\theta_e(X)$ and $\theta_e$ interchangeably.}.

\begin{equation}
\theta_e - \theta_s = \varphi_s(X_e) - \gamma_{X_e} - \varphi_s(X_s) + \gamma_{X_s}
\notag
\end{equation}

As \(X_s\) and \(X_e\) are bases from \(\mathcal{P}_{i-1}\)
(\(i-1 \ge 1\)), thus \(X_s, X_e \in \{I, R, L\}\) (Proposition
\ref{prop:IRL}). According to Figure \ref{fig:expansion_rule},
\(\gamma_{X_s}\) and \(\gamma_{X_e}\) are all zero. Then

\begin{equation}
\theta_e - \theta_s = \varphi_s(X_e) - \varphi_s(X_s) .
\notag
\end{equation}

We continue to expand the equation.

\begin{equation}
\label{eq:theta_es}
\begin{aligned}
\theta_e - \theta_s &= \varphi_s(X_e) - \varphi_s(X_s) \\
   &= \varphi_e(X_e) - \Delta(X_e) - \varphi_s(X_s) \\
   &= \varphi_e(X_{<(\pi_*)_{i-1}>,n})- \Delta(X_e) - \varphi_s(X_{<(\pi)_{i-1}>,1}) \\
   &= \varphi_e(X_{<\pi_{1*}>,n})- \Delta(X_e) - \varphi_s(X_{<\pi_1>,1}) \\
   &= \varphi_e(X_n|\pi_{1*})- \Delta(X_e) - \varphi_s(X_1|\pi_1) \\
   &= \varphi_e((X^0_n|\pi_{1*})^{\theta_n})- \Delta(X_e) - \varphi_s((X^0_1|\pi_1)^{\theta_1}) \\
   &= \theta_n + \varphi_e(X^0_n|\pi_{1*})- \Delta(X_e) - \theta_1 - \varphi_s(X^0_1|\pi_1) \\
   &= \theta_n + \varphi_e(X^{(1),0}_n)- \Delta(X_e) - \theta_1 - \varphi_s(X^{(1),0}_1) \\
\end{aligned}
\end{equation}

Explanations of the key steps are:

\begin{itemize}
\tightlist
\item
  Line 2: For a base , there is an offset denoted as $\Delta()$ between its
  exit direction $\varphi_e()$ and entry direction $\varphi_s()$ ($\Delta(X) =
  \varphi_e(X) - \varphi_s(X)$). For the three primary bases, there are
  $\Delta(I) = 0$, $\Delta(R) = -90$ and $\Delta(L) = 90$.
\item
  Line 3: $X_e$ is the last base of $\mathcal{P}_{i-1}$, and it is also the
  last base of the square unit induced by $X_n$. $X_s$ is the first base of
  $\mathcal{P}_{i-1}$, and it is also the first base of the square unit
  induced by $X_1$.
\item 
  Line 4: Using Proposition \ref{prop:entry-exit-directions}, the entry and
  the exit directions of the unit on level $i-1$ are the same as on level 1.
\item
  Line 6: If $X_n$ is associated with a rotation $\theta_n$, we apply the
  rotation on the whole level-1 expansion. This by definition of the expansion
  process. The same for $X_1$.
\item
  Line 7: The entry direction changes accordingly to the rotation of the
  curve. So we separate the rotation of the curve and the entry direction of
  the curve when $X_n$ is on the base rotation state. The same for $X_1$.
\item
  Line 8: both $\pi_{1*}$ and $\pi_1$ can be 1 or 2, we simplify the notation
  where we remove the expansion code for both notations.
\end{itemize}

Slightly modifying the results in Equation \ref{eq:theta_es}, we have

\begin{equation}
\theta_n - \theta_1 = \theta_e - \theta_s - \varphi_e(X^{(1),0}_n) + \Delta(X_e) + \varphi_s(X^{(1),0}_1) .
\end{equation}

We enumerate all combinations of \(\theta_e - \theta_s \bmod 180\) and
\(X_e\) to obtian the solution of \(\pi^\#_i\), and all combinations of
\(X_1\) and \(X_n\) as well as their two level-1 units to establish the
relations between \(\pi^\#_i\) and \(\theta_n - \theta_1\). They can be
separated into three groups:

\textit{Group 1}. \(X_1 \in \{I, R, L, U, B, P, Q\}\) where
\(\varphi_s(X_1^{(1),0})\) are all 90. The results are listed in Table
\ref{tab:rev_group1}.

\textit{Group 2}. \(X_1 = D\) where \(\varphi_s(X^{(1),0})\) are either
180 (expansion type = 1) or 0 (expansion type = 2). If we build a table,
it will be the same as Table \ref{tab:rev_group1} and only the values in
the last column will be switched, i.e., \(0 \rightarrow 90\) and
\(90 \rightarrow 0\).

\textit{Group 3}. \(X_1 = C\) where \(\varphi_s(X^{(1),0})\) are all 180
or 0, and \(n = 1\). The results are listed in Table
\ref{tab:rev_group3}. Notice since \(n =
1\), \(\theta_n - \theta_1 = 0\). Therefore we delete the first and the
fourth rows in Table \ref{tab:rev_group3}. Actually we can also prove
for a curve \(C|(\pi)_k\), if the last base is \(I\),
\(\theta_e - \theta_s \bmod 180\) can only be 0, and if the last base is
\(R\) or \(L\), \(\theta_e - \theta_s \bmod 180\) can only be 90.

\begin{table}
\centering
\begin{tabular}{cc|c|ccc|c}
\toprule
$\theta_e - \theta_s \bmod 180$ & $X_e \in \{R,L\}$ & $\pi^\#_i$ & $X_n$ & $\varphi_e(X_n^{(1),0})$ & $\Delta(X_e)$ & $\theta_n - \theta_1 \bmod 180$ \\ 
\midrule
0  & yes & $\pi_i$       & $I/U/B/D/P$ & 90/270 & 90/270 & 90  \\
0  & no  & $\hat{\pi}_i$ &             &        & 0      & 0 \\
90 & yes & $\hat{\pi}_i$ &             &        & 90/270 & 0 \\
90 & no  & $\pi_i$       &             &        & 0      & 90  \\
\midrule
0  & yes & $\pi_i$       & $R/L/Q$ & 0/180  & 90/270 & 0  \\
0  & no  & $\hat{\pi}_i$ &         &        & 0      & 90 \\
90 & yes & $\hat{\pi}_i$ &         &        & 90/270 & 90 \\
90 & no  & $\pi_i$       &         &        & 0      & 0  \\
\bottomrule
\end{tabular}
\vspace*{5mm}
\caption{\label{tab:rev_group1}Calculate $\pi_i^\#$, $X_1 \in \{I, R, L, U, B, P, Q\}$. $X_n$ is additionally separated into two groups based on the exit directions of its level-1 units. In this group, $\varphi_s(X_1^{(1),0}) = 90$.}
\end{table}

\begin{table}
\centering
\begin{tabular}{cc|c|ccc|c}
\toprule
$\theta_e - \theta_s \bmod 180$ & $X_e \in \{R,L\}$ & $\pi^\#_i$ & $X_n$ & $\varphi_e(X_n^{(1)})$ & $\Delta(X_e)$ & $\theta_n - \theta_1 \bmod 180$ \\ 
\midrule
\st{0}  & \st{yes} & \st{$\pi_i$}       & $C$ & 0/180 & \st{90/270} & \st{90}  \\
0  & no  & $\hat{\pi}_i$ &             &        & 0      & 0 \\
90 & yes & $\hat{\pi}_i$ &             &        & 90/270 & 0 \\
\st{90} & \st{no}  & \st{$\pi_i$}       &             &        & \st{0}      & \st{90}  \\
\bottomrule
\end{tabular}
\vspace*{5mm}
\caption{\label{tab:rev_group3}Calculate $\pi_i^\#$, $X_1 = C$. The first and fourth rows are deleted because they do not exist. In this group, $\varphi_s(X_1^{(1),0}) = 0$ or 180, and $n = 1$.}
\end{table}

Taking the results in Table \ref{tab:rev_group1} and
\ref{tab:rev_group3}, as well as the results in Group 2 together, the
correspondance of \(\pi^\#_i\), \(\theta_n -
\theta_1\), \(X_n\) and \(X_1\) are summarized in Table
\ref{tab:pi_solution}.

\begin{table}
\centering
\begin{tabular}{ccccl}
\toprule
$\pi_i^\#$ & $\theta_n - \theta_1 \bmod 180$ & $X_n$ & $X_1$ & Group\\ 
\midrule
$\pi_i$        & 90  & $I/U/B/D/P$ & $\notin \{C,D\}$ & Group 1\\
$\hat{\pi}_i$  & 0   & $I/U/B/D/P$ & $\notin \{C,D\}$ &\\
$\hat{\pi}_i$  & 90  & $R/L/Q$ & $\notin \{C,D\}$ &\\
$\pi_i$        & 0   & $R/L/Q$ & $\notin \{C,D\}$ &\\
\midrule
$\pi_i$        & 0  & $I/U/B/D/P$ & $D$ & Group 2\\
$\hat{\pi}_i$  & 90   & $I/U/B/D/P$ & $D$ &\\
$\hat{\pi}_i$  & 0  & $R/L/Q$ & $D$ &\\
$\pi_i$        & 90   & $R/L/Q$ & $D$ &\\
\midrule
$\hat{\pi}_i$  & 0   & $C$ & $C$ & Group 3\\
\bottomrule
\end{tabular}
\vspace*{5mm}
\caption{\label{tab:pi_solution}Final solution of $\pi_i^\#$ ($i \ge 2$).}
\end{table}

Let's use a helper function \(v()\) to represent the complex solutions
in Table \ref{tab:pi_solution}:

\begin{equation}
\pi^\#_i = v(\pi_i|\theta_n - \theta_1, X_n, X_1) \quad i \ge 2 .
\notag
\end{equation}

Note being different from \(u()\), \(v()\) additionally depends on the
base type of \(X_1\). Now we can have the final proposition of reversing
a curve:

\begin{proposition} 
\label{prop:reverse_sequence}
Reversing a curve on level $k$ initialized by a seed sequence has the
following form:

\begin{equation}
r(X_1 ... X_n|\pi_1 ... \pi_k) = X'_n ... X'_1|\pi_1^\#...\pi_k^\# .
\notag
\end{equation}

The solution of the code sequence is 

\begin{equation}
\pi_i^\# = \begin{cases}
 u(\pi_i|\theta_n - \theta_1, X_n) \quad & i = 1 \\
 v(\pi_i|\theta_n - \theta_1, X_n, X_1) \quad & 2 \le i \le k \\
\end{cases}
\notag
\end{equation}

\noindent where $\theta_1$ and $\theta_n$ are the rotations associated with
$X_1$ and $X_n$.

\end{proposition}

In particular, when the seed is a single base, Proposition
\ref{prop:reverse_sequence} can be simplified to the following
corollary.

\begin{corollary}
\label{coro:rev}
Reversing a curve on level $k$ initialized by a single seed $X$ has the
following form:

\begin{equation}
r(X|\pi_1 ... \pi_k) = X'|\pi_1^\#...\pi_k^\# .
\notag
\end{equation}

The solution of the code sequence is

\begin{equation}
\pi_1^\#...\pi_k^\# = \begin{cases}
\pi_1\pi_2...\pi_k \quad & \textnormal{if } X \in \{R, L\} \\
\hat{\pi}_1\hat{\pi}_2...\hat{\pi}_k \quad & \textnormal{if } X \in \{I, U, B, P, C\} \\
\hat{\pi}_1\pi_2...\pi_k \quad & \textnormal{if } X \in \{D, Q\} \\
\end{cases} .
\notag
\end{equation}

\end{corollary}

\begin{proof}
It can be proved by Proposition \ref{prop:reverse_sequence} by setting $X_1 = X_n$
and $\theta_n = \theta_1$.
\end{proof}

\subsection{Reversal and reflection are
redundant}\label{reversal-and-reflection-are-redundant}

When the seed is a single base, reflection and reversal are actually
redundant as they both switch the curve between clockwise and
counterclockwise orientations.

\begin{proposition}
\label{prop:redudant}
The orientation of a curve is determined by its level-1 structure. For all
curves on level $k$, let $P = \{\mathcal{P}_k\}$ be the set of curves in the
clockwise orientation, and $Q = \{\mathcal{Q}_k\}$ be the set of curves in the
counterclockwise orientation. If treating reversal $r()$ and horizontal reflection $h()$
as two mappings, then $r:P \rightarrow Q$ and $h:P \rightarrow Q$ are both bijective.
\end{proposition}

The discussion is the same if \(P\) corresponds to countryclockwise
curves and \(Q\) corresponds to clockwise curves. We omit this scenario
here.

\begin{proof}
First, it is easy to see $r(\mathcal{P}_k) \in Q$ as $r(\mathcal{P}_k)$ is counterclockwise.
For a unique curve $\mathcal{P}_k = X|\pi_1...\pi_k$, its reversal 
$r(\mathcal{P}_k) = X'|\pi_1^\#...\pi_k^\#$ is also unique because the correspondance
of the two symbolic expression is one-to-one (Equation \ref{eq:rev_base} and Corollary \ref{coro:rev}). 
From Figure \ref{fig:expansion_rule}, the
following nine level-1 units induce clockwise curves: $I_2$, $R_1$, $R_2$,
$U_1$, $B_2$, $D_1$, $P_2$, $Q_1$, $C_1$, which generate in total $9 \times
2^{k-1} \times 4 = 36 \times 2^{k-1}$ different curves in $P$,
and it in turn determines $36 \times 2^{k-1}$ different curves in
$\{r(\mathcal{P}_k)\}$ where the mapping $r()$ is bijective from
$P$ to $\{r(\mathcal{P}_k)\}$. Note the total number of 2x2
curves on level $k$ is $36 \times 2^k$ (Equation \ref{eq:total_number}) and $P$ and $Q$ are absolute complementary, then
$\{r(\mathcal{P}_k)\} = Q$, thus $r:P \rightarrow Q$ is bijective.

Also $h(\mathcal{P}_k) \in Q$ and the correspondance between
$P$ and $\{h(\mathcal{P}_k)\}$ is one-to-one. $\{h(\mathcal{P}_k)\}$
also contains $36 \times 2^{k-1}$ which makes $\{h(\mathcal{P}_k)\} = Q$. Thus
$h:P \rightarrow Q$ is also bijective.

\end{proof}

Proposition \ref{prop:redudant} indicates that, for a specific curve
\(\mathcal{Q}_k\), it can be uniquely generated by reversal of a unique
curve in \(P\) or by horizontal reflection of another unique curve in
\(P\). Next we explore the forms from these two transformations.

Write \(\mathcal{Q}_k = Y^{(1)}|(\pi)_{k-1}\) with
\(Y^{(1)} \in \{I_1, L_1, L_2,
U_2, B_1, D_2, P_1, Q_2, C_2\}\) which determine the curve in the
counterclockwise orientation. We first consider \(Y^{(1)} = I_1\)
(associated with a rotation of zero) as an example, and solve
\(r(\mathcal{P}_k) = I_1|(\pi)_{k-1}\) (with Corollary \ref{coro:rev}
and Equation \ref{eq:rev_base}).

\begin{equation}
\begin{aligned}
r(\mathcal{P}_k) &= I_1|(\pi)_{k-1} \\
\mathcal{P}_k &= r(I_1|(\pi)_{k-1}) = r(I|1(\pi)_{k-1}) \\
    &= I'|2(\hat{\pi})_{k-1} \\
    &= I^{180}|2(\hat{\pi})_{k-1} = I^{180}_2|(\hat{\pi})_{k-1} \\
\end{aligned}
\notag
\end{equation}

We solve \(h(\mathcal{P}'_k) = I_1|(\pi)_{k-1}\) (with Proposition
\ref{prop:6.5}).

\begin{equation}
\begin{aligned}
h(\mathcal{P}'_k) &= I_1|(\pi)_{k-1} \\
\mathcal{P}'_k &= h(I_1|(\pi)_{k-1}) = h(I|1(\pi)_{k-1}) \\
    &= h(I)|2(\hat{\pi})_{k-1} \\
    &= I|2(\hat{\pi})_{k-1} = I_2|(\hat{\pi})_{k-1} \\
\end{aligned}
\notag
\end{equation}

We can do it for all possible forms of \(Y^{(1)}\):

\begin{alignat*}{3}
&r(\mathcal{P}_k)   &&= \mathcal{Q}_k               &&= h(\mathcal{P}'_k)   \\
&r(I_2^{180}|(\hat{\pi})_{k-1}) &&= I_1|(\pi)_{k-1} &&= h(I_2|(\hat{\pi})_{k-1})        \\
&r(R_1^{270}|(\pi)_{k-1}) &&= L_1|(\pi)_{k-1}        &&= h(R^{180}_2|(\hat{\pi})_{k-1})         \\   
&r(R_2^{270}|(\pi)_{k-1}) &&= L_2|(\pi)_{k-1}        &&= h(R^{180}_1|(\hat{\pi})_{k-1})         \\
&r(U_1|(\hat{\pi})_{k-1}) &&= U_2|(\pi)_{k-1}       &&= h(U_1|(\hat{\pi})_{k-1})  \\
&r(B_2^{180}|(\hat{\pi})_{k-1}) &&= P_1|(\pi)_{k-1} &&= h(P_2|(\hat{\pi})_{k-1}) \\
&r(D_1^{180}|(\pi)_{k-1}) &&= Q_2|(\pi)_{k-1}       &&= h(Q_1|(\hat{\pi})_{k-1}) \\
&r(P_2^{180}|(\hat{\pi})_{k-1}) &&= B_1|(\pi)_{k-1} &&= h(B_2|(\hat{\pi})_{k-1}) \\
&r(Q_1^{180}|(\pi)_{k-1}) &&= D_2|(\pi)_{k-1}       &&= h(D_1|(\hat{\pi})_{k-1}) \\
&r(C_1|(\hat{\pi})_{k-1}) &&= C_2|(\pi)_{k-1}       &&= h(C_1|(\hat{\pi})_{k-1})   \\
\end{alignat*}

The above equations also confirm that \(r:P \rightarrow Q\) and
\(h:P \rightarrow Q\) are both bijective.

If \(Y^{(1)}\) has a rotation \(\theta\) associated, first with
Proposotion \ref{prop:6.3}, there are:

\begin{equation}
\begin{aligned}
Y^{(1),\theta}|(\pi)_{k-1} &= Y^{\theta}|\pi_1(\pi)_{k-1} \\
   &= (Y|\pi_1(\pi)_{k-1})^\theta  = \mathcal{Q}_k^\theta \\
\end{aligned} .
\notag
\end{equation}

\noindent where we assume \(\mathcal{Q}_k\) is the curve where
\(Y^{(1)}\) is associated with a rotation of zero. Rotation and reversal
on a sequence are independent (Proposition \ref{prop:trseq_2}).

\begin{equation}
\begin{aligned}
\mathcal{Q}_k^\theta &= (r(\mathcal{P}_k))^\theta \\
    &= r( (\mathcal{P}_k)^\theta) \\
\end{aligned}
\notag
\end{equation}

With Proposition \ref{prop:trseq_1} we can obtain the form with
horizontal reflection.

\begin{equation}
\begin{aligned}
\mathcal{Q}_k^\theta &= (h(\mathcal{P}'_k))^\theta \\
    &= h( (\mathcal{P}'_k)^{\theta + \alpha(\theta)} ) \\
\end{aligned}
\notag
\end{equation}

It is easy to see both \((\mathcal{P}_k)^\theta\) and
\((\mathcal{P}'_k)^{\theta + \alpha(\theta)}\) are in \(P\).

\section{Reduction}\label{reduction}

\subsection{Reduction on the curve}\label{reduction-on-the-curve}

Reduction of a curve is the reverse process of the expansion. A curve
\(\mathcal{P}_k\) induced from a seed sequence with length \(n\) on
level \(k\) is a combination of \(4^{k-1} \times n\) 2x2 units. Reducing
the curve to level \(k-1\) is to reduce each 2x2 unit into its original
single base using the rules in the diagram in Figure
\ref{fig:expansion_rule}. One important attribute in the reduction from
level \(k\) to level \(k-1\) is, the entry direction and exit direction
of each 2x2 unit are not changed when reduced to its corresponding
level-0 base. This ensures the curve after the reduction is still
well-connected (Note \ref{note}).

Denote the reduction to level \(k - 1\) as \(\mathrm{Rd}_1()\) because
the reduction is applied by depth of one, then according to the
description in the previous paragraph, we have the form of the
reduction:

\begin{equation}
\label{eq:7-1}
\begin{aligned}
\mathrm{Rd}_1(\mathcal{P}_k) &= \mathrm{Rd}_1\left( (\mathcal{S}|\pi_1...\pi_{k-1})|\pi_k \right) \\
 &= \mathrm{Rd}_1\left( X_s ... X_e | \pi_k \right) \\
 &= \mathrm{Rd}_1\left( X_{<\pi_k>,s} ... X_{<\pi_{*}>,e} \right) \\
 &= X_s ... X_e \\
 &= \mathcal{S}|\pi_1...\pi_{k-1} = \mathcal{P}_{k-1} \\
\end{aligned}
\end{equation}

\noindent where \(\mathcal{S}\) is the seed sequence, \(X_s...X_e\) is
the base sequence of the curve on level \(k-1\), and
\(X_{<\pi_k>,s} \rightarrow X_s\) is the reduction of a level-1 unit to
its corresponding base by definition.

With Equation \ref{eq:7-1}, we can have the form of reducing by any
depth \(i\), i.e., to level \(k-i\).

\begin{equation}
\label{eq:7-2}
\begin{aligned}
\mathrm{Rd}_i(\mathcal{P}_k) &= \overbrace{\mathrm{Rd}_1(...(\mathrm{Rd}_1(P_k)))}^{i\text{ }\mathrm{Rd}_1()} \\
   &= \overbrace{\mathrm{Rd}_1(...(\mathrm{Rd}_1(\mathcal{P}_{k-1})))}^{i-1\text{ }\mathrm{Rd}_1()} \\
   &= ... \\
   &= \mathrm{Rd}_1(\mathrm{Rd}_1(\mathcal{P}_{k-i+2})) \\
   &= \mathrm{Rd}_1(\mathcal{P}_{k-i+1}) \\
   &= \mathcal{P}_{k-i} \\
\end{aligned}
\notag
\end{equation}

Addtionally, we can have
\(\mathrm{Rd}_{k}(\mathcal{P}_k) = \mathcal{P}_0 =
\mathcal{S}\) (reducing the curve by the complete depth of \(k\) returns
to its seed sequence) and
\(\mathrm{Rd}_{0}(\mathcal{P}_k) = \mathcal{P}_k\) (reducing the curve
by depth zero is still the original curve).

We can say reduction of \(\mathcal{P}_k\) by depth \(i\) generates the
global structure of \(\mathcal{P}_k\) on level \(k-i\). In the following
text, if the depth is not of interest, we simplify notation
\(\mathrm{Rd}_i()\) to \(\mathrm{Rd}()\).

If a curve is represented as a list of square units, the reduction can
be applied to individual square units separately.

\begin{equation}
\label{eq:rd}
\begin{aligned}
\mathrm{Rd}(\mathcal{S}|(\pi)_k) &= \mathrm{Rd}(X_1...X_n|(\pi)_k) \\
    &= \mathrm{Rd}(X_1|(\pi)_k ... X_n|(\pi_{*,n})_k) \\
    &= \mathrm{Rd}(X_1|(\pi)_k) ... \mathrm{Rd}(X_n|(\pi_{*,n})_k) \\
\end{aligned} .
\end{equation}

\subsection{Reduction and
transformations}\label{reduction-and-transformations}

\begin{definition}
\label{def:primary_trans}
Rotation, reflection, reversal, or any combination of these three
transformations are called \textit{primary transformations}, denoted as
$f_t() = f_{t_1}(f_{t_2}(...(f_{t_*}(...))))$ where $f_{t_*}$ is an individual transformation.
\end{definition}

\begin{proposition}
\label{prop:rec_tr}
Reductions and primary transformations are independent, i.e., $\mathrm{Rd}_i(f_t(\mathcal{P}_k)) = f_t(\mathrm{Rd}_i(\mathcal{P}_k))$.
\end{proposition}

\begin{proof}
We first consider a single rotation denoted as $f_{\theta}$. Using Proposition \ref{prop:6.3}, there is:

\begin{equation}
\begin{aligned}
\mathrm{Rd}_i\left( (\mathcal{S}|\pi_1...\pi_k)^\theta \right) &= \mathrm{Rd}_i\left( \mathcal{S}^\theta|\pi_1...\pi_k \right) \\
   &= \mathcal{S}^\theta|\pi_1...\pi_{k-i} \\
\left( \mathrm{Rd}_i(\mathcal{S}|\pi_1...\pi_k )\right)^\theta &= \left( \mathcal{S}|\pi_1...\pi_{k-i} \right)^\theta \\
    &= \mathcal{S}^\theta|\pi_1...\pi_{k-i}
\end{aligned} ,
\notag
\end{equation}

\noindent thus $\mathrm{Rd}_i(f_{\theta}(\mathcal{P}_k)) =
f_{\theta}(\mathrm{Rd}_i(\mathcal{P}_k))$. Next we consider a single
reflection denoted as $f_h$. Using Proposition \ref{prop:6.5}, there is:

\begin{equation}
\begin{aligned}
\mathrm{Rd}_i\left( h(\mathcal{S}|\pi_1...\pi_k) \right) &= \mathrm{Rd}_i\left( h(\mathcal{S})|\hat{\pi}_1...\hat{\pi}_k  \right) \\
   &= h(\mathcal{S})|\hat{\pi}_1...\hat{\pi}_{k-i} \\
h\left( \mathrm{Rd}_i(\mathcal{S}|\pi_1...\pi_k) \right) &= h\left( \mathcal{S}|\pi_1...\pi_{k-i} \right) \\
    &= h(\mathcal{S})|\hat{\pi}_1...\hat{\pi}_{k-i}
\end{aligned} ,
\notag
\end{equation}

\noindent thus $\mathrm{Rd}_i(f_h(\mathcal{P}_k)) = f_h(\mathrm{Rd}_i(\mathcal{P}_k))$. Last we consider a single reversal denoted as $f_r$. Using
Proposition \ref{prop:reverse_sequence}, there is:

\begin{equation}
\begin{aligned}
\mathrm{Rd}_i\left( r(\mathcal{S}|\pi_1...\pi_k) \right) &= \mathrm{Rd}_i\left( r(\mathcal{S})|\pi^{\#_a}_1...\pi^{\#_a}_k \right) \\
   &= r(\mathcal{S})|\pi^{\#_a}_1...\pi^{\#_a}_{k-i} \\
r\left( \mathrm{Rd}_i(\mathcal{S}|\pi_1...\pi_k) \right) &= r\left( \mathcal{S}|\pi_1...\pi_{k-i} \right) \\
    &= r(\mathcal{S})|\pi^{\#_b}_1...\pi^{\#_b}_{k-i}
\end{aligned} .
\notag
\end{equation}

$\pi^{\#_a}$ and $\pi^{\#_b}$ both depend
on the same seed sequence $\mathcal{S}$, then according to Proposition \ref{prop:reverse_sequence}, $\pi^{\#_a} =
\pi^{\#_b}$. Thus $\mathrm{Rd}_i(f_r(\mathcal{P}_k)) = f_r(\mathrm{Rd}_i(\mathcal{P}_k))$.

Then we expand $f_t()$ to individual transformations with $f_{t_*} \in \{f_\theta, f_h, f_r\}$,
where in each step, we move one $f_{t_*}$ out from $\mathrm{Rd}_i()$:

\begin{equation}
\begin{aligned}
\mathrm{Rd}_i(f_t(\mathcal{P}_k)) &= \mathrm{Rd}_i(f_{t_1}(f_{t_2}(...(f_{t_*}(\mathcal{P}_k))))) \\
  &= f_{t_1}(\mathrm{Rd}_i(f_{t_2}(...(f_{t_*}(\mathcal{P}_k))))) \\
  &= ... \\
  &= f_{t_1}(f_{t_2}(...(f_{t_*}(\mathrm{Rd}_i(\mathcal{P}_k))))) \\
  &= f_t(\mathrm{Rd}_i(\mathcal{P}_k)) \\
\end{aligned} .
\notag
\end{equation}

\end{proof}

\subsection{Infer curve encoding via
reduction}\label{infer-curve-encoding-via-reduction}

Reduction of a curve can be used to reverse-infer the encoding of a
curve. For simplicity, assume \(\mathcal{P}\) is a 2x2 curve initialized
from a single base. The seed base, the level and the expansion code
sequence are all unknown. \(\mathcal{P}\) is only represented as an
ordered list of points with their \(xy\)-coordinates. The inference of
the encoding of \(\mathcal{P}\) can be applied in the following steps:

\begin{enumerate}
\tightlist
\item
   Notice $\mathrm{Rd}_k(\mathcal{P}_k) = X$. The curve should be composed of
   $2^k \times 2^k$ points. Then $k$ is assigned as the level of the curve.
   However, the value of $k$ is not necessarily to be known here because $k$ is also the
   length of the expansion code sequence which will be automatically
   determined when the inference steps are finished. The entry and exit
   directions should be manually added if they are missing. If there are
   several possible entry or exit directions, choose one combination randomly.
   In this step, the complete curve is reduced into a single point. The base
   type as well as the initial rotation can be looked up in Figure
   \ref{fig:expansion_rule}, the ``Base'' column. Note when we say
   ``reduce a unit to a point'', it means to take the average
   $xy$-coordinates of points in the unit.

   If the curve is entry-closed where the entry point is located inside the
   curve region, the base is either $B$ or $D$ but they cannot be
   distinguished on the base level. And if the curve is exit-closed where the
   exit point is located inside the curve region, the base is either $P$ or
   $Q$ but they cannot be distingshed either on the base level. For both
   scenarios, the base seed as well as its rotation can be determined on level
   1 in step 2.
\item
   Notice $\mathrm{Rd}_{k-1}(\mathcal{P}_k) = \mathcal{P}_1 = X|\pi_1$. We
   reduce the curve by depth $k-1$ to obtain $\mathcal{P}_1$. From the start
   of the curve, we replace each subunit on level $k-1$ represented as a
   $2^{k-1} \times 2^{k-1}$ square subunit to a single point, which reduces
   each of the four-quadrant subunits into a point. Visually, the reduced
   curve has a ``U-shape'' with an entry direction and an exit direction. If the
   base type of $X$ is already known from step 1, we only need to look up in
   the two level-1 expansions of $X$ in Figure \ref{fig:expansion_rule} to
   choose the code of $\pi_1$. If the seed is $B$/$D$ or $P$/$Q$ which cannot
   be determined on level 0, it can be determined in this step because their
   level-1 patterns are unique. Step 1 and step 2 can be merged into one single step
   where all types of $X|\pi_1$ can be inferred here.
\item
   Notice $\mathrm{Rd}_{k-i}(\mathcal{P}_k) = \mathcal{P}_{i} =
   \mathcal{P}_{i-1}|\pi_i$. However, we don't need to reduce the whole curve.
   With Equation \ref{eq:rd},

\begin{equation}
\begin{aligned}
\mathrm{Rd}_{k-i}(\mathcal{P}_k) &= \mathrm{Rd}_{k-i}(\mathcal{P}_{i-1}|\pi_i...\pi_k) \\
  &= \mathrm{Rd}_{k-i}(X_s...X_e|\pi_i...\pi_k) \\
  &= \mathrm{Rd}_{k-i}(X_s|\pi_i...\pi_k) ... \mathrm{Rd}_{k-i}(X_e|\pi_{i*}...\pi_{k*}) \\
  &= X_s|\pi_i ... X_e|\pi_{i*}
\end{aligned} .
\notag
\end{equation}

   In above equations, $X_s|\pi_i...\pi_k$ is the first level $k-i+1$ unit of $\mathcal{P}_k$.
   Reducing it by depth $k-i$ obtains a level-1 unit $X_s|\pi_i$. Then the value of $\pi_i$
   can be solved by looking up the shape of $X_s|\pi_i$ (the base type of $X_s$ is not of interest).
\item
   The process stops until the original curve cannot be reduced where we
   reached $\mathcal{P}_k$. We can directly look up the first 2x2 unit to get
   $\pi_k$.
\end{enumerate}

\begin{figure}
\centering{
\includegraphics[width=1\linewidth]{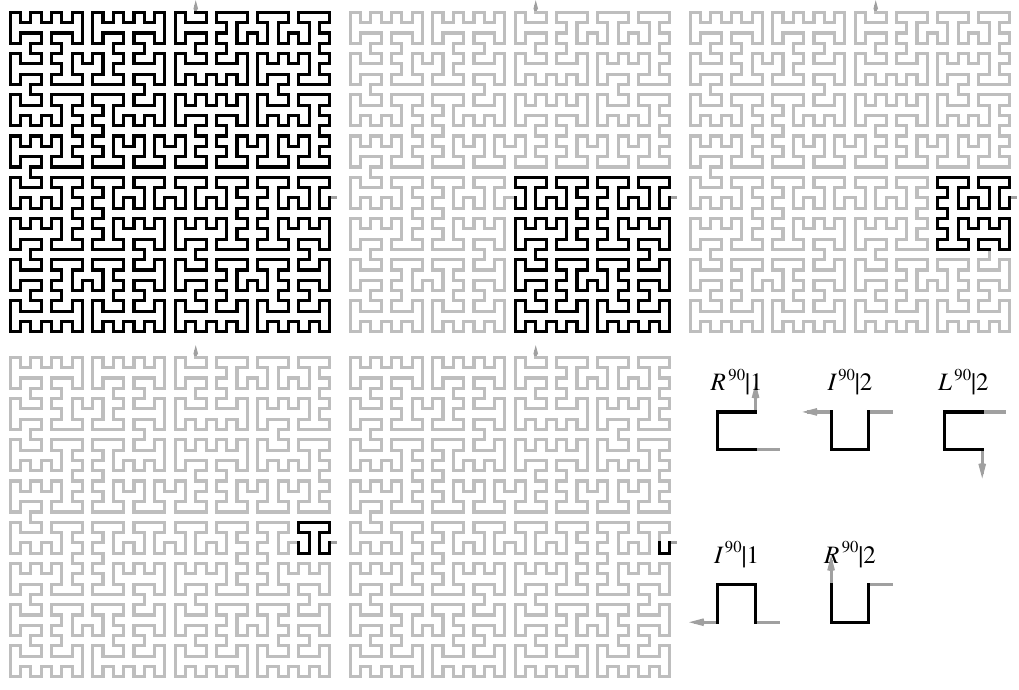}
\caption{Infer the curve encoding from its structure. In the first five panels, the first units on level 5 to 1 are highlighted in black. The last panel lists their corresponding reduced level-1 units.}\label{fig:code_infer}
}
\end{figure}

As an example, Figure \ref{fig:code_infer} illustrates the process of
inferring the symbolic expression from the curve structure. Steps are:

\begin{enumerate}
\tightlist
\item
   The curve $\mathcal{P}_k$ has an entry direction of horizontally right-in
   and an exit direction of vertically top-out. We reduce the curve into a 
   level-1 units (bottom-right panel in Figure \ref{fig:code_infer}) and have $\mathcal{P}_1 = R^{90}|1$.
\item
   We only look at the first level $k-1$ unit of $\mathcal{P}_k$ (the second panel in Figure \ref{fig:code_infer}). 
   Reduce it to a level-1 unit to have $\pi_2 = 2$.
\item
   We do it similarly to only look at the first level $k-i$ unit and we can have $\pi_3 = 2$, $\pi_4 = 1$.
\item
   Last, when $i = 4$, the first level $k-i$ is a 2x2 unit which cannot be reduced any more, thus $\pi_5 = 2$ 
   and we reach the maximal level of $\mathcal{P}_k$ ($k = 5$).
\end{enumerate}

Then the final encoding of the curve in Figure \ref{fig:code_infer} is
\(\mathcal{P}_k = R^{90}|12212\). In Section \ref{unique}, we will
introduce a simpler way for inferring the encoding of a curve which does
not require the complete structure of the curve known in advance while
the locations of the entry and exit points on the curve are already
sufficient to determine the encoding of the curve.

A little bit of more work needs to be done when inferring the encoding
of a curve induced from a seed sequence \(\mathcal{S}\). If
\(\mathcal{P}\) is composed of \(N\) points, we need to find the maximal
\(k\) that gives integer solution of \(n\) for \(4^k
\times n = N\), also each sequential block of \(4^k\) points should be
represented as a square composed of recursive quaternary partitionings.
Then \(n\) is the length of the seed sequence. Only on step 1 where we
reduce \(\mathcal{P}_k\) to \(\mathcal{S}\), the base sequence of
\(\mathcal{S}\) needs to be manually inferred, which should be simple.
Other steps are the same as using a single base as the seed introduced
in this section, where on each reduction step \(i\) we only need to
consider the first level \(k-i+1\) unit.

\section{Geometric attributes}\label{geometry}

From this section, we will study structures of 2x2 curves. We mainly
focus on the curve induced from a single seed, i.e., a square curve, but
the results can be easily extended to general 2x2 curves initialized
from seed sequences.

\subsection{Locations of entry and exit
points}\label{locations-of-entry-and-exit-points}

\begin{figure}
\centering{
\includegraphics[width=1\linewidth]{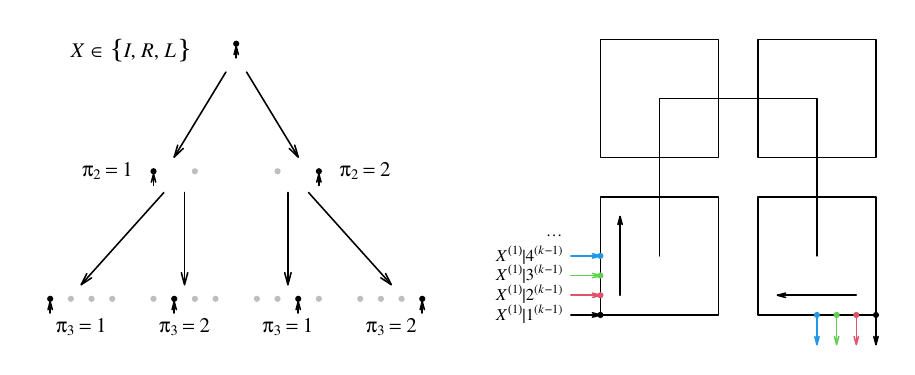}
\caption{Locations of entry points. Left: Locations in the curve expansion of $Z^0|(\pi)_k$. Right: Locations of the entry and exit points in the first and the last subunits of general $X|(\pi)_k$. The same color corresponds to the same curve.}\label{fig:expansion_code_location}
}
\end{figure}

We start from curves induced from primary bases since they form the
basic units for general 2x2 curves.

\begin{lemma}
\label{lemma:geometry_z}
For a curve $\mathcal{P}_k$ ($k \ge 1$) generated from a primary base $Z$
(associated with a rotation of zero), where $Z \in \{I, R, L\}$, there are the
following geometric attributes for the entry point of the curve:

\begin{enumerate}
\tightlist
\item
  The entry point is located on the lower left corner of the curve for
  $Z|(1)_k$, and on the lower right corner of the curve for $Z|(2)_k$.
\item
  Let the coordinate of the lower left corner be $(1, 1)$, and the length of
  the segment connecting two neighbouring points be 1, then the lower right
  corner has the coordinate $(2^k, 1)$. The entry point of curve $Z|(\pi)_k =
  Z|\delta^{(k)}$ has the coordinate of $(\delta, 1)$, where $\delta$ is the
  integer representation of the expansion code sequence on level $k$ defined
  in Equation \ref{eq:integer}.
\item 
  The entry direction is orthogonal to the side (i.e., the lower side) of the
  curve where the entry points are located for all forms of $Z|(\pi)_k$. And
  it is always vertically bottom-in.
\end{enumerate}

\end{lemma}

\begin{proof}
When $k=1$, for all level-1 units of $Z$ on base rotation state, when $\pi_1 =
1$, according to Figure \ref{fig:expansion_rule}, the entry point is located
on the lower left and when $\pi_1 = 2$ the entry point is located on the lower
right of the 2x2 grid. The entry direction is always vertically bottom-in.
Thus, the three attributes are all true.

Next we consider $k \ge 2$. The first base in $Z^{(1)}$ (i.e., a level-1 curve
induced from $Z$) only includes $I$, $R$ and $L$ all associated with rotation
of zero. Let's denote it as $W$. There are the following two properties:

First, when $Z^{(1)}$ is expanded to $Z^{(2)}$, its first base $W$ located on
its lower side will be expanded into a 2x2 unit $W^{(1)}$ which is also
located on the lower left quadrant of $Z^{(2)}$. Notice in $W^{(1)}$, $W$ is a
primary base with no rotation, thus the entry point is located on the lower
side of $W^{(1)}$ (Figure \ref{fig:expansion_rule}), which is also the lower
side of the entire $Z^{(2)}$. We can apply the same process by only looking at
the expansion of the first base on the curve and we can always conclude the
entry point is located on the lower side of the curve on any level $k$.

Second, after $k$ expansions, the first base in $\mathcal{P}_k$ is still one
of $I$/$R$/$L$ with rotation of zero. We know for $I$/$R$/$L$, their entry
directions are always vertically bottom-in. With the first property, attribute
3 is true.

Next we prove attributes 1 and 2 for $k \ge 1$ (we also include $k = 1$ here).
Assume $x$-coordinate of the entry point is $x_k$ for $\mathcal{P}_k$.
Apparently, $x_k$ depends on the expansion code sequence, then we write it as
a function $x( \pi_1...\pi_{k} )$. As mentioned, when the entry point on
$\mathcal{P}_{k-1}$ is expanded to a 2x2 unit denoted as $\mathcal{U}$, when $\pi_k = 1$, the entry
point of $\mathcal{U}$ is located on the lower left corner of $\mathcal{U}$
and when $\pi_k = 2$, the entry point is located on the lower right corner of $\mathcal{U}$
(Figure \ref{fig:expansion_code_location}, left panel). So the location of the
entry point in the expansion from level $k-1$ to level $k$ ($k \ge 1$) is:

\begin{equation}
\label{eq:coord}
x(\pi_1...\pi_k) = \begin{cases}
2\cdot x(\pi_1...\pi_{k-1}) - 1 & \quad \text{if } \pi_k = 1 \\
2\cdot x(\pi_1...\pi_{k-1}) & \quad \text{if } \pi_k = 2 \\
\end{cases}
\end{equation}

\noindent with the initial values $x(\varnothing) = 1$ when the sequence has
length of zero.

Equation \ref{eq:coord} can be merge into one line:

\begin{equation}
x(\pi_1...\pi_{k}) = 2\cdot x(\pi_1...\pi_{k-1}) + \pi_k - 2 ,
\notag
\end{equation}

\noindent and we can solve it to:

\begin{equation}
x(\pi_1...\pi_k) = 1 + \sum_{i=1}^k 2^{k-i}(\pi_i - 1)
\notag
\end{equation}

\noindent which has the same form as Equation \ref{eq:integer}. Thus the value of
$x(\pi_1...\pi_k)$ is identical to the integer representation of the curve,
i.e., $\delta$.

Then it is easy to see the entry point of $Z|(1)_k$ has a value $x = 1$, for $Z|(2)_k$ 
it has a value of $x = 2^k$, and for $Z|(\pi)_k$ it has a value of $x = \delta$. 
Thus attributes 1 and 2 are both true.

\end{proof}

Lemma \ref{lemma:geometry_z} only includes primary bases associated with
rotations of zero. For the curve initialized from any of the nine bases,
we have the following more general proposition.

\begin{proposition}
\label{prop:geometry_entry}
For a curve $\mathcal{P}_k = X|(\pi)_k = \mathcal{P}_1|\pi_2...\pi_k$ ($k \ge
2$), let's write $\mathcal{P}_k$ as a list of four subunits on level $k-1$
denoted as $\mathcal{P}_k =
\mathcal{U}_1\mathcal{U}_2\mathcal{U}_3\mathcal{U}_4$. If the level-1
expansion of $X$ is $Z_1 Z_2 Z_3 Z_4$\footnote{Note rotations are implicitly
included.}, then $\mathcal{U}_1 = Z_1|\pi_2...\pi_k$. There are the following
geometric attributes for the entry point and direction on $\mathcal{U}_1$:

\begin{enumerate}
\tightlist
\item
  When $\pi_2...\pi_k = (1)_{k-1}$, the entry point of $\mathcal{U}_1$ is
  located on a corner denoted as $a_1$, and when $\pi_2...\pi_k = (2)_{k-1}$,
  the entry point is located on the neighbouring corner of $a_1$ denoted as
  $a_2$.
\item
  Entry point of $\mathcal{U}_1$ is always located on the side determined by
  $a_1$ and $a_2$. For the integer representation $\pi_2...\pi_k \mapsto
  \delta^{(k-1)}$, $\delta-1$ is the distance to $a_1$.
\item
  The entry direction of $\mathcal{U}_1$ is orthogonal to the side determined
  by $a_1$ and $a_2$, and it comes from the outside of $\mathcal{U}_1$.
\end{enumerate}

\end{proposition}

\begin{proof}
Note $Z_1$ is from a level-1 expansion, thus $Z_1 \in \{I, R, L\}$
(Proposotion \ref{prop:IRL}). If $Z_1$ is associated with a rotation of
$\theta$, rotating the curve won't change the three attributes, where we can
simply rotate the curve by $-\theta$ to let $Z_1$ explicitly be $Z_1^0$, then
we can simply apply Lemma \ref{lemma:geometry_z} to prove it.
\end{proof}

We have similar attributes for the exit point of the curve:

\begin{corollary}
\label{coro:geometry_exit}
Using the same notations as in Proposition \ref{prop:geometry_entry}, there are
the following geometric attributes for the exit point and direction on
$\mathcal{U}_4$:

\begin{enumerate}
\tightlist
\item
  When $\pi_2...\pi_k = (1)_{k-1}$, the exit point of $\mathcal{U}_4$ is
  located on a corner of the curve denoted as $b_1$, and when $\pi_2...\pi_k =
  (2)_{k-1}$, the exit point is located on the neighbouring corner of the
  curve of $b_1$ on $\mathcal{U}_4$ denoted as $b_2$.
\item
  Exit point of $\mathcal{U}_4$ is always located on the side determined by
  $b_1$ and $b_2$. For the integer representation of $\pi_2...\pi_k \mapsto
  \delta^{(k-1)}$, $\delta-1$ is also the distance of its exit point to $b_1$
  on $\mathcal{U}_4$.
\item
  The exit direction of $\mathcal{U}_4$ is orthogonal to the side determined
  by $b_1$ and $b_2$, and it points to the outside of $\mathcal{U}_4$.
\end{enumerate}

\end{corollary}

\begin{proof}
Let's take the reversal of $\mathcal{P}_k$ denoted as $\mathcal{Q}_k$. 

\begin{equation}
\begin{aligned}
\mathcal{Q}_k &= r(\mathcal{P}_k) \\
     &= r(Z_1 Z_2 Z_3 Z_4|\pi_2...\pi_k) \\
     &= r( (Z_1|\pi_2...\pi_k)...(Z_4|\pi_{2*}...\pi_{k*}) ) \\
     &= r(Z_4|\pi_{2*}...\pi_{k*})...r(Z_1|\pi_2...\pi_k) \\
\end{aligned}
\notag
\end{equation}

In Line 3, we move the expansion code sequence to each of $Z_1$ to $Z_4$ where
$\pi_{i*}$ represents the code has not been solved yet, and in this proof its
value is not necessarily to be known. In Line 4, Reversing the whole sequence
is changed to reversing each of the four level $k-1$ subunits separately
(Proposition \ref{prop:subsequence}).

Let's write $\mathcal{V}_1 = r(Z_4|\pi_{2*}...\pi_{k*})$ as the first subunit
of $\mathcal{Q}_k$, then the encoding of $\mathcal{V}_1$ can be written as:

\begin{equation}
\begin{aligned}
\mathcal{V}_1 &= r(Z_4|\pi_{2*}...\pi_{k*}) \\
   &= Z'_4|\pi^\#_{2*}...\pi^\#_{k*} \\
\end{aligned} .
\notag
\end{equation}

Since $Z_4$ is from a level-1 extension of the seed base, $Z_4 \in \{I, R,
L\}$. Then according to Corollary \ref{coro:5.6.1}, the code sequence
$\pi_{2*}...\pi_{k*}$ is $\pi_2...\pi_k$ or $\hat{\pi}_2...\hat{\pi}_k$, and
in turn $\pi^\#_{2*}...\pi^\#_{k*}$ is also either $\pi_2...\pi_k$ or
$\hat{\pi}_2...\hat{\pi}_k$ (Corollary \ref{coro:rev}).

Note $\mathcal{V}_1$ is the reversal of $\mathcal{U}_4$, thus the entry point
of $\mathcal{V}_1$ is the exit point of $\mathcal{U}_4$. According to
Proposition \ref{prop:geometry_entry}, the following three statements are true
for $\mathcal{V}_1$ (we write the equivalent description for $\mathcal{U}_4$
in the parentheses):

\begin{enumerate}
\tightlist
\item
  When $\pi^\#_{2*}...\pi^\#_{k*} = (1)_{k-1}$, the entry point of
  $\mathcal{V}_1$ (the exit point of $\mathcal{U}_4$) is located on a corner
  denoted as $a_1$, and when $\pi^\#_{2*}...\pi^\#_{k*} = (2)_{k-1}$, the
  entry point (the exit point of $\mathcal{U}_4$) is located on the
  neighbouring corner of $a_1$ denoted as $a_2$.
\item
  Entry point of $\mathcal{V}_1$ (exit point of $\mathcal{U}_4$) is always
  located on the side determined by $a_1$ and $a_2$. For the integer
  representation of $\pi^\#_{2*}...\pi^\#_{k*} \mapsto \delta^{(k-1)}$,
  $\delta-1$ is the distance to $a_1$.
\item
  The entry direction of $\mathcal{V}_1$ (the exit direction of
  $\mathcal{U}_4$) is orthogonal to the side determined by $a_1$ and $a_2$,
  and it comes from the outside of $\mathcal{V}_1$ ($\mathcal{U}_4$).
\end{enumerate}

Since $\pi^\#_{2*}...\pi^\#_{k*}$ takes two possible values, let's discuss
them separately.

\textit{Scenario 1}: $\pi^\#_{2*}...\pi^\#_{k*} = \pi_2...\pi_k$. This
results in the above three statements the same as in this corollary if
taking $b_1 = a_1$ and $b_2 = a_2$.

\textit{Scenario 2}: $\pi^\#_{2*}...\pi^\#_{k*} = \hat{\pi}_2...\hat{\pi}_k$.
When $\pi_2...\pi_k = (1)_{k-1}$, then $\pi^\#_{2*}...\pi^\#_{k*} =
(2)_{k-1}$, which indicates $b_1 = a_2$. Similarly there is also $b_2 = a_1$.
Let the integer representation of $\pi_2...\pi_k$ be $\mu^{(k-1)}$. With
$\pi^\#_{2*}...\pi^\#_{k*} = \hat{\pi}_2...\hat{\pi}_k \mapsto
\delta^{(k-1)}$, we have $\mu = 2^{k-1} - \delta + 1$. Note $\delta - 1$ is
the distance to $a_1$/$b_2$, thus $\mu - 1 = 2^{k-1} - \delta$ is the distance
to $a_2$/$b_1$.

Attribute 3 is already proven in the equivalent text.

\end{proof}

A visualization that illustrates Proposition \ref{prop:geometry_entry}
and Corollary \ref{coro:geometry_exit} are in Figure
\ref{fig:expansion_code_location} (right panel).

\begin{remark}
Proposition \ref{prop:geometry_entry} only depends on the first subunit of
$\mathcal{P}_k$, thus Proposition \ref{prop:geometry_entry} can be extended to
a curve initialized from a seed sequence. Corollary \ref{coro:geometry_exit}
can also be extended to a curve intialized by a seed sequence, where we just
need to change the term ``$\mathcal{U}_4$'' to the ``last subunit''
in the statement.
\end{remark}

\begin{remark}
\label{remark:ee_side}
Entry points can only be located on the sides of the first subunit (including
corners) and exit points can only be located on the sides of the last subunit
of $\mathcal{P}_k$. In other words, entry and exit points cannot be located
inside the first and the last subunits.
\end{remark}

\subsection{Subunits}\label{subunits}

In the previous section, we have discussed the entry and the exit
points, but treating them separately. In this section we discuss how
they are linked on the curve (level \(\ge 2\)) via subunits.

\begin{property}
\label{prop:homo3}
The entry direction of $\mathcal{U}_1$ cannot be the reversal of its exit
direction. Similarly, the exit direction of $\mathcal{U}_4$ cannot be the
reversal of its entry direction.
\end{property}

\begin{proof}
According to Proposition \ref{prop:entry-exit-directions}, the entry and exit
directions of $\mathcal{U}_1$ are the same as $Z_1$. Since $Z_1 \in
\{I, R, L\}$, the entry direction cannot be the reversal of its exit
direction, thus so is for $\mathcal{U}_1$. Using the same method we can prove
the exit direction of $\mathcal{U}_4$ cannot be the reversal of its entry
direction.
\end{proof}

\begin{property}
\label{prop:homo1}
If the entry point is located on the corner of $\mathcal{U}_1$ which does not
attach $\mathcal{U}_2$, there are two possible choices of entry direction on
$\mathcal{U}_1$; if the entry point is located on the corner of
$\mathcal{U}_1$ which attaches $\mathcal{U}_2$, there is only one possible
entry direction on $\mathcal{U}_1$; if the entry point is not located on the
corner of $\mathcal{U}_1$, there is only one possible entry direction on
$\mathcal{U}_1$. Such property is the same for the exit point and exit
direction on $\mathcal{U}_4$.
\end{property}

\begin{proof}
According to Proposition \ref{prop:geometry_entry}, the entry direction is
orthogonal to the side of $\mathcal{U}_1$ where the entry point is located,
also the entry direction should come from the outside of $\mathcal{U}_1$. So
when the entry point is located on the corner of $\mathcal{U}_1$, there are
two sides associated with it, then possibly having two choices of entry
directions. However, according to Property \ref{prop:homo3}, when the entry
point is located on the corner which attaches $\mathcal{U}_2$, one of the two
possible entry directions which points from $\mathcal{U}_2$ is invalid because
it is a reversal of the exit direction of $\mathcal{U}_1$ (Property
\ref{prop:homo3}). When the entry point is not located on the corner of
$\mathcal{U}_1$, there is only one side for it, thus only one possible entry
direction.

With Corollary \ref{coro:geometry_exit}, we know the entry point has the same
location type as the exit point (i.e., whether it is located on the corner),
then using the same method, we can prove for the exit point and direction on
$\mathcal{U}_4$.
\end{proof}

\begin{property}
\label{prop:homo11}
The entry point can not be located on the side of $\mathcal{U}_1$ where
$\mathcal{U}_1$ and $\mathcal{U}_2$ attach (excluding the two corners of that
side). Similarly, the exit point cannot be located on the side $\mathcal{U}_4$
where $\mathcal{U}_4$ and $\mathcal{U}_3$ attach.
\end{property}

\begin{proof}
If the entry point is located on the side of $\mathcal{U}_1$ where
$\mathcal{U}_1$ and $\mathcal{U}_2$ attach, denoted as $a$, then there is only
one possible entry direction $d_1$ which is orthogonal to $a$. The entry
direction of $\mathcal{U}_2$ is also orthogonal to $a$, which makes the exit
direction of $\mathcal{U}_1$ denoted as $d_2$ is orthogonal to $a$ as well.
According to Property \ref{prop:homo3}, such scenario is not allowed. Thus the
entry point is not allowed to be located on $a$.
\end{proof}

We have defined the corners of a 2x2 unit in Section \ref{corners}.
Let's extend it to the general square units where the lower left and
upper right corners have a value of 1 and the lower right and upper left
corners have a value of 2. We first prove the following lemma:

\begin{lemma}
\label{lemma:entry_exit}
For a curve initialized by a primary base $Z$, if the entry point is located
on a corner of the curve, the exit point is located on its neighbouring corner
on the curve.
\end{lemma}

\begin{proof}
If $Z$ is associated with a rotation $\theta$, we rotate it by $-\theta$ to
let $Z$ be associated with zero rotation because rotation does not affect the
statement.

The entry point is located on the corner of the curve, implying the curve has
the encoding $Z|(1)_k$ or $Z|(2)_k$ (Lemma \ref{lemma:geometry_z}). Let's only
consider the scenario of $Z|(1)_k$. If the curve is $Z|(2)_k$, it can be
horizontally reflected to switch all expansion code to 1 (Proposition
\ref{prop:6.5}), and horizontally reflecting a primary base is still a primary
base. Reflection does not affect the statement in this lemma.

Then for the curve $\mathcal{P}_k = Z|(1)_{k}$, entry point has a coordinate
of $(1, 1)$ (Lemma \ref{lemma:geometry_z}). Exit point of $\mathcal{P}_k$ is
the entry point of its reversed curve $r(\mathcal{P}_k)$. Then, if $Z = I$,
$r(\mathcal{P}_k) = I^{180}|(2)_k$ (Equation \ref{eq:rev_base}, Corollary
\ref{coro:rev}). With Corollary \ref{coro:rev}, we know the coordinate of the
entry point of $I|(2)_k$ is $(2^k, 1)$. Then rotating $I|(2)_k$ by 180
degrees, we have the
coordinate of the entry point of $r(\mathcal{P}_k) = r(\mathcal{P}_k)$ as $(1, 2^k)$.

If $Z = R$, $r(\mathcal{P}_k) = L^{90}|(1)_k$ (Equation \ref{eq:rev_base},
Corollary \ref{coro:rev}). With Proposition \ref{prop:6.5}, we rewrite
$r(\mathcal{P}_k) = h(R^{-90}|(2)_k)$. Then the coordinate for the entry point
of $R|(2)_k$ is $(2^k, 1)$, for $R^{-90}|(2)_k$ is $(1, 1)$ and for
$r(\mathcal{P}_k) = h(R^{-90}|(2)_k)$ is $(2^k, 1)$.

If $Z = L$, $r(\mathcal{P}_k) = R^{-90}|(1)_k$ (Equation \ref{eq:rev_base},
Corollary \ref{coro:rev}). With Proposition \ref{prop:6.5}, we rewrite
$r(\mathcal{P}_k) = h(L^{90}|(2)_k)$. Then the coordinate for the entry point
of $L|(2)_k$ is $(2^k, 1)$, for $L^{90}|(2)_k$ is $(2^k, 2^k)$ and for
$r(\mathcal{P}_k) = h(L^{90}|(2)_k)$ is $(1, 2^k)$.

To summarize, when $Z = I$ or $L$, the coordinate of the exit corner is $(1,
2^k)$ which is the neighbouring corner of the entry corner and they determine
the left side of the curve. When $Z = R$, the coordinate of the exit corner is
$(2^k, 1)$ which is the neighbouring corner of the entry corner and they
determine the bottom side of the curve.

\end{proof}

\begin{property}
\label{prop:homo2}
If the entry corner has a value of $c$ on $\mathcal{U}_1$, the exit corners
$\mathcal{U}_1$ and $\mathcal{U}_4$ all have corner values of $\hat{c}$.
\end{property}

\begin{proof}
$\mathcal{U}_1$ is initialized by a primary base, according to Lemma
\ref{lemma:entry_exit}, the exit point is located on the neighbouring corner
of $\mathcal{U}_1$. Thus the exit corner of $\mathcal{U}_1$ has a corner value
of $\hat{c}$.

No matter $\mathcal{U}_2$ connects to $\mathcal{U}_1$ horizontally or
vertically, the entry point of $\mathcal{U}_2$ has an entry corner with a
value of $c$. $\mathcal{U}_2$ is also initialized by the primary base, thus
the exit corner of $\mathcal{U}_2$ is $\hat{c}$. Then finally we can have the
entry corner of $\mathcal{U}_4$ has a value of $c$ and the exit corner of
$\mathcal{U}_4$ has a value of $\hat{c}$.
\end{proof}

\begin{remark}
If the entry point is located on the corner of $\mathcal{U}_1$, we call the
curve a ``\textit{corner-induced curve}'', or else the curve is called a
``\textit{side-induced curve}''. We use this terminology throughout the
next sections.
\end{remark}

\subsection{Entry and exit points uniquely determine the
curve}\label{unique}

For \(2^k \times 2^k\) (\(k \ge 1\)) grids of points that will be
traversed by a curve on level \(k\), split the square region into four
equal quadrants. Let the quadrant where the entry point is located be
subunit 1 (\(\mathcal{U}_1\)) and the quadrant where the exit point is
located be subunit 4 (\(\mathcal{U}_4\)) which should be a neighbouring
quadrant of \(\mathcal{U}_1\). Then the other neighbouring quadrant of
\(\mathcal{U}_1\) is set to subunit 2 (\(\mathcal{U}_2\)) and the
diagonal quadrant of \(\mathcal{U}_1\) is set to subunit 3
(\(\mathcal{U}_3\)).

\begin{proposition}
\label{prop:unique}

The curve (level $\ge$ 1) is determined if the following information of the
entry and exit points is provided:

\begin{enumerate}
\tightlist
\item
   The location of the entry point. According to Remark \ref{remark:ee_side},
   the entry point can only be located on the sides of $\mathcal{U}_1$. Also
   it cannot be located on the side where $\mathcal{U}_1$ and $\mathcal{U}_2$
   attach (excluding the two end points of this side, Property \ref{prop:homo11}).
\item
   The entry direction. If the entry point is located on the corner of
   $\mathcal{U}_1$ which does not attach $\mathcal{U}_2$, then an entry
   direction must be pre-selected. If the entry point is located on the corner
   of $\mathcal{U}_1$ which attaches $\mathcal{U}_2$ or it is located on the
   side of $\mathcal{U}_1$, according to Property \ref{prop:homo1}, the entry
   direction is uniquely determined.
\item
   The exact location of the exit point is not needed. Only the side on
   $\mathcal{U}_4$ where the exit point is located is needed.
\item
   The exit direction. If the entry point is located on the corner of
   $\mathcal{U}_1$ which does not attach $\mathcal{U}_2$, this determines the
   exit point being located on the corner of $\mathcal{U}_4$ which does not
   attach $\mathcal{U}_3$, then an exit direction on $\mathcal{U}_4$ must also
   be pre-selected.
\end{enumerate}

\end{proposition}

\begin{figure}
\centering{
\includegraphics[width=0.8\linewidth]{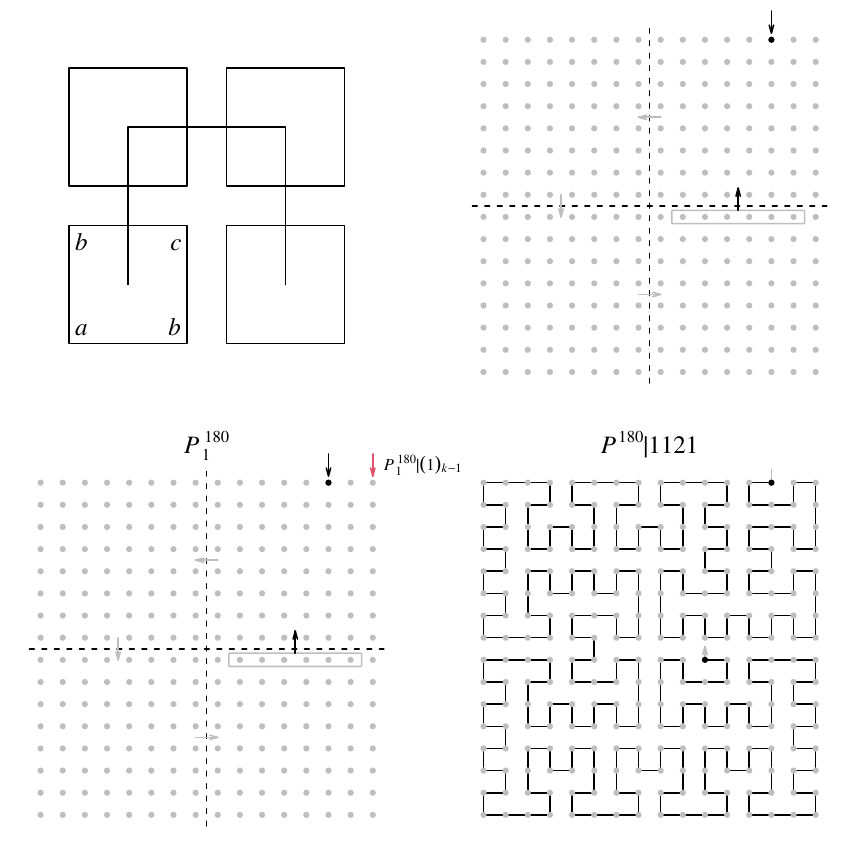}
\caption{Infer curve encoding from the entry point and the exit side.}\label{fig:curve_from_entry}
}
\end{figure}

\begin{proof}
The proof also serves as a process to determine the encoding of the curve.
First the level $k$ of the curve can be known from the dimension of the grids
of points. When $k = 1$, the encoding of $\mathcal{P}_k$ can be directly
looked up from Figure \ref{fig:expansion_rule}.

For the curve $\mathcal{P}_k = X|\pi_1...\pi_k$ ($k \ge 2$), as the entry and
exit directions, as well as the quadrants of the four subunits are all
determined, we reduce each subunit to single point to obtain the exact form of
$\mathcal{P}_1 = X|\pi_1$.

Notice the entry point is also located on $\mathcal{U}_1$, we first prepare a
table (Table \ref{tab:corner-type}) of the entry corners on $\mathcal{U}_1$
for all possible types of curves in the form of $X|\pi_1(1)_{k-1}$. We
categorize the entry corners of $\mathcal{U}_1$ into three types: a, b, and c
(Figure \ref{fig:curve_from_entry}, the first panel), where type-a corresponds
to the corners of the complete square curve, type-b corresponds to the middle side of
the square, and type-c corresponds to the inside of the square.

\begin{table}[ht]
\centering
\begin{tabular}{lc|lc|lc|lc}
\toprule
Curve & Type & Curve & Type & Curve & Type & Curve & Type \\
\midrule
$I|1(1)_{k-1}$ & a & $I|2(1)_{k-1}$ & b & $R|1(1)_{k-1}$ & a & $R|2(1)_{k-1}$ & b \\
$L|1(1)_{k-1}$ & a & $L|2(1)_{k-1}$ & b & $U|1(1)_{k-1}$ & a & $U|2(1)_{k-1}$ & b \\
$B|1(1)_{k-1}$ & c & $B|2(1)_{k-1}$ & b & $D|1(1)_{k-1}$ & b & $D|2(1)_{k-1}$ & c \\
$P|1(1)_{k-1}$ & a & $P|2(1)_{k-1}$ & b & $Q|1(1)_{k-1}$ & a & $Q|2(1)_{k-1}$ & b \\
$C|1(1)_{k-1}$ & b & $C|2(1)_{k-1}$ & c &   &   &   &  \\
\bottomrule
\end{tabular}
\vspace*{5mm}
\caption{\label{tab:corner-type}Entry corner types on $\mathcal{U}_1$. Rotations on base seeds are omitted for simplicity.}
\end{table}

Since we have already had the form of $\mathcal{P}_1 = X|\pi_1$, we look up in
Table \ref{tab:corner-type} to obtain the entry corner type of its
corresponding curve $\mathcal{Q}_k = X|\pi_1(1)_{k-1}$. With proposition
\ref{prop:geometry_entry}, we know the entry point of $\mathcal{P}_k$ denoted
as $a$ and the entry point of $\mathcal{Q}_k$ denoted as $p$ are located on
the same side of $\mathcal{U}_1$. With knowing the type of entry corner of
$\mathcal{Q}_k$ on its first subunit, the exact location of $p$ is determined (i.e. on the left or the right of $a$).
According to Proposition \ref{prop:geometry_entry}, the distance between $a$
and $p$ denoted as $d$ has the relation $d = \delta - 1$ where $\delta$ is the
integer representation of the coding sequence $\pi_2...\pi_k$. Then
$\mathcal{P}_k$ is fully determined.
\end{proof}

Figure \ref{fig:curve_from_entry} illustrates an example of the process
of identifying the curve encoding from its entry and exit points. There
are \(16
\times 16 = 2^4 \times 2^4\) grids of points, thus the level of
\(\mathcal{P}_k\) is 4. The entry point has a location of \((14, 16)\)
(we assume the lower left corner of the whole grids has a coordinate of
\((1, 1)\)), and has an entry direction of vertically top-in. The exit
point is located on the top side of \(\mathcal{U}_4\) with an exit
direction vertically top-out.

We reduce \(\mathcal{P}_k\) to \(\mathcal{P}_1\) and we know it is
\(P^{180}_1\) by looking up in Figure \ref{fig:expansion_rule}.
According to Table \ref{tab:corner-type}, \(\mathcal{U}_1\) of the curve
\(P^{180}_1|(1)_{k-1}\) has a type-a corner, which is highlighed by a
red arrow in Figure \ref{fig:curve_from_entry} with the coordinate of
\((16, 16)\). The distance of this corner to the entry point of
\(\mathcal{P}_k\) is \(16 - 14 = 2\), meaning \(\delta = 3\), thus
\(\pi_2...\pi_k = 121\), and finally \(\mathcal{P}_k =
P^{180}|1121\).

From Proposition \ref{prop:unique}, for side-induced curves, a unique
combination of entry point and exit side uniquely determine the curve.
Each side of a subunit have \(2^k - 2\) side-points. There are three
sides for the entry point on \(\mathcal{U}_1\) and three sides on
\(\mathcal{U}_4\). In the next section, we will demonstrate there are 18
corner-induced curves when the orientation of the four subunits are
fixed. Then we add up the numbers of side-induced curves (\(2^k - 2\))
and corner-induced curves (18), multiply by two types of reflections
(for clockwise and counterclockwise orientations) and four rotations.
The total number of different curves is:

\begin{equation}
((2^k - 2) \times 3 \times 3 + 18) \times 2 \times 4 = 36 \times 2^k
\notag
\end{equation}

\noindent which is also the total number of all forms of 2x2 curves
(Equation \ref{eq:total_number}).

Last, the followng equation calculates the code sequence
\(\pi_1...\pi_k\) from its integer representation \(\delta^{(k)}\)
(i.e., the reverse of Equation \ref{eq:integer}):

\begin{equation}
\begin{aligned}
\pi_k &= \left\lceil \delta / 2^{k-1} \right\rceil \\
\pi_{i} &= \left\lceil \left(\delta - \sum_{j=i}^{k-1} (\pi_{j+1}-1) \cdot 2^{j} \right) \bigg/ 2^{i-1} \right\rceil \quad \text{if } 1 \le i \le k-1 \\
\end{aligned} .
\end{equation}

\section{Homogeneous curves and shapes}\label{shapes}

In Section \ref{the-encoding-system}, we have demonstrated there are
\(36
\times 2^k\) different forms of 2x2 curves on level \(k\) initialized by
a single base, which distinguishes curves with different entry and exit
directions. However, in many current studies, the entry and exit
directions of the curve are ignored, which results in curves with the
same forms but encoded differently by our system, such as \(R_{<1>}\)
and \(I^{270}_{<2>}\) which both correspond to level-1 ``U-shape'' unit
facing bottom, starting from the lower left and ending at the lower
right. Some scenarios even treat the curves undirectional and also
ignore rotations and reflections of curves, which yields more curves
with identical shapes. In this section, we will explore families of
curves which have identical, similar or distinct structures if ignoring
their entry and exit directions, orientations, or transformations. We
only consider curves induced from a single seed base.

\subsection{Homogeneous curves}\label{homogeneous}

\begin{definition}[Homogeneous curves]
\label{def:homo}
Two curves are \textit{homogeneous} when they are only differed by their
entry or exit directions.
\end{definition}

The definition implies two homogeneous curves have the same locations of
entry and exit points, and the same path connecting them.

\begin{property}
\label{prop:homo_1}
If we express two curves $\mathcal{P}$ and $\mathcal{Q}$ as two base sequences

\begin{equation}
\begin{aligned}
\mathcal{P} &= X_1 X_2 ... X_{n-1} X_n \\
\mathcal{Q} &= Y_1 Y_2 ... Y_{n-1} Y_n \\
\end{aligned} ,
\notag
\end{equation}

\noindent $\mathcal{P}$ and $\mathcal{Q}$ are homogeneous iff $X_i = Y_i$ ($2
\le i \le n-1$) (implicitly associated rotations of $X_i$ and $Y_i$ are also
identical).
\end{property}

\begin{proof}
It is by definition that if $\mathcal{P}$ and $\mathcal{Q}$ are homogeneous,
then $X_i = Y_i$ ($2
\le i \le n-1$).

Next if $X_i = Y_i$ ($2 \le i \le n-1$), notice the second base in a sequence
has an entry direction which determines the location of the first base, then
with $X_2 = Y_2$, the exit directions and locations of $X_1$ and $Y_1$ are
identical. Similarly, the last second base in a sequence has an exit direction
which determines the location of the last base, then with $X_{n-1} = Y_{n-1}$,
the entry directions and locations of $X_n$ and $Y_n$ are also identical.
Thus, $\mathcal{P}$ and $\mathcal{Q}$ are homogeneous.
\end{proof}

When the curve is on level 0, it is represented as a single base. If the
entry and exit directions are ignored for the base, the curve is
degenerated into a single point. Thus all level-0 curves are
homogeneous.

When the curve is on level 1, we rotate all level-1 units to let them
face bottom. Then ignoring the entry and exit directions, there are two
families of homogeneous curves, one in the clockwise orientation and the
other in the counterclockwise orientation. Also considering the four
rotations, there are in total \(2 \times 4 = 8\) families of homogeneous
curves on level 1.

\begin{figure}
\centering{
\includegraphics[width=1\linewidth]{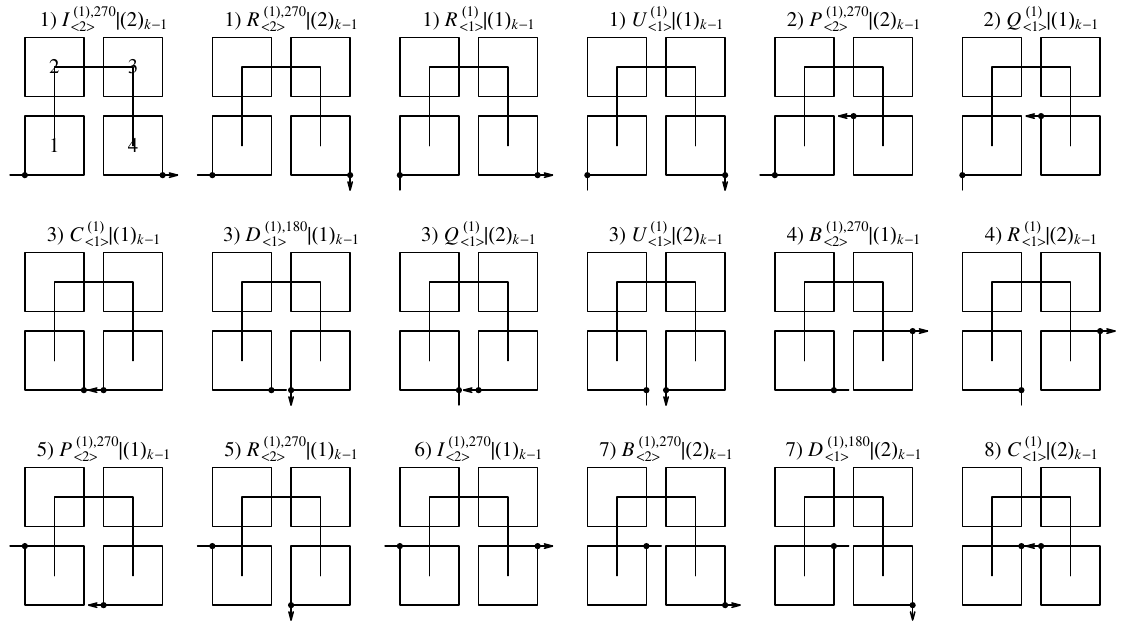}
\caption{Eight families of corner-induced curves. Each curve is encoded using level-1 unit as the seed with $k-1$ expansions.}\label{fig:homogeneous_curves}}
\end{figure}

A curve \(\mathcal{P}_k\) (\(k \ge 2\)) is composed of four subunits on
level \(k-1\) taking \(\mathcal{P}_1\)
(\(\mathcal{P}_1 = Z_1 Z_2 Z_3 Z_4\)) as its global level-1 structure.
We denote the four subunits as \(\mathcal{U}_1\), \(\mathcal{U}_2\),
\(\mathcal{U}_3\) and \(\mathcal{U}_4\). For the convenience of
discussion in the remaining sections of this article, we only consider
curves in the following state:

\begin{definition}
If $\mathcal{U}_1$, $\mathcal{U}_2$, $\mathcal{U}_3$ and $\mathcal{U}_4$ are
located in an order of lower left, upper left, upper right and lower right of
the square, $\mathcal{P}_k$ is called on \textit{the base facing state}, i.e.,
clockwise and facing downward (e.g., the first panel in Figure
\ref{fig:homogeneous_curves}).
\end{definition}

Homogeneous curves only have different entry or exit directions, then
according to Property \ref{prop:homo1}, they can only be corner-induced
curves. Property \ref{prop:homo1} implies the two lower corner of
\(\mathcal{U}_1\) can be associated with two types of entry directions
(horizontal and vertical), while the two upper corners can only be
associated with one type of entry direction (horizontal). Similarlly,
the two lower corners of \(\mathcal{U}_4\) can be associated with two
types of exit directions, and the two upper corners can only be
associated with one type of exit direction. Additionally, Property
\ref{prop:homo2} requires the entry corner and the exit corner should
have different corner values.

Now we can enumerate all corners on \(\mathcal{U}_1\) and
\(\mathcal{U}_4\), and all their valid combinations of entry and exit
corners and directions. Table \ref{tab:homo} and Figure
\ref{fig:homogeneous_curves} list the complete set of forms of curves in
the base facing states that satisfy the conditions in the previous
paragraph. These forms are classified into 8 families based on the
locations of the entry and exit points.

\begin{table}
\centering
\begin{tabular}{ccccccc}
\toprule
Family & \multicolumn{3}{c}{$\mathcal{U}_1$} & \multicolumn{3}{c}{$\mathcal{U}_4$} \\
\cmidrule(lr){2-4}\cmidrule(lr){5-7}
   & \makecell{Entry \\ location} & \makecell{Corner \\ value} & \makecell{Entry \\direction} & \makecell{Exit \\ location} & \makecell{Corner \\ value} & \makecell{Exit \\ direction} \\
\midrule
1 & lower left & 1 & h/v & lower right & 2 & h/v \\
2 & lower left & 1 & h/v & upper left & 2 & h  \\
3 & lower right & 2 & h/v & lower left & 1 & h/v  \\
4 & lower right & 2 & h/v & upper right & 1 & h  \\
5 & upper left & 2 & h & lower left & 1 & h/v  \\
6 & upper left & 2 & h & upper right & 1 & h  \\
7 & upper right & 1 & h & lower right & 2 & h/v  \\
8 & upper right & 1 & h & upper left & 2& h  \\
\bottomrule
\end{tabular}
\vspace*{5mm}
\caption{\label{tab:homo}Combinations of entry and exit locations of corner-induced curves. h: horiozntal; v: vertical.}
\end{table}

According to Property \ref{prop:geometry_entry}, corner-induced curves
have the same form of encoding
\(\mathcal{P}_k = \mathcal{P}_1|(a)_{k-1}\) (\(a \in
\{1,2\}\), i.e., from the second code are all the same). For curves in
the 8 families, the encoding of \(\mathcal{P}_1\) can be easily obtained
by reducing the four subunits to single points for curves in each family
in Figure \ref{fig:homogeneous_curves}, i.e.,
\(\mathrm{Rd}_{k-1}(\mathcal{P}_k)
=\mathcal{P}_1\) (Section \ref{reduction}). The form of each
\(\mathcal{P}_1\) is listed in the title of each curve in Figure
\ref{fig:homogeneous_curves} as well as in Table
\ref{tab:homogeneous_curves}.

Next we solve \(a\). We explicitly add the rotation to \(Z_1\), writing
\(\mathcal{P}_1\) = \(Z_1^{\theta^{(1)}} ...\) where we only consider
its first base. According to the reduction process, the entry corner of
\(\mathcal{U}_1\) is the same as the entry corner of its reduction
\(\mathrm{Rd}_{k-2}(\mathcal{U}_1) = Z_1^{\theta^{(1)}}|a\). Denote the
corner value of the entry point of \(\mathcal{U}_1\) as \(c_1\), then
the corner value of the entry point of \(Z_1^{\theta^{(1)}}|a\) is also
\(c_1\). According to Section \ref{expansion-code}, \(a\) is the entry
corner value of \(Z_1^0|a\) (rotation of zero), then with Equation
\ref{eq:tao}, the solution of \(a\) is

\begin{equation}
a = \begin{cases}
c_1 \quad & \text{if } \theta^{(1)} \bmod 180 = 0 \\
\hat{c}_1 \quad & \text{if } \theta^{(1)} \bmod 180 = 90 \\
\end{cases} .
\end{equation}

Following these calculations, the exact encodings of all corner-induced
curves are listed in Figure \ref{fig:homogeneous_curves} as well as in
Table \ref{tab:homogeneous_curves}. By applying horizontal reflection
(Proposition \ref{prop:6.5}), the reflected versions of the eight
families are also listed in Table \ref{tab:homogeneous_curves}.

\begin{table}
\centering
\renewcommand{\arraystretch}{1.2}
\begin{tabular}{clcc|l|l}
\toprule
 Family & $\mathcal{P}_1$ & $c_1$ & $a$ & $\mathcal{P}_k$ & $h(\mathcal{P}_k)$ \\
\midrule
1 & $I^{(1),270}_{<2>} = L^{270}...$ & 1 & 2 & $I^{270}|(2)_k$ & $I^{90}|(1)_k$   \\
  & $R^{(1),270}_{<2>} = L^{270}...$ & 1 & 2 &$R^{270}|(2)_k$ & $L^{90}|(1)_k$ \\
  & $R^{(1)}_{<1>} = I...$ & 1& 1 & $R|(1)_k$   & $L|(2)_k$ \\
  & $U^{(1)}_{<1>} = I...$ & 1& 1 & $U|(1)_k$   & $U|(2)_k$ \\
\midrule
2 & $P^{(1),270}_{<2>} = L^{270}...$ & 1 & 2 & $P^{270}|(2)_k$ & $P^{90}|(1)_k$  \\
  & $Q^{(1)}_{<1>} = I...$ & 1 & 1 & $Q|(1)_{k}$ & $Q|(2)_{k}$ \\
\midrule
3 & $C^{(1)}_{<1>} = R^{90}...$  & 2& 1 & $C|(1)_{k}$    & $C|(2)_{k}$ \\
  & $D^{(1),180}_{<1>} = R^{90}...$  & 2& 1 & $D^{180}|(1)_{k}$ & $D^{180}|(2)_{k}$ \\
  & $Q^{(1)}_{<1>} = I...$  & 2& 2 & $Q|1(2)_{k-1}$   & $Q|2(1)_{k-1}$ \\
  & $U^{(1)}_{<1>} = I...$ & 2 & 2 & $U|1(2)_{k-1}$   & $U|2(1)_{k-1}$ \\
\midrule
4 & $B^{(1),270}_{<2>} = R^{90}...$ & 2 & 1 & $B^{270}|2(1)_{k-1}$ & $B^{90}|1(2)_{k-1}$ \\
  & $R^{(1)}_{<1>} = I...$ & 2 & 2 & $R|1(2)_{k-1}$      & $L|2(1)_{k-1}$  \\
\midrule
5 & $P^{(1),270}_{<2>} = L^{270}...$ & 2 & 1 & $P^{270}|2(1)_{k-1}$ & $P^{90}|1(2)_{k-1}$ \\
  & $R^{(1),270}_{<2>} = L^{270}...$ & 2 & 1 & $R^{270}|2(1)_{k-1}$ &  $L^{90}|1(2)_{k-1}$  \\
\midrule
6 & $I^{(1),270}_{<2>} = L^{270}...$ & 2 & 1 & $I^{270}|2(1)_{k-1}$ & $I^{90}|1(2)_{k-1}$ \\
\midrule
7 & $B^{(1),270}_{<2>} = R^{90}...$ & 1 & 2 & $B^{270}|(2)_k$ & $B^{90}|(1)_{k}$ \\
  & $D^{(1),180}_{<1>} = R^{90}...$ & 1 & 2 & $D^{180}|1(2)_{k-1}$ &  $D^{180}|2(1)_{k-1}$ \\
\midrule
8 & $C^{(1)}_{<1>} = R^{90}...$ & 1 & 2& $C|1(2)_{k-1}$ & $C|2(1)_{k-1}$ \\
\bottomrule
\end{tabular}
\vspace*{5mm}
\caption{\label{tab:homogeneous_curves}Families of corner-induced curves. $\mathcal{P}_1$: the base structure; $c_1$: the first corner value of subunit 1; $a$: expansion code from the second expansion; $\mathcal{P}_k$: the entire curve; $h(\mathcal{P}_k)$: horizontal reflection of $\mathcal{P}_k$.}
\end{table}

The classification in Table \ref{tab:homo} and
\ref{tab:homogeneous_curves} is only based on the locations of entry and
exit points. To establish their relations to homogeneous curves, next we
prove the following proposition.

\begin{proposition}
Curves in the same family of corner-induced curves are homogeneous.
\end{proposition}

\begin{proof}
Denote $\mathcal{P}_k$ and $\mathcal{Q}_k$ are two corner-induced curves from
the same family and denote their subunits as $\mathcal{U}_i$ and
$\mathcal{V}_i$ ($i \in \{1,2,3,4\}$). First it is easy to see the entry and
exit corners of $\mathcal{U}_i$ and $\mathcal{V}_i$ are all the same.
According to Proposition \ref{prop:unique}, if the entry and exit directions
are also the same for $\mathcal{U}_i$ and $\mathcal{V}_i$, they correspond to
the same curve. This yields always $\mathcal{U}_2 = \mathcal{V}_2$ and
$\mathcal{U}_3 = \mathcal{V}_3$; if there is only one option of entry
direction on subunit 1 (e.g., Family 5), then $\mathcal{U}_1 = \mathcal{V}_1$;
and if there is only one option of exit direction on subunit 4 (e.g., Family
2), then $\mathcal{U}_4 = \mathcal{V}_4$.

We next consider when the entry directions are different on $\mathcal{U}_1$
and $\mathcal{V}_1$. We explicitly write the notation $\mathcal{U}_i$ and
$\mathcal{V}_i$ to $\mathcal{U}^{(k)}_i$ and $\mathcal{V}^{(k)}_i$ as they are
subunits from a level-$k$ curve. It is easy to see $\mathcal{U}^{(k)}_1$ and
$\mathcal{V}^{(k)}_1$ are also two corner-induced curves from the same family
(however not in the base facing state). Additionally their exit directions are
fixed and the same. Then according to the discussion in the previous
paragraph, we can conclude $\mathcal{U}^{(k-1)}_i = \mathcal{V}^{(k-1)}_i$ ($i
\in \{2,3,4\}$). We continue to split $\mathcal{U}^{(k-1)}_1$ and $\mathcal{V}^{(k-1)}_1$
to their next-level subunits. We can repeat this process and on each iteration
the last three subunits are always identical. The process is done until we
reach $\mathcal{U}^{(1)}_1$ and $\mathcal{V}^{(1)}_1$. They are two 2x2 units
with the same entry and exit corners, the same exit directions but different
entry directions. When the entry and exit corners of a 2x2 units is fixed, the
orientation regardless of its entry and exit direction is fixed (as the two
corners define the ``open side'' of the 2x2 unit). Write
$\mathcal{U}^{(1)}_1 = Z_1Z_2Z_3Z_4$ and  $\mathcal{V}^{(1)}_1 =
W_1W_2W_3W_4$. As a base can also be described as a 2-tuple of its entry and
exit directions, there is $Z_i = W_i$ for $i \in \{2, 3, 4\}$ because their
entry directions are always the same and so are their exit directions. The entry direction
of $Z_1$ is different from $W_1$ and this results in $Z_1 \ne W_1$. Note the first bases of
$\mathcal{U}^{(1)}_1$ and $\mathcal{V}^{(1)}_1$ are also the first bases on
$\mathcal{U}^{(k)}_1$ and $\mathcal{V}^{(k)}_1$. Thus if $\mathcal{P}_k$ and
$\mathcal{Q}_k$ have different entry directions, only the first base in their
base sequences are different.

We can perform similar analysis on the case when the exit directions are
different on $\mathcal{U}_4$ and $\mathcal{V}_4$. We can conclude only the
last bases in their base sequences are different.

Putting together, if $\mathcal{P}_k$ and $\mathcal{Q}_k$ are from the same
family of corner-induced curves, it is only possible that the first or the
last base are different. Then according to Property \ref{prop:homo_1},
$\mathcal{P}_k$ and $\mathcal{Q}_k$ are homogeneous curves.

\end{proof}

Family 6 and 8 only contain one type of curve, the number of curves is
not enough to form a family. As rotations and reflections are already
enough to generate the full set of level-\(k\) curves (Proposition
\ref{prop:redudant}), by also considering the four rotations, there are
\((8-2) \times 2 \times 4 =
48\) families of homogeneous curves for \(\mathcal{P}_k\).

\begin{corollary}
\label{coro:homo_first_last}
Related to Property \ref{prop:homo_1}, if two homogeneous curves $\mathcal{P}$
and $\mathcal{Q}$ have the same entry direction, then $X_1 = Y_1$; if they
have different entry directions, then $X_1 \ne Y_1$ with values of $X_1 = I$,
$Y_1 \in \{R, L\}$ or $X_1 \in \{R,L\}$, $Y_1 = I$. If $\mathcal{P}$ and
$\mathcal{Q}$ have the same exit direction, then $X_n = Y_n$; if they have
different exit directions, then $X_n \ne Y_n$ with values of $X_n = I$, $Y_n
\in \{R, L\}$ or $X_n \in \{R,L\}$, $Y_n = I$.
\end{corollary}

\begin{proof}
Denote $X_1 = (\varphi_{s, X_1}, \varphi_{e, X_1})$ and $Y_1 = (\varphi_{s, Y_1}, \varphi_{e, Y_1})$
where each of both is represented as a 2-tuple of its entry direction and exit direction.
It is always $\varphi_{e, X_1} = \varphi_{e, Y_1}$ if $\mathcal{P}_k$ and $\mathcal{Q}_k$ are
homogeneous. With the condition $\varphi_{s, X_1} = \varphi_{s, Y_1}$, there is $X_1 = Y_1$.

According to Figure \ref{fig:homogeneous_curves}, if two homogeneous curves
$\mathcal{P}$ and $\mathcal{Q}$ have different entry directions, the
difference between the two entry directions is 90. Note the exit directions of
$X_1$ and $Y_1$ are the same and $X_1, Y_1 \in \{I, R, L\}$. Then only $X_1 =
I$, $Y_1 \in \{R, L\}$ or $X_1 \in \{R,L\}$, $Y_1 = I$ satisfies.

Denote $X_n = (\varphi_{s, X_n}, \varphi_{e, X_n})$ and $Y_n = (\varphi_{s, Y_n}, \varphi_{e, Y_n})$.
It is always $\varphi_{s, X_n} = \varphi_{s, Y_n}$. With the condition $\varphi_{e, X_n} = \varphi_{e, Y_n}$, there is $X_n = Y_n$.

If $\mathcal{P}$ and $\mathcal{Q}$ have different exit directions, the
difference between the two exit directions is 90. Note the entry directions of
$X_n$ and $Y_n$ are the same and $X_n, Y_n \in \{I, R, L\}$. Then only $X_n =
I$, $Y_n \in \{R, L\}$ or $X_n \in \{R,L\}$, $Y_n = I$ satisfies.
\end{proof}

\begin{corollary}
\label{coro:homo_entry}
Let $\mathcal{P}_k = \mathcal{P}_1|(a)_{k-1}$ and $\mathcal{Q}_k =
\mathcal{Q}_1|(b)_{k-1}$ be homogeneous. If $\mathcal{P}_k$ and
$\mathcal{Q}_k$ have the same entry direction, then $a = b$, if they have
different entry directions, then $a \ne b$ (or $\hat{a} = b$).
\end{corollary}

\begin{proof}
For curves in the base facing states, it can be directly seen from Figure
\ref{fig:homogeneous_curves}. Rotations and reflections change code in
$(a)_{k-1}$ and $(b)_{k-1}$ simultaneously, then the statement in this
corollary is always true.

We can prove it in another way. Since $\mathcal{P}_k$ and $\mathcal{Q}_k$ are
homogeneous, $\mathcal{P}_2 = \mathcal{P}_1|a$ and $\mathcal{Q}_2 =
\mathcal{Q}_1|b$ are also homogeneous. Write $\mathcal{P}_1|a =
Z_1Z_2Z_3Z_4|a$ and $\mathcal{Q}_1|b = W_1W_2W_3W_4|b$. If $\mathcal{P}_2$ and
$\mathcal{Q}_2$ have the same entry direction, their first 2x2 units are
identical, i.e., $Z_1|a = W_1|b$. Then According to Definition \ref{eq:def},
$Z_1 = W_1$ and $a = b$. If $\mathcal{P}_2$ and $\mathcal{Q}_2$ have different
entry directions, with Corollary \ref{coro:homo_first_last}, $Z_1 = I$, $W_1
\in \{R, L\}$ or $Z_1 \in \{R,L\}$, $W_1 = I$. Also the last bases in $Z_1|a$
and $W_1|b$ are identical. With these requirements, from Figure
\ref{fig:expansion_rule}, only pairs of $I_1$/$L_2$ or $I_2$/$R_1$ satisfy for
$Z_1|a$ and $W_1|b$. This results in $a \ne b$.
\end{proof}

\begin{note}
In Proposition \ref{prop:unique} which uniquely determines the curve encoding
from its entry and exit points, when the entry point is located on the corner
of subunit 1 or the exit point is located on the corner of subunit 4, the
entry direction or the exit direction should be preselected if there are
mulitple options. Different selection gives different encodings of curves.
According to this section, they are actually homogeneous curves, which are
only differed by the entry or exit direction of the complete curves, but the
internal structures are identical.
\end{note}

Last, if two curves \(\mathcal{P}\) and \(\mathcal{Q}\) are homogeneous,
we denote \(\mathcal{P} = \mathcal{H}(\mathcal{Q})\). Apparently, it is
also \(\mathcal{Q} =
\mathcal{H}(\mathcal{P})\).

\begin{proposition}
\label{prop:homo_tr}
Let $f_t()$ be primary transformations (Definition \ref{def:primary_trans}),
then

\begin{equation}
f_t(\mathcal{H}(\mathcal{P})) = \mathcal{H}(f_t(\mathcal{P})) .
\notag
\end{equation}

\end{proposition}

\begin{proof}
We write $\mathcal{P} = *X_2 ... X_{n-1}*$ where we denote the first and the
last base as ``$*$'' since they are not used when evaluating the
homogeneity of curves. We also use ``$*$'' for any transformation on them. Then
it is obvious:

\begin{equation}
\mathcal{H}(*X_2 ... X_{n-1}*) = *X_2 ... X_{n-1}*
\notag
\end{equation}

Note when two curves have the same expression $*X_2 ... X_{n-1}*$, we cannot
conclude they are identical curves while we could only say they are homogeneous.

If $f_t()$ is a single rotation or a single reflection,

\begin{equation}
\begin{aligned}
f_t(\mathcal{H}(\mathcal{P})) &= f_t(\mathcal{H}(*X_2 ... X_{n-1}*)) \\
  &= f_t(*X_2 ... X_{n-1}*) \\
  &= *f_t(X_2)...f_t(X_{n-1})* \\
f_t(\mathcal{P}) &= f_t(*X_2 ... X_{n-1}*) \\
  &= *f_t(X_2) ... f_t(X_{n-1})* \\
\end{aligned} .
\notag
\end{equation}

Thus $f_t(\mathcal{H}(\mathcal{P}))$ and $f_t(\mathcal{P})$ are homogeneous
curves for rotation and reflection, i.e., $f_t(\mathcal{H}(\mathcal{P})) =
\mathcal{H}(f_t(\mathcal{P}))$ (Property \ref{prop:homo_1}). Next we consider
$f_t()$ as a single reversal:

\begin{equation}
\begin{aligned}
r(\mathcal{H}(\mathcal{P})) &= r(\mathcal{H}(*X_2 ... X_{n-1}*)) \\
  &= r(*X_2 ... X_{n-1}*) \\
  &= *r(X_{n-1})...r(X_{2})* \\
r(\mathcal{P}) &= r(*X_2 ... X_{n-1}*) \\
  &= *r(X_{n-1}) ... r(X_{2})* \\
\end{aligned} .
\notag
\end{equation}

We can also have $r(\mathcal{H}(\mathcal{P}))$ and $r(\mathcal{P})$ are
homogeneous, i.e., $r(\mathcal{H}(\mathcal{P})) =
\mathcal{H}(r(\mathcal{P}))$.

Using the same method as in the proof for Proposition \ref{prop:rec_tr}, we
can prove this statement is true for any combination of rotation, reflection
and reversal.

\end{proof}

\subsection{Identical shapes}\label{identical-shapes}

Homogeneous curves are still distinguished by their rotations and
orientations. They can be further simplified to only considering their
``shapes''.

\begin{definition}[Identical shapes]
\label{def:shape}
For two curves, ignoring their entry and exit directions, if rotation,
reflection, reversal or combinations of these transformations make them
completely overlapped, they are called to have the same shape.
\end{definition}

\begin{note}
Definition \ref{def:shape} implies that two curves $\mathcal{P}$ and $\mathcal{Q}$ have
the same shape if there exist primary transformations $f_t()$ that make
$\mathcal{P} = f_t(\mathcal{Q})$ or $\mathcal{H}(\mathcal{P}) =
f_t(\mathcal{Q})$.
\end{note}

It is easy to see, all level-0 curves have the same shape as a point,
and all level-1 curves have the same ``U-shape''.

We still consider curves (level \(\ge\) 2) in their base facing states
as in Figure \ref{fig:homogeneous_curves}. Other forms of curves can be
transformed to them by rotations and reflections. Based on the
definition, they have the same shapes.

\subsubsection{Corner-induced curves}\label{corner-induced-curves}

Curves in each of the eight families in Figure
\ref{fig:homogeneous_curves} share the same shape. Family 2 is a
horizontal reflection of the reversed curve in Family 7, and Family 4 is
a horizontal reflection of the reversed curve in Family 5. So Family 7
has the same shape as Family 2, and Family 5 has the same shape as
Family 4. Then we have the first six shapes from the eight famillies
where family 7 is merged with family 2, and family 5 is merged with
family 4. We can see the six families of curves have different shapes
because the entry or exit points are located differently. We take the
first curve in each family (i.e., in the base facing state) as the
inducing curve and the full sets for the six shapes are listed in Table
\ref{tab:corner-induced}. Note the full set of a curve also contains the
horizontally reflected versions of the corresponding curves. The
inducing curve can be any of the curves in the corresponding family
associated with any rotation.

\begin{table}
\centering
\begin{tabular}{clccp{5.9cm}c}
\toprule
Group & Inducing curve & Family & $n$ & Full set & $n_\mathrm{total}$  \\
\midrule
1 & $I^{270}|(2)_k$ & 1 & 4 & $I|(1)_k$, $I|(2)_k$, $R|(1)_k$, $R|(2)_k$, \newline $L|(1)_k$, $L|(2)_k$, $U|(1)_k$, $U|(2)_k$ & 32 \\
2 & $P^{270}|(2)_k$ & 2, 7 &4 & $B|(1)_k$, $B|(2)_k$, $D|2(1)_{k-1}$, $D|1(2)_{k-1}$,\newline $P|(1)_k$, $P|(2)_k$, $Q|(1)_k$, $Q|(2)_k$  & 32 \\
3 & $C|(1)_{k}$   &  3  &4 & $U|2(1)_{k-1}$, $U|1(2)_{k-1}$, $D|(1)_k$, $D|(2)_k$,\newline $Q|2(1)_{k-1}$, $Q|1(2)_{k-1}$, $C|(1)_k$, $C|(2)_k$  & 32 \\
4 & $B^{270}|2(1)_{k-1}$ & 4, 5 & 4& $R|2(1)_{k-1}$, $R|1(2)_{k-1}$, $L|2(1)_{k-1}$, $L|1(2)_{k-1}$,\newline $B|2(1)_{k-1}$, $B|1(2)_{k-1}$, $P|2(1)_{k-1}$, $P|1(2)_{k-1}$ & 32 \\
5 & $I^{270}|2(1)_{k-1}$ & 6 & 1 & $I|2(1)_{k-1}$, $I|1(2)_{k-1}$ & 8 \\
6 & $C|1(2)_k$ & 8 & 1 & $C|2(1)_{k-1}$, $C|1(2)_{k-1}$ & 8 \\
\bottomrule
\end{tabular}
\vspace*{5mm}
\caption{\label{tab:corner-induced}The six groups of corner-induced curves that have the same shapes. $n$: number of curves in Figure \ref{fig:homogeneous_curves}. Full set: the full set of curves in the corresponding family and their horizontal reflections. The initial rotation of base seed are all set to zero. $n_\mathrm{total}$: total number of curves by considering rotations and reflections ($n \times 4 \times 2$).}
\end{table}

\subsubsection{Side-induced curves}\label{side-induced-curves}

There are also side-induced curves (level \(\ge\) 3) where entry points
are not located on the corners of subunit 1. This type of curves can be
represented as \(\mathcal{P}_1|(\omega)_{k-1}\) where \((\omega)_{k-1}\)
is a code sequence of length \(k-1\) where at least two code have
different values (Proposition \ref{prop:geometry_entry}).

\begin{figure}
\centering{
\includegraphics[width=1\linewidth]{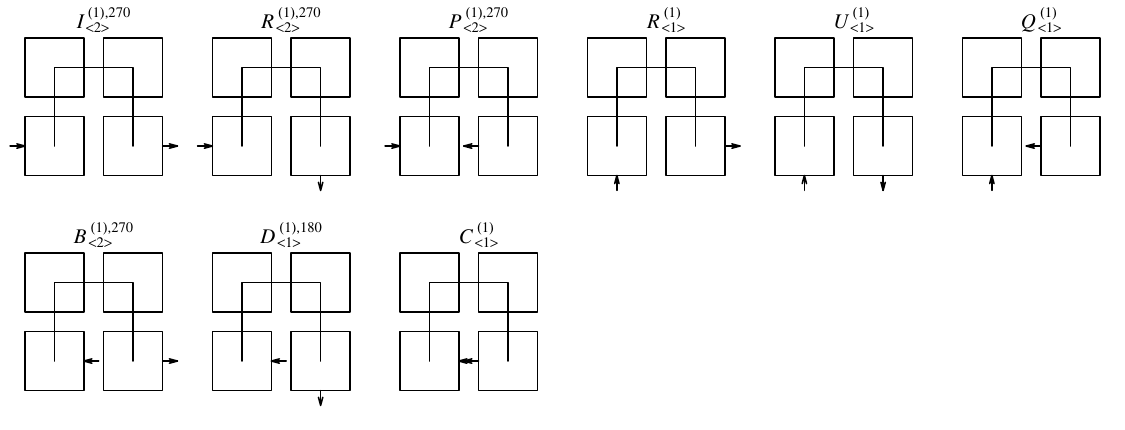}
\caption{Locations of the entry and exit points on side-induced curves. Code above each curve is its global structure on level 1.}\label{fig:side_induced_curves}}
\end{figure}

On subunit 1, the entry point can be located on the left, the bottom or
the right side, but it cannot be on the top side because this is where
subunit 1 connects to subunit 2 (Property \ref{prop:homo11}). Similarly,
the exit point can only be located on the left, the bottom or the right
side of subunit 4. In Figure \ref{fig:side_induced_curves}, all possible
combinations of the sides of entry point and exit point are listed.

Code of \(\mathcal{P}_1\) can be inferred by reducing \(\mathcal{P}_k\)
into a 2x2 unit. Among these nine forms in Figure
\ref{fig:side_induced_curves}, \(R^{(1),270}_{<2>}\) is a horizontal
reflection of the reversal of \(R^{(1)}_{<1>}\), \(P^{(1),270}_{<2>}\)
is a horizontal reflection of the reversal of \(B^{(1),270}_{<2>}\), and
\(Q^{(1)}_{<1>}\) is a horizontal reflection of the reversal of
\(D^{(1),180}_{<1>}\). Thus, there are six groups of global structures
for side-induced curves listed in Table \ref{tab:side-induced}.

\begin{table}
\centering
\begin{tabular}{cllclc}
\toprule
Group & $\mathcal{P}_1$ & $\mathcal{P}_k$ & $n$ & $h(\mathcal{P}_k)$ & $n_\mathrm{total}$  \\
\midrule
1 & $I^{(1),270}_{<2>}$ & $I^{270}|2(\omega)_{k-1}$ & 1 & $I^{90}|1(\hat{\omega})_{k-1}$ & 8 \\
2 & $R^{(1),270}_{<2>}$ & $R^{270}|2(\omega)_{k-1}$ & 2 & $L^{90}|1(\hat{\omega})_{k-1}$ & 16 \\
  & $R^{(1)}_{<1>}$ & $R|1(\hat{\omega})_{k-1}$ &   & $L|2(\omega)_{k-1}$ &   \\
3 & $P^{(1),270}_{<2>}$ & $P^{270}|2(\omega)_{k-1}$ & 2 & $P^{90}|1(\hat{\omega})_{k-1}$ & 16 \\
  & $B^{(1),270}_{<2>}$ & $B^{270}|2(\omega)_{k-1}$ &   & $B^{90}|1(\hat{\omega})_{k-1}$ &   \\
4 & $U^{(1)}_{<1>}$ & $U|1(\omega)_{k-1}$ & 1& $U|2(\hat{\omega})_{k-1}$ & 8 \\
5 & $Q^{(1)}_{<1>}$ & $Q|1(\omega)_{k-1}$ & 2& $Q|2(\hat{\omega})_{k-1}$ & 16 \\
  & $D^{(1),180}_{<1>}$ & $D^{180}|1(\hat{\omega})_{k-1}$ & & $D^{180}|2(\omega)_{k-1}$ &  \\
6 & $C^{(1)}_{<1>}$ & $C|1(\omega)_{k-1}$ & 1 & $C|2(\hat{\omega})_{k-1}$ & 8 \\
\bottomrule
\end{tabular}
\vspace*{5mm}
\caption{\label{tab:side-induced}The six groups of side-induced curves characterized by their level-1 global structures. $\mathcal{P}_1$: the base structure; $\mathcal{P}_k$: the entire curve; $h(\mathcal{P}_k)$: horizontal reflection of $\mathcal{P}_k$; $n$: number of curves in the group; $n_\mathrm{total}$: total number of curves by considering rotations and reflections ($n \times 4 \times 2$). In Group 2, $R|1(\hat{\omega})_{k-1}$ is a horizontal reflection of the reversal of $R^{270}|2(\omega)_{k-1}$. In Group 3, $B^{270}|2(\omega)_{k-1} = h(r(P^{270}|2(\omega)_{k-1}))$. In Group 5, $D^{180}|1(\hat{\omega})_{k-1} = h(r(Q|1(\omega)_{k-1}))$.}
\end{table}

In each group, it is easy to see, for the curve
\(\mathcal{P}_1|(\omega)_{k-1}\), a different sequence of
\((\omega)_{k-1}\) generates a different shape of the curve (fixing the
form of \(\mathcal{P}_1\)) because it corresponds to a different integer
representation \(\delta^{(k-1)}\) thus a different location of the entry
point on subunit 1 (Proposition \ref{prop:geometry_entry}). Then, for a
given level-1 seed \(\mathcal{P}_1\), there are in total \(2^{k-1}-2\)
forms of side-induced curves\footnote{Note
there are in total $2^{k-1}$ curves induced by $\mathcal{P}_1$ where 2 of them
are corner-induced.}, thus they generate \(2^{k-1}-2\) different shapes.

\subsubsection{Put together}\label{put-together}

According to Remark \ref{remark:ee_side}, Corner-induced and
side-induced curves are the only two types of curves. For curves on
level \(k\) (\(k \ge 2\)), there are six shapes from the corner-induced
curves, and \(6 \times (2^{k-1} -
2)\) shapes from the side-induced curves. Putting together, we have the
final number of different shapes of curves on level \(k\):

\begin{equation}
\label{eq:shapes_nn}
\begin{cases}
6 + 6 \times (2^{k-1}-2) &\quad k \ge 2\\ 
1 &\quad k \in \{0, 1\}\\ 
\end{cases} .
\end{equation}

\begin{figure}
\centering{
\includegraphics[width=1\linewidth]{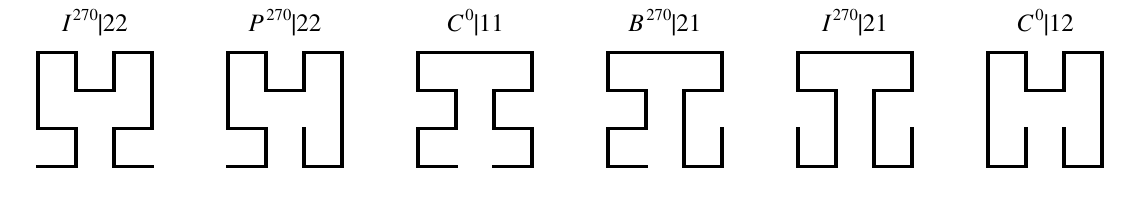}
\caption{All six level-2 shapes.}\label{fig:level_2_shapes}.}
\end{figure}

\begin{figure}
\centering{
\includegraphics[width=1\linewidth]{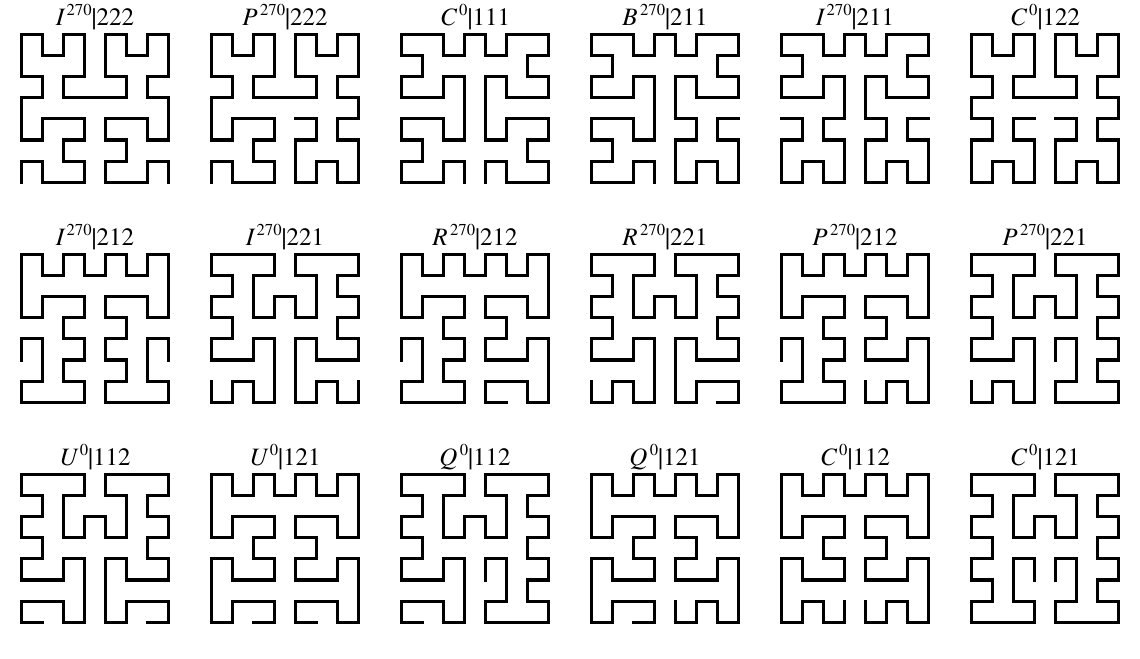}
\caption{All 18 level-3 shapes. The first row contains shapes from corner-induced curves. The second and third rows contain shapes from side-induced curves.}\label{fig:level_3_shapes}
}
\end{figure}

Figure \ref{fig:level_2_shapes} lists all shapes of curves on level 2.
Note the level-2 curve only has corner-induced shapes. The six curves in
Figure \ref{fig:level_2_shapes} are generated by the six inducing curves
in Table \ref{tab:corner-induced}. Figure \ref{fig:level_3_shapes} lists
all 18 shapes for curves on level 3 where the first row contains the six
corner-induced shapes according to Table \ref{tab:corner-induced}, and
the second and third rows contain the 12 side-induced shapes according
to Table \ref{tab:side-induced}. Note the inducing curves can be any one
from the full set of inducing curves of the corresponding shape group.
Table \ref{tab:corner-induced} and \ref{tab:side-induced} can be used to
generate the full set of shapes for curves on any level \(k\).

Let's add the number of curves for each shape from Table
\ref{tab:corner-induced} and \ref{tab:side-induced} (the
\(n_\mathrm{total}\) column):

\begin{equation}
(32+32+32+32+8+8) + (8+16+16+8+16+8)\times(2^{k-1} - 2) = 36 \times 2^k
\notag
\end{equation}

\noindent which is exactly the number of all forms of a level-\(k\)
curve (Equation \ref{eq:total_number}). This implies the shape analysis
includes all forms of \(\mathcal{P}_k\).

\subsubsection{Hierarchical shape generation}\label{hc_shape}

Shapes on level \(k\) can be generated from a certain shape on level
\(k-1\). Taking the first shape in Figure \ref{fig:level_2_shapes} which
corresponds to corner-induced shape group 1 in Table
\ref{tab:corner-induced} as an example, it generates four shapes on
level 3. The encoding of this level-2 shape can be \(I|22\), \(R|22\),
\(L|22\) and \(U|22\) (ignore other versions after rotations and
reflections). There are the following four shapes on level 3: \(I|222\)
which is still a corner-induced curve, \(I|221\), \(R|221\) and
\(U|221\) (in Figure \ref{fig:level_3_shapes}, its reflected version
\(U|112\) is used) which are side-induced curves and according to Table
\ref{tab:side-induced}. Since the seed bases are different, the three
ones have different shapes. \(L|221\) is excluded because it has the
same shape as \(R|221\).

\begin{figure}
\centering{
\includegraphics[width=1\linewidth]{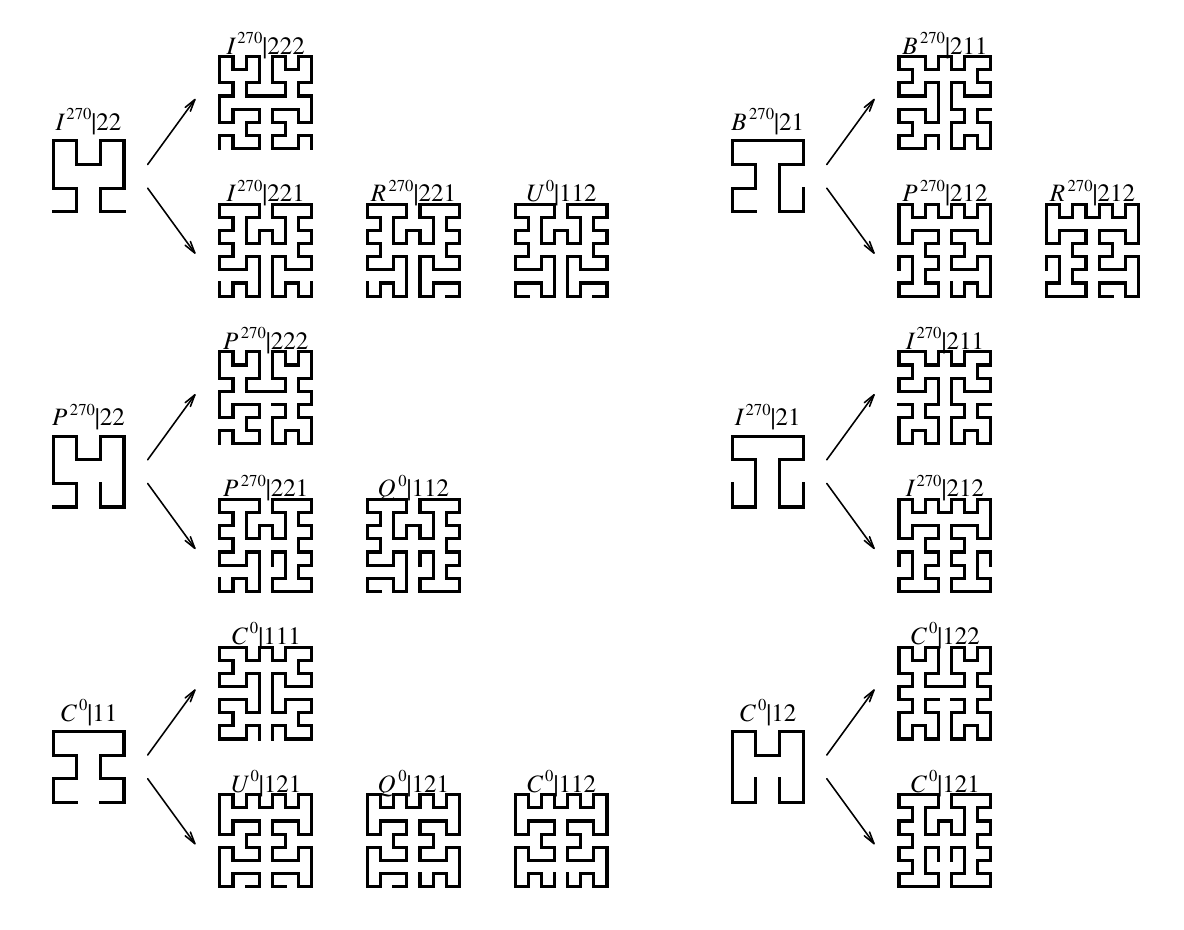}
\caption{Hierarchical generation of shapes from level 2 to level 3.}\label{fig:shape_hierarchical}
}
\end{figure}

Denote a level-2 curve from a certain corner-induced shape group as
\(\mathcal{C}^{(2)}\), it can induces two types of shapes on level 3:
corner-induced and side-induced. As shown in the following diagram,

\begin{center}
\begin{tikzpicture}
    \node(formula){$\mathcal{C}^{(2)}$};
    \node(solution1) [above right =-0.5em and 2em of formula]{$\mathcal{C}^{(3)}$};
    \node(solution2) [below right =-0.5em and 2em of formula]{$\mathcal{D}^{(3)} = \mathcal{C}^{(2)}|\omega$};
    \draw [->] (formula.east) --  (solution1.west);
    \draw [->] (formula.east) --  (solution2.west);
\end{tikzpicture}
\end{center}

\noindent \(\mathcal{C}^{(3)}\) is a corner-induced curve from the same
shape group as \(\mathcal{C}^{(2)}\), and
\(\mathcal{D}^{(3)} = \mathcal{C}^{(2)}|\omega\) is a side-induced curve
where \(\omega\) has a different code from its preceding code. Since
there are multiple encodings for \(\mathcal{C}^{(2)}\), there might be
multiple shapes for \(\mathcal{D}^{(3)}\) as well, depending on the seed
of the curve. Note \(R\)/\(L\) generate side-induced curves in the same
shape group, and so are \(B\)/\(P\) and \(D\)/\(Q\). The list of
possible seeds for side-induced curves on level 3 induced from
corresponding level-2 shape is in Table \ref{tab:shape_seed_2}. The full
set of shape expansion from level 2 to level 3 is listed in Figure
\ref{fig:shape_hierarchical}.

\begin{table}
\centering
\begin{tabular}{clc}
\toprule
Shape group of $\mathcal{C}^{(2)}$ & Seeds for $\mathcal{C}^{(2)}|\omega$ & $h_g$ \\
\midrule
1 & $I$, $R$/$L$, $U$ & 3 \\
2 & $B$/$P$, $D$/$Q$  & 2 \\
3 & $U$, $D$/$Q$, $C$ & 3 \\
4 & $R$/$L$, $B$/$P$  & 2 \\
5 & $I$ & 1 \\
6 & $C$ & 1 \\
\bottomrule
\end{tabular}
\vspace*{5mm}
\caption{\label{tab:shape_seed_2}Seeds for side-induced shape on level 3. $h_g$: number of shapes of $\mathcal{C}^{(2)}|\omega$ for a specific shape group of $\mathcal{C}^{(2)}$.}
\end{table}

Starting from \(\mathcal{C}^{(2)}\), when expanding from level \(i-1\)
to level \(i\) (\(i-1 \ge 3\)), there are always two types of shapes on
level \(i-1\): corner-induced \(\mathcal{C}^{(i-1)}\) and several
side-induced \(\mathcal{D}^{(i-1)}\). They are expanded to level \(i\)
in the following ways:

\begin{center}
\begin{tikzpicture}
    \node(formula){$\mathcal{C}^{(i-1)}$};
    \node(solution1) [above right =-0.5em and 2em of formula]{$\mathcal{C}^{(i)}$};
    \node(solution2) [below right =-0.5em and 2em of formula]{$\mathcal{D}^{(i)} = \mathcal{C}^{(i-1)}|\omega$, \quad $h_g$ \textnormal{forms}};
    \draw [->] (formula.east) --  (solution1.west);
    \draw [->] (formula.east) --  (solution2.west);
\end{tikzpicture}
\end{center}

\noindent and

\begin{center}
\begin{tikzpicture}
    \node(formula){$\mathcal{D}^{(i-1)}$};
    \node(solution1) [above right =-0.5em and 2em of formula]{$\mathcal{D}^{(i-1)}|1$};
    \node(solution2) [below right =-0.5em and 2em of formula]{$\mathcal{D}^{(i-1)}|2$};
    \draw [->] (formula.east) --  (solution1.west);
    \draw [->] (formula.east) --  (solution2.west);
\end{tikzpicture}
\end{center}

We can easily see one of \(\mathcal{D}^{(i-1)}|1\) and
\(\mathcal{D}^{(i-1)}|2\) have all its level-2 units as Hilbert units
and the other one has all its level-2 units as \(\beta\Omega\)-units,
then they have different shapes. \(\mathcal{C}^{(i-1)}|\omega\),
\(\mathcal{D}^{(i-1)}|1\) and \(\mathcal{D}^{(i-1)}|2\) are all denoted
as \(\mathcal{D}^{(i)}\) for the next-level expansion for simplicity as
they are all side-induced.

Denote \(n_i\) as the number of shapes on level \(i\)
(\(3 \le i \le k\)). On level \(i-1\), there is only one corner-induce
shapes, and all other \(n_{i-1}-1\) shapes are side-induced. A
corner-induced shape generates \(1 + h_g\) shapes on level \(i\), and
each side-induced curve generates two shapes on level \(i\). Then we
have the relation \(n_i = (1 + h_g) + 2\times (n_{i-1} - 1)\) with the
initial value \(n_2 = 1\) if we fix the initial level-2 shape group. We
can obtain the final solution on level \(k\):
\(n_k = 1 + h_g \times (2^{k-2} - 1)\). Adding all six inducing groups
on \(\mathcal{C}^{(2)}\), we can finally have:

\begin{equation}
\sum_{g \in \{1,...,6\}} (1 + h_g \times (2^{k-2} - 1)) = 6 \times 2^{k-1} - 6
\notag
\end{equation}

\noindent which is exactly the number of all different shapes for curves
on level \(\ge 2\), as in Equation \ref{eq:shapes_nn}.

Last, notably, also as shown in Figure \ref{fig:shape_hierarchical},
\(h_g\) shapes in the form of \(\mathcal{C}^{(i-1)}|\omega\) induced
from shape \(\mathcal{C}^{(i-1)}\) in group 1-4 have identical shapes
except the first or the last 2x2 unit. They are called
``partially identical shapes'' which will be introduced in Section
\ref{partially-identical-shapes}.

\subsubsection{Other attributes}\label{other-attributes}

\begin{proposition}
\label{prop:shape_reduce}

There are the following two contrapositive statements related to the shape of
a curve and its reduced forms. For two curves $\mathcal{P}_k$ and
$\mathcal{Q}_k$ ($k \ge 2$),

\begin{enumerate}
\tightlist
\item
  If they have the same shape, denoted as $\mathcal{S}(\mathcal{P}_k) =
  \mathcal{S}(\mathcal{Q}_k)$, their reductions also have the same shape.

\begin{equation}
\mathcal{S}(\mathrm{Rd}_i(\mathcal{P}_k)) = \mathcal{S}(\mathrm{Rd}_i(\mathcal{Q}_k)) \quad 1 \le i \le k-2
\notag
\end{equation}

\item
  If $\mathcal{S}(\mathcal{P}_k) \ne \mathcal{S}(\mathcal{Q}_k)$, their
  expansion with the same number of code have different
  shapes.

\begin{equation}
\mathcal{S}(\mathcal{P}_k|(\pi)_{l}) \ne \mathcal{S}(\mathcal{Q}_k|(\sigma)_{l}) \quad l \ge 1
\notag
\end{equation}

\end{enumerate}

\end{proposition}

\begin{proof}
We only prove the first statement. Let $f_{t_1}()$ be primary transformations
that transform $\mathcal{P}_k$ to its base facing state and $f_{t_2}()$ be
primary transformations that transform $\mathcal{Q}_k$ to its base facing
state. First let's perform the two transformations:

\begin{equation}
\begin{aligned}
\mathcal{R}_k &= f_{t_1}(\mathcal{P}_k) \\
\mathcal{T}_k &= f_{t_2}(\mathcal{Q}_k) \\
\end{aligned} .
\notag
\end{equation}

Apparently there are $\mathcal{S}(\mathcal{R}_k) = \mathcal{S}(\mathcal{P}_k)$
and $\mathcal{S}(\mathcal{T}_k) = \mathcal{S}(\mathcal{Q}_k)$. With the
condition $\mathcal{S}(\mathcal{P}_k) = \mathcal{S}(\mathcal{Q}_k)$, there is
$\mathcal{S}(\mathcal{R}_k) = \mathcal{S}(\mathcal{T}_k)$. Let's write
$\mathcal{R}_k$ and $\mathcal{T}_k$ as expansions taking level-1 units as
seeds:

\begin{equation}
\begin{aligned}
\mathcal{R}_k &= \mathcal{R}_1|(\pi)_{k-1} \\
\mathcal{T}_k &= \mathcal{T}_1|(\sigma)_{k-1} \\
\end{aligned} .
\notag
\end{equation}

According to both Table \ref{tab:corner-induced} and \ref{tab:side-induced},
for two curves in the same shape group, if reducing the expansion code
sequences $(\pi)_{k-1}$ and $(\sigma)_{k-1}$ by the same amount to a length
$\ge$ 1, they are still in the same shape group, i.e.,

\begin{equation}
\mathcal{S}(\mathrm{Rd}_i(\mathcal{R}_k)) = \mathcal{S}(\mathrm{Rd}_i(\mathcal{T}_k)) \quad 1 \le i \le k-2 .
\notag
\end{equation}

According to Proposition \ref{prop:rec_tr} (for Line 2 and 5 in the following
equations), we expand $\mathcal{P}_k$ and $\mathcal{Q}_k$ separately:

\begin{equation}
\begin{aligned}
\mathcal{S}(\mathrm{Rd}_i(\mathcal{P}_k)) &= \mathcal{S}(f_{t_1}(\mathrm{Rd}_i(\mathcal{P}_k))) \\
  &= \mathcal{S}(\mathrm{Rd}_i(f_{t_1}(\mathcal{P}_k))) \\
  &= \mathcal{S}(\mathrm{Rd}_i(\mathcal{R}_k)) \\
\mathcal{S}(\mathrm{Rd}_i(\mathcal{Q}_k)) &= \mathcal{S}(f_{t_2}(\mathrm{Rd}_i(\mathcal{Q}_k))) \\
  &= \mathcal{S}(\mathrm{Rd}_i(f_{t_2}(\mathcal{Q}_k))) \\
  &= \mathcal{S}(\mathrm{Rd}_i(\mathcal{T}_k)) \\
\end{aligned} .
\notag
\end{equation}

Then $\mathcal{S}(\mathrm{Rd}_i(\mathcal{P}_k)) = \mathcal{S}(\mathrm{Rd}_i(\mathcal{Q}_k))$.

\end{proof}

\begin{lemma}
\label{lemma:diff_last2}
If $\mathcal{P}_k = X|\pi_1...aa$ and $\mathcal{Q}_k = Y|\sigma_1...a\hat{a}$
where $a = 1$ or 2, then $\mathcal{P}_k$ and $\mathcal{Q}_k$ have different
shapes.
\end{lemma}

\begin{proof}
When $k \ge 3$, $\mathcal{P}_k$ is composed of a list of Hilbert units and
$\mathcal{Q}_k$ is composed of a list of $\beta\Omega$-units, thus
$\mathcal{P}_k$ and $\mathcal{Q}_k$ have different shapes. When $k = 2$, we
can easily see $X|aa$ and $Y|a\hat{a}$ have different shapes with Figure
\ref{fig:level_2_shapes}.
\end{proof}

\begin{proposition}
\label{prop:diff_2code}
Let $\mathcal{P}_k = X|(\pi)_k$ and $\mathcal{Q}_k = Y|(\sigma)_k$. If there
exist $i$ and $j$ ($2 \le i < j \le k$) where $\pi_i = \sigma_i$ and $\pi_j  =
\hat{\sigma}_j$, then $\mathcal{P}_k$ and $\mathcal{Q}_k$ have different
shapes.
\end{proposition}

\begin{proof}
If such $i$ and $j$ exist, there must exist two
neighbouring code $i'$ and $i'+1$ that makes $\pi_{i'} = \sigma_{i'}$ and
$\pi_{i'+1}  = \hat{\sigma}_{i'+1}$ ($i' \ge i$, $i'+1 \le j$). Then according
to Lemma \ref{lemma:diff_last2}, $\mathcal{P}_{i'+1}$ and $\mathcal{Q}_{i'+1}$
have different shapes. In turn according to Proposition
\ref{prop:shape_reduce}, $\mathcal{P}_k$ and $\mathcal{Q}_k$ have different
shapes.
\end{proof}

\begin{proposition}
\label{prop:side-identical}
If $\mathcal{P}_k$ and $\mathcal{Q}_k$ are two side-induced curves in the same
shape group, there exists primary transformations $f_t()$ that makes

\begin{equation}
\mathcal{P}_k = f_t(\mathcal{Q}_k) .
\notag
\end{equation}

If $\mathcal{P}_k$ and $\mathcal{Q}_k$ are two corner-induced curves in the same
shape group, then

\begin{equation}
\mathcal{H}(\mathcal{P}_k) = f_t(\mathcal{Q}_k) \text{ or } \mathcal{P}_k = f_t(\mathcal{Q}_k).
\notag
\end{equation}

\begin{proof}
According to the definition of identical shapes, there is a $f_t()$ that makes $f_t(\mathcal{Q}_k)$ completely overlaps with
$\mathcal{P}_k$. If the direction of two curves are still mutually reversed,
we additionally add $r()$ to $f_t()$. If $\mathcal{P}_k$ and
$f_t(\mathcal{Q}_k)$ are side-induced curves, there is only one possible entry
direction and one possible exit direction for both (Proposition
\ref{prop:homo2}), then $\mathcal{P}_k = f_t(\mathcal{Q}_k)$. If
$\mathcal{P}_k$ and $f_t(\mathcal{Q}_k)$ are corner-induced curves, there
might be multiple combinations of entry and exit directions for both, then
$\mathcal{H}(\mathcal{P}_k) = f_t(\mathcal{Q}_k)$ or $\mathcal{P}_k = f_t(\mathcal{Q}_k)$
\end{proof}

\end{proposition}

\begin{proposition}
\label{prop:shape_expand}
Let $\mathcal{S}(\mathcal{P}_i) = \mathcal{S}(\mathcal{Q}_i)$ ($i \ge 2$).
Write $\mathcal{P}_k = X|\pi_1...\pi_k$ and $\mathcal{Q}_k = Y|\sigma_1...\sigma_k$. 
$\mathcal{S}(\mathcal{P}_k) = \mathcal{S}(\mathcal{Q}_k)$ ($k > i$), iff

\begin{equation}
\label{eq:shape_expand}
\sigma_{i+1}...\sigma_k = \begin{cases}
\pi_{i+1}...\pi_k & \quad \textnormal{if } \sigma_i = \pi_i \\
\hat{\pi}_{i+1}...\hat{\pi}_k & \quad \textnormal{if } \sigma_i = \hat{\pi}_i \\
\end{cases} .
\end{equation}

If $\mathcal{P}_i$ and $\mathcal{Q}_i$ are corner-induced but $\mathcal{P}_k$
and $\mathcal{Q}_k$ are side-induced, additionally we require $X$ and $Y$ to
be valid seeds to induce side-induced curves in the same shapes, i.e., $X, Y =
I$, $X, Y = U$, $X, Y = C$, $X, Y \in \{R, L\}$, $X, Y \in \{B, P\}$, or $X, Y
\in \{D, Q\}$.

\end{proposition}

\begin{proof}

First we prove $\Rightarrow$. If $\mathcal{S}(\mathcal{P}_k) =
\mathcal{S}(\mathcal{Q}_k)$, then the two code sequences
$\pi_2...\pi_i\pi_{i+1}...\pi_k$ and
$\sigma_2...\sigma_i\sigma_{i+1}...\sigma_k$ are also either the same or
complementary no matter they are corner- or side-induced (Table \ref{tab:corner-induced} and \ref{tab:side-induced}). Then it is obvious that Equation \ref{eq:shape_expand} is true.
Additionally, if $\mathcal{P}_k$ and $\mathcal{Q}_k$ are side-induced, their seed should
come from the same side-induced shape group.

Next we prove $\Leftarrow$. With the condition $\mathcal{S}(\mathcal{P}_i) = \mathcal{S}(\mathcal{Q}_i)$, there is $\sigma_2...\sigma_i = \pi_2...\pi_i$ or $\sigma_2...\sigma_i = \hat{\pi}_2...\hat{\pi}_i$. Together with the condition in Equation \ref{eq:shape_expand}, there is $\sigma_2...\sigma_k = \pi_2...\pi_k$ or $\sigma_2...\sigma_k = \hat{\pi}_2...\hat{\pi}_k$. We consider a second curve 

\begin{equation}
\mathcal{Q}'_k = \begin{cases}
\mathcal{Q}_k = \mathcal{Q}_1|\pi_2...\pi_k \quad & \textnormal{if } \sigma_2...\sigma_k = \pi_2...\pi_k\\
h(\mathcal{Q}_k) = h(\mathcal{Q}_1)|\pi_2...\pi_k = \mathcal{Q}'_1|\pi_2...\pi_k \quad & \textnormal{if } \sigma_2...\sigma_k = \hat{\pi}_2...\hat{\pi}_k\\
\end{cases} .
\notag
\end{equation}

Apparently, it is also $\mathcal{S}(\mathcal{P}_i) = \mathcal{S}(\mathcal{Q}'_i)$. 
If $\mathcal{P}_i$/$\mathcal{Q}'_i$/$\mathcal{P}_k$/$\mathcal{Q}'_k$ are all corner-induced or all side-induced, with $\mathcal{S}(\mathcal{P}_i) = \mathcal{S}(\mathcal{Q}'_i)$, their level-1 seeds $\mathcal{P}_1$ and $\mathcal{Q}'_1$ induce the same shape group, then $\mathcal{S}(\mathcal{P}_k) = \mathcal{S}(\mathcal{Q}'_k)$ because their code sequences from the second base are identical so they are also in the same corner-induced or side-induced shape group. If $\mathcal{P}_i$/$\mathcal{Q}_i$
are corner-induced while $\mathcal{P}_k$/$\mathcal{Q}_k$ are side-induced, we only need to additionally ensure $\mathcal{P}_1$ and $\mathcal{Q}'_1$ also induce the same side-induced shape groups.

With $\mathcal{S}(\mathcal{P}_k) = \mathcal{S}(\mathcal{Q}'_k)$, there is
$\mathcal{S}(\mathcal{P}_k) = \mathcal{S}(\mathcal{Q}_k)$.
\end{proof}

\subsection{Partially identical
curves}\label{partially-identical-curves}

Next let's consider a type of curves which has a loose requirement on
shapes.

\subsubsection{Differed by level-1
units}\label{differed-by-level-1-units}

\begin{definition}
For a curves $\mathcal{P}_k$ ($k \ge 2$) without considering its entry and
exit directions, if its first or last 2x2 unit can be adjusted to generate
$\mathcal{Q}_k$, then $f_{t_1}(\mathcal{P}_k)$ and $f_{t_2}(\mathcal{Q}_k)$
where $f_{t_1}()$ and $f_{t_2}()$ are two arbitrary primary transformations
have partially identical shapes only differed by the first or the last 2x2
unit.
\end{definition}

\begin{figure}
\centering{
\includegraphics[width=0.8\linewidth]{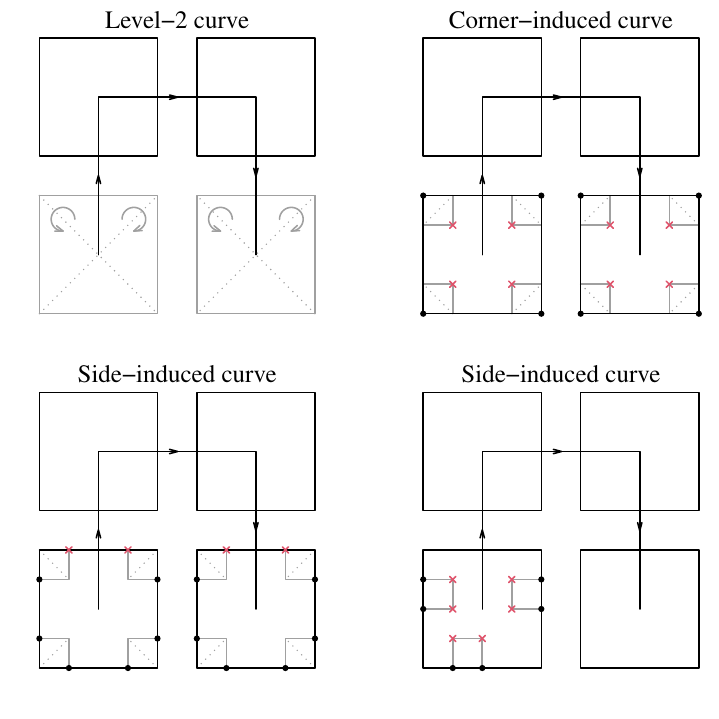}
\caption{Partially identical shapes. First panel: adjust the first and the last 2x2 units on a level-2 curve. Second-fourth panels: adjust the first 2x2 units on corner-induced and side-induced curves. Solid points are all possible entry points. Dashed lines represent the diagonal lines for reflecting the 2x2 units which are determined by Note \ref{note:pp}. Red crosses represent the location of the entry point after the adjustment which makes the curves invalid. In the fourth panel, diagonal lines are not illustrated.}\label{fig:partially_identical_shapes_structure}
}
\end{figure}

\begin{note}
\label{note:pp}
The exit corner $p_s$ of the first 2x2 unit $\mathcal{U}_s$ is fixed. To
adjust $\mathcal{U}_s$ means to reflect $\mathcal{U}_s$ by the diagonal line
determined by $p_s$. For the last 2x2 unit $\mathcal{U}_e$, its entry corner
$p_e$ is also fixed, then to adjust $\mathcal{U}_e$ is to reflect
$\mathcal{U}_e$ by the diagonal line determined by $p_e$ (Figure
\ref{fig:partially_identical_shapes_structure}, the first panel). Entry
and exit directions are not considered in the adjustment.
\end{note}

In this section, if adjusting the first or the last 2x2 unit of
\(\mathcal{P}_k\) still generates a valid curve, we specifically call
\(\mathcal{P}_k\) \textit{partially shapable}. In the following text, we
consider \(\mathcal{P}_k\) in the base facing state and explore the
conditions that make \(\mathcal{P}_k\) partially shapable.

When \(k = 2\), the first and the last subunits of \(\mathcal{P}_k\) are
2x2 units. According to the first panel in Figure
\ref{fig:partially_identical_shapes_structure}, diagonally reflecting
the first 2x2 unit still makes the entry point located on one of its
corners, and so is for the exit point and the last 2x2 unit. However, to
make the adjusted curve valid, the entry or the exit direction of
\(\mathcal{P}_2\) cannot be horizontal on the bottom corners, or else
after the adjustment, the entry direction of the first or the exit
direction of the last 2x2 unit will be vertical on its upper-corners
which makes the adjusted curve invalid (Property \ref{prop:homo3}). Then
there are the following level-2 curves of which the first 2x2 unit is
adjustable (from Figure \ref{fig:homogeneous_curves}):

\begin{equation}
\begin{aligned}
& R|11, U|11, Q|11, Q|12, U|12, R|12, \\ 
& P^{270}|21, R^{270}|21, I^{270}|21, B^{270}|22, D^{180}|12, C|12
\end{aligned}
\notag
\end{equation}

And there are the following level-2 curves of which the last 2x2 unit is
adjustable.

\begin{equation}
\begin{aligned}
& R^{270}|22, U|11, P^{270}, Q|11, D^{180}|11, U|12, \\ 
& B^{270}|21, R|12, R^{270}|21, I^{270}|21, D^{180}|12, C|12
\end{aligned}
\notag
\end{equation}

When \(k \ge 3\), all corner-induced curves are not partially shapable
because adjusting the first 2x2 unit always makes the reflected entry
point located inside subunit 1, which is not allowed (Remark
\ref{remark:ee_side} and the second panel in Figure
\ref{fig:partially_identical_shapes_structure}). So is for the last 2x2
unit.

For side-induced curves, when the first 2x2 unit is located on the
corner part of subunit 1, then if it is located on the two top corner
part of subunit 1, the reflected entry points will be located on the
side where subunit 1 and 2 attach, which makes the curve invalid
(Property \ref{prop:homo11}). It can only be located on the two bottom
corner parts (the third panel in Figure
\ref{fig:partially_identical_shapes_structure}). There is the same
requirement for the last 2x2 unit. When the first 2x2 unit is not
located on the corners of subunit 1, then all their reflected entry
points will be located inside subunit 1 which is not allowed (the fourth
panel in Figure \ref{fig:partially_identical_shapes_structure}).

Now we have the only type of curve for \(\mathcal{P}_k\) which is
partially shapable where its first or last 2x2 unit is located on the
bottom corner of its subunit 1 or 4. For \(k \ge 3\), if
\(\mathcal{Q}_k\) is partially shaped from \(\mathcal{P}_k\) by only
reflecting its first and/or last 2x2 units, there are the following
results: 1. \(\mathcal{P}_{k-1}\) and \(\mathcal{Q}_{k-1}\) are
corner-induced and in the same shape with entry corner on the bottom of
subunit 1 and/or exit corner on the bottom of subunit 4. Then with
Figure \ref{fig:homogeneous_curves} and Table \ref{tab:corner-induced},
\(\mathcal{P}_{k-1}\) and \(\mathcal{Q}_{k-1}\) should be from the same
corner-induced shape group from group 1-4; 2. \(\mathcal{P}_k\) and
\(\mathcal{Q}_k\) are side-induced with different shapes, thus in the
encoding \(\mathcal{P}_k = X|\pi_1(a)_{k-2}\hat{a}\) and
\(\mathcal{Q}_k = Y|\sigma_1(b)_{k-2}\hat{b}\).

Here we only discuss \(\mathcal{P}_k\) and \(\mathcal{Q}_k\) in the base
facing state. Then result 1 becomes \(\mathcal{P}_{k-1}\) and
\(\mathcal{Q}_{k-1}\) should be from the same homogeneous family 1, 2/7,
3, or 4/5. Note curves in homogeneous family 7 can be generated from
family 5 via primary transformations and so are for family 4 and 5. We
finally rephrase result 1 to: \(\mathcal{P}_{k-1}\) and
\(\mathcal{Q}_{k-1}\) or after certain primary transformations are from
the same homogeneous family 1-4.

\begin{proposition}
\label{prop:partial}
$\mathcal{P}_k$ and $\mathcal{Q}_k$ ($k \ge 3$) are two curves in the base facing state. They have partially identical shapes 
if and only if
\begin{enumerate}
\tightlist
\item
  There exist two primary transformations $f_{t_1}()$ and $f_{t_2}()$, so that $f_{t_1}(\mathcal{P}_{k-1})$ and $f_{t_2}(\mathcal{Q}_{k-1})$ are homogeneous curves from family 1-4.
\item
  $\mathcal{P}_k = X|\pi_1(a)_{k-2}\hat{a}$ and $\mathcal{Q}_k = Y|\sigma_1(b)_{k-2}\hat{b}$.
\end{enumerate}

\end{proposition}

\begin{proof}

We have already proven $\Rightarrow$ in the previous discussion. Here we only prove $\Leftarrow$.

First, if $\mathcal{P}_{k-1}$ and $\mathcal{Q}_{k-1}$ are homogeneous curves from family 1-4, write them as

\begin{equation}
\begin{aligned}
\mathcal{P}_{k-1} = X_1X_2...X_{n-1}X_n \\
\mathcal{Q}_{k-1} = Y_1Y_2...Y_{n-1}Y_n \\
\end{aligned}
\notag
\end{equation}

\noindent where $n = 4^{k-1}$ and $X_j = Y_j$ for all $2 \le j \le n-1$, there are two scenarios to consider.

\textit{Scenario 1}. If $\mathcal{P}_{k-1}$
and $\mathcal{Q}_{k-1}$ have different entry directions, according to Corollary \ref{coro:homo_entry},
$X_1 = I$, $Y_1 \in \{R,L\}$ or $X_1 \in \{R,L\}$, $Y_1 = I$, and according to Corollary \ref{coro:homo_first_last},
$a \ne b$ or $\hat{a} = b$. Then with Proposition \ref{prop:3.5},

\begin{equation}
\begin{aligned}
\mathcal{P}_{k} &= \mathcal{P}_{k-1}|\hat{a}  = X_{<\hat{a}>,1}X_{<\hat{a}>,2}... \\
\mathcal{Q}_{k} &= \mathcal{Q}_{k-1}|a = Y_{< a>,1}Y_{<\hat{a}>,2}... \\
\end{aligned} \quad \textnormal{if } X_1 = I, Y_1 \in \{R,L\}
\notag
\end{equation}

\noindent or

\begin{equation}
\begin{aligned}
\mathcal{P}_{k} &= X_{<\hat{a}>,1}X_{< a >,2}... \\
\mathcal{Q}_{k} &= Y_{< a>,1}Y_{< a >,2}... \\
\end{aligned} \quad \textnormal{if } X_1 \in \{R,L\}, Y_1 = I .
\notag
\end{equation}

Both indicate the second 2x2 units in $\mathcal{P}_k$ and $\mathcal{Q}_k$ are identical.
With Proposition \ref{prop:3.5}, expansion code of a base in a sequence is determined
by its preceding base, then since $X_j = Y_j$ for all $2 \le j \le n-1$, from the second to the last second 2x2 unit are all identical 
in $\mathcal{P}_{k}$ and $\mathcal{Q}_{k}$.
The first
2x2 units $X_{< \hat{a}>,1}$ and $Y_{< a>,1}$ in the two curves have the same exit point and exit direction. Since the entry directions
of $X_{< \hat{a}>,1}$ and $Y_{< a>,1}$ have a difference of $\pm 90$, then their entry points can only be located
on the two different neighbouring corners of the exit point, thus in different facings. Thus $\mathcal{P}_{k}$ and $\mathcal{Q}_{k}$ have partially
identical shapes.

\textit{Scenario 2}. If $\mathcal{P}_{k-1}$ and $\mathcal{Q}_{k-1}$ have the
same entry direction but different exit directions. Then $X_1 = Y_1$, $a = b$ (Corollary
\ref{coro:homo_entry}),
$X_n = I$, $Y_n \in \{R,L\}$ or $X_n \in \{R,L\}$, $Y_n = I$, (Corollary \ref{coro:homo_first_last}). It is easy to see that
the subsequences from the first to the last second 2x2 unit are all
identical. The last 2x2 unit in $\mathcal{P}_k$ is $X_{< a_*>, n}$ and in
$\mathcal{Q}_k$ is $Y_{< a_*>, n}$. The two 2x2 units have the same entry
point and entry direction. Since the exit directions of them have a difference
of $\pm 90$, the last 2x2 units are in different facings. Thus
$\mathcal{P}_{k}$ and $\mathcal{Q}_{k}$ also have partially identical shapes.

Next for the general case, let $\mathcal{P}'_k = f_{t_1}(\mathcal{P}_k)$
and $\mathcal{Q}'_k = f_{t_2}(\mathcal{P}_k)$ which makes $\mathcal{P}'_k$
and $\mathcal{Q}'_k$ homogeneous from family 1-4. Primary transformations
change the expansion code simultaneously from the second code, thus we can write

\begin{equation}
\begin{aligned}
\mathcal{P}'_k &= f_{t_1}(X|\pi_1(a)_{k-2}\hat{a}) = f_{t_1}(X)|\pi'_1(c)_{k-2}\hat{c} \\
\mathcal{Q}'_k &= f_{t_2}(Y|\sigma_1(b)_{k-2}\hat{b}) = f_{t_2}(Y)|\sigma'_1(d)_{k-2}\hat{d} \\
\end{aligned}
\notag
\end{equation}

\noindent where $\pi'_1$ and $\sigma'_1$ are the code after transformations $f_{t_1}$
and $f_{t_2}$, and $c$ and $d$ are two new variables to represent expansion code.
According to the discussion we have already made, $\mathcal{P}'_k$ and $\mathcal{Q}'_k$ have partially identical shapes. As $\mathcal{P}_k$ has the same shape as $\mathcal{P}'_k$
and $\mathcal{Q}_k$ has the same shape as $\mathcal{Q}'_k$, then $\mathcal{P}_k$ and $\mathcal{Q}_k$ have partially identical shapes.

\end{proof}

Proposition \ref{prop:partial} indicates there are four groups of curves
in partially identical shapes that are induced from homogeneous family
1-4. According to the discussion that has been made, if
\(\mathcal{P}_{k-1}\) and \(\mathcal{Q}_{k-1}\) have different entry
directions, then the first 2x2 units of \(\mathcal{P}_{k}\) and
\(\mathcal{Q}_{k}\) are in different facings; if \(\mathcal{P}_{k-1}\)
and \(\mathcal{Q}_{k-1}\) have different exit directions, the last 2x2
units of \(\mathcal{P}_{k}\) and \(\mathcal{Q}_{k}\) are in different
facings. Thus a different combination of entry direction and exit
direction of \(\mathcal{P}_{k-1}\) determines a different shape. The
full list of the four groups of curves is listed in Table
\ref{tab:partial_identical_shapes_1}, where group 1 and 3 both include 4
shapes, and group 2 and 4 both include 2 shapes.

Last, for curves on level 2, there are two groups of curves in partially
identical shapes. The first group includes curves in corner-induced
group 1, 2, 6, and the second group includes curves in corner-induced
group 3, 4, 5 (Figure \ref{fig:level_2_shapes}).

\begin{table}
\centering
\begin{tabular}{cllcllr}
\toprule
Family & $\mathcal{P}_k$ & $h(\mathcal{P}_k)$ & \makecell{Other \\ family} & $\mathcal{P}_k$  & $h(\mathcal{P}_k)$ & $n_\mathrm{total}$ \\
\midrule
1 & $I^{270}|(2)_{k-1}1$ & $I^{90}|(1)_{k-1}2$ & &  & & 8  \\
  & $R^{270}|(2)_{k-1}1$ & $L^{90}|(1)_{k-1}2$ & &  & & 8 \\
  & $R|(1)_{k-1}2$ & $L|(2)_{k-1}1$ & & & & 8 \\
  & $U|(1)_{k-1}2$ & $U|(2)_{k-1}1$ & & & & 8 \\
\midrule
2 & $P^{270}|(2)_{k-1}1$ & $P^{90}|(1)_{k-1}2$  & 7 & $B^{270}|(2)_{k-1}1$ & $B^{90}|(1)_{k-1}2$ & 16 \\
  & $Q|(1)_{k-1}2$       & $Q|(2)_{k-1}1$       &   & $D^{180}|1(2)_{k-2}1$ & $D^{180}|2(1)_{k-2}2$ & 16\\
\midrule
3 & $C|(1)_{k-1}2$ & $C|(2)_{k-1}1$ & &  & & 8 \\
  & $D^{180}|(1)_{k-1}2$ & $D^{180}|(2)_{k-1}1$ & & & & 8 \\
  & $Q|1(2)_{k-2}1$ & $Q|2(1)_{k-2}2$ & & & & 8 \\
  & $U|1(2)_{k-2}1$ & $U|2(1)_{k-2}2$ & &  & & 8 \\
\midrule
4 & $B^{270}|2(1)_{k-2}2$ & $B^{90}|1(2)_{k-2}1$ & 5 & $P^{270}|2(1)_{k-2}2$ & $P^{90}|1(2)_{k-2}1$ & 16\\
  & $R|1(2)_{k-2}1$ & $L|2(1)_{k-2}2$ & & $R^{270}|2(1)_{k-2}2$ & $L^{90}|1(2)_{k-2}1$ & 16\\
\bottomrule
\end{tabular}
\vspace*{5mm}
\caption{\label{tab:partial_identical_shapes_1}Groups of curves (level $\ge 3$) in partially identical shapes. Each row contains curves in the same shapes. $n_\mathrm{total}$: total number of curves by considering four rotations.}
\end{table}

!!! the second and the third ones are symmetric !!!

\subsubsection{\texorpdfstring{Differed by level \(k-1\)
units}{Differed by level k-1 units}}\label{differed-by-level-k-1-units}

All homogeneous families are level \(k-1\) units.

\subsubsection{\texorpdfstring{Differed by level-\(i\) (\(i \ge 2\))
units}{Differed by level-i (i \textbackslash ge 2) units}}\label{differed-by-level-i-i-ge-2-units}

We have only discussed one-level expansion from \(\mathcal{P}_{k-1}\)
and \(\mathcal{Q}_{k-1}\) to generate partially identical shapes. Next
we discuss more general cases. Let \(\mathcal{P}_{k-i}\) and
\(\mathcal{Q}_{k-i}\) (\(k-i \ge 2\), \(i \ge 1\)) be from the same
homogeneous family of family 1-4 while \(\mathcal{P}_{k-i+1}\) and
\(\mathcal{Q}_{k-i+1}\) not. Write

\begin{equation}
\begin{aligned}
\mathcal{P}_{k} &= \mathcal{P}_{k-i}|(\pi)_{k-i+1...k} \\
    &= X_1X_2...X_{n-1}X_n|(\pi)_{k-i+1...k} \\
    &= \mathcal{U}_1\mathcal{U}_2...\mathcal{U}_{n-1}\mathcal{U}_n \\
\mathcal{Q}_{k} &= \mathcal{Q}_{k-i}|(\sigma)_{k-i+1...k} \\
    &= Y_1Y_2...Y_{n-1}Y_n|(\sigma)_{k-i+1...k} \\
    &= \mathcal{V}_1\mathcal{V}_2...\mathcal{V}_{n-1}\mathcal{V}_n \\
\end{aligned}
\notag
\end{equation}

\noindent where \(\mathcal{U}_j\) and \(\mathcal{V}_j\) are level-\(i\)
units. With conditions \(\mathcal{U}_j = \mathcal{V}_j\)
(\(2 \le j \le n - 1\)), we want to find the solution of
\((\sigma)_{k-i+1...k}\) based on \((\pi)_{k-i+1...k}\).

Expansion code for \(X_1\) is \((\pi)_{k-i+1...k}\) and for \(Y_1\) is
\((\sigma)_{k-i+1...k}\). Expansion code for \(X_2\) and \(Y_2\) can be
calculated based on the type of \(X_1\) and \(Y_1\) (Corollary
\ref{coro:5.6.2}).

\begin{equation}
\label{eq:ppp}
\begin{aligned}
(\pi_{*,2})_{k-i+1...k} = \begin{cases}
(\pi)_{k-i+1...k} & \quad \textnormal{if } X_1 = I \\
(\hat{\pi})_{k-i+1...k} & \quad \textnormal{if } X_1 \in \{R, L\} \\
\end{cases}
\\
(\sigma_{*,2})_{k-i+1...k} = \begin{cases}
(\sigma)_{k-i+1...k} & \quad \textnormal{if } Y_1 = I \\
(\hat{\sigma})_{k-i+1...k} & \quad \textnormal{if } Y_1 \in \{R, L\} \\
\end{cases}
\end{aligned}
\end{equation}

With the condition \(\mathcal{U}_2 = \mathcal{V}_2\), there is
\((\pi_{*,2})_{k-i+1...k} = (\sigma_{*,2})_{k-i+1...k}\). When the entry
directions of \(\mathcal{P}_{k-i}\) and \(\mathcal{Q}_{k-i}\) are the
same, then \(X_1 = Y_1\) (Corollary \ref{coro:homo_entry}), with
Equation \ref{eq:ppp}, there is
\((\sigma)_{k-i+1...k} = (\pi)_{k-i+1...k}\). When the entry directions
of \(\mathcal{P}_{k-i}\) and \(\mathcal{Q}_{k-i}\) are different, then
\(X_1 \ne Y_1\). According to Corollary \ref{coro:homo_first_last},
\(X_1\) and \(Y_1\) cannot be \(R\)/\(L\) at the same time. Then with
Equation \ref{eq:ppp}, we obtain
\((\sigma)_{k-i+1...k} = (\hat{\pi})_{k-i+1...k}\) in this category. We
write the solution of \((\sigma)_{k-i+1...k}\) as:

\begin{equation}
(\sigma)_{k-i+1...k} = \begin{cases}
(\pi)_{k-i+1...k} & \quad \textnormal{if } \varphi_s(\mathcal{P}_{k-i}) = \varphi_s(\mathcal{Q}_{k-i}) \\
(\hat{\pi})_{k-i+1...k} & \quad \textnormal{if } \varphi_s(\mathcal{P}_{k-i}) \ne \varphi_s(\mathcal{Q}_{k-i}) \\
\end{cases} .
\notag
\end{equation}

\noindent where \(\varphi_s()\) represents the entry direction of a
curve.

Now we write \(\mathcal{P}_{k-i} = \mathcal{P}_1|(a)_{2...k-i}\) and
\(\mathcal{Q}_{k-i} = \mathcal{Q}_1|(b)_{2...k-i}\) since they are
corner-induced curves. With Corollary \ref{coro:homo_entry}, given
\(\mathcal{P}_k = \mathcal{P}_1|(a)_{2...k-i}(\pi)_{k-i+1...k}\),
\(\mathcal{Q}_k\) can be expressed as:

\begin{equation}
\label{eq:partial_q}
\mathcal{Q}_k = \begin{cases}
\mathcal{Q}_1|(a)_{2...k-i}(\pi)_{k-i+1...k} & \quad \textnormal{if } \varphi_s(\mathcal{P}_{k-i}) = \varphi_s(\mathcal{Q}_{k-i}) \\
\mathcal{Q}_1|(\hat{a})_{2...k-i}(\hat{\pi})_{k-i+1...k} & \quad \textnormal{if } \varphi_s(\mathcal{P}_{k-i}) \ne \varphi_s(\mathcal{Q}_{k-i}) \\
\end{cases} .
\end{equation}

Condition that \(\mathcal{P}_{k-i}\) and \(\mathcal{Q}_{k-i}\) are
homogeneous while \(\mathcal{P}_{k-i+1}\) and \(\mathcal{Q}_{k-i+1}\)
are not implies at least one of \(\mathcal{P}_{k-i+1}\) and
\(\mathcal{Q}_{k-i+1}\) are not corner-induced. If
\(\mathcal{P}_{k-i+1}\) is not corner-induced, then
\(\pi_{k-i+1} = \hat{a}\). If \(\mathcal{P}_{k-i+1}\) is corner-induced
and \(\mathcal{Q}_{k-i+1}\) is not corner-induced, then
\(\pi_{k-i+1} = a\), but this results in that \(\mathcal{Q}_{k-i+1}\) is
corner-induced which has conflict with the assumption. Thus we have the
only solution here \(\pi_{k-i+1} = \hat{a}\) which makes both
\(\mathcal{P}_{k-i+1}\) and \(\mathcal{Q}_{k-i+1}\) side-induced.

We summarize the discussion to the next proposition.

\begin{proposition}
Let $\mathcal{P}_{k-i} = X|\pi_1(a)_{k-i-1}$ and $\mathcal{Q}_{k-i} =
Y|\sigma_1(b)_{k-i-1}$ ($k-i \ge 2$, $i \ge 2$) be two curves from the same homogeneous family of 
family 1-4. For a curve $\mathcal{P}_k = \mathcal{P}_{k-i}|(\pi)_{k-i+1...k}$ and a
second curve $\mathcal{Q}_k = \mathcal{Q}_{k-i}|(\sigma)_{k-i+1...k}$, if the
following requirements are satisfied: 

\begin{enumerate}
\tightlist
\item
  $\pi_{k-i+1} = \hat{a}$,
\item
\begin{equation}
(\sigma)_{k-i+1...k} = \begin{cases}
(\pi)_{k-i+1...k} & \quad \textnormal{if } a = b \\
(\hat{\pi})_{k-i+1...k} & \quad \textnormal{if } \hat{a} = b \\
\end{cases} ,
\notag
\end{equation}
\end{enumerate}

\noindent then $f_{t_1}(\mathcal{P}_k)$ and
$f_{t_2}(\mathcal{Q}_k)$ have partially identical shapes only differed by the first (if
the entry directions are different) or the last (if the exit directions are
different) level-$i$ units where $f_{t_1}()$ and $f_{t_2}()$ are two primary transformations.

\end{proposition}

\begin{proof}
It has already been proven by previous discussions.
We only need to translate Equation \ref{eq:partial_q} to requirement 2. With Corollary \ref{coro:homo_entry},
when $\varphi_s(\mathcal{P}_{k-i}) = \varphi_s(\mathcal{Q}_{k-i})$, then $a = b$, and when 
$\varphi_s(\mathcal{P}_{k-i}) \ne \varphi_s(\mathcal{Q}_{k-i})$, then $\hat{a} = b$.
\end{proof}

\begin{figure}
\centering{
\includegraphics[width=1\linewidth]{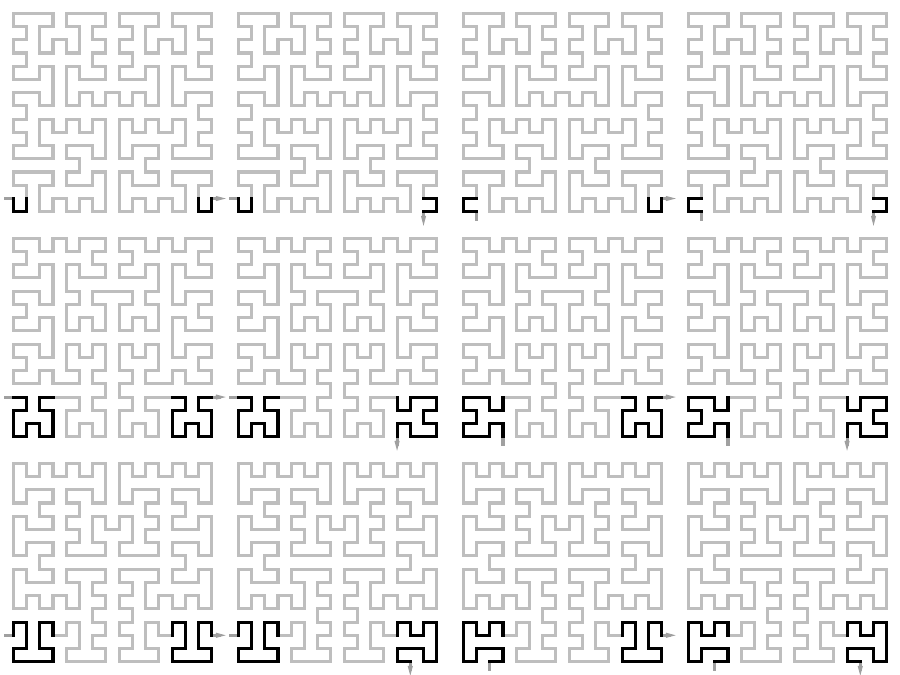}
\caption{Partially identical shapes. $\mathcal{G}(1, 1)$, $\mathcal{G}(1, 11)$, $\mathcal{G}(1, 12)$.}\label{fig:partially_identical_shapes}
}
\end{figure}

The grouping of partially identical curves differed by level-\(i\)
(\(i \ge 2\)) is not only determined by which homogeneous family they
are induced from, but also the code sequence \(\pi_{k-i+1...k}\). The
first code \(\pi_{k-i+1}\) is determined by its ``homogeneous seed'',
however the code \(a\) changes between 1 and 2 depending on which curve
in the homogeneous family and which primary transformation applied to
it. To standardize the notation, let
\(\mathcal{P}_{k-i} = X|\pi_1(\kappa_g)_{k-i-1}\) be the ``inducing
curve'' for the corresponding homogeneous family \(g\) listed in Table
\ref{tab:corner-induced}, then \(\pi_{k-i+1} = \hat{\kappa}_g\). With
the remaining code \(\pi_{k-i+2...k}\), a unique group of partially
identical curves is determined. We denote each group as
\(\mathcal{G}(g, \hat{\kappa}_g\pi_{k-i+2...k})\) where
\(g \in \{1, 2, 3, 4\}\) and
\(\kappa_1 = 2, \kappa_2 = 2, \kappa_3 = 1, \kappa_4 = 1\). We can also
simplify the notation to \(\mathcal{G}(g, \hat{\kappa}_g(\pi)_{i-1})\).
The scenario of \(i = 1\), i.e., partially identical shapes only
differed by level-1 unit, can also be integrated in to this notation as
\(\mathcal{G}(g, \hat{\kappa}_g)\). Table
\ref{tab:partial_identical_shapes_2} lists the completel groups of
partially identical shapes and Figure
\ref{fig:partially_identical_shapes} illustrates the three groups of
\(\mathcal{G}(1, 1)\), \(\mathcal{G}(1, 11)\), \(\mathcal{G}(1, 12)\).

\begin{proposition}
Let $\mathcal{P}_k$ be from $\mathcal{G}(g, \hat{\kappa}_g(\pi)_{i-1})$ and $\mathcal{Q}_k$ be from $\mathcal{G}(g, \hat{\kappa}_g(\sigma)_{i-1})$. Express $\mathcal{P}_k$ as a list of level-$i$ units $\mathcal{U}_1...\mathcal{U}_n$ and express $\mathcal{Q}_k$ as a list of level-$i$ units $\mathcal{V}_1...\mathcal{V}_n$ ($n = 4^{k-i}$). If $(\pi)_{i-1} \ne (\sigma)_{i-1}$, then $\mathcal{S}(\mathcal{U}_j) \ne \mathcal{S}(\mathcal{V}_j)$ for all $1 \le j \le n$.
\end{proposition}

\begin{proof}
The two curves are induced from the same homogeneous seed. We only consider $\mathcal{P}_{k-i} = \mathcal{Q}_{k-i}$ as the ``inducing curve'' from the corresponding family. General cases can be generated by primary transformations, but it does not affect the statement
in this proposition.

Now we can write 

\begin{equation}
\begin{aligned}
\mathcal{P}_k &= \mathcal{P}_{k-i}|\hat{\kappa}(\pi)_{i-1} \\
   &= X_1...X_n|\hat{\kappa}(\pi)_{i-1} \\
   &= \mathcal{U}_1...\mathcal{U}_n \\
\mathcal{Q}_k &= \mathcal{P}_{k-i}|\hat{\kappa}(\sigma)_{i-1} \\
   &= X_1...X_n|\hat{\kappa}(\pi)_{i-1} \\
   &= \mathcal{V}_1...\mathcal{V}_n \\
\end{aligned}
\notag
\end{equation}

\noindent where $\mathcal{U}_* = X_*|\hat{\kappa}_*(\pi_*)_{i-1}$ and $\mathcal{V}_* = X_*|\hat{\kappa}_*(\sigma_*)_{i-1}$.
When $i-1 \ge 2$, there are three cases that cause $(\pi)_{i-1} \ne (\sigma)_{i-1}$. 1. One of $\mathcal{U}_*$ and $\mathcal{V}_*$
a corner-induced curve and the other a side-induced curve, then of course they are in different shapes. 2. Both $\mathcal{U}_*$ and $\mathcal{V}_*$ are side-induced curves, then with $X_* \in \{I, R, L\}$ and the first code being the same, with Table \ref{tab:side-induced}, they are always in different shapes. 3. Both $\mathcal{U}_*$ and $\mathcal{V}_*$ are corner-induced curves,
then $(\pi)_{i-1} = (\hat{\sigma})_{i-1}$, then with $X_* \in \{I, R, L\}$ and the first code being the same, with Table \ref{tab:corner-induced}, they are in different shapes.

When $i = 2$, $\mathcal{U}_* = X_*|\hat{\kappa}_*(\pi_*)$ and $\mathcal{V}_* = X_*|\hat{\kappa}_*(\sigma_*)$. With $\pi_* \ne \sigma_*$,
one of $\mathcal{U}_*$ and $\mathcal{V}_*$ is always a Hilbert unit and the other is always a $\beta$- or $\Omega$-unit, then always in different shapes.

\end{proof}

\begin{corollary}
Let $\mathcal{P}_k$ be from $\mathcal{G}(g_1, \hat{\kappa}_g(\pi)_{i-1})$ and $\mathcal{Q}_k$ be from $\mathcal{G}(g_2, \hat{\kappa}_g(\sigma)_{i-1})$. Denote their second level $k-1$ subunit as $\mathcal{U}_2$ and $\mathcal{V}_2$. $\mathcal{S}(\mathcal{U}_2) = \mathcal{S}(\mathcal{V}_2)$ only when $g_1 = g_2$ and $(\pi)_{i-1} = (\sigma)_{i-1}$.
\end{corollary}

\begin{corollary}
If a curve $\mathcal{P}_k$ is not from any homogeneous family, but its global
structure $\mathcal{P}_i$ ($i < k$ where $i$ is the largest value that
satisfies) is, then $\mathcal{P}_k$ must come from a partially identical group.
\end{corollary}

\begin{table}
\centering
\begin{tabular}{cclcl}
\toprule
Group & Family & $\mathcal{P}_k$ & Other family & $\mathcal{P}'_k$  \\
\midrule
$\mathcal{G}(1, 1(\pi)_{i-1})$ & 1 & $I^{270}|(2)_{k-i-1}1(\pi)_{i-1}$  & &  \\
 &  & $R^{270}|(2)_{k-i-1}1(\pi)_{i-1}$ &  & \\
 & & $R|(1)_{k-i-1}2(\hat{\pi})_{i-1}$ & & \\
 &  & $U|(1)_{k-i-1}2(\hat{\pi})_{i-1}$ & & \\
\midrule
$\mathcal{G}(2, 1(\pi)_{i-1})$ & 2 & $P^{270}|(2)_{k-i-1}1(\pi)_{i-1}$ & 7 & $B^{270}|(2)_{k-i-1}1(\pi)_{i-1}$ \\
 & & $Q|(1)_{k-i-1}2(\hat{\pi})_{i-1}$       &   & $D^{180}|1(2)_{k-i-2}1(\pi)_{i-1}$ \\
\midrule
$\mathcal{G}(3, 2(\pi)_{i-1})$ & 3 & $C|(1)_{k-i-1}2(\pi)_{i-1}$ & & \\
 & & $D^{180}|(1)_{k-i-1}2(\pi)_{i-1}$ & & \\
 & & $Q|1(2)_{k-i-2}1(\hat{\pi})_{i-1}$ & & \\
 & & $U|1(2)_{k-i-2}1(\hat{\pi})_{i-1}$ & & \\
\midrule
$\mathcal{G}(4, 2(\pi)_{i-1})$ & 4 & $B^{270}|2(1)_{k-i-2}2(\pi)_{i-1}$ & 5 & $P^{270}|2(1)_{k-i-2}2(\pi)_{i-1}$\\
 & & $R|1(2)_{k-i-2}1(\hat{\pi})_{i-1}$ & & $R^{270}|2(1)_{k-i-2}2(\pi)_{i-1}$\\
\bottomrule
\end{tabular}
\vspace*{5mm}
\caption{\label{tab:partial_identical_shapes_2}Groups of curves with partially identical shapes differed by the first or the last level-$i$ units. Each row contains curves in the same shapes. Curves after rotations or reflections are not listed in the table.}
\end{table}

\subsection{Completely distinct
shapes}\label{completely-distinct-shapes}

\begin{definition}[Completely distinct shapes]
\label{def:diff}
If square units in the same size (with corresponding level $\ge$ 2) on the
same location of $\mathcal{P}_k$ and $\mathcal{Q}_k$ ($k \ge 2$) are always in
different shapes and this statement is always true for all $\mathcal{P}_k$
and $\mathcal{Q}_k$'s reductions until level 2, then $\mathcal{P}_k$ and
$\mathcal{Q}_k$ are called to have completely distinct shapes.
\end{definition}

A unit on level \textgreater{} 2 is expanded from a level-2 unit.
According to Proposition \ref{prop:shape_reduce}, if the two level-2
units are in different shapes, corresponding higher-level units expanded
from them are also in different shapes. This yields the following
proposition.

\begin{proposition}
\label{prop:diff_shapes}
For $\mathcal{P}_k$ and $\mathcal{Q}_k$, let $\mathcal{P}_i$ and
$\mathcal{Q}_i$ be them or their reductions. Express $\mathcal{P}_i$ and
$\mathcal{Q}_i$ as lists of level-2 units. If units on the same locations of
$\mathcal{P}_i$ and $\mathcal{Q}_i$ always have different shapes, and it is
true for all $2 \le i \le k$, then $\mathcal{P}_k$ and $\mathcal{Q}_k$ have
completely distinct shapes.

\end{proposition}

\begin{proof}

The set of units in Proposition \ref{prop:diff_shapes} denoted as $A$ is a
subset from those denoted as $S$ in Definition \ref{def:diff}. We first
prove the extra units from $S$ can be generated by expanding units in $A$.

Denote $A = \{A_i\} = \bigcup_{i=2}^k A_i$ where set $A_i$ contains level-2 units for all $\mathcal{P}_i$. We write it as

\begin{equation}
A_i = \{X|\pi_1...\pi_{i-2} \parallel \pi_{i-1}\pi_i\} = \{\mathcal{U}^{(i),2}\}
\notag
\end{equation}

\noindent where we use the notation ``$\parallel$'' to denote the curve
as a set of level-2 units. We denote these units simply as
$\mathcal{U}^{(i),2}$ as they are level-2 units on a level-$i$ curve. More
generally,

\begin{equation}
\{X|\pi_1...\pi_{i-d} \parallel \pi_{i-d+1}...\pi_i\} = \{\mathcal{U}^{(i),d}\}
\notag
\end{equation}

\noindent represents a list of level-$d$ ($2 \le d \le i$) units of $\mathcal{P}_i$.
Note each level-$d$ unit is expanded from the corresponding level-2 units on
$\mathcal{P}_{i-d+2}$ (i.e., $\{X|\pi_1...\pi_{i-d} \parallel
\pi_{i-d+1}\pi_{i-d+2}\}$), then we denote $\mathcal{U}^{(i),d} =
\mathrm{Expand}(\mathcal{U}^{(i-d+2),2}, d - 2)$ where $d-2$ represents the number
of expansions from $\mathcal{P}_{i-d+2}$ to $\mathcal{P}_{i}$.

The full set on level-$i$ denoted as $S_i$ can be written as:

\begin{equation}
\begin{aligned}
S_i =\{ &X|\pi_1...\pi_{i-2} \parallel \pi_{i-1}\pi_i, \\
   &X|\pi_1...\pi_{i-3} \parallel \pi_{i-2}\pi_{i-1}\pi_i, \\
   &..., \\
   &X \parallel \pi_1...\pi_i \}
\end{aligned}
\notag
\end{equation}

\noindent where from the second line the units can be written as expansions of
corresponding $\mathcal{U}^{(*),2}$:

\begin{equation}
\begin{aligned}
S_i &=\{ \mathcal{U}^{(i),2}, \mathrm{Expand}(\mathcal{U}^{(i-1),2},1), ..., \mathrm{Expand}(\mathcal{U}^{(2),2},i-2) \} \\
  &= \bigcup_{j = 0}^{i-2} \mathrm{Expand}(\mathcal{U}^{(i - j),2}, j) \\
\end{aligned} .
\notag
\end{equation}

\noindent and the full set $S$:

\begin{equation}
S = \bigcup_{i = 2}^k \bigcup_{j = 0}^{i-2} \mathrm{Expand}(\mathcal{U}^{(i - j),2}, j) .
\notag
\end{equation}

The superscript $i-j$ ranges within $[2, k]$, thus $S$ can be constructed by $A$ (when $j = 0$) and expansion units from $A$ (when $j \ge 1$). 

From the condition of this proposition, the pairwise unit
$\mathcal{V}^{(i-j),2}$ of $\mathcal{Q}_k$ ($2 \le i \le k$, $0 \le j \le i-2$) is always different
from $\mathcal{U}^{(i-j),2}$ on $\mathcal{P}_k$. According to Proposition \ref{prop:shape_reduce},
then all pairwise $\mathrm{Expand}(\mathcal{V}^{(i-j),2}, j)$ and
$\mathrm{Expand}(\mathcal{U}^{(i-j),2}, j)$ are also always
different. Then according to Definition \ref{def:diff}, $\mathcal{P}_k$ and
$\mathcal{Q}_k$ have completely dictinct shapes.

\end{proof}

For a curve \(\mathcal{P}_k = X|\pi_1...\pi_k\) (\(k \ge 2\)), we next
explore the form of a second curve
\(\mathcal{Q}_k = Y|\sigma_1...\sigma_k\) that has a completely distinct
shape from \(\mathcal{P}_k\). Let's go through their reductions
\(\mathcal{P}_i\) and \(\mathcal{Q}_i\) from \(i = 2\).

When \(i = 2\),
\(\mathcal{S}(X|\pi_1\pi_2) \ne \mathcal{S}(Y|\sigma_1\sigma_2)\). This
is the initial criterion.

When \(i = 3\),
\(\mathcal{P}_3 = X|\pi_1\pi_2\pi_3 = X|\pi_1|\pi_2\pi_3\) and
\(\mathcal{Q}_3 = Y|\sigma_1\sigma_2\sigma_3 = Y|\sigma_1|\sigma_2\sigma_3\).
According to Definition \ref{def:diff}, their \(j\)-th level-2 units
(there are four level-2 subunits for each, \(1 \le j \le 4\)) should
always have different shapes. We enumerate all possible combinations of
\(\pi_2\pi_3\) and \(\sigma_2\sigma_3\).

\begin{enumerate}
\tightlist
\item
  When $\pi_2 = \pi_3$ and $\sigma_2 = \sigma_3$, all level-2 units are
  Hilbert units, thus in the same shape.
\item
  When $\pi_2 \ne \pi_3$ and $\sigma_2 \ne \sigma_3$, note in $X|\pi_1$ and
  $Y|\sigma_1$, the second bases are always $R$/$L$ (turning right or left in
  the U-shape), which makes the second level-2 units all $\beta$-units, thus
  in the same shape.
\item
  When $\pi_2 = \pi_3$ and $\sigma_2 \ne \sigma_3$, all level-2 units in
  $\mathcal{P}_3$ are Hilbert units while in $\mathcal{Q}_3$ are all
  $\beta$-units and $\Omega$-units, thus always in different shapes.
\item 
  When $\pi_2 \ne \pi_3$ and $\sigma_2 = \sigma_3$, all level-2 units in
  $\mathcal{P}_3$ are $\beta$-units and $\Omega$-units while in
  $\mathcal{Q}_3$ are all Hilbert units, thus always in different shapes.
\end{enumerate}

Thus, combinations 3 and 4 make level-2 units on the same positions
always different on \(\mathcal{P}_3\) and \(\mathcal{Q}_3\).

When \(4 \le i \le k\),
\(\mathcal{P}_i = \mathcal{P}_{i-2}|\pi_{i-1}\pi_i\) and
\(\mathcal{Q}_i = \mathcal{Q}_{i-2}|\sigma_{i-1}\sigma_i\). Similarly,
we can go all combinations of \(\pi_{i-1}\pi_i\) and
\(\sigma_{i-1}\sigma_i\), and we can have the scenarios
\(\pi_{i-1} = \pi_i\) and \(\sigma_{i-1} \ne \sigma_i\), or \(\pi_{i-1}
\ne \pi_i\) and \(\sigma_{i-1} = \sigma_i\) make level-2 units of
\(\mathcal{P}_i\) and \(\mathcal{Q}_i\) on the same positions always
different.

To summarize, we have the following proposition:

\begin{proposition}
\label{prop:diff}
For a curve $\mathcal{P}_k = X|\pi_1...\pi_k$ ($k \ge 2$), a second curve
$\mathcal{Q}_k = Y|\sigma_1...\sigma_k$ has a completely distinct shape
from $\mathcal{P}_k$ iff the following two conditions are satisfied:

\begin{enumerate}
\tightlist
\item
  $\mathcal{S}(Y|\sigma_1\sigma_2) \ne \mathcal{S}(X|\pi_1\pi_2)$.
\item
  When $3 \le i \le k$, 
\begin{equation}
\begin{aligned}
\sigma_{i}  &= \begin{cases}
  \hat{\sigma}_{i-1} & \quad \textnormal{if } \pi_{i} = \pi_{i-1} \\
  \sigma_{i-1} & \quad \textnormal{if } \pi_{i} = \hat{\pi}_{i-1} \\
\end{cases}
\end{aligned} .
\notag
\end{equation}  
\end{enumerate}

\end{proposition}

\begin{proof}
First we prove $\Rightarrow$. If $\mathcal{P}_k$ has a completely distinct
shape from $\mathcal{Q}_k$, the two conditions are exactly the results that we
have discussed previously.

Next we prove $\Leftarrow$. If condition 2 are satisfied, then for all $3
\le i \le k$, the lowest level-2 units on the same locations of
$\mathcal{P}_i$ and $\mathcal{Q}_i$ are always in different shapes as one
curve only contains Hilbert units and the other only contains
$\beta\Omega$-units. Together with condition 1 which implies $\mathcal{P}_2$
has a different shape from $\mathcal{Q}_2$, then according to Proposition
\ref{prop:diff_shapes}, $\mathcal{P}_k$ and $\mathcal{Q}_k$ have completely
distinct shapes.
\end{proof}

\begin{figure}
\centering{
\includegraphics[width=0.8\linewidth]{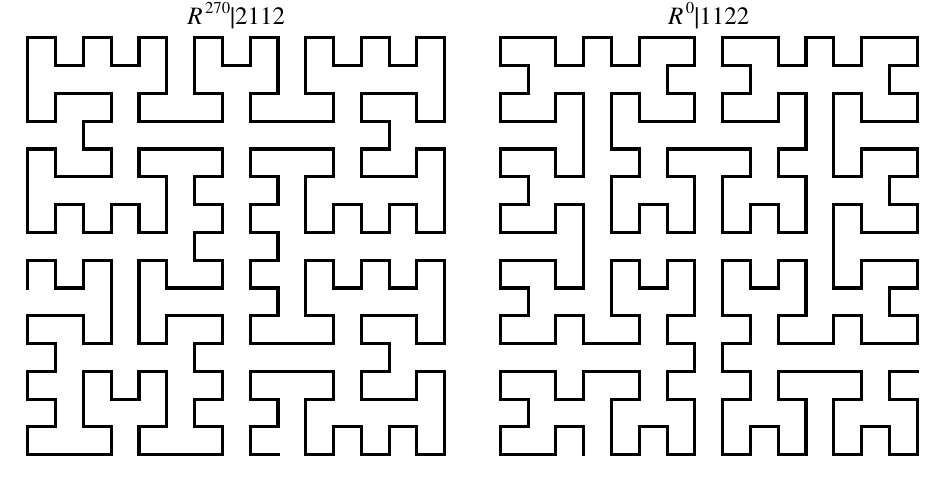}
\caption{Two curves in completely distinct shapes.}
\label{fig:different_curve}
}
\end{figure}

As an example, Figure \ref{fig:different_curve} illustrates two curves
in completely distinct shapes (\(R|2112\) and \(R|1122\)).

\section{\texorpdfstring{The Hilbert curve and the
\(\beta\Omega\)-curve}{The Hilbert curve and the \textbackslash beta\textbackslash Omega-curve}}\label{def-hc}

In this section, we provide definitions of the Hilbert curve and the
\(\beta\Omega\)-curve from the aspect of their specific structures. We
will not distinguish rotations associated with the seed bases since
rotations won't affect the statements in this section. We require these
two types of curves to have level \(\ge\) 2. Note again, we study these
curves after finite iterations.

\subsection{The Hilbert curve}\label{the-hilbert-curve}

\begin{definition}[Hilbert curve]
\label{def:hc}
For a curve $\mathcal{P}_k$, if the lowest level-2 units of $\mathcal{P}_{i} =
\mathrm{Rd}_{k-i}(\mathcal{P}_k)$ are always Hilbert units (Section
\ref{global-structure-and-local-unit}) for all $2 \le i \le k$, then
$\mathcal{P}_k$ is called a Hilbert curve.
\end{definition}

\begin{proposition}
\label{prop:hc1}
$\mathcal{P}_k$ is a Hilbert curve iff $\mathcal{P}_k = X|(a)_k$  where
$X \in \{I, R, L, U\}$, $a \in \{1,2\}$.
\end{proposition}

\begin{proof}

The reduction $\mathcal{P}_i$ ($i \ge 3$) is composed
of Hilbert units if and only if $\pi_i = \pi_{i-1} = a$ (Proposition
\ref{prop:unit}). 
Then all the reductions in $3 \le i \le k$ are all composed of Hilbert units
if and only if $\mathcal{P}_k = X|\pi_1(a)_{k-1}$.

When $k = 2$, according to the first shape group in Table
\ref{tab:corner-induced} and the first curve in Figure
\ref{fig:level_2_shapes}, $\mathcal{P}_2 = X|(a)_2$ where $X \in \{I, R, L,
U\}$ and $a \in \{1, 2\}$ is the only form of the Hilbert unit.

Then according the definition, $\mathcal{P}_k$ is a Hilbert curve if and only
if $\mathcal{P}_k = X|(a)_k$  where $X \in \{I, R, L, U\}$, $a \in \{1,2\}$.

\end{proof}

\begin{remark}
\label{remark:hc}
All possible forms of Hilbert curves on the same level $k$ have the same shape.
\end{remark}

\begin{proof}
The encodings of the Hilbert curves indicate they are only from the shape group
1 of corner-induced curves (Table \ref{tab:corner-induced}). Thus all Hilbert
curves on level $k$ in different encodings have the same shape.
\end{proof}

\subsection{The Hilbert variant}\label{the-hilbert-variant}

\begin{definition}[Hilbert variant]
For a curve $\mathcal{P}_k$, if the lowest level-2 units of $\mathcal{P}_{i} =
\mathrm{Rd}_{k-i}(\mathcal{P}_k)$ are always Hilbert units for all $2 < l+2
\le i \le k$, while the lowest level-2 units of $\mathcal{P}_{l+1}$ are not
Hilbert units, $\mathcal{P}_k$ is called an \textit{order-$l$ Hilbert variant}
($l \ge 1$).
\end{definition}

\begin{proposition}
\label{prop:hc2}
$\mathcal{P}_k$ is an order-$l$ Hilbert variant iff $\mathcal{P}_k =
X|\pi_1...\pi_l(a)_{k-l}$ ($k - l \ge 2$) with the following requirements:

\begin{enumerate}
\tightlist
\item
  If $l \ge 2$, then $\pi_l \ne a$ and there is no restriction on the type of $X$.
\item
  If $l = 1$ and $\pi_l \ne a$, then there is no restriction on the type of $X$.
\item
  If $l = 1$ and $\pi_l = a$, then $X \in \{B, D, P, Q, C\}$. 
\end{enumerate}

If $\mathcal{P}_k$ is expressed as a list of $4^l$ level $k-l$ subunits, then
each subunit is a Hilbert curve.
\end{proposition}

\begin{proof}
When $l \ge 2$, we can reduce $\mathcal{P}_k$ to $\mathcal{P}_{l+1} =
X|\pi_1...\pi_la$ level-by-level, where in previous reduction steps, the last
two code are always $aa$, thus all the lowest level-2 units are Hilbert units.
Now we look at $\mathcal{P}_{l+1}$ ($l+1 \ge 3$). In this category,
$\mathcal{P}_{l+1}$ is not composed of Hilbert units if and only if $\pi_l \ne
a$ (Proposition \ref{prop:unit}). Thus, when $l \ge 2$, $\mathcal{P}_k$ is an
order-$l$ Hilbert variants iff $\pi_l \ne a$.

When $l = 1$, similarly, we can reduce $\mathcal{P}_k$ to $\mathcal{P}_{2} =
X|\pi_1a$. According to Table \ref{tab:corner-induced} and Figure
\ref{fig:level_2_shapes}, $\mathcal{P}_2$ is not a Hilbert unit if and only
if $\pi_l \ne a$, or $\pi_l = a$ and $X \in \{B, D, P, Q, C\}$. Thus in this
category, $\mathcal{P}_k$ is an order-$l$ Hilbert variant iff requirement 2
or 3 is satisfied.

Let's write $\mathcal{P}_k$ as a list of level $k-l$ subunits: $\mathcal{P}_k
= X_{<\pi_1...\pi_l>}|(a)_{k-l} = Z...|(a)_{k-l} = \mathcal{U}_1 ... \mathcal{U}_e$ where
$\mathcal{U}_1 = Z|(a)_{k-l}$ and $\mathcal{U}_* = Z_{*}|(a_*)_{k-l}$. With $l
\ge 1$, we have $Z, Z_* \in \{I, R, L\}$. With Corollary \ref{coro:5.6.1},
$(a_*)_{k-l}$ is either $(1)_{k-l}$ or $(2)_{k-l}$. Then with Proposition
\ref{prop:hc1}, $\mathcal{U}_1$ and $\mathcal{U}_*$ are all Hilbert curves on
level $k-l$.
\end{proof}

\begin{figure}
\centering{
\includegraphics[width=1\linewidth]{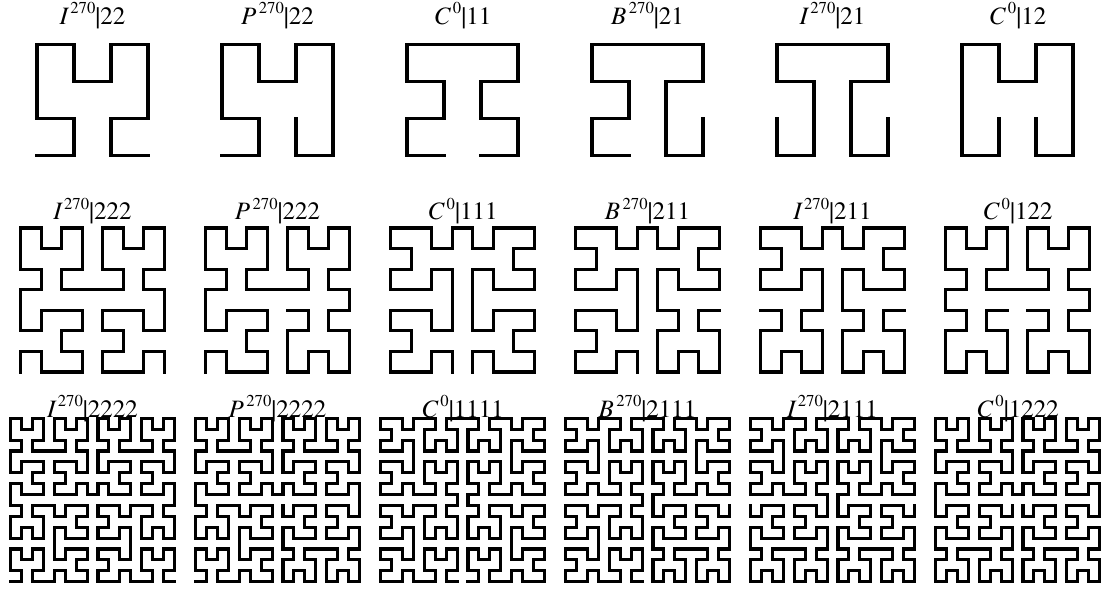}
\caption{Hilbert curve and type-$V_1$ to $V_5$ order-1 Hilbert variants on level 2, 3 and 4.}\label{fig:hilbert_variants}
}
\end{figure}

\subsubsection{Order-1 Hilbert variant}\label{order-1-hilbert-variant}

\begin{proposition}
\label{prop:hc_od1}
The union of Hilbert curves and order-1 Hilbert variants compose the full set of
corner-induced curves.
\end{proposition}

\begin{proof}
For the curve $\mathcal{P}_k = X|\pi_1(a)_{k-1}$, when $\pi_1 \ne a$,
$\mathcal{P}_k$ is a order-1 Hilbert variant; when $\pi_1 = a$ and $X \in \{B,
D, P, Q, C\}$, $\mathcal{P}_k$ is a order-1 Hilbert variant; and when $\pi_1 =
a$ and $X \in \{I, R, L, U\}$, $\mathcal{P}_k$ is a Hilbert curve. Thus the
union of Hilbert curves and order-1 Hilbert variants compose the full set of
corner-induced curves.
\end{proof}

According to Table \ref{tab:corner-induced}, all corner-induced curves
are classified into six shape groups, where shape group 1 only includes
Hilbert curves and other five groups include order-1 Hilbert variants.
They are named type-\(V_1\) to type-\(V_5\) in Table
\ref{tab:Liu-variants}.

\citet{LIU2004741} studied the structure of 2x2 curves and concluded
that there are six variants of general Hilbert curves, including the
standard Hilbert curve, the Moore curve and four other variants termed
as \(L_1\) to \(L_4\). Our analysis revealed that the Moore curve and
Liu-variants \(L_1\) to \(L_4\) are actually order-1 Hilbert variants in
different shapes. The shape groups of the Liu-variants, their
correspondance to the classification of order-1 Hilbert variants are
listed in Table \ref{tab:Liu-variants}, with their corresponding curves
illustrated in Figure \ref{fig:hilbert_variants}.

According to Proposition \ref{prop:hc2}, the Hilbert curve as well as
order-1 Hilbert variants can be expressed as a list of four Hilbert
curves on level \(k-1\) (\(k \ge 3\)). Their structures are determined
by their level-2 global structures \(X|\pi_1 a\) and the facing of four
subunits of \(\mathcal{P}_k\) are also determiend by the facings of four
2x2 units in \(X|\pi_1 a\). Then the construction of the Hilbert curve
and order-1 Hilbert variants can be expressed in the copy-paste mode
(Section \ref{introduction}) by pasting the four level \(k-1\) Hilbert
curve and positioning them in their specific facings. The facing of the
four subunits are listed in Table \ref{tab:Liu-variants}. If the curve
is considered as directional, reflections might need to be applied on
some of the subunits.

\begin{table}
\centering
\begin{tabular}{clrrl}
\toprule
Shape group & $\mathcal{P}_k$ & Type & Liu-variant & Facing of four subunits \\
\midrule
1  &  $I^{270}|(2)_{k}$ & Hilbert & Hilbert & left, down, down, right \\
2  &  $P^{270}(2)_{k}$ & $V_1$ & $L_3$ & left, down, down, up \\
3  &  $C|(1)_{k}$ & $V_2$ & Moore  & right, right, left, left \\
4  &  $B^{270}|2(1)_{k-1}$ & $V_3$ & $L_4$  & right, right, left, up  \\
5  &  $I^{270}|2(1)_{k-1}$ & $V_4$ & $L_2$ & up, right, left, up \\
6  &  $C|1(2)_{k-1}$ & $V_5$ & $L_1$ & up, down, down up \\
\bottomrule
\end{tabular}
\vspace*{5mm}
\caption{\label{tab:Liu-variants}The shape groups of the Hilbert curve and order-1 Hilbert variants, as well as the classification by Liu. The first curve in each shape group is selected for the ``$\mathcal{P}_k$'' column in the table. Shape groups are from Table \ref{tab:corner-induced}. Note curves in each group can be transformed by rotations, reflections and reversals, then the values in the last columns should be adjusted accordingly.}
\end{table}

\begin{proposition}
\label{prop:all_hc_units}
If $\mathcal{P}_k$ is a Hilbert curve or an order-1 Hilbert variant, then its
unit on any location with level $2 \le l < k$ is always a Hilbert curve.
\end{proposition}

\begin{proof}
We write

\begin{equation}
\begin{aligned}
\mathcal{P}_k &= X|\pi_1(a)_{k-1} \\
       &= X|\pi_1(a)_{k-l-1}|a_{k-l}...a_{k-1} \\
       &= Z_1...Z_*...|a_{k-l}...a_{k-1} \\
       &= Z_1|a_{k-l}...a_{k-1}\ ...\ Z_*|a_{k-l*}...a_{k-1*}\ ... \\
       &= \mathcal{U}_1...\mathcal{U}_{*}... \\
\end{aligned} .
\notag
\end{equation}

$X|\pi_1(a)_{k-l-1}$ is a curve with level $\ge 1$, thus $Z_1...Z_*...$ is
only composed of $I$/$R$/$L$ (Proposition \ref{prop:IRL}). With Corollary
\ref{coro:5.6.1}, $a_{k-l*}...a_{k-1*} = a_{k-l}...a_{k-1}$ or
$a_{k-l*}...a_{k-1*} = \hat{a}_{k-l}...\hat{a}_{k-1}$. For both scenarios, all
code in $a_{k-l*}...a_{k-1*}$ are all the same. Thus $\mathcal{U}_1$ and
$\mathcal{U}_*$ are all Hilbert curves on level $l$.
\end{proof}

\subsection{\texorpdfstring{The
\(\beta\Omega\)-curve}{The \textbackslash beta\textbackslash Omega-curve}}\label{bo-curve}

The definition and further description of the \(\beta\Omega\)-curve is
very similar as the Hilbert curve.

\begin{definition}[$\beta\Omega$-curve]
For a curve $\mathcal{P}_k$, if the lowest level-2 units of $\mathcal{P}_i =
\mathrm{Rd}_{k-i}(\mathcal{P}_k)$ are always $\beta$-units and $\Omega$-units
(Section \ref{global-structure-and-local-unit}) for all $2 \le i \le k$, then
$\mathcal{P}_k$ is called a $\beta\Omega$-curve.
\end{definition}

In this and next sections, we use the notation \((a_1...a_k)\) for a
sequence where digits 1 and 2 appear alternatively, i.e.,
\(a_i = \hat{a}_{i-1}\) (\(2 \le
i \le k\)). And we explicitly use \((1212...)\) and \((2121...)\) (at
least two explicit digits) for such cases.

\begin{proposition}
\label{prop:betaomega}
$\mathcal{P}_k$ is a $\beta\Omega$-curve iff $\mathcal{P}_k = X|(a_1...a_k)$
where $X \in \{I, R, L, B, P\}$.
\end{proposition}

\begin{proof}
The reduction $\mathcal{P}_i$ ($i \ge 3$) is composed of $\beta\Omega$-units
if and only if $a_i = \hat{a}_{i-1}$ (Proposition \ref{prop:unit}). Then all
the reductions in $3 \le i \le k$ are all composed of $\beta\Omega$-units if
and only if $\mathcal{P}_k = X|\pi_1(a_2...a_k)$.

When $k = 2$, according to the shape groups 4 and 5 from Table
\ref{tab:corner-induced} (also curves 4 and 5 in Figure
\ref{fig:level_2_shapes}), the form $\mathcal{P}_2 = X|a_1 a_2$ where $X \in
\{I, R, L, B, P\}$ and $a_2 = \hat{a}_1$ is the only form of the $\beta\Omega$-unit.

Then according the definition, $\mathcal{P}_k$ is a $\beta\Omega$-curve if and
only if $\mathcal{P}_k = X|(a_1...a_k)$ where $X \in
\{I, R, L, B, P\}$.

\end{proof}

\begin{figure}
\centering{
\includegraphics[width=0.8\linewidth]{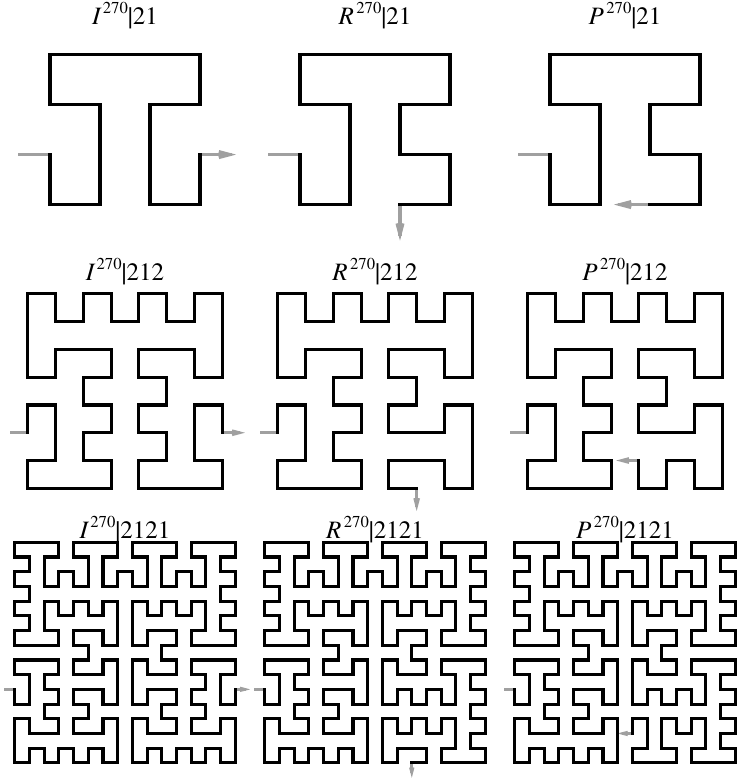}
\caption{Three shapes of the $\beta\Omega$-curves on level 2, 3, and 4. Entry and exit directions are additionally added to distinguish the second and the third shapes.}\label{fig:beta_omega_curve}
}
\end{figure}

When \(k \ge 3\), it is easy to see the \(\beta\Omega\)-curve is a
side-induced curve since the last two code are always different in the
expansion code sequence. Then according to Table \ref{tab:side-induced},
on the same level \(k\), all forms of the \(\beta\Omega\)-curves have
three possible shapes according to their level-1 units, listed in Table
\ref{tab:beta-omega-type} and Figure \ref{fig:beta_omega_curve}. We name
the first type of \(\beta\Omega\)-curves as type-\(O\) because
\(\mathcal{P}_2\) has an \(\Omega\)-shape, and the other two types as
type-\(B_1\) and type-\(B_2\) because their \(\mathcal{P}_2\) have
\(\beta\)-shapes.

\begin{table}
\centering
\begin{tabular}{clcll}
\toprule
Shape group & $\mathcal{P}_k$ & Type & Type of the four subunits & Facing of the four subunits\\
\midrule
1  & $I^{270}|(2121...)$ & $O$ & $B_1B_1B_1B_1$ & up, right, left, up\\
2  & $R^{270}|(2121...)$ & $B_1$ & $B_1B_1B_1O$ & up right left left \\
3  & $P^{270}|(2121...)$ & $B_2$ & $B_1B_1B_1B_1$ & up, right, left, left \\
\bottomrule
\end{tabular}
\vspace*{5mm}
\caption{\label{tab:beta-omega-type}The three types of $\beta\Omega$-curves. The first curve in each shape group is selected for the ``$\mathcal{P}_k$'' column in the table. The shape groups are from Table \ref{tab:side-induced}. Note curves in each group can be transformed by rotations, reflections and reversals, then the values in the last columns should be adjusted accordingly.}
\end{table}

It is easy to see if \(\mathcal{P}_k\) (\(k \ge 3\)) is a
\(\beta\Omega\)-curve, its four subunits are also
\(\beta\Omega\)-curves. Taking \(I^{270}|(2121...)\) (type-\(O\) from
shape group 1) as an example, its level-1 curve \(I^{270}|2\) only
contains bases \(R\) and \(L\), thus \(I^{270}|(2121...)\) is composed
of four type-\(B_1\) \(\beta\Omega\)-curves on level \(k-1\).
Additionally the facings of the four subunits are determined by
\(I^{270}|21\) which are up, right, left and up. Then the construction
of the type-\(O\) \(\beta\Omega\)-curve can be expressed in the
copy-paste mode by pasting four copies of type-\(B_1\)
\(\beta\Omega\)-curves on level \(k-1\) and positioning them in their
specific facings. The types of their subunits as well as the facings are
listed in the last two columns in Table \ref{tab:beta-omega-type}. If
the curve is considered as directional, reflections might need to be
applied on some of the subunits.

Type-\(B_2\) \(\beta\Omega\)-curve cannot be used as subunits to
construct higher-level \(\beta\Omega\)-curve under the copy-paste mode
because the entry and exit directions of a Type-\(B_2\) have a
difference of 180 while a valid \(\beta\)-subunit always have a
difference of 90 between its entry and exit directions.

\subsection{\texorpdfstring{The
\(\beta\Omega\)-variant}{The \textbackslash beta\textbackslash Omega-variant}}\label{the-betaomega-variant}

\begin{definition}[$\beta\Omega$-variant]
For a curve $\mathcal{P}_k$, if the lowest level-2 units of $\mathcal{P}_i =
\mathrm{Rd}_{k-i}(\mathcal{P}_k)$ are always $\beta$-units and $\Omega$-units
for all $2 < l+2 \le i \le k$, while the lowest level-2 units of
$\mathcal{P}_{l+1}$ are not $\beta$-units or $\Omega$-units, $\mathcal{P}_k$
is called an \textit{order-$l$ $\beta\Omega$-variant} ($l \ge 1$).
\end{definition}

\begin{proposition}
\label{prop:betaomega-variants}
$\mathcal{P}_k$ is an order-$l$ $\beta\Omega$-variant iff $\mathcal{P}_k =
X|\pi_1...\pi_l(a_{l+1}...a_k)$ ($k-l \ge 2$) with 
the following requirements:

\begin{enumerate}
\tightlist
\item
  If $l \ge 2$, then $\pi_l = a_{l+1}$ and there is no restriction on the type of $X$.
\item
  If $l = 1$ and $\pi_l = a_{l+1}$, there is no restriction on the type of $X$.
\item
  If $l = 1$ and $\pi_l \ne a_{l+1}$, then $X \in \{U, D, Q, C\}$. 
\end{enumerate}

If $\mathcal{P}_k$
is expressed as a list of $4^l$ level $k-l$ subunits, then each subunit is a
$\beta\Omega$-curve.
\end{proposition}

\begin{proof}
When $l \ge 2$, we can reduce $\mathcal{P}_k$ to $\mathcal{P}_{l+1} =
X|\pi_1...\pi_la_{l+1}$ level-by-level, where in previous reduction steps, the last
two code are always different, thus all the lowest level-2 units are $\beta\Omega$-units.
Now we look at $\mathcal{P}_{l+1}$ ($l+1 \ge 3$). In this category,
$\mathcal{P}_{l+1}$ is not composed of $\beta\Omega$-units if and only if $\pi_l =
a_{l+1}$ (Proposition \ref{prop:unit}). Thus, when $l \ge 2$, $\mathcal{P}_k$ is an
order-$l$ $\beta\Omega$-variants iff $\pi_l = a_{l+1}$.

When $l = 1$, similarly, we can reduce $\mathcal{P}_k$ to $\mathcal{P}_{2} =
X|\pi_1a_2$. According to Table \ref{tab:corner-induced} and Figure
\ref{fig:level_2_shapes}, $\mathcal{P}_2$ is not a $\beta\Omega$-unit if and only
if $\pi_l = a_2$, or $\pi_l \ne a$ and $X \in \{U, D, Q, C\}$. Thus in this
category, $\mathcal{P}_k$ is an order-$l$ $\beta\Omega$-variant iff requirement 2
or 3 is satisfied.

Let's write $\mathcal{P}_k$ as a list of level $k-l$ subunits: $\mathcal{P}_k
= X_{<\pi_1...\pi_l>}|(a_{l+1}...a_k) = Z...|(a_{l+1}...a_k) = \mathcal{U}_1
... \mathcal{U}_e$ where $\mathcal{U}_1 = Z|(a_{l+1}...a_k)$ and
$\mathcal{U}_* = Z_{*}|(a_{l+1*}...a_{k*})$. With $l
\ge 1$, we have $Z, Z_* \in \{I, R, L\}$. With Corollary \ref{coro:5.6.1},
$(a_{l+1*}...a_{k*})$ is either $(a_{l+1}...a_k)$ or
$(\hat{a}_{l+1}...\hat{a}_k)$. Then with Proposition \ref{prop:betaomega},
$\mathcal{U}_1$ and $\mathcal{U}_*$ are all $\beta\Omega$-curves on level
$k-l$.
\end{proof}

\subsubsection{\texorpdfstring{Order-1
\(\beta\Omega\)-variant}{Order-1 \textbackslash beta\textbackslash Omega-variant}}\label{order-1-betaomega-variant}

\begin{figure}
\centering{
\includegraphics[width=1\linewidth]{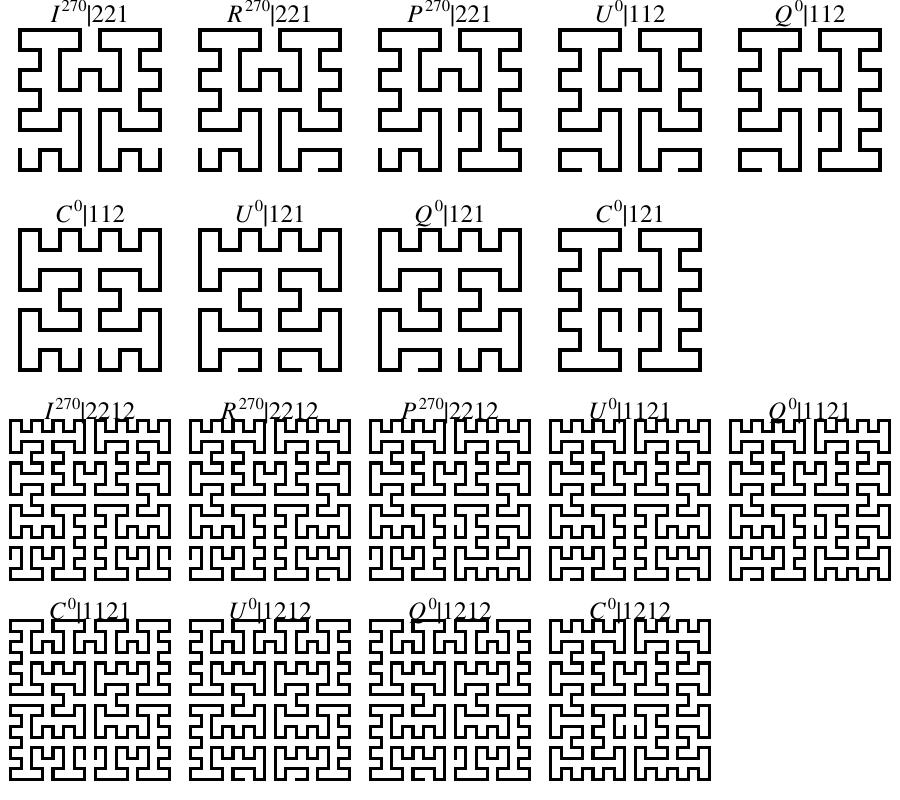}
\caption{The nine types of order-1 $\beta\Omega$-variants on level 3 and 4.}\label{fig:beta_omega_variants}
}
\end{figure}

According to Proposition \ref{prop:betaomega-variants}, all order-1
\(\beta\Omega\)-variants are \(X|a_2(a_2...a_k)\) and
\(Y|\hat{a}_2(a_2...a_k)\) where \(Y \in \{U, D, Q, C\}\). All order-1
\(\beta\Omega\)-variants on the same level \(k\) (\(k \ge 3\)) have nine
possible different shapes listed in Table \ref{tab:beta-omega-variants}
and illustrated in Figure \ref{fig:beta_omega_variants}. We term these
nine types type-\(V_1\) to type-\(V_9\).

The four subunits of order-1 \(\beta\Omega\)-variants are all
\(\beta\Omega\)-curves on level \(k-1\). Then the construction of
order-1 \(\beta\Omega\)-variants can also be expressed in the copy-paste
mode where each subunit is a specific type of \(\beta\Omega\)-curve and
is positioned in its specific facing. The types of the subunits and
their facings for each order-1 \(\beta\Omega\)-variant are listed in the
last two columns in Table \ref{tab:beta-omega-variants}.

In Equation \ref{eq:betaomega} (Section \ref{special-curves}), we give
one encoding for the \(\beta\Omega\)-curve of which the structure is
often used in literatures. Here we can see the curve in the encoding is
actually an order-1 \(\beta\Omega\)-variant in type \(V_6\).

\begin{table}
\centering
\begin{tabular}{clcll}
\toprule
Shape group & $\mathcal{P}_k$ & Type & Type of the four subunits & Facing of the four subunits  \\
\midrule
1  & $I^{270}|2(212...)$ & $V_1$ & $B_1B_1B_1B_1$ & left, down, down, right \\
2  & $R^{270}|2(212...)$ & $V_2$ & $B_1B_1B_1O$ & left, down, down, right \\
3  & $P^{270}|2(212...)$ & $V_3$ & $B_1B_1B_1B_1$ & left, down, down, up \\
4  & $U|1(121...)$ & $V_4$ & $OB_1B_1O$ & left, down, down, right\\
5  & $Q|1(121...)$ & $V_5$ & $OB_1B_1B_1$ & left, down, down, up\\
6  & $C|1(121...)$ & $V_6$ & $B_1B_1B_1B_1$ & right, right, left, left \\
\midrule
4  & $U|1(212...)$ & $V_7$ & $OB_1B_1O$ & right, right, left, left \\
5  & $Q|1(212...)$ & $V_8$ & $OB_1B_1B_1$ & right, right, left, left \\
6  & $C|1(212...)$ & $V_9$ & $B_1B_1B_1B_1$ & up, down, down, up\\
\bottomrule
\end{tabular}
\vspace*{5mm}
\caption{\label{tab:beta-omega-variants}The nine types of order-1 $\beta\Omega$-variants. The first curve in each shape group is selected in the table. The shape groups are from Table \ref{tab:side-induced}. Note curves in each group can be transformed by rotations, reflections and reversals, then the values in the last columns should be adjusted accordingly. Base $D$ induces curves in the same shape as $Q$, thus it is not listed in the table.}
\end{table}

\begin{proposition}
If $\mathcal{P}_k$ is a $\beta\Omega$-curve or an order-1 $\beta\Omega$-variant, then its
unit on any location with level $2 \le l < k$ is always a $\beta\Omega$-curve.
\end{proposition}

\begin{proof}
We write

\begin{equation}
\begin{aligned}
\mathcal{P}_k &= X|\pi_1(ab...)_{[k-1]} \\
       &= X|\pi_1(ab...)_{[k-l-1]}|(ab...)_{[l]} \\
       &= Z_1...Z_*...|(ab...)_{[l]} \\
       &= Z_1|(ab...)_{[l]}\ ...\ Z_*|(a_*b_*...)_{[l]}\ ... \\
       &= \mathcal{U}_1...\mathcal{U}_{*}... \\
\end{aligned} .
\notag
\end{equation}

$X|\pi_1(ab...)_{[k-l-1]}$ is a curve with level $\ge 1$, thus $Z_1...Z_*...$ is
only composed of $I$/$R$/$L$ (Proposition \ref{prop:IRL}). With Corollary
\ref{coro:5.6.1}, $(a_*b_*...)_{[l]} = (ab...)_{[l]}$ or
$(a_*b_*...)_{[l]} = (\hat{a}\hat{b}...)_{[l]}$. For both scenarios, all
code in $(a_*b_*...)_{[l]}$ change alternatively. Thus $\mathcal{U}_1$ and
$\mathcal{U}_*$ are all $\beta\Omega$-curves on level $l$.
\end{proof}

\subsection{\texorpdfstring{Relations of Hilbert curves, Hilbert
variants, \(\beta\Omega\)-curves, and
\(\beta\Omega\)-variants}{Relations of Hilbert curves, Hilbert variants, \textbackslash beta\textbackslash Omega-curves, and \textbackslash beta\textbackslash Omega-variants}}\label{relations-of-hilbert-curves-hilbert-variants-betaomega-curves-and-betaomega-variants}

\begin{proposition}
\label{prop:hc_bo_full}
The full set of 2x2 curves (level $\ge 3$) are composed of Hilbert curves, Hilbert variants, $\beta\Omega$-curves and $\beta\Omega$-variants.
\end{proposition}

\begin{proof}
If the last $i$ ($2 \le i \le k-1$) code of $\mathcal{P}_k$ are all the same, written as $\mathcal{P}_k = X|(\pi)_{k-i-1}b(a)_{i}$ ($a \ne b$). With Proposition \ref{prop:hc2}, if $k-i-1 \ge 1$, $\mathcal{P}_k$ is an order-$i$ Hilbert variant.

If $\mathcal{P}_k = X|(a)_k$, when $X \in \{B, D, P, Q, C\}$, $\mathcal{P}_k$ is an order-1 Hilbert variant; and when $X \in \{I, R, L, U\}$, $\mathcal{P}_k$ is a Hilbert curve.

If the code sequence of $\mathcal{P}_k$ is ended with $(ab...)_{[i]}$\footnote{Subscript ``$[i]$'' represents the length of the sequence.} ($2 \le i \le k-1$), 
written as $\mathcal{P}_k = X|(\pi)_{k-i-1}a(ab...)_{[i]}$. With Proposition \ref{prop:betaomega-variants}, 
if $k-i-1 \ge 1$, $\mathcal{P}_k$ is an order-$i$ $\beta\Omega$-variant.

If $\mathcal{P}_k = X|(ab...)_{[k]}$, when $X \in \{U, D, Q, C\}$, $\mathcal{P}_k$ is an order-1 $\beta\Omega$-variant; and when $X \in \{I, R, L, B, O\}$, $\mathcal{P}_k$ is a $\beta\Omega$-curve.
\end{proof}

\begin{proposition}
The Hilbert curves have completely distinct shapes from the
$\beta\Omega$-curves.
\end{proposition}

\begin{proof}
Any reduction $\mathcal{P}_i$ ($2 \le i \le k$) of a Hilbert curve
$\mathcal{P}_k$ is still a Hilbert curve, and any reduction $\mathcal{Q}_i$
($2 \le i \le k$) of a $\beta\Omega$-curve $\mathcal{Q}_k$ is still a
$\beta\Omega$-curve curve. Then $\mathcal{P}_i$ is always composed of Hilbert
units and $\mathcal{Q}_i$ is always composed of $\beta\Omega$-units. According
to Proposition \ref{prop:diff_shapes}, $\mathcal{P}_k$ has complete distinct
shapes from $\mathcal{Q}_k$.
\end{proof}

Let \(\mathcal{P}_k\) be a Hilbert curve or an order-1 Hilbert variant,
\(\mathcal{Q}_k\) be a \(\beta\Omega\)-curve or an order-1
\(\beta\Omega\)-variant. In Proposition \ref{prop:diff}, condition 2 is
always satisfied. Then if
\(\mathcal{S}(\mathcal{P}_2) \ne \mathcal{S}(\mathcal{Q}_2)\),
\(\mathcal{P}_k\) and \(\mathcal{Q}_k\) have completely distinct shapes.
Table \ref{tab:hc_compare} categorizes all Hilbert curve/variants and
\(\beta\Omega\)-curves/variants based on their level-2 shapes where on
each row, curves have the same level-2 shape. Then a Hilbert
curve/variant has a completely distinct shape from a
\(\beta\Omega\)-curve/variant if they are form different rows in Table
\ref{tab:hc_compare}.

\begin{table}
\centering
\begin{tabular}{ccc}
\toprule
Hilbert curve / variants & $\beta\Omega$-curves / variants & Level-2 shape  \\
\midrule
Hilbert & $V_1$/$V_2$/$V_4$ & $I|22$ \\
$V_1$   & $V_3$/$V_5$ & $P|22$ \\
$V_2$   & $V_6$/$V_7$/$V_8$ & $C|11$ \\
$V_3$   & $\beta\Omega$-$B_1$/$\beta\Omega$-$B_2$ & $B|21$\\
$V_4$   & $\beta\Omega$-$O$ & $I|21$\\
$V_5$   & $V_9$ & $C|12$ \\
\bottomrule
\end{tabular}
\vspace*{5mm}
\caption{\label{tab:hc_compare}Categorize Hilbert curve/variants and $\beta\Omega$-curves/variants based on their level-2 shapes.}
\end{table}

\subsection{Hierarchical generation}\label{hierarchical-generation}

According to Section \ref{hc_shape}, the full set of shapes of 2x2
curves can be generated hierarchically. As Proposition
\ref{prop:hc_bo_full} indicates, Hibert curves, Hilbert variants,
\(\beta\Omega\)-curves and \(\beta\Omega\)-variants also compose the
full set of 2x2 curves. Then, if we only look at the shapes of the four
types of curves, the full set of them can be also be generated in a
hierarchical procedure.

\begin{figure}
\centering{
\includegraphics[width=1\linewidth]{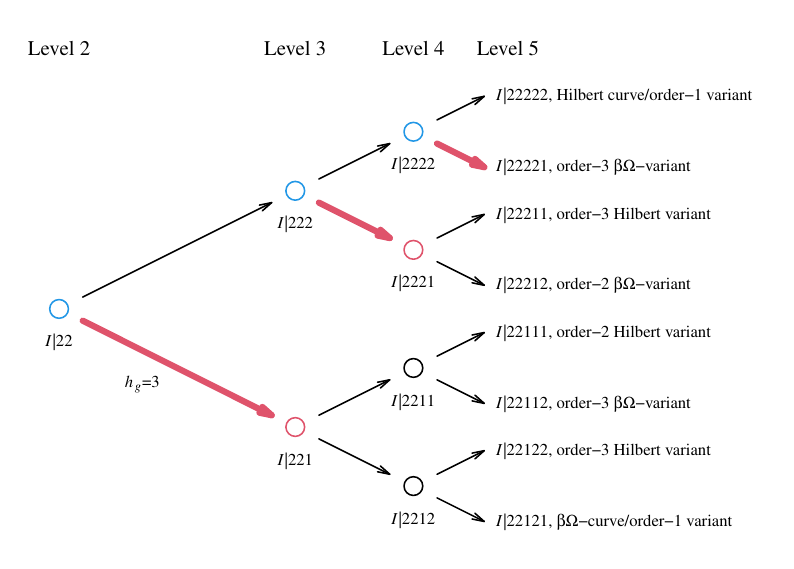}
\caption{Hierarchical generation of Hilbert curves/variants and $\beta\Omega$-curves/variants. The diagram illustrates the generation from $I|22$ (corner-induced shape group 1). Red arrow represents there are $h_g$ forms of side-induced curves generated from a corner-induced curve in the previous level.}\label{fig:hc_bo_generation}
}
\end{figure}

The hierarchical generation starts from a certain level-2 shape group.
Figure \ref{fig:hc_bo_generation} illustrates the hierarchical
generation of curves to level 5 in the shape group of \(I|22\) (shape
group 1, Table \ref{tab:corner-induced}). In the diagram, from level
\(i-1\) to level \(i\), if the last expansion code is not changed in the
curve, we use an up-right arrow \(\nearrow\) to link them; if the code
changes, we use a down-right arrow \(\searrow\). Then in the diagram,
corner-induced curves on any level is always located on the top border
line. According to Section \ref{hc_shape}, when a corner-induced level
is expanded to the next level as a side-induced curve (i.e., changing
the last code), there are \(h_g\) different forms (Table
\ref{tab:shape_seed_2}) depending on which level-2 shape group it is
generated from. In the diagram, we explicitly use a thick red arrow to
represent the branch under the red arrow is just one of \(h_g\) forms.

In the diagram, the first curve on level \(i\) (i.e., all with up-right
arrows on its generation path) is always a Hilbert curve (if it is
generated from shape group 1) or an order-1 Hilbert variant (if it is
generated from shape group 2-6). The last curve (i.e., all with
down-right arrows on its generation path) is always a
\(\beta\Omega\)-curve (if generated from shape group 4-5) or an order-1
\(\beta\Omega\)-variants (if generated from shape group 1-3 and 6)
(Table \ref{tab:hc_compare}). Denote \(H^{(i)}(j)\) and \(B^{(i)}(j)\)
as an order-\(j\) Hilbert variant and an order-\(j\)
\(\beta\Omega\)-variant on level \(i\) where \(1
\le j \le i - 2\). \(H^{(i)}(1)\) represent an order-\(1\) Hilbert
variant or a Hilbert curve, depending which shape group it is from.
Similarly \(B^{(i)}(1)\) represent an order-\(1\)
\(\beta\Omega\)-variant or a \(\beta\Omega\)-curve. The generation from
level \(i-1\) to level \(i\) can be summarized into the following
diagrams. When \(i = 2\), we replace \(H^{(i-1)}(j)\) and
\(B^{(i-1)}(j)\) with \(G^{(2)}\) as they correspond to the level-2 base
shape in this group.

\begin{center}

\begin{tikzpicture}
    \node(formula){$H^{(i-1)}(j)$};
    \node(solution1) [above right =-0.5em and 2em of formula]{$H^{(i)}(j)$};
    \node(solution2) [below right =-0.5em and 2em of formula]{$B^{(i)}(i-2)$};
    \draw [->] (formula.east) --  (solution1.west);
    \draw [->] (formula.east) --  (solution2.west);
\end{tikzpicture}

\begin{tikzpicture}
    \node(formula){$B^{(i-1)}(j)$};
    \node(solution1) [above right =-0.5em and 2em of formula]{$H^{(i)}(i-2)$};
    \node(solution2) [below right =-0.5em and 2em of formula]{$B^{(i)}(j)$};
    \draw [->] (formula.east) --  (solution1.west);
    \draw [->] (formula.east) --  (solution2.west);
\end{tikzpicture}

\end{center}

Let's reformat these two diagrams to:

\begin{equation}
\label{eq:hc_bo_generation}
\begin{aligned}
& H^{(i)}(j) \leftarrow H^{(i-1)}(j)  \quad & 1 \le j \le i-3 \\
& H^{(i)}(i-2) \leftarrow B^{(i-1)}(j)  \quad & 1 \le j \le i-3 \\
& B^{(i)}(j) \leftarrow B^{(i-1)}(j)  \quad & 1 \le j \le i-3 \\
& B^{(i)}(i-2) \leftarrow H^{(i-1)}(j)  \quad & 1 \le j \le i-3 \\
\end{aligned} .
\end{equation}

With Equation \ref{eq:hc_bo_generation}, we can study how a order-\(i\)
Hilbert or \(\beta\Omega\)-variant is generated. There is only one
unique path in the hierarchical diagram to generate each of
\(H^{(k)}(1)\), \(B^{(k)}(1)\), \(H^{(k)}(2)\), \(B^{(k)}(2)\)
(\(k \ge 3\)).

\begin{alignat*}{11}
&H^{(k)}(1) &\leftarrow &H^{(k-1)}(1) &\leftarrow &... &\leftarrow &H^{(4)}(1) &\leftarrow &H^{(3)}(1) &\leftarrow &G^{(2)}\\ 
&B^{(k)}(1) &\leftarrow &B^{(k-1)}(1) &\leftarrow &... &\leftarrow &B^{(4)}(1) &\leftarrow &B^{(3)}(1) &\leftarrow &G^{(2)} \\
&H^{(k)}(2) &\leftarrow &H^{(k-1)}(2) &\leftarrow &... &\leftarrow &H^{(4)}(2) &\leftarrow &B^{(3)}(1) &\leftarrow &G^{(2)} \\ 
&B^{(k)}(2) &\leftarrow &B^{(k-1)}(2) &\leftarrow &... &\leftarrow &B^{(4)}(2) &\leftarrow &H^{(3)}(1) &\leftarrow &G^{(2)} \\
\end{alignat*}

\noindent where \(G^{(2)}\) represent the level-2 shape. Next take
\(H^{(k)}(3)\) for example:

\begin{equation}
\begin{aligned}
& H^{(k)}(3) \leftarrow H^{(k-1)}(3) \leftarrow ... \leftarrow H^{(5)}(3) \\ 
\end{aligned} .
\notag
\end{equation}

Now \(H^{(5)}(3)\) has two options to be generated into:

\begin{equation}
\begin{aligned}
& H^{(5)}(3) \leftarrow B^{(4)}(2) \\ 
& H^{(5)}(3) \leftarrow B^{(4)}(1) \\ 
\end{aligned} ,
\notag
\end{equation}

\noindent which makes two different paths to generate \(H^{(k)}(3)\).
More generally, for \(H^{(k)}(i)\) (\(3 \le i \le k-2\)), the generation
is

\begin{equation}
H^{(k)}(i) \leftarrow H^{(k-1)}(i) \leftarrow ... H^{(i+2)}(i)
\notag
\end{equation}

\noindent and

\begin{equation}
\begin{aligned}
H^{(i+2)}(i) & \leftarrow B^{(i+1)}(1) \\
& ... \\
H^{(i+2)}(i) & \leftarrow B^{(i+1)}(i-1) \\
\end{aligned} .
\notag
\end{equation}

The procedure is the same if using \(B^{(k)}(i)\). Denote \(n(k, i)\) as
the number of paths to generate \(H^{(k)}(i)\) or \(B^{(k)}(i)\), then

\begin{equation}
\left\{\begin{aligned}
n(k, i) &= n(i+1, 1) + n(i+1, 2) + ...  + n(i+1, i-1) \quad 3 \le i \le k-2 \\
n(k, 1) &= 1 \\
n(k, 2) &= 1 \\
\end{aligned}\right.
\notag
\end{equation}

\noindent which is identical to

\begin{equation}
\left\{\begin{aligned}
n(k, i) &= n(i+1, 1) + n(i+1, 2) + ...  + n(i+1, i-1) \quad 2 \le i \le k-2 \\
n(k, 1) &= 1 \\
\end{aligned}\right. .
\notag
\end{equation}

Note \(n(k, i) = n(i+1, i)\), then we can solve the previous equations
to

\begin{equation}
n(k, i) = \begin{cases}
2^{i-2} & \quad 2 \le i \le k-2 \\
1 & \quad i = 1 \\
\end{cases} .
\notag
\end{equation}

When the curve is side-induced, the number of total forms should be
multiplied by \(h_g\).

\begin{equation}
n(k, i) = \begin{cases}
h_g \times 2^{i-2} & \quad 2 \le i \le k-2 \textnormal{ for both } H^{(k)}(i) \textnormal{ and } B^{(k)}(i)\\
1 & \quad \textnormal{for } H^{(k)}(1) \\
h_g & \quad \textnormal{for } B^{(k)}(1) \\
\end{cases}
\notag
\end{equation}

Each curve is generated via a unique path thus in unique shape. The
total number of shapes of \(H^{(k)}(i)\) and \(B^{(k)}(i)\)
(\(1 \le i \le k-2\)) is
\(1 + h_g + 2 \times \sum h_g \times 2^{i-2} = 1 + h_g \times 2^{k-2} - h_g\).
Then the total number of shapes of Hilbert variants and
\(\beta\Omega\)-varints generated from all the six level-2 shape groups
is

\begin{equation}
\sum_{g \in \{1,...,6\}}(1 + h_g \times 2^{k-2} - h_g) = 6 \times 2^{k-1} - 6
\notag
\end{equation}

\noindent which is identical to the total number of shapes in Equation
\ref{eq:shapes_nn} and Section \ref{hc_shape} for \(k \ge 2\).

\section{Other structures}\label{structures}

In this section, we only consider a curve initialized from a single
base.

\subsection{Recursive curves}\label{recursive-curves}

In many studies, the space-filling curve is described to have
self-similarity where structure of the curve is recursively inherited
from its lower levels.

\begin{definition}[Recursive curve]
For a reduction $\mathcal{P}_i = \mathrm{Rd}_{k-i}(\mathcal{P}_k)$, if the
shapes of its four subunits on level $i-1$ as well as its depth-1 reduction
are always the same, i.e., $\mathcal{S}(\mathcal{U}_1) =
\mathcal{S}(\mathcal{U}_2) = \mathcal{S}(\mathcal{U}_3) =
\mathcal{S}(\mathcal{U}_4) = \mathcal{S}(\mathcal{P}_{i-1})$, for all $3 \le i \le k$,
then $\mathcal{P}_k$ is recursive.
\end{definition}

In the definition, the scenario of \(i = 2\) is excluded because
\(\mathcal{U}_*\) and \(\mathcal{P}_1\) are always in the same
\textit{U}-shapes. This definition is similar to Definition 7.1 in
\citet{bader}.

\begin{proposition}
\label{prop:9.1}
There are two types of recursive curves: 1. the Hilbert curve on any level $k
\ge 2$, and 2. a level-3 curve $X|121$ or $X|212$ where $X \in \{B, P\}$.
\end{proposition}

\begin{proof}
We first look at the corner-induced curves. With Proposition
\ref{prop:hc_od1}, corner-induced curves are composed of
Hilbert curves and order-1 Hilbert variants. For the reduction of a Hilbert
curve $\mathcal{P}_i = X|(a)_i$ ($X \in \{I, R, L, U\}$, $3 \le i \le k$), its
reduction $\mathcal{P}_{i-1} = X|(a)_{i-1}$, and its four subunits
$Z_*|(a_*)_{i-1}$ ($Z_* \in \{I, R, L\}$, $a_* = 1$ or 2) are all Hilbert
curves on level $i-1$. So they are always in the same shape (Remark
\ref{remark:hc}). Thus the Hilbert curve is a resursive curve. For order-1
Hilbert variant, with Proposition \ref{prop:hc2}, when reducing
$\mathcal{P}_k$ to $\mathcal{P}_3$, all its four level-2 units are Hilbert
units, but its further depth-1 reduction $\mathcal{P}_2$ is not a Hilbert
unit. This makes $\mathcal{S}(\mathcal{U}_*) \ne \mathcal{S}(\mathcal{P}_2)$
on $\mathcal{P}_3$, then the order-1 Hilbert variants are not recursive.

Next we consider the side-reduced curve $X|\pi_1(\omega)_{i-1}$. Since at
least two neighbouring code are different, we use 12 as an example. The proof
for the scenario of 21 is basically the same. The original curve is written
as:

\begin{equation}
\label{eq:recursive_big}
X|\pi_1(\pi_*)_{k_1}12(...) \quad \pi_1 \in \{1, 2\}, k_1 \ge 0
\end{equation}

\noindent where $(\pi_*)_{k_1}$ is a sequence of arbitrary code of length $k_1$ and
$(...)$ is also a sequence of arbitrary code. We only consider its reduced
version denoted as $\mathcal{P}_r$:

\begin{equation}
\mathcal{P}_r = X|\pi_1(\pi_*)_{k_1}12
\notag
\end{equation}

If we write $X|\pi_1 = Z_1 Z_2 Z_3 Z_4$, the first subunit of $\mathcal{P}_r$
is $\mathcal{V}_1 = Z_1|(\pi_*)_{k_1}12$, and its depth-1 reduction is
$\mathcal{P}_{r-1} = \mathrm{Rd}_1(\mathcal{P}_r) = X|\pi_1(\pi_*)_{k_1}1$.
Let's check whether $\mathcal{P}_r$ is recursive. There are two scenarios.

\textit{Scenario 1}. If $k_1 \ge 1$, $\mathcal{V}_1$ is side-induced because
its code sequence $(\pi_*)_{k_1}12$ has a length $\ge$ 3 and the last two
code are different. Also note $Z_1 \in \{I, R, L\}$, then if $\mathcal{V}_1$
and $\mathcal{P}_{r-1}$ have the same shape, they should be all from shape
group 1 or 2 in Table \ref{tab:side-induced}. For the other subunits
$\mathcal{V}_2$, $\mathcal{V}_3$ and $\mathcal{V}_4$, their code sequences are
either the same as $\mathcal{V}_1$ or the complement (Corollary
\ref{coro:5.6.1}), so they are also side-induced curves and they should also
come from shape group 1 or 2 accordingly. Then there are two possible
combinations of values for $Z_*$ and $X$ if $\mathcal{V}_*$ and
$\mathcal{P}_{r-1}$ have the same shape:

\begin{equation}
\label{eq:rec_base}
\begin{cases}
Z_* &= I \\
X &= I \\
\end{cases}
\quad
\textnormal{ or }
\quad
\begin{cases}
Z_* & \in \{R, L\} \\
X & \in \{R, L\} \\
\end{cases} .
\end{equation}

As $X = Z_1Z_2Z_3Z_4$, when $X = I$, there must be $R$/$L$ in $Z_*$; and when
$X \in \{R, L\}$, there must be $I$ in $Z_*$. Thus the conditions in Equation
\ref{eq:rec_base} are impossible and for scenario 1, and $\mathcal{P}_r$ is not
recursive.

\textit{Scenario 2}. If $k_1 = 0$, then $\mathcal{V}_1 = Z_1|12$ and
$\mathcal{P}_{r-1} = X|\pi_11$. We first exclude the scenario $\pi_1 = 1$
because $Z_1|12$ is a $\beta$- or $\Omega$-unit but $X|11$ always comes from
shape group 1-3 in Table \ref{tab:corner-induced} or the first three shapes in
Figure \ref{fig:level_2_shapes}, never a $\beta\Omega$-unit. So we only
discuss $\mathcal{P}_{r-1} = X|21$. Notice when a certain $Z_* \in \{R, L\}$,
$Z_*|12$ or $Z_*|21$ is a $\beta$-unit; and when $Z_* = I$, $Z_*|12$ or
$Z_*|21$ is a $\Omega$-unit. The two types of units have different shapes. It
is impossible that all four $Z_*$ are $I$, then we restrict to $Z_* \in \{R,
L\}$. We then look up in all level-1 expansion rules in Figure
\ref{fig:expansion_rule}, only $I^{(1)}$, $B^{(1)}$, $P^{(1)}$ and $C^{(1)}$
are composed of $R$/$L$, which makes $\mathcal{P}_3$ being represented as a
list of $\beta$-units. In them, we additionally exclude $I^{(1)}$ and
$C^{(1)}$ because for these two scenario $\mathcal{P}_2$ does not have the
$\beta$-unit shape.

Now we have the only recursive form for side-induced curves: $\mathcal{P}_r =
X|212$ ($X \in \{B, P\}$. Of course there is another form $X|121$ but we omit
the discussion here), but only on level 3. Next we go back to Equation
\ref{eq:recursive_big} and rewrite $\mathcal{P}_k$ as

\begin{equation}
\begin{aligned}
\mathcal{P}_k &= X|212(...)_{k_2} \quad X \in \{B, P\}, k_2 \ge 1 \\
\mathcal{P}_{k-1} &= X|212(...)_{k_2 - 1} \\
\mathcal{V}_1 &= W_1|12(...)_{k_2} \\
\end{aligned}
\notag
\end{equation}

\noindent where $W_1$ is the first base of $X|2$ and $(...)_{k_2}$ is a sequence of code
of length $k_2$. $\mathcal{P}_{k-1}$ is a side-induced curve because the
second and the third code are different. If $(...)_{k_2} = (2)_{k_2}$ which
makes $\mathcal{V}_1$ a corner-induced curve, apparently $\mathcal{P}_{k-1}$
has a different shape from $\mathcal{V}_1$. If there are at least two code
different in $2(...)_{k_2}$ which makes $\mathcal{V}_1$ also a side-induced
curve, since $W_1 \in \{I, R, L\}$ and $X \in \{B, P\}$, $\mathcal{P}_{k-1}$
and $\mathcal{V}_1$ are not in the same shape groups (Table
\ref{tab:side-induced}). Thus $\mathcal{P}_k$ is not recursive from level 4 in
this category.

\end{proof}

\subsection{Subunit identically shaped
curves}\label{subunit-identically-shaped-curves}

\begin{definition}
For a reduction $\mathcal{P}_i$, if the
shapes of its four subunits on level $i-1$ are always the same, i.e.,
$\mathcal{S}(\mathcal{U}_{1}) = \mathcal{S}(\mathcal{U}_{2}) =
\mathcal{S}(\mathcal{U}_{3}) = \mathcal{S}(\mathcal{U}_{4})$ for every $3 \le i
\le k$, then $\mathcal{P}_k$ is called a subunit identically shaped curve.
\end{definition}

Compared to the recursive curve, a subunit identically shaped curve does
not require \(\mathcal{U}_*\) to have the same shape as
\(\mathcal{P}_{i-1}\). We discuss corner-induced curves and side-induced
curves separately.

\begin{proposition}
All corner-induced curves are subunit identically shaped.
\end{proposition}

\begin{proof}
A Corner-induced curve $\mathcal{P}_k$ has four Hilbert curves as its four
subunits (Proposition \ref{prop:all_hc_units}), thus with four identically
shaped subunits. $\mathcal{P}_i$ is still a corner-induced curve for all $3
\le i \le k$, Thus $\mathcal{P}_k$ is a subunit identically shaped curve.
\end{proof}

\begin{lemma}
\label{lemma:RL_shape}

\begin{enumerate}
\tightlist
\item
  $R|(\pi)_k$, $R|(\hat{\pi})_k$, $L|(\pi)_k$, $L|(\hat{\pi})_k$ ($k \ge 2$) are always in the same shape.
\item
  Let $\mathcal{P}_k$ be one of the four forms. If $\mathcal{Q}_k$ is
  initialized from $R$/$L$ and has the same shape as $\mathcal{P}_k$, then
  $\mathcal{Q}_k$ should also be one of the four forms.
\end{enumerate}

Rotations are ommited in the two statements.
\end{lemma}

\begin{proof}
First we prove statement 1. There are the following relations:

\begin{equation}
\begin{aligned}
\mathcal{S}(L|(\hat{\pi})_k) &= \mathcal{S}(h(R|(\pi)_k)) \\
\mathcal{S}(R|(\hat{\pi})_k) &= \mathcal{S}(h(r(R|(\pi)_k))) \\
\mathcal{S}(L|(\pi)_k) &= \mathcal{S}(h(R|(\hat{\pi})_k)) \\
\end{aligned}
\notag
\end{equation}

So the four types of curves are always in the same shape.

Next we prove statement 2. Let $\mathcal{P}_k = X|(\pi)_k$ and $\mathcal{Q}_k = Y|(\sigma)_k$. 
According to Proposition \ref{prop:diff_2code}, if $\mathcal{P}_k$ and $\mathcal{Q}_k$
are in the same shape, then for any $2 \le i < j \le k$, it is always $\pi_i = \sigma_i$, $\pi_j = \sigma_j$,
or $\pi_i = \hat{\sigma}_i$, $\pi_j = \hat{\sigma}_j$. This results in $\pi_2...\pi_k = \sigma_2...\sigma_k$ or
$\pi_2...\pi_k = \hat{\sigma}_2...\hat{\sigma}_k$. We also require $\mathcal{P}_2$ and $\mathcal{Q}_2$ in the
same shape. Denote both $R$ and $L$ as $W$. With $\mathcal{P}_2 = W|\pi_1\pi_2$ and $\mathcal{Q}_2 = W|\sigma_1\sigma_2$,
from Table \ref{tab:corner-induced}, $\pi_1\pi_2 = \sigma_1\sigma_2$ or  $\pi_1\pi_2 = \hat{\sigma}_1\hat{\sigma}_2$.
Then $\sigma_1...\sigma_k = \pi_1...\pi_k$ or $\sigma_1...\sigma_k = \hat{\pi}_1...\hat{\pi}_k$.

The two statements can also be validated directly from Table \ref{tab:corner-induced} and \ref{tab:side-induced}.
\end{proof}

\begin{lemma}
\label{lemma:RL_shape2}
Write $\mathcal{P}_k = Z_1Z_2Z_3Z_4|\pi_2...\pi_k = \mathcal{U}_1\mathcal{U}_2\mathcal{U}_3\mathcal{U}_4$.
If $Z_* \in \{L, R\}$, then $\mathcal{U}_*$ are in the same shape.
\end{lemma}

\begin{proof}
In $\mathcal{U}_* = Z_*|(\pi_{2*}...\pi_{k*})$, the code sequence $(\pi_{2*}...\pi_{k*})$ is
either $\pi_2...\pi_k$ or its complement $\hat{\pi}_2...\hat{\pi}_k$ (Corollary \ref{coro:5.6.1}),
then according to Lemma \ref{lemma:RL_shape}, $\mathcal{U}_*$ are in the same shape.
\end{proof}

\begin{proposition}
\label{prop:subunit_ident_side}
The side-induced curves that are subunit identically shaped should have the following
form:

\begin{equation}
X|\pi_1(\omega)_{k-1} \quad X \in \{I, B, P, C\}, k \ge 3 .
\notag
\end{equation}

\end{proposition}

\begin{proof}

$\mathcal{P}_k = X^{(1)}|(\omega)_{k-1} = Z_1 Z_2 Z_3 Z_4|(\omega)_{k-1}$, we
write the four subunit as (the first code in the $\omega$-sequence is moved
out and denoted explicitly as $\omega_2$ or $\omega_{2*}$):

\begin{equation}
\mathcal{U}_* = Z_*|\omega_{2*}(\omega_*)_{k-2} .
\notag
\end{equation}

If code in $(\omega_*)_{k-2}$ ($k \ge 3$) are all the same and they are only
different from $\omega_{2*}$, then $\mathcal{U}_*$ are all
corner-induced curves. With the two constraints of $\omega_{2*}$ being different
from $(\omega_*)_{k-2}$ and $Z_* \in \{I, R, L\}$, from Table
\ref{tab:corner-induced}, $\mathcal{U}_*$ can only take values from
$R|1(2)_{k-2}$/$L|2(1)_{k-2}$/$R|2(1)_{k-2}$/$L|1(2)_{k-2}$ (Group 4) or
$I|2(1)_{k-2}$/$I|1(2)_{k-2}$ (Group 5). Note $Z_*$ is the level-1 expansion
of $X$, then it is not possible that all $Z_*$ are $I$. So $Z_*$ should only
contain $R$/$L$, this results in $X \in \{I, B, P, C\}$.

If at least two code are different in $(\omega_*)_{k-2}$ ($k \ge 4$), then
$\mathcal{U}_*$ are all side-induced curves. If
they are in the same shape group, according to Table \ref{tab:side-induced},
all $Z_*$ should be all $I$ or all $R$/$L$. Note $Z_*$ is the level-1 expansion of
$X$, then it is not possible that all $Z_*$ are $I$. So $Z_*$ should only
contain $R$/$L$. This also makes $X \in \{I, B, P, C\}$. 

Then according to Lemma \ref{lemma:RL_shape2}, the four subunits of $\mathcal{P}_k$ always have
the same shape.

The reduction $\mathcal{P}_i$ is also side-induced for all $3 \le i \le k$, with its four subunits
always in the same shape. Thus the side-induced curve $\mathcal{P}_k$ is subunit identically shaped
when $X \in \{I, B, P, C\}$.

\end{proof}

\subsection{Subunit differently shaped and completely non-recursive
curves}\label{diffcurve}

\begin{definition}
If for the reduction $\mathcal{P}_i$, at least two shapes of $\mathcal{U}_1$, $\mathcal{U}_2$,
$\mathcal{U}_3$ and $\mathcal{U}_4$ are different for
every $3 \le i \le k$, then $\mathcal{P}_k$ is called a subunit differently shaped curve.
\end{definition}

Let's explore the form of curves that is subunit differently shaped.
With Proposition \ref{prop:all_hc_units}, if \(\mathcal{P}_i\) is a
corner-induced curve, its four subunits are all Hilbert curves on level
\(i-1\) in the same shape. Thus, any reduction of \(\mathcal{P}_k\)
cannot be a Hilbert curve. Then we first restrict \(\mathcal{P}_k\) to
the form \(X|\pi_1(ab)\pi_{4}...\pi_{k}\) (if \(k = 3\), then
\(\pi_{4}...\pi_{k}\) is an empty sequence) where the second and the
third code should be different or complementary.

Next write \(\mathcal{P}_k = X|\pi_1(ab)\pi_{4}...\pi_{k} =
Z_1Z_2Z_3Z_4|(ab)\pi_{4}...\pi_{k} =
\mathcal{U}_1\mathcal{U}_2\mathcal{U}_3\mathcal{U}_4\). If
\(\mathcal{U}_*\) are corner-induced curves, i.e.,
\(\pi_4...\pi_k = (b)_{k-3}\), to make at least two of
\(\mathcal{U}_* = Z_*|a_*(b_*)_{k-2}\) to have different shapes (note it
is also \(a_* \ne b_*\)), \(Z_*\) should contain both \(I\) and
\(R\)/\(L\) (shape group 5 and 4 in Table \ref{tab:corner-induced}),
then \(X \in \{R, L, U, D, Q\}\). If \(\mathcal{U}_*\) are side-induced
curves, we write \(\mathcal{U}_* =
Z_*|a_*(\omega_*)_{k-2}\), similarly, \(Z_*\) should also contain both
\(I\) and \(R\)/\(L\), then also \(X \in \{R, L, U, D, Q\}\).

\begin{proposition}
\label{prop:no_ident}
$\mathcal{P}_k$ ($k \ge 3$) is subunit differently shaped if
$\mathcal{P}_k = X|\pi_1a\hat{a}$ ($k = 3$) or $\mathcal{P}_k =
X|\pi_1a\hat{a}\pi_4...\pi_k$ ($k \ge 4$) where $X \in \{R, L, U, D, Q\}$.
\end{proposition}

\begin{proof}
According to the previous discussion, the curve $\mathcal{P}_k = X|\pi_1a\hat{a}\pi_4...\pi_k$ has subunits in
different shapes if $X \in \{R, L, U, D, Q\}$ for $k \ge 4$. Then the reduction $\mathcal{P}_i$ for any $4 \le i \le k$
has subunits in different shapes. Reduction to level 3 $\mathcal{P}_3 = X|\pi_1a\hat{a}$ also has subunits 
in different shapes. Thus $\mathcal{P}_k$ ($k \ge 3$) is subunit differently shaped. 

\end{proof}

Next we make a stronger statement.

\begin{definition}
If for the reduction $\mathcal{P}_i$, at least two shapes of $\mathcal{U}_1$,
$\mathcal{U}_2$, $\mathcal{U}_3$ and $\mathcal{U}_4$ are different and
$\mathcal{S}(\mathcal{P}_{i-1}) \ne \mathcal{S}(\mathcal{U}_j)$ (for all $1 \le j
\le 4$) for every $3 \le i \le k$, then $\mathcal{P}_k$ is completely non-recursive
or has completely no self-similarity.
\end{definition}

To explore the form of \(\mathcal{P}_k\), we directly start with
\(\mathcal{P}_k
= X|\pi_1(ab)\pi_{4}...\pi_{k} = Z_1Z_2Z_3Z_4|(ab)\pi_{4}...\pi_{k} =
\mathcal{U}_1\mathcal{U}_2\mathcal{U}_3\mathcal{U}_4\) with
\(X \in \{R, L, U,
D, Q\}\) from Proposition \ref{prop:no_ident}.

\textit{Scenario 1}. \(\mathcal{U}_*\) is corner-induce (\(k \ge 3\)).
We write

\begin{equation}
\begin{aligned}
\mathcal{U}_* &= Z_*|a_*(b_*)_{k-2} \\
\mathcal{P}_{k-1} &= X|\pi_1a(b)_{k-3} \\
\end{aligned} .
\notag
\end{equation}

The second and the third code in \(\mathcal{U}_*\) are \(b_*b_*\) (two
identical code) and in \(\mathcal{P}_{k-1}\) are \(ab\) (two different
code). According to Proposition \ref{prop:diff_2code}, the four
\(\mathcal{U}_*\) and \(\mathcal{P}_{k-1}\) always have different shapes
for reduction \(\mathcal{P}_i\) with \(4 \le i \le k\). When \(i = 3\),

\begin{equation}
\begin{aligned}
\mathcal{U}_* &= Z_*|a_*b_* \\
\mathcal{P}_{2} &= X|\pi_1a \\
\end{aligned} .
\notag
\end{equation}

Notice since \(X \in \{R, L, U, D, Q\}\), \(Z_*\) contains both \(I\)
and \(R\)/\(L\) which results in that \(\mathcal{U}_*\) includes both
\(\beta\)- and \(\Omega\)-units. Then \(\mathcal{P}_2 = X|\pi_1a\)
cannot be a \(\beta\)- or \(\Omega\)-unit. This results

\begin{equation}
\mathcal{P}_2 = \begin{cases}
X|aa \quad & \textnormal{if } X \in \{R, L\} \\
X|\pi_1a \quad & \textnormal{if } X \in \{U, D, Q\} \\
\end{cases} .
\notag
\end{equation}

Now we have the first form of \(\mathcal{P}_k\) (\(k \ge 3\)) if its
subunits are corner-induced:

\begin{equation}
\label{eq:no_recur1}
\mathcal{P}_k = \begin{cases}
X|aa(b)_{k-2} \quad & \textnormal{if } X \in \{R, L\} \\
X|\pi_1a(b)_{k-2} \quad & \textnormal{if } X \in \{U, D, Q\} \\
\end{cases} .
\end{equation}

\textit{Scenario 2}. \(\mathcal{U}_*\) is side-induced (\(k \ge 4\)). We
write \(\mathcal{U}_* = Z_*|a_*b_*\pi_{4*}...\pi_{k*}\) where at least
two code are different in \(b_*\pi_{4*}...\pi_{k*}\). The form of
\(\mathcal{P}_{k-1}\) is

\begin{equation}
\mathcal{P}_{k-1} = X|\pi_1ab\pi_4...\pi_{k-1} ,
\notag
\end{equation}

\noindent and obvious \(\mathcal{P}_{k-1}\) is also side-induced because
the second and the third code are different. If \(X \in \{U, D, Q\}\),
it is always
\(\mathcal{S}(\mathcal{U}_*) \ne \mathcal{S}(\mathcal{P}_{i-1})\) for
\(\mathcal{P}_i\) till \(i = 4\) because they always come from different
side-induced shape groups with different set of initial seed. When
reducing to \(i = 3\), \(\mathcal{P}_{i-1}\) and four \(\mathcal{U}_*\)
are all corner-induced. With Equation \ref{eq:no_recur1},
\(\mathcal{P}_{i-1}\) and four \(\mathcal{U}_*\) always have different
shapes. Then we have the second form of \(\mathcal{P}_k\) in this
subcategory:

\begin{equation}
\label{eq:no_recur2}
\mathcal{P}_k = X|\pi_1ab\pi_4...\pi_k \quad \textnormal{if } X \in \{U, D, Q\}, k \ge 4 .
\end{equation}

If \(X \in \{R, L\}\) whose level-1 expansion contains both \(I\) and
\(R\)/\(L\). If \(Z_1\) or \(Z_4\) is \(I\), then corresponding
\(\mathcal{U}_1\) or \(\mathcal{U}_4\) has a different shape from
\(\mathcal{P}_{k-1}\) because the latter is from shape group initialized
from \(R\)/\(L\). If \(Z_1\) or \(Z_4\) is \(R\)/\(L\), they have the
same shape as \(\mathcal{U}_2\) or \(\mathcal{U}_3\) because their code
sequences are the same. Additionally \(\mathcal{U}_2\) always has the
same shape as \(\mathcal{U}_3\) (Lemma \ref{lemma:RL_shape2}). Then we
only need to consider
\(\mathcal{S}(\mathcal{P}_{k-1}) \ne \mathcal{S}(\mathcal{U}_2)\). There
are two possible forms of \(\mathcal{U}_2\) depending on \(X|\pi_1\):

\begin{equation}
\mathcal{U}_2 = \begin{cases}
Z_2|ab\pi_4...\pi_k \\
Z_2|\hat{a}\hat{b}\hat{\pi}_4...\hat{\pi}_k \\
\end{cases} .
\notag
\end{equation}

We first consider the opposite case,
\(\mathcal{S}(\mathcal{P}_{k-1}) = \mathcal{S}(\mathcal{U}_2)\). As in
this category, \(X, Z_2 \in \{R, L\}\), with Lemma \ref{lemma:RL_shape},
it is only possible

\begin{equation}
\begin{aligned}
& \pi_1ab\pi_4...\pi_k = ab\pi_4...\pi_k  \\
\textnormal{ or } & \pi_1ab\pi_4...\pi_k = \hat{a}\hat{b}\hat{\pi}_4...\hat{\pi}_k \\
\end{aligned} .
\notag
\end{equation}

This gives the solution (note \(b = \hat{a}\))

\begin{equation}
\pi_4...\pi_k = \begin{cases}
(\hat{a}a...)_{[k-3]} \quad & \textnormal{if } \pi_1 = a \\
(a\hat{a}...)_{[k-3]} \quad & \textnormal{if } \pi_1 = \hat{a} \\
\end{cases} .
\notag
\end{equation}

And the negation

\begin{equation}
\pi_4...\pi_k \ne \begin{cases}
(\hat{a}a...)_{[k-3]} \quad & \textnormal{if } \pi_1 = a \\
(a\hat{a}...)_{[k-3]} \quad & \textnormal{if } \pi_1 = \hat{a} \\
\end{cases}
\notag
\end{equation}

\noindent ensures for reduction \(\mathcal{P}_i\) (\(4 \le i \le k\)),
\(\mathcal{S}(\mathcal{P}_{i-1}) \ne \mathcal{S}(\mathcal{U}_2)\). When
reducing to \(i = 3\), \(\mathcal{P}_{i-1}\) and four \(\mathcal{U}_*\)
are all corner-induced. With Equation \ref{eq:no_recur1}, it only allows
\(\pi_1 = a\). Then we have the third form of \(\mathcal{P}_k\) in this
subcategory:

\begin{equation}
\label{eq:no_recur3}
\begin{aligned}
\mathcal{P}_k = X|aa\hat{a}\pi_4...\pi_k \quad & \textnormal{if } X \in \{R, L\}, \pi_4...\pi_k \ne (\hat{a}a...)_{[k-3]} \\
    & \textnormal{ and at least two code are different in } \hat{a}\pi_4...\pi_k \\
\end{aligned}
\end{equation}

We sum Equation \ref{eq:no_recur1}, \ref{eq:no_recur2} and
\ref{eq:no_recur3} up to the following proposition.

\begin{proposition}
\label{prop:non_recur}
$\mathcal{P}_k$ being completely non-recursive should have the following form.

\begin{equation}
\mathcal{P}_k = \begin{cases}
X|aa\hat{a}\pi_4...\pi_k \quad & \textnormal{if } X \in \{R, L\}, \pi_4...\pi_k \ne (\hat{a}a...)_{[k-3]}\\
X|\pi_1a\hat{a}\pi_4...\pi_k \quad & \textnormal{if } X \in \{U, D, Q\} \\
\end{cases}
\notag
\end{equation}

\end{proposition}

\begin{figure}
\centering{
\includegraphics[width=0.8\linewidth]{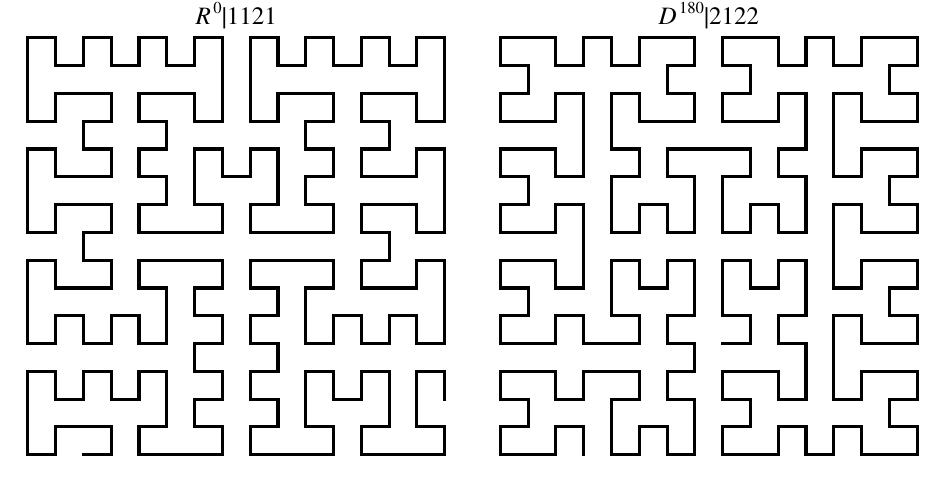}
\caption{Two examples of completely non-recursive curves.}\label{fig:non_recursive}
}
\end{figure}

Figure \ref{fig:non_recursive} lists two example curves for the two
groups in Proposition \ref{prop:non_recur}. \(\beta\Omega\)-curves are
not completely non-recursive, i.e., they show self-similarity on certain
levels. When \(X \in
\{I, B, P\}\), \(\mathcal{P}_k\) are subunit identically shaped
(Proposition \ref{prop:subunit_ident_side}). When \(X \in \{R, L\}\),
\(\mathcal{P}_{k-1}\) always has the same shape as \(\mathcal{U}_2\) and
\(\mathcal{U}_3\). For all the order-1 \(\beta\Omega\)-variants, only
type-\(V_2\), \(V_4\), \(V_5\), \(V_7\) and \(V_8\) are completely
non-recursive. Other types, i.e., \(V_1\), \(V_3\), \(V_6\) and \(V_9\),
are subunit identically shaped with seed \(I\), \(P\) and \(U\).

\subsection{Symmetric curves}\label{symmetric-curves}

\begin{definition}

Let $\mathcal{P}_k$ be a curve in the base facing state. 
Write $\mathcal{P}_k$ as a list of four level
$k-1$ subunits ($k \ge 2$): $\mathcal{P}_k  =
\mathcal{U}_1\mathcal{U}_2\mathcal{U}_3\mathcal{U}_4$. There are the following
three types of symmetries:

\begin{enumerate}
\tightlist
\item
  If $\mathcal{H}(\mathcal{U}_1) = h(r(\mathcal{U}_4))$ and $\mathcal{U}_2 =
  h(r(\mathcal{U}_3))$, then $\mathcal{P}_k$ is type-A symmetric.
\item
  If $\mathcal{H}(\mathcal{U}_1) = v(r(\mathcal{U}_2))$ and
  $\mathcal{H}(\mathcal{U}_4) = v(r(\mathcal{U}_3))$, then $\mathcal{P}_k$ is
  type-B symmetric.
\item
  If $\mathcal{P}_k$ is both type-A and type-B symmetric, then it is called type-AB
  symmetric.
\end{enumerate}

\end{definition}

\(f_t(\mathcal{P}_k)\) has the same symmetry type as \(\mathcal{P}_k\)
where \(f_t()\) is arbitrary combinations of rotations, reflections and
reversals. When \(\mathcal{P}_k\) faces downward, type-A symmetry
corresponds to horizontal symmetry and type-B corresponds to vertical
symmetry.

\begin{remark}
If the curve is type-B symmetric, it is also type-A symmetric, thus type-AB
symmetric.
\end{remark}

\begin{proof}
We first consider a curve $\mathcal{P}_k$ in the base facing state.

On level-1 reduction of $\mathcal{P}_k$, the second base is always $R$ (bottom-in, right-out) and the third
base is always $R^{270}$ (left-in, bottom-out). If $\mathcal{U}_2 =
R|(\pi_*)_{k-1}$, according to Lemma \ref{lemma:code1}, $\mathcal{U}_3 =
R^{270}|(\hat{\pi}_*)_{k-1}$. Then we apply Corollary \ref{coro:rev} to get
$r(\mathcal{U}_3) = L|(\hat{\pi}_*)_{k-1}$, and finally we apply Proposition
\ref{prop:6.5} to get that it is always $h(r(\mathcal{U}_3)) = R|(\pi_*)_{k-1} = \mathcal{U}_2$.

$\mathcal{P}_k$ is type-B symmetric, then this means (by applying Corollary
\ref{coro:rev} and Proposition \ref{prop:6.5}):

\begin{equation}
\begin{aligned}
\mathcal{H}(\mathcal{U}_1) &= v(r(\mathcal{U}_2)) \\
              &= v(r(R|(\pi_*)_{k-1})) \\
              &= R^{90}|(\hat{\pi}_*)_{k-1}
\end{aligned} ,
\notag
\end{equation}

\noindent and 

\begin{equation}
\begin{aligned}
\mathcal{H}(\mathcal{U}_4) &= v(r(\mathcal{U}_3)) \\
              &= v(r(R^{270}|(\hat{\pi}_*)_{k-1})) \\
              &= R^{180}|(\pi_*)_{k-1} \\
\end{aligned} .
\notag
\end{equation}

Applying reversal and horizontal reflection to $\mathcal{H}(\mathcal{U}_4)$:

\begin{equation}
\begin{aligned}
h(r(\mathcal{H}(\mathcal{U}_4))) &= h(r( R^{180}|(\pi_*)_{k-1} )) \\
        &= R^{90}|(\hat{\pi}_*)_{k-1} \\
\end{aligned} ,
\notag
\end{equation}

\noindent we have

\begin{equation}
\mathcal{H}(\mathcal{U}_1) = h(r(\mathcal{H}(\mathcal{U}_4))) .
\notag
\end{equation}

Finally according to Proposition \ref{prop:homo_tr}, 

\begin{equation}
\begin{aligned}
\mathcal{H}(\mathcal{U}_1) &= h(r(\mathcal{H}(\mathcal{U}_4))) \\
  &= \mathcal{H}(h(r(\mathcal{U}_4))) \\
  &= h(r(\mathcal{U}_4)) \\
\end{aligned} .
\notag
\end{equation}

The statement of this Remark is also true for $f_t(\mathcal{P}_k)$.
\end{proof}

\subsubsection{Type-A symmetric curves}\label{type-a-symmetric-curves}

If \(\mathcal{P}_k\) is type-A symmetric, the facings of
\(\mathcal{U}_1\) and \(\mathcal{U}_4\) should also be symmetric. Note a
curve has the same facing as its level-1 ``U-shape'' unit. We reduce
\(\mathcal{P}_k\) to level 2 which is composed of four 2x2 units in
U-shapes. Since \(\mathcal{U}_2\) always has the symmetric facing of
\(\mathcal{U}_3\), we only need to require \(\mathcal{U}_1\) to have the
symmetric facing of \(\mathcal{U}_4\), where they should face all upward
(note both downward facing is not valid for the curve in its base facing
state), or one leftward and the other rightward. According to Figure
\ref{fig:level_2_shapes} and Table \ref{tab:corner-induced}, the level-2
structure that \(\mathcal{U}_1\) and \(\mathcal{U}_4\) have symmetric
facings are in shape groups 1, 3, 5, 6, which correspond to homogeneous
family 1, 3, 6, 8. The inducing level-2 seeds from the four homogeneous
families (Table \ref{tab:homogeneous_curves}) as well as the complete
encodings of \(\mathcal{P}_k\) are listed in Table \ref{tab:symmetric-a}
where we only include curves in the base facing state and other forms
can be obtained simply by rotation and reflection.

\begin{table}
\centering
\begin{tabular}{crrrc}
\toprule
\makecell{Shape group \\ of $\mathcal{P}_2$} & $\mathcal{P}_k$  &  $\mathcal{U}_1$ & $h(r(\mathcal{U}_4))$  & Type-A \\
\midrule
1  &  $I^{270}|22|(\pi)_{k-2}$ & $L^{270}|2(\pi)_{k-2}$ & $L^{270}|2(\pi)_{k-2}$ & always \\
   &  $R^{270}|22|(\pi)_{k-2}$ & $L^{270}|2(\pi)_{k-2}$ & $I|1(\hat{\pi})_{k-2}$ & when all $\pi=2$ \\
   &  $R|11|(\pi)_{k-2}$ & $I|1(\pi)_{k-2}$ &  $L^{270}|2(\hat{\pi})_{k-2}$ & when all $\pi=1$ \\
   &  $U|11|(\pi)_{k-2}$ & $I|1(\pi)_{k-2}$ &  $I|1(\pi)_{k-2}$ & always \\
\midrule
3  &  $C|11|(\pi)_{k-2}$ & $R^{90}|1(\pi)_{k-2}$ & $R^{90}|1(\pi)_{k-2}$ & always \\
   &  $D^{180}|11|(\pi)_{k-2}$ & $R^{90}|1(\pi)_{k-2}$ & $I|2(\hat{\pi})_{k-2}$ & when all $\pi=1$ \\
   &  $Q|12|(\pi)_{k-2}$ & $I|2(\pi)_{k-2}$ &  $R^{90}|1(\hat{\pi})_{k-2}$ & when all $\pi=2$ \\
   &  $U|12|(\pi)_{k-2}$ & $I|2(\pi)_{k-2}$ &  $I|2(\pi)_{k-2}$ & always \\
\midrule
5  &  $I^{270}|21|(\pi)_{k-2}$ & $L^{270}|1(\pi)_{k-2}$ &  $L^{270}|1(\pi)_{k-2}$ & always \\
\midrule
6  &  $C|12|(\pi)_{k-2}$ & $R^{90}|2(\pi)_{k-2}$ &  $R^{90}|2(\pi)_{k-2}$ & always\\
\bottomrule
\end{tabular}
\vspace*{5mm}
\caption{\label{tab:symmetric-a}Type-A symmetric curves. Taking $I^{270}|22$ as an example (the first row),  $\mathcal{P}_k = I^{270}|22(\pi)_{k-2} = L^{270}RR^{270}L^{180}|2(\pi)_{k-2}$. 
Then its first subunit is $\mathcal{U}_1 = L^{270}|2(\pi)_{k-2}$ and the fourth subunit $\mathcal{U}_4 = L^{180}|1(\hat{\pi})_{k-2}$ (Corollary \ref{coro:5.6.1}). Then $r(\mathcal{U}_4) = R^{90}|1(\hat{\pi})_{k-2}$ (Corollary \ref{coro:rev}) and $h(r(\mathcal{U}_4)) = L^{270}|2(\pi)_{k-2}$ (Proposition \ref{prop:6.5}). We explicitly write $I^{270}|22(\pi)_{k-2}$ as $I^{270}|22|(\pi)_{k-2}$ to emphasize its level-2 global structure.}
\end{table}

The last column in Table \ref{tab:symmetric-a} gives the condition where
the corresponding \(\mathcal{P}_k\) is type-A symmetric, i.e.,
\(\mathcal{H}(\mathcal{U}_1) = h(r(\mathcal{U}_4))\). By also
considering their reflections and rotations, the form of type-A
symmetric curves can be summarized as follows.

\begin{proposition}
The forms for type-A symmetric curves $\mathcal{P}_k = X|(\pi)_k$ are :

\begin{enumerate}
\tightlist
\item
  If $X \in \{I, U, C\}$, then $\mathcal{P}_k$ is always type-A symmetric.
\item
  If $X \in \{R, L, D\}$, then $\mathcal{P}_k = X|(a)_k$.
\item 
  If $X = Q$, then $\mathcal{P}_k = X|a(\hat{a})_{k-1}$.
\end{enumerate}

where $a = 1$ or 2 and $X$ can be associated with any rotation.

\end{proposition}

\subsubsection{Type-AB symmetric curves}\label{type-ab-symmetric-curves}

To simplify the calculation, we first write the following remark:

\begin{remark}
If a curve in the base facing state is type-A symmetric, i.e., subunits 1 and 4 are horizontally
symmetric, subunits 2 and 3 are horizontally symmetric, if
$\mathcal{H}(\mathcal{U}_1) = v(r(\mathcal{U}_2))$, i.e., subunits 1 and 2 are
verticall symmetric, then subunits 3 and 4 are also vertically symmetric, thus
the curve is type-AB symmetric.
\end{remark}

Using the same method, if a curve is type-B symmetric, \(\mathcal{U}_1\)
and \(\mathcal{U}_2\) should face upward/downward and so is for
\(\mathcal{U}_4\) and \(\mathcal{U}_3\). Then the level-2 structures are
shape 3 and 6 in Figure \ref{fig:level_2_shapes} which corresponds to
group 3 and 6 in Table \ref{tab:corner-induced}. We only need to
validate whether \(\mathcal{H}(\mathcal{U}_1) = v(r(\mathcal{U}_2))\)
for type-AB symmetric curves. The results are in Table
\ref{tab:symmetric-ab}.

\begin{table}
\centering
\begin{tabular}{rrrrcc}
\toprule
\makecell{Shape group \\ of $\mathcal{P}_2$}  & $\mathcal{P}_k$ & $\mathcal{U}_1$ & $v(r(\mathcal{U}_2))$ & Type-A & Type-AB \\
\midrule
3 & $C|11|(\pi)_{k-2}$ & $R^{90}|1(\pi)_{k-2}$ & $R^{90}|1(\pi)_{k-2}$ & always & always \\
  &$D^{180}|11|(\pi)_{k-2}$ & $R^{90}|1(\pi)_{k-2}$ & $R^{90}|1(\pi)_{k-2}$ & when all $\pi=1$ & when all $\pi=1$ \\
  & $Q|12|(\pi)_{k-2}$ & $I|2(\pi)_{k-2}$ &  $R^{90}|1(\hat{\pi})_{k-2}$ & when all $\pi=2$ & when all $\pi=2$ \\
  & $U|12|(\pi)_{k-2}$ & $I|2(\pi)_{k-2}$ &  $R^{90}|1(\hat{\pi})_{k-2}$ & always & when all $\pi=2$ \\
\midrule
6 & $C|12|(\pi)_{k-2}$ & $R^{90}|2(\pi)_{k-2}$ &  $R^{90}|2(\pi)_{k-2}$ & always & always\\
\bottomrule
\end{tabular}
\vspace*{5mm}
\caption{\label{tab:symmetric-ab}Type-AB symmetric curves. The transformation of vertical reflection $v()$ is based on Section \ref{other-types-of-reflections}. The conditions in the ``Type-AB'' column is based on the equality of $\mathcal{H}(\mathcal{U}_1) = v(r(\mathcal{U}_2))$ and conditions in the ``Type-A'' column is from Table \ref{tab:symmetric-a}.}
\end{table}

By also considering their reflections and rotations, the form of type-AB
symmetric curves can be summarized as follows.

\begin{proposition}
The forms for type-AB symmetric curves $\mathcal{P}_k = X|(\pi)_k$ are :

\begin{enumerate}
\tightlist
\item
  If $X = C$, then $\mathcal{P}_k$ is always type-AB symmetric.
\item
  If $X = D$, then $\mathcal{P}_k = X|(a)_k$.
\item 
  If $X \in \{U, Q\}$, then $\mathcal{P}_k = X|a(\hat{a})_{k-1}$.
\end{enumerate}

\noindent where $a = 1$ or 2 and $X$ can be associated with any rotation.

\end{proposition}

\subsection{Closed curves}\label{closed-curves}

Let the length of the unit segment on the curve be 1. Then if a curve is
closed, the distance between the entry point and the exit point is 1, so
that an additional horizontal or vertical segment can connect the two
points.

From Figure \ref{fig:homogeneous_curves}, corner-induced curves in
family 3 and 8 are closed. From Figure \ref{fig:side_induced_curves} and
Table \ref{tab:side-induced}, side-induced curves in shape group 6
induced by \(C_1\) can possibly be closed curves because entry point is
located on the right side of subunit 1 denoted as \(p\) and exit point
is located on the left side of subunit 4 denoted as \(q\). In the first
curve in the second row of Figure \ref{fig:homogeneous_curves} which
corresponds to \(C_1|(1)_{k-1}\), the entry point is located on the
lower right of subunit 1 denoted as \(a\) and the exit point is located
on the lower left of subunit 4 denoted as \(b\). According to
Proposition \ref{prop:geometry_entry} and Corollary
\ref{coro:geometry_exit}, we know for the side-induced curve
\(C_1|(\omega)_{k-1} = C_1|\delta^{(k-1)}\), its entry point denoted as
\(a'\) has a distance to \(a\) of \(\delta - 1\) and its exit point
denoted as \(b'\) has a distance of \(\delta - 1\) to \(b\). This
results \(a'\) and \(b'\) move parallely on \(p\) and \(q\), and the
distance between \(a'\) and \(b'\) is always 1.

\begin{proposition}
By also considering the reflections, the following curves

\begin{equation}
\begin{aligned}
& C|(\pi)_k \\
& D|(1)_k, D|(2)_k \\
& Q|1(2)_{k-1}, Q|2(1)_{k-1} \\
& U|1(2)_{k-1}, U|2(1)_{k-1} \\
\end{aligned}
\notag
\end{equation}

\noindent associated with any rotation are closed.
\end{proposition}

Among them, \(C|(1)_k\), \(C|(2)_k\), \(D|(1)_k\), \(D|(2)_k\),
\(Q|1(2)_{k-1}\), \(Q|2(1)_{k-1}\), \(U|1(2)_{k-1}\), \(U|2(1)_{k-1}\)
(Family 3 in Table \ref{tab:corner-induced}) are the Moore curves,
\(C|2(1)_{k-1}\), \(C|1(2)_{k-1}\) (Family 8 in Table
\ref{tab:corner-induced}) are type-\(V_6\) order-1 Hilbert variants
(Table \ref{tab:Liu-variants}), \(C|1(1212...)\), \(C|2(2121...)\) are
type-\(V_7\) order-1 \(\beta\Omega\)-variants (Table
\ref{tab:beta-omega-variants}), and \(C|1(2121...)\), \(C|2(1212...)\)
are type-\(V_9\) order-1 \(\beta\Omega\)-variants (Table
\ref{tab:beta-omega-variants}).

\subsection{Summarize}\label{summarize}

Structural attributes introduced in this section for the Hilbert curve,
order-1 Hilbert variants, the \(\beta\Omega\)-curve and order-1
\(\beta\Omega\)-variants are summarized in Table
\ref{tab:structure_summary}.

\begin{sidewaystable}
\centering
\begin{tabular}{lclccccccc}
\toprule
Curves & 
Type & 
$\mathcal{P}_k$ & 
Recur. & 
\makecell{Subunit \\ identically \\ shaped} & 
\makecell{Subunit \\ differently \\ shaped} & 
\makecell{Completely \\ non- \\ recursive} & 
\makecell{Type-A \\ symmetric} & 
\makecell{Type-AB \\ symmetric} & 
Closed \\
\midrule
Hilbert curve         & Hilbert       & $I|(2)_k$      & yes & yes &     &     & yes &     &     \\
\midrule
Hilbert variant       & $V_1$         & $P|(2)_k$      &     & yes &     &     &     &     &     \\
                      & $V_2$ (Moore) & $C|(1)_k$      &     & yes &     &     & yes & yes & yes \\
                      & $V_3$         & $B|2(1)_{k-1}$ &     & yes &     &     &     &     &     \\
                      & $V_4$         & $I|2(1)_{k-1}$ &     & yes &     &     & yes &     &     \\
                      & $V_5$         & $C|1(2)_{k-1}$ &     & yes &     &     & yes & yes & yes \\
\midrule
$\beta\Omega$-curve   & $O$           & $I|(21...)$    &     & yes &     &     & yes &     &     \\
                      & $B_1$         & $R|(21...)$    &     &     & yes &     &     &     &     \\
                      & $B_2$         & $P|(21...)$    &     & yes &     &     &     &     &     \\
\midrule
$\beta\Omega$-variant & $V_1$         & $I|2(21...)$   &     & yes &     &     & yes &     &     \\
                      & $V_2$         & $R|2(21...)$   &     &     & yes & yes &     &     &     \\
                      & $V_3$         & $P|2(21...)$   &     & yes &     &     &     &     &     \\
                      & $V_4$         & $U|1(12...)$   &     &     & yes & yes & yes &     &     \\
                      & $V_5$         & $Q|1(12...)$   &     &     & yes & yes &     &     &     \\
                      & $V_6$         & $C|1(12...)$   &     & yes &     &     & yes & yes & yes \\
                      & $V_7$         & $U|(121...)$   &     &     & yes & yes & yes &     &     \\
                      & $V_8$         & $Q|(121...)$   &     &     & yes & yes &     &     &     \\
                      & $V_9$         & $C|(121...)$   &     & yes &     &     & yes & yes & yes \\
\bottomrule
\end{tabular}
\vspace*{5mm}
\caption{\label{tab:structure_summary}Summary of the structural attributes of the Hilbert curve, order-1 Hilbert variants, the $\beta\Omega$-curve and order-1 $\beta\Omega$-variants. Column $\mathcal{P}_k$ is from Table \ref{tab:Liu-variants}, \ref{tab:beta-omega-type} and \ref{tab:beta-omega-variants} where rotations are ommited.}
\end{sidewaystable}

\section{Arithmetic representation}\label{arithmetic}

In this section we discuss the calculation of the coordinates of the
curve.

\subsection{Sequential}\label{sequential}

For a base \(X^\theta\), denote its \(xy\)-coordinate as \(\mathbf{v}\)
and let the length of the unit segment be 1, then the coordinate of its
next base is \(\mathbf{v} + R(\theta)t(X)\) where \(t(X)\) is the offset
of \(X\) to its next base in its base rotation state, which can be
inferred from its exit direction:

\begin{equation}
t(X) = \begin{cases}
(0, 1) & \quad \textnormal{if } X \in \{I, B, D\} \\
(1, 0) & \quad \textnormal{if } X = R \\
(-1, 0) & \quad \textnormal{if } X = L \\
(0, -1) & \quad \textnormal{if } X = U \\
\end{cases}
\notag
\end{equation}

\noindent and \(R(\theta)\) is the rotation matrix:

\begin{equation}
R(\theta) = \left[\begin{matrix}
\cos \theta & - \sin \theta \\
\sin \theta & \cos \theta \\
\end{matrix}\right] .
\notag
\end{equation}

Let's denote the offset of the complete base as
\(p(X^\theta) = R(\theta)t(X)\). Then if the base sequence of
\(\mathcal{P}_k\) is already known and the coordinate of the entry base
is \(\mathbf{v}_1 = (x, y)\), the coordinate of the \(i\)-th base is

\begin{equation}
\label{eq:arith1}
\mathbf{v}_i = \mathbf{v}_1 + \sum_{j = 1}^{i-1} p(X_j^{\theta_j}) .
\end{equation}

When \(k \ge 1\), \(\mathcal{P}_k\) is only composed of primary bases.
Then there are only three possible values of \(t(X)\) and four possible
values of \(R(\theta)\). We can precompute the value of \(p(X^\theta)\)
for these 12 combinations of \(X\) and \(\theta\), and we define a new
offset table \(p'(X^\theta)\), then Equation \ref{eq:arith1} can be
simplied to

\begin{equation}
\label{eq:arith2}
\mathbf{v}_i = \mathbf{v}_1 + \sum_{j = 1}^{i-1} p'(X_{j}^{\theta_j}) 
\end{equation}

\noindent to get rid of \(i-1\) matrix multiplications.

\subsection{Individual bases}\label{individual-bases}

\begin{figure}
\centering{
\includegraphics[width=0.9\linewidth]{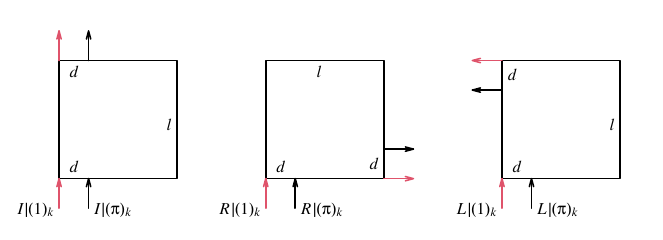}
\caption{Entry and exit points on subunits induced from $I$, $R$ and $L$. $d$ is the distance between the entry points of $Z|(1)_k$ and $Z|(\pi)_k$ ($Z \in \{I, R, L\}$), $l$ is the side length of the square.}
\label{fig:arithmetic}
}
\end{figure}

Equation \ref{eq:arith2} is convenient when calculating coordinates of
the whole curve sequentially, but it is not efficient for calculating
the coordinate of only one single base in the curve because the
coordinates of all its preceding bases need to be calculated in advance,
thus the time complexity is exponential to the level \(k\). In this
section, we discuss an efficient way to calculate the coordinate of the
\(i\)-th base (\(2 \le i \le 4^k\)) that only has a linear time
complexity to \(k\). We consider the curve initialized from a single
seed base.

\subsubsection{Method 1}\label{method-1}

\(\mathcal{P}_k\) (\(k \ge 1\)) is only composed of \(I\), \(R\), \(L\),
and any subunit of it on any level is also only induced from these three
primary bases. Let a subunit \(\mathcal{U} = Z|(\pi)_k\) where
\(Z \in \{I, R, L\}\), the following equation calculates the offset
between the entry point of a subunit and that of its next subunit
denoted as \(p()\) (black arrows in Figure \ref{fig:arithmetic}):

\begin{equation}
\label{eq:unit_offset}
p() = \begin{cases}
R(\theta)\left( 
          \left[\begin{matrix} 0 \\ l \\ \end{matrix} \right] + 
          \left[\begin{matrix} 0 \\ 1 \\ \end{matrix} \right] 
        \right)  = 
    R(\theta)\left[\begin{matrix} 0 \\ 2^k \end{matrix} \right]
 & \quad \textnormal{if } Z = I \\
R(\theta)\left( 
          \left[\begin{matrix} l - d \\ 0 \\ \end{matrix} \right] + 
          \left[\begin{matrix} 0 \\ d \\ \end{matrix} \right] +
          \left[\begin{matrix} 1 \\ 0 \\ \end{matrix} \right]
        \right) = 
    R(\theta)\left[\begin{matrix} 2^k - \delta^{(k)} + 1 \\ \delta^{(k)} - 1 \end{matrix}\right]
 & \quad \textnormal{if } Z = R \\
R(\theta)\left( 
          \left[\begin{matrix} -d \\ 0 \\ \end{matrix} \right] + 
          \left[\begin{matrix} 0 \\ l-d  \\ \end{matrix} \right] +
          \left[\begin{matrix} -1 \\ 0 \\ \end{matrix} \right]
        \right) = 
    R(\theta)\left[ \begin{matrix} - \delta^{(k)} \\ 2^k - \delta^{(k)} \end{matrix}\right]
  & \quad \textnormal{if } Z = L \\
\end{cases} .
\end{equation}

In it, \(R()\) is the rotation matrix, \(\theta\) is the rotation
associated with \(Z\), \(l\) is the side length of the square curve and
\(d\) is the distance between the entry point of \(\mathcal{U}\) and
\(Z|(1)_k\) (Figure \ref{fig:arithmetic}). According to Proposition
\ref{prop:geometry_entry}, there are \(l = 2^k - 1\) and
\(d = \delta^{(k)} - 1\) where \(\delta^{(k)}\) is the integer
representation of the code sequence of the subunit. As \(p()\) depends
on the type of \(Z\), its rotation and its expansion code sequence, we
write \(p()\) parametrically as \(p(Z, \theta, (\pi)_k)\) or the
single-parameter form \(p(\mathcal{U})\). When \(\mathcal{U}\) is
reduced to a single point, i.e., \(k = 0\) and \(\delta(0)=1\), \(p()\)
is the same as in the previous section. When the unit is a single point,
we denote \(p(Z^\theta)\) or \(p(Z, \theta, \varnothing)\) as the code
sequence is empty.

Now let's go back to the problem. For a curve \(\mathcal{P}_k =
X|\pi_1...\pi_k\) of which the encoding is already known, the coordinate
of its entry point is \(\mathbf{v}\), then the calculation of the
coordinate of the \(n\)-th point on the curve denoted as \(t\) is
applied in the following steps.

\emph{The preparation step}. We first transform the index \(n\) to its
quaternary form \(n\mapsto q_1 ... q_k\) (\(1 \le n \le 4^k\),
\(q_* \in \{1,2,3,4\}\)) where \(q_i\) represents the subunit index on
the level \(k-i+1\) curve.

\begin{equation}
\begin{aligned}
q_1 &= \lceil n/4^{k-1} \rceil \\
q_i &= \Bigl\lceil \left( n - \sum_{j = 1}^{i-1} (q_j -1)\cdot 4^{k-j} \right)\Big/4^{k-i} \Bigr\rceil \quad\quad 2 \le i \le k \\
\end{aligned}
\notag
\end{equation}

\emph{Step 1}. Let's start from level \(k\). We write
\(\mathcal{P}_k = \mathcal{U}_1
\mathcal{U}_2 \mathcal{U}_3 \mathcal{U}_4\) where
\(X|\pi_1 = Z_1 Z_2 Z_3 Z_4\),
\(\mathcal{U}_1 = Z_1|(\pi)_{2...k}\)\footnote{In this section $(\pi)_{a...b} =
\pi_a...\pi_b$.} and \(\mathcal{U}_i = Z_i|s((\pi)_{2...k}|\theta_i -
\theta_1)\) (\(i \ge 2\), Corollary \ref{coro:5.6.1}) where \(\theta_i\)
is the rotation associated with \(Z_i\). For simplicity, we also write
\(\mathcal{U}_1 = Z_1|s((\pi)_{2...k}|\theta_1 -\theta_1) = (\pi)_{2...k}\).

The entry point of \(\mathcal{P}_k\) is also the entry point of
\(\mathcal{U}_1\). The first quaternary index \(q_1\) implies that the
point \(t\) is located on the \(q_1\)-th subunit of \(\mathcal{P}_k\),
then we calculate the coordinate of the entry point of
\(\mathcal{U}_{q_1}\) denoted as \(\mathbf{v}_{q_1}\) according to
Equation \ref{eq:unit_offset} as:

\begin{equation}
\mathbf{v}_{q_1} = \mathbf{v} + \sum_{i=1}^{q_1 - 1} p\left(Z_i, \theta_i, s((\pi)_{2...k}|\theta_i-\theta_1)\right) .
\notag
\end{equation}

From this step, we will use different forms of notations, as we will
reach point \(t\) with the hierarchical indices of \(q_1...q_k\). We
change the notations of \(\mathcal{U}_{q_1}\) to
\(\mathcal{U}^{(q_1)}\), \(\mathbf{v}_{q_1}\) to \(\mathbf{v}^{(q_1)}\),
\(Z_{q_1}\) to \(X^{(q_1)}\). The code sequence for \(X^{(q_1)}\) is
\(s((\pi)_{2...k}|\theta_{q_1} - \theta_1)\) and we denote it to
\((\pi)^{(q_1)}_{2...k}\).

\emph{Step 2}. Now we are on subunit \(\mathcal{U}^{(q_1)} =
X^{(q_1)}|(\pi)^{(q_1)}_{2...k}\) of which the encoding is known, also
the coordinate \(\mathbf{v}^{(q_1)}\) of its entry point is also known
(all have been calculated form the prevous step). Let \(\pi^{(q_1)}_2\)
be the first code in \((\pi)^{(q_1)}_{2...k}\), and
\((\pi)^{(q_1)}_{3...k}\) be the remaining code sequence, then we write
\(X^{(q_1)}|\pi_2^{(q_1)} = Z^{(q_1)}_1 Z^{(q_1)}_2
Z^{(q_1)}_3 Z^{(q_1)}_4\) where associated rotations are
\(\theta^{(q_1)}_i\). The second quaternary index \(q_2\) implies that
point \(t\) is located on the \(q_2\)-th subunit of
\(\mathcal{U}^{(q_1)}\) denoted as \(\mathcal{U}^{(q_1q_2)}\), then
applying the same method as in the first step, we can obtain the
coordinate of the entry point of \(\mathcal{U}^{(q_1q_2)}\), denoted as
\(\mathbf{v}^{(q_1q_2)}\):

\begin{equation}
\mathbf{v}^{(q_1q_2)} = \mathbf{v}^{(q_1)} + \sum_{i=1}^{q_2 - 1} p\left(Z^{(q_1)}_i, \theta^{(q_1)}_i, s((\pi)^{(q_1)}_{3...k}|\theta^{(q_1)}_i - \theta^{(q_1)}_1)\right) .
\notag
\end{equation}

\emph{Step 3 to Step \(k\)}. Similarly, we can denote \(X^{(q_1q_2)} =
Z^{(q_1q_2)}_{q_2}\) and its code sequence \((\pi)^{(q_1q_2)}_{3...k} =
s((\pi)^{(q_1)}_{3...k}|\theta^{(q_1)}_{q_2} - \theta^{(q_1)}_1)\). We
know the point \(t\) is located on \(\mathcal{U}^{(q_1q_2q_3)}\) and we
can use the same method to calculate the coordinate
\(\mathbf{v}^{(q_1q_2q_3)}\) of its entry point.

Generally, for \(m+1 \le k\),

\begin{equation}
\label{eq:vqm}
\mathbf{v}^{(q_1...q_{m+1})} = \mathbf{v}^{(q_1...q_m)} + \sum_{i=1}^{q_{m+1} - 1} p\left(Z^{(q_1...q_m)}_i, \theta^{(q_1...q_m)}_i, s((\pi)^{(q_1...q_m)}_{{m+2}...k}|\theta^{(q_1...q_m)}_i - \theta^{(q_1...q_m)}_1)\right) .
\end{equation}

\noindent where the values of \(\mathbf{v}^{(q_1...q_m)}\),
\(X^{(q_1...q_m)}\) and \((\pi)^{(q_1...q_m)}_{{m+2}...k}\) are already
known from the previous step.

Let's consider the number of calculations taking the worst case where
\(t\) is the last point of the curve. On each step of traversing down
the hierarchical index, there are the following calculations:

\begin{enumerate}
\tightlist
\item
  Expand $X^{(q_1...q_i)}|\pi^{(q_1...q_i)}_{i+1} = Z^{(q_1...q_i)}_1 Z^{(q_1...q_i)}_2 Z^{(q_1...q_i)}_3 Z^{(q_1...q_i)}_4$.
\item
  For subunit 2-4, use Corollary \ref{coro:5.6.1} or Corollary \ref{coro:5.6.2} to calculate their expansion code sequences.
\item
  Apply Equation \ref{eq:unit_offset} to calculate $p()$ for subunit 1, 2, 3.
\item
  Add all offsets to the entry location to obtain the entry location of the next unit.
\end{enumerate}

The number of calculations on each iteration is roughly a constant, thus
the time complexity is linear to the level \(k\).

\begin{figure}
\centering{
\includegraphics[width=0.7\linewidth]{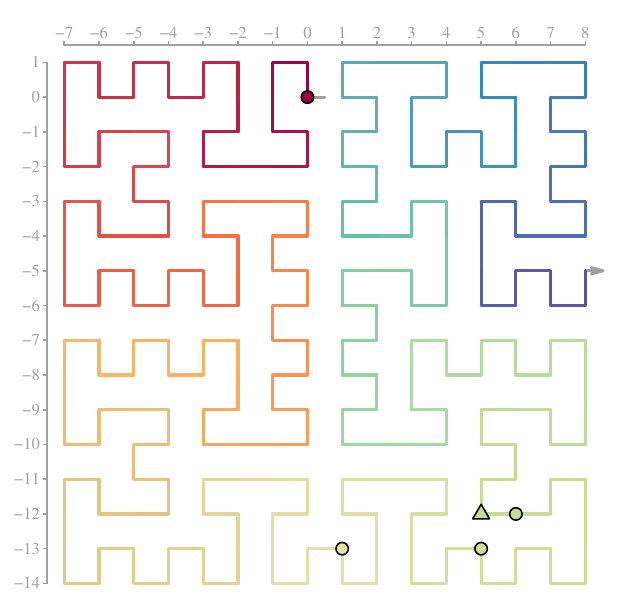}
\caption{Calculate the coordinate of point-158 (the triangle point) in $B^{270}|1221$. The 4 round points are the entry points of subunits on corresponding levels. They have coordinates of $\mathbf{v}$, $\mathbf{v}^{(3)}$, $\mathbf{v}^{(32)}$ and $\mathbf{v}^{(324)}$.}\label{fig:arithmetic_example}
}
\end{figure}

As an example (Figure \ref{fig:arithmetic_example}), for the curve
\(\mathcal{P}_k = X|\pi_1...\pi_k = B^{270}|1221\) (level \(k=4\) with
total 256 points), let the entry coordinate be \((0, 0)\), we calculate
the coordinate of point with index 158 on the curve. We have
\(\mathbf{v} = (0,0)\), \(q_1q_2q_3q_4
= 3242\).

\begin{itemize}
\tightlist
\item
  Step 1. \(X|\pi_1 = B^{270}|1 = L^{90}L^{180}L^{270}R\). With
  Corollary \ref{coro:5.6.1}, we have the four subunits
  \(\mathcal{U}_1 = L^{90}|221\), \(\mathcal{U}_2 = L^{180}|112\),
  \(\mathcal{U}_3 = L^{270}|221\) and \(\mathcal{U}_4 = R|112\). With
  \(q_1 = 3\), then the location of entry point of
  \(\mathcal{U}^{(q_1)}\) is (In Line 2, we simplified the notation
  \(s((\pi)_{2...k}|\theta_i-\theta_1)\) to \(s_i\)):
\end{itemize}

\begin{equation}
\begin{aligned}
\mathbf{v}^{(q_1)} &= \mathbf{v} + \sum_{i=1}^{q_1 - 1} p\left(Z_i, \theta_i, s((\pi)_{2...k}|\theta_i-\theta_1)\right) \\
    &= \mathbf{v} + \sum_{i=1}^{q_1 - 1} p\left(Z_i, \theta_i, s_i\right) \\
\mathbf{v}^{(3)} &= \mathbf{v} + p(L, 90, 221) + p(L, 180, 112) \\
     &= \left[\begin{matrix}0\\0\end{matrix}\right] + R(90)\left[\begin{matrix}- 7 \\ 2^3 - 7 \end{matrix}\right] + R(180)\left[\begin{matrix} -2 \\ 2^3 - 2 \end{matrix}\right] \\
     &= \left[\begin{matrix}0\\0\end{matrix}\right] + \left[\begin{matrix}0 & -1\\ 1 & 0 \end{matrix}\right]\left[\begin{matrix}-7\\ 1 \end{matrix}\right] + \left[\begin{matrix}-1 & 0\\ 0 & -1 \end{matrix}\right]\left[\begin{matrix}-2\\ 6 \end{matrix}\right] \\
     &= \left[\begin{matrix}1\\-13\end{matrix}\right]
\end{aligned}
\notag
\end{equation}

\begin{itemize}
\tightlist
\item
  Step 2. we have
  \(\mathcal{U}^{(q_1)} = \mathcal{U}^{(3)} = L^{270}|221\) from the
  previous step. The four subunits of \(\mathcal{U}^{(3)}\) are
  \(I^{270}|21\), \(L^{270}|21\), \(L|12\) and \(R^{90}|21\), then the
  location of entry point of \(\mathcal{U}^{(q_1q_2)}\) (\(q_2 = 2\))
  is:
\end{itemize}

\begin{equation}
\begin{aligned}
\mathbf{v}^{(q_1q_2)} &= \mathbf{v}^{(q_1)} + \sum_{i=1}^{q_2 - 1} p\left(Z^{(q_1)}_i, \theta^{(q_1)}_i, s_i^{(q_1)}\right) \\
\mathbf{v}^{(32)} &= \mathbf{v}^{(3)} + p(I, 270, 21)  \\
     &= \left[\begin{matrix}1\\-13\end{matrix}\right] + R(270)\left[\begin{matrix}0 \\ 2^2 \end{matrix}\right] \\
     &= \left[\begin{matrix}1\\-13\end{matrix}\right] + \left[\begin{matrix}0 & 1\\ -1 & 0 \end{matrix}\right]\left[\begin{matrix}0\\ 4 \end{matrix}\right] \\
     &= \left[\begin{matrix}5\\-13\end{matrix}\right]
\end{aligned}
\notag
\end{equation}

\begin{itemize}
\tightlist
\item
  Step 3. we have
  \(\mathcal{U}^{(q_1q_2)} = \mathcal{U}^{(32)} = L^{270}|21\). Its four
  subunits are \(I^{270}|1\), \(L^{270}|1\), \(L|2\) and \(R^{90}|1\),
  then the location of entry point of \(\mathcal{U}^{(q_1q_2q_3)}\)
  (\(q_3 = 4\)) is:
\end{itemize}

\begin{equation}
\begin{aligned}
\mathbf{v}^{(q_1q_2q_3)} &= \mathbf{v}^{q_1q_2} + \sum_{i=1}^{q_3 - 1} p\left(Z^{(q_1q_2)}_i, \theta^{(q_1q_2)}_i, s_i^{(q_1q_2)}\right) \\
\mathbf{v}^{(324)} &= \mathbf{v}^{(32)} + p(I, 270, 1) + p(L, 270, 1) + p(L, 0, 2)  \\
     &= \left[\begin{matrix}5\\-13\end{matrix}\right] + R(270)\left[\begin{matrix}0 \\ 2^1 \end{matrix}\right] + R(270)\left[\begin{matrix}-1 \\2^1- 1 \end{matrix}\right] + R(0)\left[\begin{matrix}-2 \\ 2^1 - 2 \end{matrix}\right] \\
     &= \left[\begin{matrix}5\\-13\end{matrix}\right] + \left[\begin{matrix}0 & 1\\ -1 & 0 \end{matrix}\right]\left[\begin{matrix}0\\ 2 \end{matrix}\right] + \left[\begin{matrix}0 & 1\\ -1 & 0 \end{matrix}\right]\left[\begin{matrix}-1\\ 1 \end{matrix}\right] +\left[\begin{matrix}1 & 0\\ 0 & 1 \end{matrix}\right]\left[\begin{matrix}-2\\ 0 \end{matrix}\right] \\
     &= \left[\begin{matrix}6\\-12\end{matrix}\right]
\end{aligned}
\notag
\end{equation}

\begin{itemize}
\tightlist
\item
  Step 4. Last we reach the last index \(q_4 = 2\). The unit
  \(\mathcal{U}^{(q_1q_2q_3)} = \mathcal{U}^{(324)} = R^{90}|1 =
  I^{90}R^{90}RL^{270}\). Then with Equation \ref{eq:arith1}:
\end{itemize}

\begin{equation}
\begin{aligned}
\mathbf{v}^{(q_1q_2q_3q_4)} &= \mathbf{v}^{(q_1q_2q_3)} + \sum_{i=1}^{q_4-1}R(\theta_i^{(q_1q_2q_3)})t(Z_i^{(q_1q_2q_3)}) \\
\mathbf{v}^{(3242)} &= \mathbf{v}^{(324)} + R(90)t(I) \\
    &= \left[\begin{matrix}6\\-12\end{matrix}\right] + \left[\begin{matrix}0 & -1\\ 1 & 0 \end{matrix}\right]\left[\begin{matrix}0\\ 1 \end{matrix}\right] \\
    &= \left[\begin{matrix}5\\-12\end{matrix}\right]
\end{aligned} 
\notag
\end{equation}

The coordinate of \(t\) can be validated by applying the sequential
method in Section \ref{sequential}.

\subsubsection{Method 2}\label{method-2}

Equation \ref{eq:vqm} can be rewritten as:

\begin{equation}
\mathbf{v}^{(q_1...q_{m+1})} = \mathbf{v}^{(q_1...q_{m})} + \sum_{i = 1}^{q_{m+1}-1} p(\mathcal{U}_i^{(q_1...q_m)})
\notag
\end{equation}

\noindent then, \(\mathbf{v}^{(q_1...q_k)}\) can be expanded as:

\begin{equation}
\label{eq:vqm2}
\begin{aligned}
\mathbf{v}^{(q_1...q_{k})} &= \mathbf{v} + \sum_{j = 1}^{q_1-1} p(\mathcal{U}_j) + \sum_{j = 1}^{q_2-1} p(\mathcal{U}_j^{(q_1)}) + ... + \sum_{j = 1}^{q_{k}-1} p(\mathcal{U}_j^{(q_1...q_{k-1})}) \\
   &= \mathbf{v} + \sum_{i=1}^{k} \sum_{j = i}^{q_i-1} p(\mathcal{U}_j^{(q_1...q_{i-1})}) \\
\end{aligned}
\end{equation}

When \(i = 1\), we denote \(\mathcal{U}^{(p_1...p_{i-1})} =
\mathcal{U}^{(\varnothing)} = \mathcal{U}\), i.e., the subunit of the
complete curve \(\mathcal{P}_k\). Note
\(\mathcal{U}^{(q_1...q_{i-1})}_j\) is the \(j\)-th subunit of
\(\mathcal{U}^{(q_1...q_{i-1})}\) which is the \(q_{i-1}\)-th subunit of
\(\mathcal{U}^{(q_1...q_{i-2})}\). Then all forms of
\(\mathcal{U}^{(q_1...q_{i-1})}\) are determined recursively from
\(P_k\).

For the previous example, instead of moving from subunits, we can first
calculate all necessary forms of the subunit on every level as in Table
\ref{tab:arith}. Then according to Equation \ref{eq:vqm2}:

\begin{equation}
\begin{aligned}
\mathbf{v}^{(3242)} &= \mathbf{v} + p(L^{90}|221) + p(L^{180}|112) + p(I^{270}|21) + \\
    & p(I^{270}|1) + p(L^{270}|1) + p(L|2) + p(I^{90}) \\
   &= \left[\begin{matrix}5\\-12\end{matrix}\right] \\
\end{aligned}
\notag
\end{equation}

\begin{table}
\centering
\begin{tabular}{llcllll}
\toprule
Curve & Level-1 expansion & $q$ & $\mathcal{U}_1$ & $\mathcal{U}_2$  & $\mathcal{U}_3$ &  $\mathcal{U}_4$ \\
\midrule
$\mathcal{P}_4 = B^{270}|1221$ & $B^{270}|1 = L^{90}L^{180}L^{270}R$ & $q_1 = 3$  & $L^{90}|221$ & $L^{180}|112$ & $L^{270}|221$ & - \\
$\mathcal{U}^{(q_1)}=L^{270}|221$ & $L^{270}|2=I^{270}L^{270}LR^{90}$ & $q_2 = 2$ & $I^{270}|21$ & $L^{270}|21$ & - & - \\
$\mathcal{U}^{(q_1q_2)}=L^{270}|21$ & $L^{270}|2=I^{270}L^{270}LR^{90}$ & $q_3 = 4$ & $I^{270}|1$ & $L^{270}|1$ & $L|2$ & $R^{90}|1$ \\
$\mathcal{U}^{(q_1q_2q_3)}=R^{90}|1$ & $R^{90}|1=I^{90}R^{90}RL^{270}$ & $q_4 = 2$ & $I^{90}$ & $R^{90}$ & - & - \\
\bottomrule
\end{tabular}
\vspace*{5mm}
\caption{Encodings of subunits on every level. On each level, we only need to calculate the encodings for the first to the $q_i$-th subunits.}
\label{tab:arith}
\end{table}

If the seed is a base sequence \(\mathcal{P}_k = X_1...X_w|(\pi)_k\),
note \(\mathcal{P}_k\) is represented as a list of \(w\) square curves,
we first calculated which square curve the point \(t\) is located on.
The index \(c\) of the square curve can be calculated as
\(c = \left\lceil n/4^k \right\rceil\) where \(n\) is the index of \(t\)
on the entire curve. Let's denote this square curve as
\(\mathcal{Q}_{k,[c]} = X_c|s((\pi)_k|\theta_c - \theta_1)\) where
\(\theta_c\) and \(\theta_1\) are the rotations associated with \(X_c\)
and \(X_1\). Next we calculate the coordinate of the entry point of
\(\mathcal{Q}_{k,[c]}\), denoted as \(\mathbf{v}_c\):

\begin{equation}
\mathbf{v}_c = \mathbf{v} + \sum_{i-1}^{c-1} p(\mathcal{Q}_{k,[i]})
\notag
\end{equation}

\noindent where \(\mathbf{v}\) is the coordinate of the entry point of
the entire curve. As \(X_1\) is also possible from \(\{U, B, P\}\), it
can be used in the same way as \(X_1 =I\) when calculating \(p()\). We
also need to calculate the index of \(t\) only on
\(\mathcal{Q}_{k,[c]}\) as \(n' = n - (c-1)\times 4^k\). Then with
\(\mathcal{Q}_{k,[c]}\), \(\mathbf{v}_c\) and \(n'\), we can use the
method for single square curve proposed in this section to calculate the
coordinate of \(t\).

\subsubsection{Method 3}\label{method-3}

The previosu two methods can be thought of as ``entry-point-based'', we
can also do a center-based method, where each level of \(\mathcal{U}\)
and when it is reduced to a single point, it is the location.

for a 2x2 unit, let \(\mathbf{c}\) as the center and
\(\mathcal{c}^(1)_q\) as the \(i\)-th center of the four quandants, let
\(\mathbf{v}\) as a vector from c to c\_i, then \(c + v\) is the vector
from origin to c\_q.

\begin{equation}
v = v_1 + \sum v_q
\end{equation}

\subsection{Obtain index on the curve}\label{obtain-index-on-the-curve}

Next we consider the reversed problem. With knowing the coordinate of a
point \(t\) in the two-dimensional space, we want to calculate its
sequential index \(n\) on the curve. \(n\) can be transformed from its
quaternary form \(q_1...q_k\):

\begin{equation}
\label{eq:quad2n}
n = 1 + \sum_{i=1}^{k}((q_i - 1) \times 4^{k-i}) .
\end{equation}

Thus, we only need to calculate the quaternary index of \(t\) on the
curve.

\subsubsection{Method 1}\label{method-1-1}

\begin{figure}
\centering{
\includegraphics[width=0.4\linewidth]{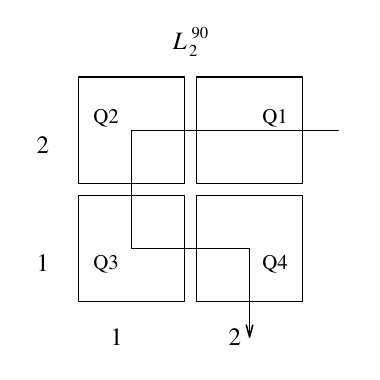}
\caption{Quadrants and quaternary indicies.}\label{fig:quadrant}
}
\end{figure}

On each level, the four quarters of the curve are represented as four
quadrants. However, the correspondance between them changes for
different bases in different rotations. We first build a list which
contains the correspondance between quaternary index and quadrants for
every \(X^{(1),\theta}\) (the level-1 curve determines the orientation
of the four quadrants). The correspondance is represented as a 2x2
matrix, e.g.,
\(Q(L_2^{90}) = \begin{bmatrix} 2 & 1 \\ 3 & 4 \end{bmatrix}\) (Figure
\ref{fig:quadrant}) where row and column indicies correspond to the
indicies of the quadrants (indicies on the sides in Figure
\ref{fig:quadrant}) and the values in the matrix correspond to the
quaternary indicies of the curve. With knowing the index of quadrants,
the quaternary index is determined, which we denote as
\(q = Q(X^{(1),\theta}, i, j)\), e.g., \(Q(L^{90,\theta}_2, 2, 1) = 4\).

The calculation of \(q_1...q_k\) can be calculated by recursively
partitioning the curve. Let the bottom left corner have a coordinate
\((x_1, y_1)\) and the top right corner have a coordinate
\((x_2, y_2)\). The coordinate of the point \(t\) is \((a, b)\).

\emph{Step 1}.
\(\mathcal{P}_k = X|\pi_1...\pi_k = X_{<\pi_1>}|\pi_2...\pi_k\). The
four quadrants of \(\mathcal{P}_k\) are determined by \(X_{<\pi_1>}\).
We first calculate which quadrant the point \(t\) is located on. The
values of \(i\) and \(j\) are in \(\{1, 2\}\).

\begin{equation}
\begin{aligned}
i &= \Bigl\lceil 2 \cdot \frac{a - x_1 + 0.5}{x_2 - x_1 + 1} \Bigr\rceil \\
j &= \Bigl\lceil 2 \cdot \frac{b - y_1 + 0.5}{y_2 - y_1 + 1} \Bigr\rceil \\
\end{aligned}
\notag
\end{equation}

We add an offset of \(-0.5\) both to the \(xy\)-coordinate of the bottom
left corner, and an offset of 0.5 both to the \(xy\)-coordinate of the
top right corner.

The quaternary index of \(\mathcal{P}_k\) where \(t\) is in is
calculated from the precomplied list as \(q_1 = Q(X_{<\pi_1>}, i, j)\).

\emph{Step 2}. In the previous step, \(t\) is located on the \(q_1\)-th
quarter of \(\mathcal{P}_k\). Write
\(\mathcal{P}_k = \mathcal{U}_1\mathcal{U}_2\mathcal{U}_3\mathcal{U}_4\),
then \(t\) is located on \(\mathcal{U}_{q_1}\). If
\(X|\pi_1 = Z_1Z_2Z_3Z_4\), then
\(\mathcal{U}_{q_1} = Z_{q_1}|s((\pi)_{2...k}|\theta_{q_1} - \theta_1)\)
where \(\theta_{q_1}\) and \(\theta_1\) are rotations associated with
\(Z_{q_1}\) and \(Z_1\). Using the same notation as in the previous
section, we write
\(\mathcal{U}^{(q_1)} = X^{(q_1)}|(\pi)^{(q_1)}_{2...k}\).

Since now we are on \(\mathcal{U}^{(q_1)}\), we calculate the
coordinates of its bottom left and top right corners.

\begin{equation}
\begin{aligned}
x_1^{(q_1)} &= x_1 + I(i = 2) \cdot \frac{x_2 - x_1 + 1}{2} \\
y_1^{(q_1)} &= y_1 + I(j = 2) \cdot \frac{y_2 - y_1 + 1}{2} \\
x_2^{(q_1)} &= x_2 - I(i = 1) \cdot \frac{x_2 - x_1 + 1}{2} \\
y_2^{(q_1)} &= y_2 - I(j = 1) \cdot \frac{y_2 - y_1 + 1}{2} \\
\end{aligned}
\notag
\end{equation}

Similarly, \(\mathcal{U}^{(q_1)}\) has four quadrants determined by
\(X^{(q_1)}|\pi_2^{(q_1)}\). The quaternary index on the next level can
be calculated as:

\begin{equation}
\begin{aligned}
i^{(q_1)} &= \Bigl\lceil 2 \cdot \frac{a - x_1^{(q_1)} + 0.5}{x_2^{(q_1)} - x_1^{(q_1)} + 1} \Bigr\rceil \\
j^{(q_1)} &= \Bigl\lceil 2 \cdot \frac{b - y_1^{(q_1)} + 0.5}{y_2^{(q_1)} - y_1^{(q_1)} + 1} \Bigr\rceil \\
q_2 &= Q(X^{(q_1)}|\pi_2^{(q_1)}, i^{(q_1)}, j^{(q_1)}) \\
\end{aligned} .
\notag
\end{equation}

\emph{Step 3 to step k}. To calculate \(q_{m+1}\) (\(m+1 \le k\)), we
always first obtain the unit on level \(m\) where point \(t\) is located
on:
\(\mathcal{U}^{(q_1...q_m)} = X^{(q_1...q_m)}|(\pi)^{(q_1...q_m)}_{m+1...k}\).
Then calculate the coordinates of the two corners.

\begin{equation}
\begin{aligned}
x_1^{(q_1...q_m)} &= x_1^{(q_1...q_{m-1})} + I(i^{(q_1...q_{m-1})} = 2) \cdot \frac{x_2^{(q_1...q_{m-1})} - x_1^{(q_1...q_{m-1})} + 1}{2} \\
y_1^{(q_1...q_m)} &= y_1^{(q_1...q_{m-1})} + I(j^{(q_1...q_{m-1})} = 2) \cdot \frac{y_2^{(q_1...q_{m-1})} - y_1^{(q_1...q_{m-1})} + 1}{2} \\
x_2^{(q_1...q_m)} &= x_2^{(q_1...q_{m-1})} - I(i^{(q_1...q_{m-1})} = 1) \cdot \frac{x_2^{(q_1...q_{m-1})} - x_1^{(q_1...q_{m-1})} + 1}{2} \\
y_2^{(q_1...q_m)} &= y_2^{(q_1...q_{m-1})} - I(j^{(q_1...q_{m-1})} = 1) \cdot \frac{y_2^{(q_1...q_{m-1})} - y_1^{(q_1...q_{m-1})} + 1}{2} \\
\end{aligned}
\notag
\end{equation}

We then calculate the quadrant index of \(\mathcal{U}^{(q_1...q_m)}\).

\begin{equation}
\begin{aligned}
i^{(q_1...q_{m})} &= \Bigl\lceil 2 \cdot \frac{a - x_1^{(q_1...q_m)} + 0.5}{x_2^{(q_1...q_m)} - x_1^{(q_1...q_m)} + 1} \Bigr\rceil \\
j^{(q_1...q_{m})} &= \Bigl\lceil 2 \cdot \frac{b - y_1^{(q_1...q_m)} + 0.5}{y_2^{(q_1...q_m)} - y_1^{(q_1...q_m)} + 1} \Bigr\rceil \\
\end{aligned}
\notag
\end{equation}

And finally obtain the quaternary index.

\begin{equation}
q_{m+1} = Q(Z^{(q_1...q_m)}|\pi_2^{(q_1...q_m)}, i^{(q_1...q_{m})}, j^{(q_1...q_{m})})
\notag
\end{equation}

We use the same example from the previous section to demonstrate the
calculation. We set \((a, b) = (5, -12)\), \((x_1, y_1) = (-7, -14)\),
\((x_2, y_2) = (8, 1)\), and \(\mathcal{P}_k = B^{270}|1221\).

\emph{Step 1}.

\begin{equation}
\begin{aligned}
i &= \Bigl\lceil 2 \cdot \frac{5 - (-7)) + 0.5}{8 - (-7) + 1} \Bigr\rceil = 2 \\
j &= \Bigl\lceil 2 \cdot \frac{-12 - (-14) + 0.5}{1 - (-14) + 1} \Bigr\rceil = 1 \\
\end{aligned}
\notag
\end{equation}

We obtain \(q_1 = Q(B^{270}|1, 2, 1) = 3\).

\emph{Step 2}. \(t\) is also in \(\mathcal{U}^{(q_1)} = \mathcal{U}_3\).
With the form of \(\mathcal{P}_k\), we know
\(\mathcal{U}^{(3)} = L^{270}|221\). We first calculate the coordinates
of the two corners:

\begin{equation}
\begin{aligned}
x_1^{(3)} &= x_1 + I(i = 2) \cdot \frac{x_2 - x_1 + 1}{2} = -7 + \frac{8 - (-7) + 1}{2} = 1\\
y_1^{(3)} &= y_1 + I(j = 2) \cdot \frac{y_2 - y_1 + 1}{2} = -14\\
x_2^{(3)} &= x_2 - I(i = 1) \cdot \frac{x_2 - x_1 + 1}{2} = 8\\
y_2^{(3)} &= y_2 - I(j = 1) \cdot \frac{y_2 - y_1 + 1}{2} = 1 - \frac{1-(-14)+1}{2} = -7\\
\end{aligned}
\notag
\end{equation}

The quadrant index on \(U^{(q_1)}\) is

\begin{equation}
\begin{aligned}
i^{(3)} &= \Bigl\lceil 2 \cdot \frac{a - x_1^{(3)} + 0.5}{x_2^{(3)} - x_1^{(3)} + 1} \Bigr\rceil = \Bigl\lceil 2 \cdot \frac{5 - 1 + 0.5}{8 - 1 + 1} \Bigr\rceil = 2\\
j^{(3)} &= \Bigl\lceil 2 \cdot \frac{b - y_1^{(3)} + 0.5}{y_2^{(3)} - y_1^{(3)} + 1} \Bigr\rceil = \Bigl\lceil 2 \cdot \frac{-12 - (-14) + 0.5}{-7 - (-14) + 1} \Bigr\rceil = 1\\
\end{aligned}
\notag
\end{equation}

We obtain \(q_2 = Q(L^{270}|2, 2, 1) = 2\).

\emph{Step 3}. The \(q_2\)-th subunit of \(\mathcal{U}^{(3)}\) is
\(\mathcal{U}^{(32)} = L^{270}|21\). The coordinates of its two corners
are:

\begin{equation}
\begin{aligned}
x_1^{(32)} &= x_1^{(3)} + I(i^{(3)} = 2) \cdot \frac{x_2^{(3)} - x_1^{(3)} + 1}{2} = 1 + \frac{8-1+1}{2} = 5 \\
y_1^{(32)} &= y_1^{(3)} + I(j^{(3)} = 2) \cdot \frac{y_2^{(3)} - y_1^{(3)} + 1}{2} = -14\\
x_2^{(32)} &= x_2^{(3)} - I(i^{(3)} = 1) \cdot \frac{x_2^{(3)} - x_1^{(3)} + 1}{2} = 8\\
y_2^{(32)} &= y_2^{(3)} - I(j^{(3)} = 1) \cdot \frac{y_2^{(3)} - y_1^{(3)} + 1}{2} = -7 - \frac{-7-(-14)+1}{2} = -11\\
\end{aligned}
\notag
\end{equation}

The quadrant index on \(U^{(q_1q_2)}\) is

\begin{equation}
\begin{aligned}
i^{(32)} &= \Bigl\lceil 2 \cdot \frac{a - x_1^{(32)} + 0.5}{x_2^{(32)} - x_1^{(32)} + 1} \Bigr\rceil = \Bigl\lceil 2 \cdot \frac{5 - 5 + 0.5}{8 - 5 + 1} \Bigr\rceil = 1\\
j^{(32)} &= \Bigl\lceil 2 \cdot \frac{b - y_1^{(32)} + 0.5}{y_2^{(32)} - y_1^{(32)} + 1} \Bigr\rceil = \Bigl\lceil 2 \cdot \frac{-12 - (-14) + 0.5}{-11 - (-14) + 1} \Bigr\rceil = 2\\
\end{aligned}
\notag
\end{equation}

We obtain \(q_3 = Q(L^{270}|2, 1, 2) = 4\).

\emph{Step 4}. The \(q_3\)-th subunit of \(\mathcal{U}^{(32)}\) is
\(\mathcal{U}^{(324)} = R^{90}|1\). The coordinates of its two corners
are:

\begin{equation}
\begin{aligned}
x_1^{(324)} &= x_1^{(32)} + I(i^{(32)} = 2) \cdot \frac{x_2^{(32)} - x_1^{(32)} + 1}{2} = 5 \\
y_1^{(324)} &= y_1^{(32)} + I(j^{(32)} = 2) \cdot \frac{y_2^{(32)} - y_1^{(32)} + 1}{2} = -14 + \frac{-11 -(-14) + 1}{2} = -12\\
x_2^{(324)} &= x_2^{(32)} - I(i^{(32)} = 1) \cdot \frac{x_2^{(32)} - x_1^{(32)} + 1}{2} = 8 - \frac{8-5+1}{2} = 6\\
y_2^{(324)} &= y_2^{(32)} - I(j^{(32)} = 1) \cdot \frac{y_2^{(32)} - y_1^{(32)} + 1}{2} = -11\\
\end{aligned}
\notag
\end{equation}

The quadrant index on \(U^{(q_1q_2q_3)}\) is

\begin{equation}
\begin{aligned}
i^{(324)} &= \Bigl\lceil 2 \cdot \frac{a - x_1^{(324)} + 0.5}{x_2^{(324)} - x_1^{(324)} + 1} \Bigr\rceil = \Bigl\lceil 2 \cdot \frac{5 - 5 + 0.5}{6 - 5 + 1} \Bigr\rceil = 1\\
j^{(324)} &= \Bigl\lceil 2 \cdot \frac{b - y_1^{(324)} + 0.5}{y_2^{(324)} - y_1^{(324)} + 1} \Bigr\rceil = \Bigl\lceil 2 \cdot \frac{-12 - (-12) + 0.5}{-11 - (-12) + 1} \Bigr\rceil = 1\\
\end{aligned}
\notag
\end{equation}

We obtain \(q_4 = Q(R^{90}|1, 1, 1) = 2\).

\(\mathcal{U}^{(q_1q_2q_3q_4)}\) is a single point, thus
\(q_1q_2q_3q_4 = 3242\) is the quaternary form of \(n\). Then with
Equation \ref{eq:quad2n}, we have \(n = 158\).

\subsubsection{Method 2}\label{method-2-1}

The point \(t\) is always located on a corner of a square on a certain
level \(i\) (\(1 \le i \le k\)) of the curve \(\mathcal{P}_k\). The
calculation of \(q_1...q_k\) only needs to be applied till \(q_i\) and
the remaining quaternary indicies can be directly obtained from the code
sequence of \(\mathcal{P}_k\).

Let's denote \(t^{(k)}\) as the point \(t\) on \(\mathcal{P}_k\). Since
\(\mathcal{P}_k\) is composed of a list of \(2^i \times 2^i\)
(\(1 \le i \le k\)) units if we look at the curve on different levels,
\(t^{(k)}\) is at least a corner of a lowest 2x2 unit.

\begin{proposition} 
\label{prop:p1_p2_sim}
For $t^{(k)}$ on $\mathcal{P}_k = X|(\pi)_k$, $\mathcal{P}_i$ ($1 \le i < k$) is a reduction
where $t^{(i)}$ on $\mathcal{P}_i$ is still a corner of a square unit. 
The unit where $t^{(k)}$ is located on $\mathcal{P}_k$
is reduced to the single point $t^{(i)}$ on $\mathcal{P}_i$.
Then $q_1...q_i$ for $t^{(i)}$ on $\mathcal{P}_i$ are inferred by Method 1, 
and the remaining quaternary indicies $q_{i+1}...q_k$ are calculated as:

\begin{equation}
\label{eq:qqq}
\begin{aligned}
q_j = \begin{cases}
q_i & \quad \textnormal{if } \pi_j = \pi_{j-1} \\
\hat{q}_i & \quad \textnormal{if } \pi_j = \hat{\pi}_{j-1} \\
\end{cases}
\end{aligned}
\notag
\end{equation}

\noindent where $i+1 \le j \le k$ and $\hat{q}$ is the complement of $q$ defined as:

\begin{equation}
\begin{aligned}
\hat{q} = \begin{cases}
2 & \quad \textnormal{if } q = 3 \\
3 & \quad \textnormal{if } q = 2 \\
2 & \quad \textnormal{if } q = 1 \\
3 & \quad \textnormal{if } q = 4 \\
\end{cases}
\end{aligned}
\notag
\end{equation}

\end{proposition}

To prove it, we start from the simplest form.

\begin{lemma}
\label{lemma:p1_p2_sim_1}
Proposition \ref{prop:p1_p2_sim} is true for the special case $\mathcal{P}_2 = X|\pi_1\pi_2$ (i.e. $k = 2$) and $X \in \{I, R, L\}$.
\end{lemma}

\begin{figure}
\centering{
\includegraphics[width=1\linewidth]{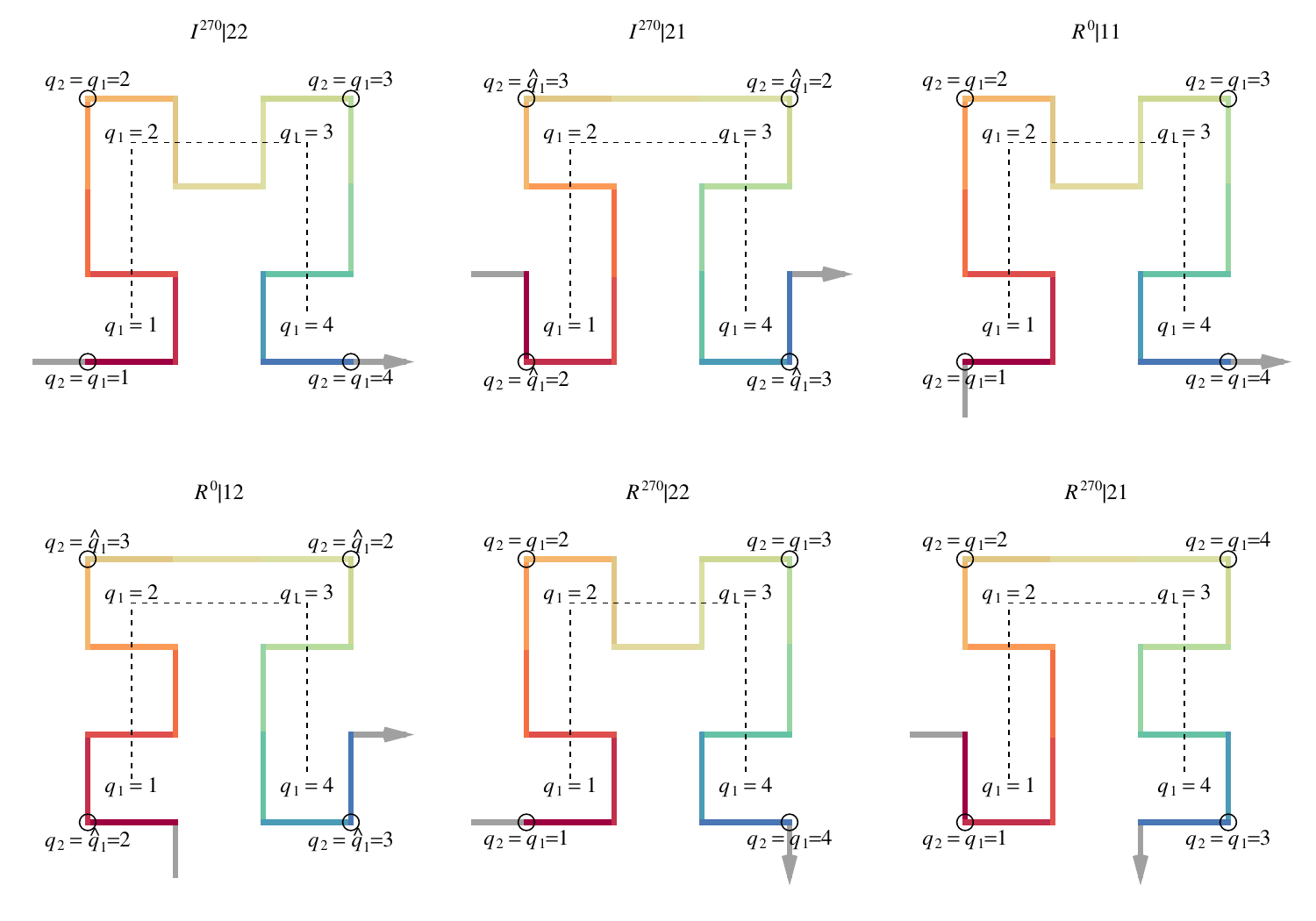}
\label{fig:quat_index_level_2}
}
\end{figure}

\begin{proof}
We consider $\mathcal{P}_2$ in the base facing state. All curves are listed in Figure \ref{fig:quat_index_level_2}, which demonstrates Lemma \ref{lemma:p1_p2_sim_1} is true
for this set of curves. 

The reflection of $\mathcal{P}_2$ won't change quaternary indicies of points but changes all code in the code sequence to the complement (Proposition \ref{prop:6.5}). However,
this won't change the condition in Equation \ref{eq:qqq}. Thus Lemma \ref{lemma:p1_p2_sim_1} is true
for the reflections. 

Last rotations won't change quaternary indicies as well as the code sequence, then for the
complete set of $\mathcal{P}_2$, Lemma \ref{lemma:p1_p2_sim_1} is true.
\end{proof}

Next we look at the case where \(i = k - 1\).

\begin{lemma}
\label{lemma:p1_p2_sim_2}
Proposition \ref{prop:p1_p2_sim} is true for the special case $i = k-1$.
\end{lemma}

\begin{proof}
The level-2 unit where $t^{(k)}$ is located and the level-1 unit where $t^{(k-1)}$ is located
are all induced from the base of $t^{(k-2)}$. We write the level $k-2$ curve as
$\mathcal{P}_{k-2} = X|\pi_1...\pi_2$ and expand the code sequence to every base on $\mathcal{P}_{k-1}$:

\begin{equation}
\mathcal{P}_{k-2} = X_1^{(k-2)}|\pi_1...\pi_{k-2} \cdot ... X_i^{(k-2)}|f(\pi_1...\pi_{k-2}) \cdot ...
\notag
\end{equation}

\noindent where $f()$ returns the original code sequence or the complement according to Corollary \ref{coro:5.6.1} since $X_i^{(k-2)} \in \{I, R, L\}$. Let's write
$f(\pi_1...\pi_{k-2}) = \sigma_1...\sigma_{k-2}$, then according to \ref{prop:p1_p2_sim},
Lemma \ref{lemma:p1_p2_sim_2} is true for evey of the base $X_i^{(k-2)}$ using $\sigma$ as
the conditions. since all $\sigma$ changes simultanuously, $\sigma_j = \sigma_i \equiv \pi_j = \pi_i$ and $\sigma_j \ne \sigma_i \equiv \pi_j \ne \pi_i$. Then Lemma \ref{lemma:p1_p2_sim_2} is true by taking $\pi$ as the condition.
\end{proof}

Last the proof of Proposition \ref{prop:p1_p2_sim}:

\begin{proof}
We can repeatedly apply Lemma \ref{lemma:p1_p2_sim_2}, until the minimal $i$ where $t^{(i)}$
is a corner of a 2x2 unit.
\end{proof}

The minimal \(i\) that satifies Proposition \ref{prop:p1_p2_sim} can be
calculated as

\begin{equation}
\min \left\{ i \le k \mid 2^{k-i} \mid (a-x_1+1) \text{ and } 2^{k-i} \mid (b-y_1+1) \right\}
\end{equation}

\noindent where \((a, b)\) is the coordinate of \(t^{(k)}\) and
\((x_1, y_1)\) is the coordinate of the bottom left corner of
\(\mathcal{P}_k\).

The method is efficient when \(i\) is small as all scenarios of
\(q_1...q_i\) can be pre-computated and the remaining quaternary
indicies can be simply calculated. And the original Method 1 depends
specific forms on lower levels which have no uniform form.

\begin{figure}
\centering{
\includegraphics[width=1\linewidth]{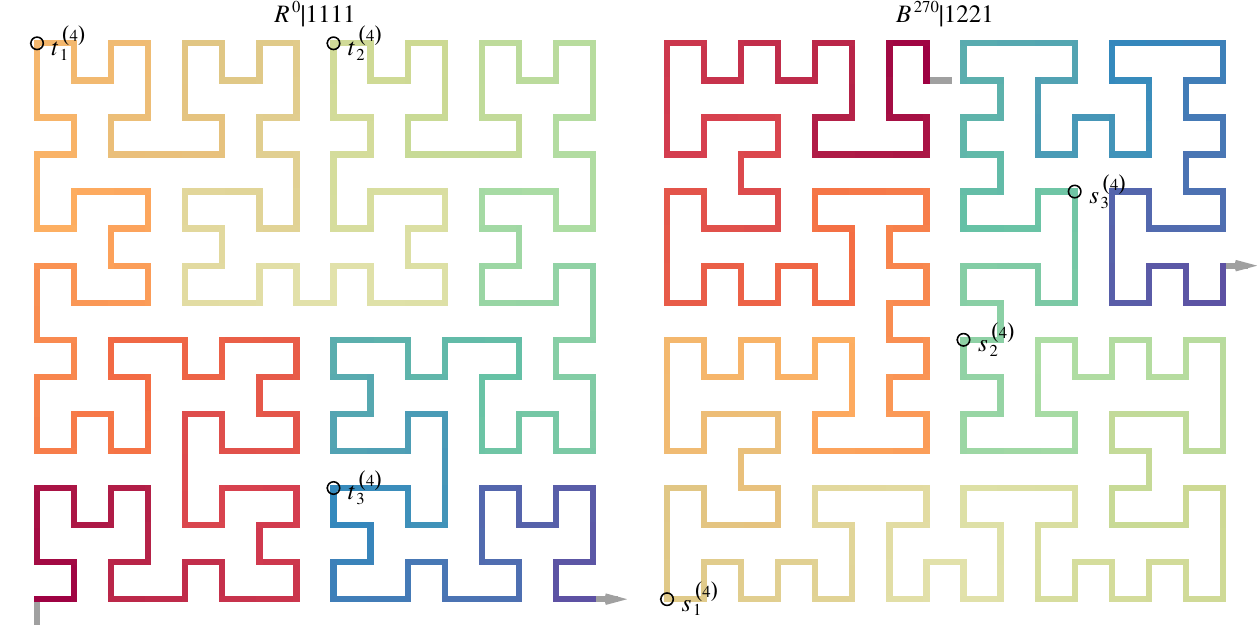}
\label{fig:quat_index_example}
}
\end{figure}

Figure \ref{fig:quat_index_example} illustrates two examples of
\(\mathcal{P}_4\). In Figure \ref{fig:quat_index_example}A, the lowest
reduced curve where \(t_1\) is still a corner point is
\(\mathcal{P}_1\), and the quaternary index of \(t_1^{(1)}\) is 2,
i.e.~\(q_1 = 2\). The remaining three quaternary indicies can be
directly calculated as \(q_2q_3q_4 = 222\)

\begin{table}
\centering
\begin{tabular}{rlllll}
\toprule
$\mathcal{P}_4$ & point       & minimal $i$ & $q_1...q_i$ & $q_{i+1}...q_{4}$ & n \\
\midrule
$R|1111$       & $t^{(4)}_1$ &    1        & 2          & 222             & xx \\
               & $t^{(4)}_2$ &    2        & 32         & 22              & xx \\
               & $t^{(4)}_3$ &    3        & 423        & 3               & xx \\
\midrule
$B|1221$       & $s^{(4)}_1$ &    1        & 2          & 332             & xx \\
               & $s^{(4)}_2$ &    2        & 34         & 43              & xx \\
               & $s^{(4)}_3$ &    3        & 442        & 3               & xx \\
\bottomrule
\end{tabular}
\vspace*{5mm}
\label{tab:quat_index_example}
\caption{.}
\end{table}

\section{Conclusion}\label{conclusion}

In this work, we presented a new framework for constructing and
representing 2x2 space-filling curves, which is built upon two essential
components: the full set of rules of level 0-to-1 expansions and the
encoding system. Based on it, we established comprehensive theories for
studying the construction, expansion, transformation and structures of
2x2 curves. The 2x2 curve is the simplest form of the general nxn
(\(n\)-by-\(n\), \(n
\ge 2\)) curves. However, the framework proposed in this work can be a
conceptual foundation for extension studies on more complex nxn curves.

\renewcommand\refname{References}
\bibliography{space-filling-curve.bib}

\end{document}